\documentstyle[preprint,aps,rmp,epsfig,tighten,floats]{revtex}

\def\ms {\overline{\rm MS}}
\def\dis{{\rm DIS}_{\gamma}}
\def\eps{\varepsilon}
\def\d  {{\rm d}}
\def\M  {{\cal M}}
\def\O  {{\cal O}}
\def\Q  {{\cal Q}}

\def\yytqb{\gamma\gamma\rightarrow q\bar{q}}
\def\yqtgq{\gamma q\rightarrow gq}

\def\ygtqb{\gamma g\rightarrow q\bar{q}}
\def\qptqp{qq'\rightarrow qq'}

\def\qqtqq{qq\rightarrow qq}

\def\qbtgg{q\bar{q}\rightarrow gg}

\def\ggtgg{gg\rightarrow gg}

\def\yythh{\gamma\gamma\rightarrow h\bar{h}}
\def\ygthh{\gamma g    \rightarrow h\bar{h}}
\def\qbthh{q\bar{q}\rightarrow h\bar{h}}
\def\ggthh{gg      \rightarrow h\bar{h}}

\def\OOSZE{{\langle O_8^{J/\Psi}(^1S_0)\rangle}}
\def\OTSOO{{\langle O_1^{J/\Psi}(^3S_1)\rangle}}
\def\OTSOE{{\langle O_8^{J/\Psi}(^3S_1)\rangle}}

\def\OTPJE{{\langle O_8^{J/\Psi}(^3P_J)\rangle}}

\def\yytQy{\gamma\gamma\rightarrow \OTSOO\gamma}
\def\ygtQg{\gamma g    \rightarrow \OTSOO g}
\def\ggtQg{ g     g    \rightarrow \OTSOO g}

\def\lp {\left.   }
\def\rp {\right.  }
\def\lr {\left(   }
\def\rr {\right)  }
\def\le {\left[   }
\def\re {\right]  }
\def\lg {\left\{  }
\def\rg {\right\} }

\def\tux{\frac{t}{u}}

\def\utx{\frac{u}{t}}

\def\tus{\frac{tu}{s^2}}

\def\ede{\frac{1}{\varepsilon}}

\def\beq{\begin{equation}}
\def\eeq{\end{equation}}
\def\bea{\begin{eqnarray}}
\def\eea{\end{eqnarray}}

\begin{document}

\preprint{DESY 02-086}

\title{Theory of hard photoproduction}

\author{Michael Klasen\thanks{Electronic address: michael.klasen@desy.de}}

\address{II.\ Institut f\"ur Theoretische Physik, Universit\"at Hamburg,
         D--22761 Hamburg, Germany}

\date{\today}

\maketitle

\begin{abstract}

The present theoretical knowledge about photons and hard photoproduction
processes,
{\it i.e.} the production of jets, light and heavy hadrons, quarkonia, and
prompt photons in photon-photon and photon-hadron collisions, is reviewed.
Virtual and polarized photons and prompt photon production in hadron
collisions are also discussed.
The most important leading and next-to-leading order QCD results are compiled
in analytic form. A large variety of numerical predictions
is compared to data from TRISTAN, LEP, and HERA and extended to future
electron and muon colliders. The sources of all relevant results are collected
in a rich bibliography.

\end{abstract}

\vfill
\begin{center}\large\bf
Scheduled for publication in Reviews of Modern Physics.
\end{center}
\vfill

\pagebreak
\tableofcontents
\pagebreak


\section{Introduction}
\label{sec:intro}

The photon is a fascinating particle. It is the fundamental gauge boson of
Quantum Electrodynamics and as such one of the best studied elementary
particles of the Standard Model of particle physics. However, in high-energy
reactions the photon exhibits a complex hadronic structure, which is far less
well understood. Investigations took off with fixed-target experiments and
soft vector-meson production more than 40 years ago, which established the
Vector Meson Dominance model of fluctuations between photons and vector mesons.
With the advent of Quantum Chromodynamics (QCD) as the theory of strong
interactions in the 1970s, the pointlike nature of the photon, its coupling to
quarks, and the perturbative evolution of the photon structure moved to the
center of interest. For some time it seemed possible to actually calculate the
hadronic structure of the photon. Unfortunately it turned out shortly later,
that this was only possible at asymptotically large energies, which could not
be reached with the available accelerators. In the mid-80s, the photon beam
energies were high enough to prove that there were two different types of
photon interactions. The photon could interact directly with the quarks and
gluons in the hadronic target, but it could also resolve into a hadronic
structure and the partonic constituents of the photon could participate in the
hard scattering. At about the same time, the two processes were found to be
theoretically related. The pointlike photon structure was shown to be singular
and in need of regularization by a non-perturbative hadronic boundary
condition, which had to be determined from theoretical models or deep-inelastic
electron-photon experiments at the $e^+e^-$ colliders PETRA and PEP.

During the last 15 years, a tremendous experimental and theoretical effort has
been invested into refining the picture of the photon, testing its structure,
and using the photon as a tool for studying the production and properties of
jets, light and heavy hadrons, quarkonia, and prompt photons. Experimentally
this was facilitated by the construction and operation of the $e^+e^-$
colliders TRISTAN and LEP with center-of-mass energies between 58 and
210 GeV and the HERA $ep$ collider with a center-of-mass energy of 300
GeV, which has just been upgraded for high-luminosity operation. The
experimental results on the structure and interactions of the photon have been
discussed at many general and the topical ``PHOTON'' conferences, and
they have recently been reviewed by several authors (Erdmann, 1997; Abramowicz
and Caldwell, 1999; Nisius, 2000; Krawczyk, Staszel, and Zembrzuski, 2001).
The experimental advances were
paralleled by similar theoretical improvements. Until the mid-80s,
determinations of the photon structure and calculations of photoproduction
processes were performed only in leading order (LO) of perturbative QCD, and a
review of the LO formalism exists for prompt photon, jet, and particle
production in hadron collisions (Owens, 1987). Since then, several 
photon parameterizations and many photoproduction calculations
have been performed in next-to-leading order (NLO). In comparisons with
experimental data, these calculations have provided stringent tests on QCD
and precise information about the photon structure and fragmentation and about
the formation of jets and hadrons (for short reviews see Klasen, 1997a, 1999b;
Kniehl, 1997; Kramer, 1996, 1998, 1999).

While LO predictions are straightforward to calculate, they are unfortunately
not very precise and can only be used to estimate hard photoproduction cross
sections within factors of two. The reason for this is that LO QCD cross
sections depend strongly on the renormalization scale in the strong
coupling and on the factorization scales in the photon and proton parton
densities, which are commonly varied by factors of two around the physical
scale to estimate the relative theoretical error.
Specific problems exist for jet and heavy-flavor production.
In LO every jet corresponds to just one parton, and jet algorithms cannot be
implemented. Only beyond LO this becomes possible by combining two or more
partons into a single jet. Heavy flavors can be treated as massive or massless
particles, and only in NLO the logarithmic mass and scale dependences in fixed
and variable flavor number schemes can be made explicit.
The large luminosities and trigger rates at modern high-energy colliders and
detectors lead to great improvements in the statistical accuracy of the
experimental data, which exceeds the theoretical precision of LO calculations.
In NLO the uncertainties can be reduced to a reasonable level (a few percent).
This precision is generally limited to scattering processes with four external
legs, and it has now been reached for almost all photoproduction processes.
Exceptions are the real photoproduction of three jets, two hadrons or photons,
and various quarkonium states, and the virtual or polarized photoproduction of
hadrons and prompt photons. In some cases NLO precision may be insufficient to
describe the experimental data as will be discussed later. It would be
necessary to work at next-to-next-to-leading order, but this precision has so
far been feasible only for inclusive processes like the total rates of
$e^+e^-$ and $ep$ scattering into hadrons.

It is the aim of this Review to give a complete theoretical description of
hard photoproduction processes, ranging from the generation of real, slightly
virtual, and also polarized photons with lepton beams, over the current
knowledge about the photon structure, to the methods and applications
of NLO QCD calculations for jet, hadron, and prompt photon photoproduction.
While the theoretical tools cannot be described
in full detail, the main lines of argument and techniques are explained
conceptually, and the most important results are stated explicitly.
The predictions are compared only to the most recent and precise data
from TRISTAN, LEP, and HERA in order to reach up-to-date conclusions about
the applicability of QCD and the properties of jets, hadrons, and photons.
Selected predictions are made for future accelerators like linear $e^+e^-$
and circular muon colliders or a future $ep$ machine in order to provide
a look ahead, and a rich bibliography guides the reader to all of the
relevant theoretical and
experimental literature. In this sense, this Review will hopefully serve
as a compendium of the current state of the art in hard photoproduction.


\pagebreak
\section{Photon spectra}
\label{sec:phospec}
\setcounter{equation}{0}

In the first generation of photoproduction experiments, real photons with
energies below 60 GeV were generated by pion decays or electron bremsstrahlung
and scattered off nuclear targets in order to study soft particle production
and total cross sections (Paul, 1992). Measurements of hard photoproduction of
photons and mesons began with the CERN experiment NA14 at beam energies between
50 and 150 GeV (Auge {\it et al.}, 1986). Later energies up to 400 GeV
were reached at Fermilab. They led to a first observation of jets in
photoproduction (Adams {\it et al.}, 1994), but were still too small for
definite tests of QCD. Center-of-mass energies of 200 to 300 GeV have recently
been reached at the $e^+e^-$ colliders PETRA, PEP,
TRISTAN, and LEP
and at the $ep$ collider HERA. Here spacelike, almost real bremsstrahlung
photons are exchanged during the hard collision. At future linear $e^+e^-$
colliders large particle bunch densities are needed to reach high
luminosities. Then additional beamstrahlung photons will be created before the
hard interaction by the coherent action of the
electromagnetic field of one bunch on the opposite one. If the electron beams
are collided with additional high-energy laser beams, real photons can be
produced through Compton scattering. Thus three different mechanisms can
contribute to photon scattering at high-energy colliders: bremsstrahlung,
beamstrahlung,
and laser backscattering. In this Section, the corresponding photon energy
spectra will be discussed.

\subsection{Bremsstrahlung}
\label{sec:bremsstrahlung}

If the outgoing and incoming leptons in a hard scattering process are almost
collinear, the calculation of the corresponding cross section can be
considerably simplified by using the Weizs\"acker-Williams or Equivalent Photon
approximation (for a review see Budnev {\it et al.}, 1974). Current
conservation and the small photon
virtuality lead to a factorization of the lepton scattering cross section
into a broad-band photon spectrum in the lepton and the hard photon
scattering cross section. Already in the 1920s, Fermi (1924) discovered the
equivalence between
the perturbation of distant atoms by the field of charged particles flying
by and the one due to incident electromagnetic radiation.
His semi-classical treatment was then extended to high-energy electrodynamics
by Weizs\"acker (1934) and Williams (1934)
independently, who used a Fourier analysis to unravel the
predominance of transverse over longitudinal photons radiated from a
relativistic charged particle. In the fifties, Curtis (1956) and
Dalitz and Yennie (1957) gave the first field-theoretical derivations
and applied the approximation to meson production in electron-nucleon
collisions.

In the Weizs\"acker-Williams approximation,
the energy spectrum of the exchanged photons is given by
\bea
 f_{\gamma/l}^{\rm brems}(x)&=&\frac{\alpha}{2\pi}\left[
 \frac{1+(1-x)^2}{x}\ln\frac{Q^2_{\max}(1-x)}{m_l^2 x^2}\rp\nonumber \\
 &+&\lp 2 m_l^2 x\left(\frac{1}{Q^2_{\max}}-\frac{1-x}{m_l^2 x^2}\right)
 \right].
 \label{eq:unpol_brems}
\eea
The subleading non-logarithmic terms (Kessler, 1975; Frixione {\it et al.},
1993) modify the cross section typically by 5\%. $\alpha=1/137$ is the
electromagnetic fine-structure constant, $x=E_\gamma/E_l\,(l=e,\mu)$ is the
energy fraction transferred from the lepton to the photon, $m_l$ is the lepton
mass, and $Q^2_{\max}=E_l^2\, (1-x)\,\theta^2_{\max}$ is the maximal photon
virtuality for lepton scattering angles below $\theta_{\max}$. This angle
can be determined by tagging the outgoing lepton in the forward
direction or by requiring that it is lost in the beam pipe (anti-tagging).
When no information about the scattered lepton is available, one has
to integrate over the whole phase space, thus allowing large transverse
momenta and endangering the factorization property of the cross section.
Bremsstrahlung photons can be generated by electron and muon beams, but for
the latter the photon density is smaller by approximately a factor of two due
to the larger muon mass (Klasen, 1997b).

\subsection{Beamstrahlung}
\label{sec:beamstrahlung}

\begin{figure}
 \begin{center}
  \epsfig{file=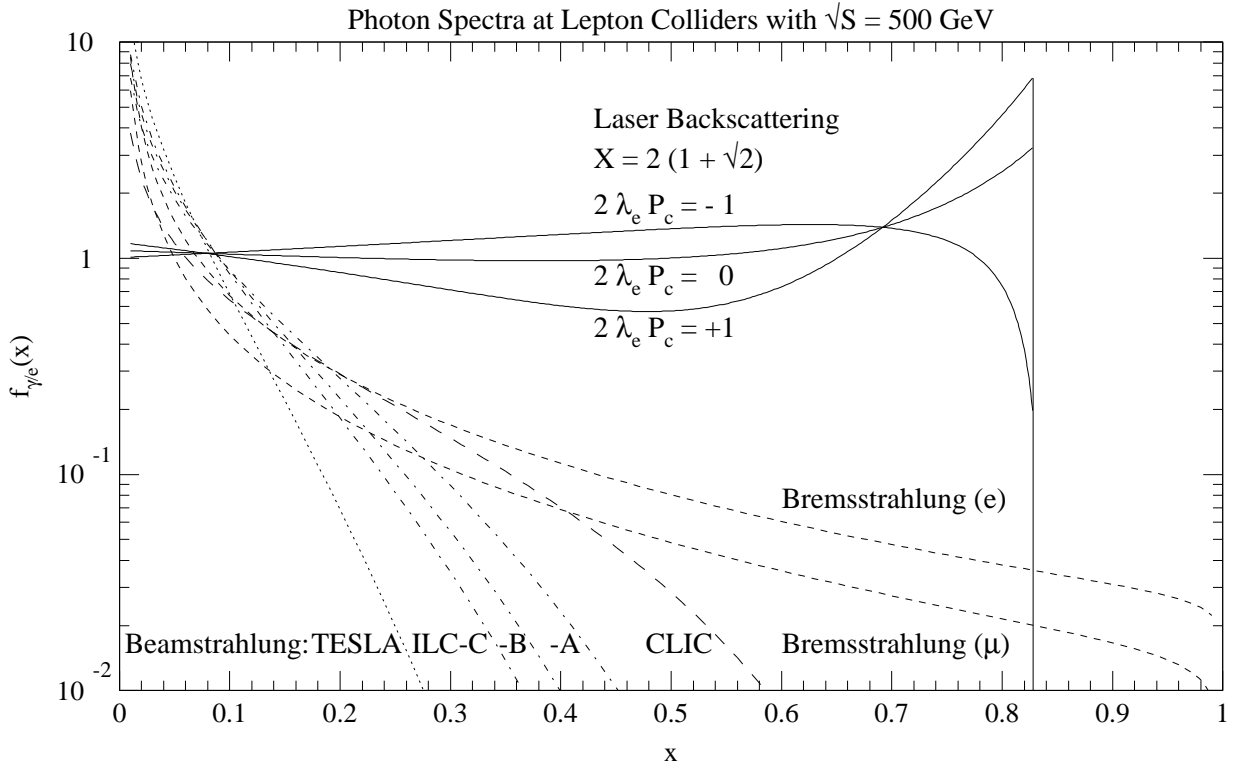,width=\linewidth}
 \caption{\label{fig:01}
 Photon energy spectra at lepton colliders with $\sqrt{S}=500$ GeV. The
 bremsstrahlung spectrum depends logarithmically on $\sqrt{S}$ and is roughly
 two times larger for electrons than for muons. The electron beamstrahlung
 spectra depend strongly on the design parameters and rise with $\sqrt{S}$. 
 The laser backscattering spectrum is completely independent of $\sqrt{S}$,
 if the the center-of-mass energy of the electron-laser photon collision is
 kept fixed, but it depends on the polarizations of the electron $\lambda_e$
 and of the laser photon $P_c$.}
 \end{center}
\end{figure}

Future circular $e^+e^-$ colliders with center-of-mass energies above $\sqrt{S}
=500$ GeV would suffer from very high synchrotron radiation. They must
therefore have a linear design and dense particle bunches in order to still
obtain large luminosities. Inside the opposite bunch, electrons and positrons
experience transverse acceleration and radiate beamstrahlung.
The spectrum is controlled by
the beamstrahlung parameter
\beq
 \Upsilon = {5 r_e^2 E_e N\over 6\alpha\sigma_z(\sigma_x+\sigma_y)m_e},
\eeq
which is proportional to the effective
electromagnetic field of the bunches
and depends on the classical electron radius $r_e = \alpha/m_e = 2.818
\cdot 10^{-15}$ m, the beam energy $E_e$, the total number of particles in a
bunch $N$, and on the r.m.s.\ sizes of the Gaussian beam $\sigma_x,\,\sigma_y,
\,\sigma_z$. For not too large $\Upsilon$ $(\Upsilon\leq 5)$, Chen (1992)
derived the approximate spectrum
\bea
\label{eq:appbeam}
 f_{\gamma/e}^{\rm beam}(x) &=& \frac{1}{\Gamma\left(\frac{1}{3}\right)}
  \left(\frac{2}{3\Upsilon}\right)^\frac{1}{3} x^{-\frac{2}{3}}
  (1-x)^{-\frac{1}{3}} e^{-2 x/[3\Upsilon (1-x)]}\nonumber\\
 & \times & \left\{\frac{1-\sqrt{\frac{\Upsilon}{24}}}{g(x)}\left[ 1-\frac{1}
  {g(x) N_\gamma}\left(1-e^{-g(x) N_\gamma}\right)\right]\rp\nonumber\\
 &+& \lp\sqrt{\frac{\Upsilon}{24}} \left[1-\frac{1}{N_\gamma} \left( 
  1-e^{-N_\gamma}\right)\right]\right\} 
\eea
with
\beq
 g(x) = 1 - \frac{1}{2} \left[(1+x)\sqrt{1+\Upsilon^{\frac{2}{3}}}+1-x\right]
 (1-x)^{\frac{2}{3}}.
\eeq
The average number of photons radiated per electron throughout the
collision is 
\beq
 N_\gamma = {5 \alpha^2 \sigma_z m_e\Upsilon\over 2 r_e E_e
 \sqrt{1+\Upsilon^{\frac{2}{3}}}}.
\eeq
Current design parameters for possible future $e^+e^-$ colliders
are listed in Tab.\ \ref{tab:colliders}, and the
spectra corresponding to the $\sqrt{S}=500$ GeV designs are shown in
Fig.\ \ref{fig:01}.
This figure also displays the bremsstrahlung spectra for electrons and muons,
which have been integrated
over the photon virtuality up to an upper bound $Q^2_{\max}=4E_l^2(1-x)$ for
untagged outgoing leptons. Both the bremsstrahlung and
beamstrahlung spectra are soft, but for particular collider designs
beamstrahlung can be more important
%
\begin{table}
\begin{center}
\begin{tabular}{|c|ccccc|}
  Collider                        & TESLA& ILC-A & ILC-B & ILC-C & CLIC \\
  Last update                     & 3/01 & 4/00  & 4/00  & 4/00  & 3/01 \\
\hline
\hline
  $\sqrt{S}$ (GeV)                & 500  & 535   & 515   & 500   & 500  \\
  Particles/Bunch ($10^{10}$)     & 2    & 0.75  & 0.95  & 1.1   & 0.4  \\
  $\sigma_x$ (nm)                 & 553  & 277   & 330   & 365   & 202  \\
  $\sigma_y$ (nm)                 & 5    & 3.4   & 4.9   & 7.6   & 2.5  \\
  $\sigma_z$ ($\mu$m)             & 300  & 90    & 120   & 145   & 30   \\
  $\Upsilon$                      & 0.05 & 0.14  & 0.11  & 0.09  & 0.28 \\
\hline
  $\sqrt{S}$ (GeV)                & 800  & 1046  & 1008  & 978   & 1000 \\
  Particles/Bunch ($10^{10}$)     & 1.4  & 0.75  & 0.95  & 1.1   & 0.4  \\
  $\sigma_x$ (nm)                 & 391  & 197   & 235   & 260   & 115  \\
  $\sigma_y$ (nm)                 & 2.8  & 2.7   & 3.9   & 5.4   & 1.75 \\
  $\sigma_z$ ($\mu$m)             & 300  & 90    & 120   & 145   & 30   \\
  $\Upsilon$                      & 0.10 & 0.38  & 0.30  & 0.25  & 0.98
\end{tabular}
\end{center}
\caption{\label{tab:colliders}Current design parameters for possible future
 linear $e^+e^-$ colliders (Brinkmann {\it et al.}, 2001; Adolphsen {\it et
 al.}, 2000; Kamitani and Rinolfi, 2001).}
\end{table}
%
%
%
than bremsstrahlung over a wide
range in $x$.
Beamstrahlung is completely negligible at muon colliders, since the
parameter $\Upsilon_\mu=m_e^3/m_\mu^3\cdot\Upsilon_e$,
which controls the spectrum mostly through the first exponential
function in Eq.\ (\ref{eq:appbeam}),
is about five orders of magnitude smaller than for electrons (Klasen, 1997b).

\subsection{Laser backscattering}
\label{sec:laserback}

Real photons of very high energy can be produced, if laser photons are
backscattered off electrons. The laser backscattering spectrum (Ginzburg
{\it et al.}, 1984)
\bea
 f_{\gamma/e}^{\rm laser}(\!\!&x&\!\!) = \frac{1}{N_c+2 \lambda_e P_c N_c'} \left[
    1-x +\frac{1}{1-x}-\frac{4 x}{X (1-x)} \right. \nonumber \\
 &+& \left. \frac{4 x^2- 2 \lambda_e P_c x (2-x)X
 [x (X+2)-X]} {X^2 (1-x)^2} \right],
\eea
where
\beq
 N_c = \left[1-\frac{4}{X}-\frac{8}{X^2}\right] \ln(1+X)+\frac{1}{2}
 + \frac{8}{X}-\frac{1}{2 (1+X)^2}
\eeq
and
\beq
 N_c' = \left[\left(1+\frac{2}{X}\right)\ln(1+X)-\frac{5}{2}+\frac{1}{1+X}
 -\frac{1}{2(1+X)^2}\right]
\eeq
are related to the total Compton cross section, depends on the center-of-mass
energy of the electron-laser photon collision $s_{e\gamma}$ through the
parameter $X = s_{e\gamma}/m_e^2-1$. The
optimal value of $X$ is determined by the threshold for the process
$\gamma\gamma\rightarrow e^+e^-$ and is $X=2(1+\sqrt{2})\simeq 4.83$
(Telnov, 1990). If this value is kept fixed, the laser backscattering
spectrum becomes independent of $\sqrt{S}$.  A large fraction of the photons
is then produced close to the kinematic limit $x < x_{\max} = X/(1+X)
\simeq 0.828$, so that one obtains an almost monochromatic ``photon collider.''
The monochromaticity of the produced
photons can be improved further if the helicities of the electron
$\lambda_e$ and of the laser photon $P_c$ satisfy the condition
$2 \lambda_e P_c = -1$ (see Fig.\ \ref{fig:01}).
For electron beams, the optimal pulse
energies ($\sim 1$~J) and repetition rates ($\sim 1$~kHz) can be provided by
current laser technology. For muon beams, one would need much higher flash
energies ($\sim 1$~GJ), which are currently not feasible (Klasen, 1997b).
High-energy interactions of laser-backscattered photons have yet to be
experimentally observed, and the simulation tools for realistic photon
spectra including low-energy tails and non-linear effects have to be refined in
order to reliably estimate the luminosities and event rates for photon
colliders.


\pagebreak
\section{Photon structure}
\label{sec:phostr}
\setcounter{equation}{0}

The first generation of photoproduction experiments on nuclear targets
revealed striking similarities with purely hadronic processes like pion
scattering. The total cross sections showed strong resonances below
center-of-mass energies of 3 GeV and flat behavior above, and the elastic
scattering cross sections fell strongly with the momentum transfer
(Bauer {\it et al.}, 1978). The most striking feature, however, was the
copious production of vector mesons, particularly of the $\rho$ meson, which
led Stodolsky (1964) to postulate an analogy between the isovector
electromagnetic current and the $\rho$ field operator and a
relation between diffractive $\gamma p$ and elastic $\rho p$ cross sections
(Ross and Stodolsky, 1966). This analogy was subsequently extended in the
Vector Meson Dominance (VMD) model, where the photon was viewed as a
superposition of $\rho$, $\omega$, and $\phi$ mesons with a small, if not
negligible, pointlike contribution (Stodolsky, 1967; Joos, 1967).
In the Generalized Vector Dominance (GVD) model (Sakurai and Schildknecht,
1972), all mesons, which carried the
same quantum numbers as the photon $(J^{PC}=1^{--})$, were included as
contributions to the photon structure, {\it e.g.} also the heavier $J/\Psi$
meson.

The hadronic picture of the photon was revolutionized with the advent of QCD.
Witten (1977) showed, that the photon structure had an anomalous (pointlike)
component, which could be understood as a short-time fluctuation into
quark-antiquark pairs and gluons and could be calculated perturbatively.
QCD also predicted direct photon scattering off quarks and gluons, which was
subsequently observed in fixed-target experiments at higher energies and
momentum transfers (Auge {\it et al.}, 1986). Thus it became clear that the
photon had a pointlike and a hadronic component, that they
were important at different momentum transfers, and that they induced photon
and parton scattering reactions, respectively (Brodsky {\it et al.}, 1978,
1979; Lewellyn-Smith, 1978). 
Much of the present knowledge about the hadronic structure of the photon has
been obtained from measurements of the photon structure functions $F_2^\gamma$
and $F_L^\gamma$ in the deep-inelastic scattering process $e(k)\gamma(p)\to
e(k')X$ at $e^+e^-$ colliders with
\beq
 {\d^2\sigma\over\d x\d Q^2}={2\pi\alpha^2\over xQ^4}
 \le\lr 1+(1-y)^2\rr F_2^\gamma(x,Q^2)-y^2F_L^\gamma(x,Q^2)\re,
\eeq
where the flux of quasi-real photons is given by the Equivalent Photon
approximation, Eq.\ (\ref{eq:unpol_brems}), and where the kinematic variables
$Q^2=-q^2=-(k-k')^2,\,x=Q^2/(2\,p\cdot q),$ and $y=(p\cdot q)/(p\cdot k)$ are
experimentally determined.

\subsection{Evolution equations}

In NLO of QCD, an initial-state photon can split into a
massless quark-antiquark pair, which then interacts in the hard scattering, and
a collinear singularity is encountered.
In dimensional regularization ('t Hooft and Veltman, 1972; Bollini and
Giambiagi, 1972a, 1972b; Gastmans and Meuldermans, 1973; Marciano and Sirlin,
1975; Marciano, 1975; Leibbrandt, 1975), where
scattering matrix elements and phase-space factors are evaluated in $n=4-2\eps$
dimensions, the singularity manifests itself as a $1/\eps$-pole multiplying
the space-like photon-quark splitting function $P_{q\leftarrow\gamma}(x)$ in
\bea
 \overline{\Gamma}_{q\leftarrow\gamma}(x,M_f^2)&=&\delta_{q\gamma}\,\delta(1-x)
 -\frac{1}
 {\eps}\frac{\alpha}{2\pi}\frac{\Gamma(1-\eps)}{\Gamma(1-2\eps)}\lr
 \frac{4\pi\mu^2}{M_f^2}\rr^\eps \nonumber \\
 &\times& P_{q\leftarrow\gamma}(x) 
 +{\cal O}(\eps,\alpha^2,\alpha\alpha_s)\nonumber \\
 &=& \delta_{q\gamma}\,\delta(1-x)
 \!-\!\le \frac{1}{\eps}\!-\!\gamma_E\!+\!\ln(4\pi)\!+\!\ln\frac{\mu^2}
 {M_f^2}\re\nonumber \\
 &\times&\frac{\alpha}{2\pi}P_{q\leftarrow\gamma}(x)
 +{\cal O}(\eps,\alpha^2,\alpha\alpha_s).
 \label{eq:realphotsplit}
\eea
$x$ is the longitudinal momentum fraction of the quark in the photon,
$\gamma_E=0.5772 \dots$ is the Euler constant, and the scale $\mu$ has been
introduced to preserve the dimension of physical quantities. The factorization
theorem (Amati, Petronzio, and Veneziano, 1978a, 1978b; Ellis {\it et al.},
1979; Collins, Soper, and Sterman, 1988)
ensures, that the collinear singularity appearing in the transition function
$\overline{\Gamma}$ is universal, {\it i.e.} independent of the LO
scattering process, and can be absorbed into a renormalized
quark density in the photon
\beq
 f_{q/\gamma}(x,M_f^2)=\overline{f}_{q/\gamma}(x)+\le
 \overline{\Gamma}_{q\leftarrow\gamma}(M_f^2)\otimes\overline{f}_{\gamma/\gamma}\re (x)
 \label{eq:fqyren}
\eeq
at a factorization scale $M_f$. Thus logarithmic dependences on the artificial
scale $M_f$ are induced in $f_{q/\gamma}$ and in the NLO partonic scattering
cross section, which cancel up to higher orders in the perturbative expansion.
In the modified minimal subtraction ($\ms$) scheme (Bardeen {\it et al.}, 1978)
no additional finite terms are subtracted, since these depend generally
on the hard scattering process. In Eq.\ (\ref{eq:fqyren}) the $x$-space
convolution is defined as
\beq 
 \le \overline{\Gamma}_{j\leftarrow i} \otimes \overline{f}_{i/\gamma}\re (x)
 = \int\limits_{x}^1\frac{\mbox{d}
 y}{y}\overline{\Gamma}_{j\leftarrow i}\lr\frac{x}{y}\rr \overline{f}_{i/\gamma}(y),
\eeq
$\overline{f}_{q/\gamma}(x)=\delta_{q\gamma}\delta(1-x)+{\cal O}(\alpha)$
is the bare quark density, and $\overline{f}_{\gamma/\gamma}(x)=\delta(1-x)+
{\cal O}(\alpha^2)$ is the bare photon density, which gets only renormalized at
next-to-next-to-leading order (NNLO) in $\alpha$. A renormalized gluon density
in the photon $f_{g/\gamma}(x,M_f^2)$ is generated in a similar fashion by
gluon radiation from a quark-antiquark pair at ${\cal O}(\alpha\alpha_s)$,
where $\alpha_s(\mu^2)$ is the running coupling constant of quarks and gluons.
Beyond LO, the direct photon scattering processes are thus intimately connected
to the resolved processes. In this Section, we review the NLO QCD evolution
equations, boundary conditions, factorization schemes, and hadronic
solutions of the photon structure function.

The evolution of the parton densities in the photon with the scale
$M_f^2$ can be calculated in perturbation theory by taking the
logarithmic derivatives of $f_{q/\gamma},\,f_{g/\gamma},$ and
$f_{\gamma/\gamma}$ with respect to $M_f^2$. This leads to a coupled system
of integro-differential equations (De Witt {\it et al.}, 1979)
\bea
 \label{eq:evol_eq}
 \frac{\mbox{d}f_{q     /\gamma}(Q^2)}{\mbox{d}\ln Q^2} &=&
   \frac{\alpha       }{2\pi} P_{q\leftarrow\gamma}\otimes f_{\gamma/\gamma}
  (Q^2)
  +\frac{\alpha_s(Q^2)}{2\pi}\\&\times&
  \left[ P_{q\leftarrow q} \otimes f_{q/\gamma}(Q^2)
  +      P_{q\leftarrow g} \otimes f_{g/\gamma}(Q^2) \right] ,\nonumber\\
 \frac{\mbox{d}f_{g     /\gamma}(Q^2)}{\mbox{d}\ln Q^2} &=&
   \frac{\alpha       }{2\pi} P_{g\leftarrow\gamma}\otimes f_{\gamma/\gamma}
  (Q^2)
  +\frac{\alpha_s(Q^2)}{2\pi}\nonumber\\&\times&
  \left[ P_{g\leftarrow q} \otimes f_{q/\gamma}(Q^2)
  +      P_{g\leftarrow g} \otimes f_{g/\gamma}(Q^2) \right] ,\nonumber\\
 \frac{\mbox{d}f_{\gamma/\gamma}(Q^2)}{\mbox{d}\ln Q^2} &=&
   \frac{\alpha       }{2\pi} P_{\gamma\leftarrow\gamma}\otimes f_{\gamma/
  \gamma}(Q^2)
  +\frac{\alpha}{2\pi}\nonumber\\&\times&
  \left[ P_{\gamma\leftarrow q} \otimes f_{q/\gamma}(Q^2)
  +      P_{\gamma\leftarrow g} \otimes f_{g/\gamma}(Q^2) \right]. \nonumber
\eea
These evolution equations differ from the well-known hadronic case by the
inhomogeneous terms of $\O(\alpha)$, which arise from the pointlike coupling of
photons to quarks. Eqs.\ (\ref{eq:evol_eq})
are given for a single quark flavor $q$, but $2N_f$ light
quarks and antiquarks are easily accommodated by summing over the index $q$
from one to $2N_f$ whenever it appears twice.
The renormalization scale $\mu$ and the factorization scale $M_f$ have
been identified with the physical scale $Q$. The $x$-space convolution
$P_{j\leftarrow i} \otimes f_{i/\gamma}(Q^2)$
reduces to a simple product $P_{j\leftarrow i}(n) f_{i/\gamma}(n,Q^2)$ in
Mellin $n$-space, where the $n^{\rm th}$ moment is defined as
$f(n)=\int_0^1{\rm d}xx^{n-1}f(x)$.
In moment space, analytical solutions to the evolution equations can be found,
but they have to be transformed back to $x$-space for physical cross section
predictions. Alternatively, the evolution equations can be solved directly in
$x$-space by iteration, but this requires a careful separation of LO and NLO
terms in order to avoid spurious higher order terms. An original ansatz by
Rossi (1984) was subsequently generalized by Da Luz Vieira and Storrow (1991)
in order to allow for non-zero input parton densities.

The LO and NLO splitting functions $P_{j\leftarrow i}$ have been
calculated by Curci, Furmanski, and Petronzio (1980) for the flavor-nonsinglet
case, by Furmanski and Petronzio (1980) for the coupled gluon and
flavor-singlet cases, and also by Floratos, Kounnas, and Lacaze (1981). Here
we review only the LO results (Altarelli and Parisi, 1977)
\bea
  P_{q\leftarrow q} (x) & = &
    C_F \left[ \frac{1+x^2}{(1-x)}_+ + \frac{3}{2} \delta (1-x)
    \right] +{\cal O}(\alpha_s), \nonumber\\
  P_{g\leftarrow q} (x) & = &
    C_F \left[ \frac{1+(1-x)^2}{x} \right] +{\cal O}(\alpha_s), \nonumber\\
  P_{q\leftarrow g} (x) & = &
    T_R \left[ x^2+(1-x)^2 \right]+{\cal O}(\alpha_s), \nonumber \\
  P_{g\leftarrow g} (x) & = &
    2 N_C \left[ \frac{1}{(1-x)}_++\frac{1}{x}+x(1-x)-2 \re\nonumber \\
    &+& \left[ \frac{11}{6}N_C-\frac{1}{3}N_f\right] \delta (1-x)+{\cal O}
    (\alpha_s),
\label{eq:altparsplit}
\eea
where the $+$-distributions are defined as usual with a test function $f(x)$
in the integral
\beq
 \int_0^1\d xf(x)g_+(x)=\int_0^1\d x [ f(x)-f(1)] g(x).
\eeq
The photon-quark splitting function can be obtained in LO
from $P_{q\leftarrow g}$
by the transformation $P_{q\leftarrow\gamma} = 2N_Ce_q^2P_{q\leftarrow g}$.
$C_F=(N_C^2-1)/(2N_C)$ and $N_C=3$ are the SU(3) color factors, $T_R=1/2$,
$N_f$ is the number of active flavors, and $e_q$ is the fractional quark
charge. Since there is no direct coupling of photons to
gluons, the photon-gluon splitting function enters only in NLO. It can
be obtained from the NLO gluon-gluon splitting function by replacing the
appropriate color factors and dropping the part proportional to $\delta(1-x)$
(Fontannaz and Pilon, 1992)
\bea
  P_{g\leftarrow\gamma} (x) &=& \frac{\alpha_s(Q^2)}{2\pi} 
 \frac{e_q^2N_CC_F}{2}\le
  -16+8x+\frac{20}{3}x^2+\frac{4}{3x}\rp\nonumber\\
 &-&\lp (6+10x)\ln x-2(1+x)\ln^2x\re.
\eea

\subsection{Boundary conditions and factorization schemes}

The general solutions of the evolution equations Eqs.\ (\ref{eq:evol_eq})
are given by the sums of
pointlike (``anomalous'') and hadronic contributions $f_{i/\gamma}(Q^2) =
f^{\rm pl}_{i/\gamma}(Q^2)+f^{\rm had}_{i/\gamma}(Q^2)$. Due to the pointlike
coupling of the photon to quarks, the former can be calculated perturbatively
at asymptotically large scales $Q^2$ and large Bjorken-$x$ in LO (Witten,
1977) and NLO (Bardeen and Buras, 1979). The pointlike solution is of the form
\beq
 f^{\rm pl}_{i/\gamma}(Q^2)=\frac{\alpha}{2\pi}\le
 \frac{4\pi}{\alpha_s(Q^2)}a_i+b_i+{\cal O}(\alpha_s)\re,
 \label{eq:asymphot}
\eeq
where $a_i$ and $b_i$ are known
analytic functions in moment space. Scaling is already violated at LO,
${\cal O} (\alpha/\alpha_s(Q^2))$, and the pointlike solution dominates
at large $Q^2$. As the parton-parton cross sections are of
${\cal O} (\alpha_s^2)$, the resolved photon contributions are of the same
order ${\cal O}(\alpha\alpha_s)$ as the direct photon-parton
contributions. The pointlike solutions turn out to be singular at low $x$ and
moderate $Q^2$ (Duke and Owens, 1980) and have to be regularized by
boundary conditions $f^{\rm had}_{i/\gamma}(Q_0^2)$ at some low starting scale
$Q_0^2$ (Gl\"uck and Reya, 1983).

By combining the parton distributions functions $f_{q/\gamma}(Q^2)$ and
$f_{g/\gamma}(Q^2)$ with the appropriate Wilson coefficients
(Bardeen {\it et al.}, 1978)
\begin{eqnarray}
C_q(x) & = & C_F\! \left[ \frac{1+x^2}{1-x}\! \left( \ln\!
   \frac{1-x}{x}\!-\!\frac{3}{4}\right) \!+\! \frac{1}{4}\, 
     (9+5x)\right]_+\!\!,\!\!\\
C_g(x) & = & T_R \left[ \left(x^2+(1-x)^2\right) \ln\, \frac{1-x}{x} 
         + 8x(1-x)-1 \right], \nonumber
\end{eqnarray}
and (Bardeen and Buras, 1979)
\bea
C_{\gamma}(x)&=&2N_C\, C_g(x)=3\, \left[ \left( x^2+(1-x)^2\right)
   \,\ln \, \frac{1-x}{x}\rp\nonumber\\
 &+&\lp 8x(1-x)-1 \right],
\label{eq:cgamma}
\eea
one obtains the NLO photon structure function in the $\ms$ scheme
\bea
 F_2^{\gamma}(Q^2)&=&\sum_q 2 x e_q^2\lg f_{q/\gamma}(Q^2)+\frac{\alpha_s(Q^2)}
 {2\pi}\le C_{q}\otimes f_{q/\gamma}(Q^2)\rp\rp\nonumber\\
 &+&\lp\lp C_{g}\otimes f_{g/\gamma}(Q^2)\re
 +\frac{\alpha}{2\pi}e_q^2C_{\gamma}\rg.
 \label{eq:f2gamma}
\eea
The factor of two arises from the fact that $f_{q/\gamma}(Q^2)=
f_{\overline{q}/\gamma}(Q^2)$ due to charge conjugation invariance.
Gl\"uck, Reya, and Vogt (1992a) observed that the direct term $C_{\gamma}(x)$
contains a term $\ln(1-x)$, which diverges in the large-$x$ region and is
better absorbed in the quark distributions $f^{\dis}_{q/\gamma}(Q^2) =
f^{\ms}_{q/\gamma}(Q^2)+\alpha/(2\pi)e_q^2 C_{\gamma}$. This also affects
the NLO photon splitting functions
\begin{eqnarray}
P_{q\leftarrow\gamma}^{\dis} & = & P_{q\leftarrow\gamma}^{\ms} -
		e_q^2\, P_{q\leftarrow q} \otimes C_{\gamma}\nonumber,\\
P_{g\leftarrow\gamma}^{\dis} & = & P_{g\leftarrow\gamma}^{\ms} -
		2\, \sum_q  e_q^2\, P_{g\leftarrow q}\otimes C_{\gamma}\, .
\label{eq:dissplit}
\end{eqnarray}
The definition of the gluon densities in this new $\dis$ factorization scheme
remains unchanged.
Equivalently in the $\ms$ scheme one can absorb $C_\gamma$ (Gordon and Storrow,
1992a, 1997) or its process-independent part (Aurenche, Fontannaz, and
Guillet, 1994a) into pointlike initial quark distributions.

Heavy quarks with mass $m_h$ and velocity squared $\beta^2=1-4m_h^2x/
[(1-x)Q^2]$ contribute to the photon structure function through the
Bethe-Heitler process $\gamma^\ast(Q^2)\gamma\rightarrow h\overline{h}$
with (Budnev {\it et al.}, 1974; Hill and Ross, 1979)
\bea
 F_{2,h}^{\gamma}(x,Q^2)&=&3 x e_h^4 \frac{\alpha}{\pi}\lg\beta\le
 8x(1-x)\!-\!1\!-\!x(1-x)\frac{4m_h^2}{Q^2}\re\rp\nonumber \\
 &+&\!\!\lp\le x^2\!+
 (1-x)^2\!+x(1-3x)\frac{4m_h^2}{Q^2}-x^2\frac{8m_h^4}{Q^4}\re\rp\nonumber\\
 &\times&\!\lp\ln\frac{1+\beta} {1-\beta}\rg,
 \label{eq:f2bh}
\eea
if the available hadronic energy squared $W^2=Q^2(1-x)/x$ is larger
than the production threshold $4m_h^2$. The NLO corrections to Eq.\
(\ref{eq:f2bh}) are at most 20\% (Laenen {\it et al.}, 1994) and usually
neglected by authors of photonic parton distributions, but a resolved
contribution from the process $\gamma^\ast(Q^2)g\rightarrow h\overline{h}$
\beq
 F_{2,h}^g(x,Q^2) = \frac{\alpha_s(Q^2)}{2N_Ce_h^2\alpha} F_{2,h}^{\gamma}(Q^2)
 \otimes f_{g/\gamma}(Q^2)
 \label{eq:f2bhres}
\eeq
is sometimes included. Far
above threshold heavy quarks are treated as light flavors with boundary
conditions $f_{h/\gamma}(m_h^2)=f_{\overline{h}/\gamma}(m_h^2)=0$.

\subsection{Hadronic solutions}
\label{sec:hadsol}

Since the $x$-dependence of the boundary conditions $f_{i/\gamma}(x,Q_0^2)$
cannot be calculated in perturbation theory, it is necessary to make
theoretical assumptions. Usually one takes a form similar to $f_{i/\gamma}
(x,Q_0^2)=Nx^\alpha (1-x)^\beta$ and fits the normalization $N$ and the
exponents $\alpha$ and $\beta$ to experimental
data. However, only one particular combination of photonic parton densities,
{\it i.e.} $F_2^{\gamma}(x,Q^2)$, is well constrained by experimental data
from PETRA, PEP, TRISTAN, and LEP in the ranges $0.001\leq x\leq 0.9$ and $0.24
\leq Q^2 \leq 390$ GeV$^2$ (Nisius, 2000), and this combination is dominated
by the up-quark density. Thus one has to rely further on models like Vector
Meson Dominance (see the review by Bauer {\it et al.}, 1978), which relates
the photon to the $\rho$, $\omega$, and $\phi$ mesons with
%
\begin{table}
\begin{center}
\begin{tabular}{|c|ccc|}
 Vector Meson V & $m_V$/MeV & $\Gamma^V_{e^+e^-}$/keV & $f_V^2/(4 \pi)$ \\
\hline
\hline
 $\rho$   &  769.3   $\pm$ 0.8   & 6.77~ $\pm$ 0.32      & 2.02 \\
 $\omega$ &  782.57  $\pm$ 0.12  & 0.60~ $\pm$ 0.02      & 23.2 \\
 $\phi$   & 1019.417 $\pm$ 0.014 & 1.297 $\pm$ 0.04      & 13.96 \\
\end{tabular}
\end{center}
\caption{\label{tab:deccons}Masses, leptonic decay widths, and decay constants
 in the zero-width approximation $\Gamma^V_{e^+e^-}=4\pi\alpha^2
 m_V/(3f_V^2)$ for the $\rho$, $\omega$, and $\phi$ vector mesons
 (Groom {\it et al.}, 2000).}
\end{table}
%
 $J^{PC}=1^{--}$
\bea
 |\gamma\rangle &=& \sum_{V=\rho,\omega,\phi}\frac{e}{f_V}|V\rangle =
 \sqrt{\frac{e^2}{f_\rho^2}+\frac{e^2}{f_\omega^2}}(e_u^2+e_d^2)^{-1/2}
 \nonumber\\
 &\times& (e_u|u\bar{u}\rangle+e_d|d\bar{d}\rangle)
 +\frac{e}{f_\phi}|s\bar{s}\rangle
 \label{eq:vmd}
\eea
and allows for a successful phenomenological description of experimental data
on the photoproduction of vector mesons.
A flavor SU(3)-symmetric superposition requires $f_\rho/f_\omega=1/3$ and
$f_\rho/f_\phi=-\sqrt{2}/3$ in fairly good agreement with experimental
measurements ($1/3.4$ and $-\sqrt{1.3}/3$, see Tab.\ \ref{tab:deccons}).
A coherent superposition of up-
and down-quarks, which is favorable
at large scales, is obtained with $e_u=2/3$ and
$e_d=-1/3$, while an incoherent superposition is obtained with $e_u=e_d=1$.

For safely high starting scales $Q_0^2\geq 2$ GeV$^2$, the pure VMD ansatz
turns out to be insufficient to describe the $F_2^\gamma$ data at larger $Q^2$
and has to be supplemented by an additional hard component. For quarks it
can be naturally provided by the quark-box diagram with four external photons,
but this is unfortunately
not viable for the gluon. Two solutions are possible. The first option is to
retain a relatively large $Q_0$, $Q_0 \geq 1$ GeV, fit the quark densities to
$F_2^\gamma$ data, and estimate the gluon input (Drees and Grassie [DG], 1985;
Abramowicz, Charchula,
and Levy [LAC], 1991; Gordon and Storrow [GS, GS96], 1992a, 1997; Hagiwara
{\it et al.} [WHIT], 1995; Schuler and Sj\"ostrand [SaS~2], 1995; Abramowicz,
Gurvich, and Levy [GAL], 1998). The second option is to retain the pure VMD
ansatz and start the evolution at a very low scale $Q_0\simeq 0.5...0.7$ GeV
(Gl\"uck, Reya, and Vogt [GRV], 1992b; Aurenche {\it et al.} [ACFGP], 1992;
Aurenche, Fontannaz, and Guillet [AFG], 1994a; Schuler and
Sj\"ostrand [SaS~1], 1995; Gl\"uck, Reya, and Schienbein [GRSc], 1999b).
Unfortunately the parton distributions in the vector mesons
are unknown, so that in practice one has to resort to those in the pseudoscalar
pion. The various parameterizations are summarized in Tab.\ \ref{tab:vmdinput}.

Recently, an energy-momentum sum rule for the photon has been derived
(Schuler and Sj\"ostrand, 1995; Frankfurt and Gurvich, 1996)
\beq
 \int_0^1{\rm d}x x\left[ \Sigma(x,Q^2)+f_{g/\gamma}(x,Q^2)
 +f_{\gamma/\gamma}(x,Q^2)\right] = 1,
 \label{eq:sumrule}
\eeq
which further constrains the boundary conditions. $\Sigma(x,Q^2)=
\sum_{i=1}^{N_f}2f_{q/\gamma}(x,Q^2)$ is the singlet quark distribution.
Integrating the third evolution equation in Eq.\ (\ref{eq:evol_eq}) one
obtains the solution
\beq
 f_{\gamma/\gamma}(x,Q^2)=\delta(1-x)\left[1-\frac{\alpha}{6\pi}\left(
 \sum_q2N_Ce_q^2\ln\frac{Q^2}{Q_0^2}+c\right)\right],
\eeq
%
\begin{table}
\begin{center}
\begin{tabular}{|ccc|ccccc|}
 Group & Year & Set & $Q_0^2$ & Factor.\        & VMD & $N_f$ &
   $\Lambda_{\ms}^{N_f=4}$ \\
       &      &     & (GeV$^2$) &        Scheme & Model & &  (MeV) \\
\hline
\hline
 DG & 1985 & 1-3 & 1               & LO            & -     & 3-5 &
   400 \\
 LAC & 1991& 1-3 & 4,4,1           & LO            & -     & 4     &
   200 \\
 GS & 1992 & LO  & 5.30            & LO            & incoherent & 5  &
   200 \\
   &    & NLO & 5.30            & $\ms$         & incoherent & 5  &
   200 \\
 GRV & 1992  & LO  & 0.25            & LO            & incoherent & 5     &
   200 \\
   &    & NLO & 0.30            & $\dis$        & incoherent & 5     &
   200 \\
 ACFGP & 1992 & NLO & 0.25          & $\ms$         &   coherent  & 4     &
   200 \\
 AFG &1994& NLO & 0.50            & $\ms$         &   coherent  & 4     &
   200 \\
 WHIT &1995& 1-6 & 4.00            & LO            & -     & 4     &
   400 \\
 SaS&1995& 1D/M&0.36            & LO&  coherent & 5     &
   200 \\
   &    & 2D/M&4.00            & LO&  coherent & 5     &
   200 \\
 GS96&1997& LO  & 3.00            & LO            & incoherent & 5  &
   200 \\
   &    & NLO & 3.00            & $\ms$         & incoherent & 5  &
   200 \\
 GAL&1998& LO  & 4               & LO            & - & 4 &
   200 \\
 GRSc&1999& LO  & 0.26            & LO            &   coherent & 3     &
   204 \\
   &    & NLO & 0.40            & $\dis$        &   coherent & 3     &
   299 \\
\end{tabular}
\end{center}
\caption{\label{tab:vmdinput}Parameterizations of the parton densities in the
 photon. The sets SaS~1M and 2M are defined by absorbing the $C_\gamma$
 coefficient in Eq.\ (\ref{eq:cgamma}) into the LO quark distributions.
 The coherence of the VMD model is determined by the coefficients $e_u$ and
 $e_d$ in Eq.\ (\ref{eq:vmd}).
 Most parameterizations fit $F_2^\gamma$ (except ACFGP, AFG, and
 GRSc) and the pion densities (except DG, LAC, WHIT, SaS, and GAL) and add
 the direct Bethe-Heitler contribution for heavy quarks in Eq.\
 (\ref{eq:f2bh}) (except DG, LAC, and GAL).
 WHIT, SaS, and GRSc also add the resolved contribution in Eq.\
 (\ref{eq:f2bhres}).}
\end{table}
%
which can be directly inserted in Eq.\ (\ref{eq:sumrule}) with the result
\bea
  \int_0^1{\rm d}x x\left[ \Sigma(x,Q^2)\rp &+&\lp f_{g/\gamma}(x,Q^2)\right] =
 \nonumber \\
 &=& \frac{\alpha}{6\pi}\left(
 \sum_q2N_Ce_q^2\ln\frac{Q^2}{Q_0^2}+c\right).
 \label{eq:sumrule2}
\eea
So the quark and gluon momentum fractions rise logarithmically with $Q^2$.
For the sum rule to be of practical use, the unknown integration constant $c$
has to be related, {\it e.g.}, to the total hadronic cross section in $e^+e^-$
annihilation via a dispersion relation in the photon virtuality.
Schuler and Sj\"ostrand (1995) obtained $c/\pi=\sum_{V=\rho,\omega,\phi}4\pi/
f_V^2\simeq 0.55$ at $Q_0^2\simeq 0.36$ GeV$^2$, while the numerical values in
Tab.\ \ref{tab:deccons} give a slightly larger value of $c/\pi\simeq 0.61$.

\begin{figure}
 \begin{center}
  \epsfig{file=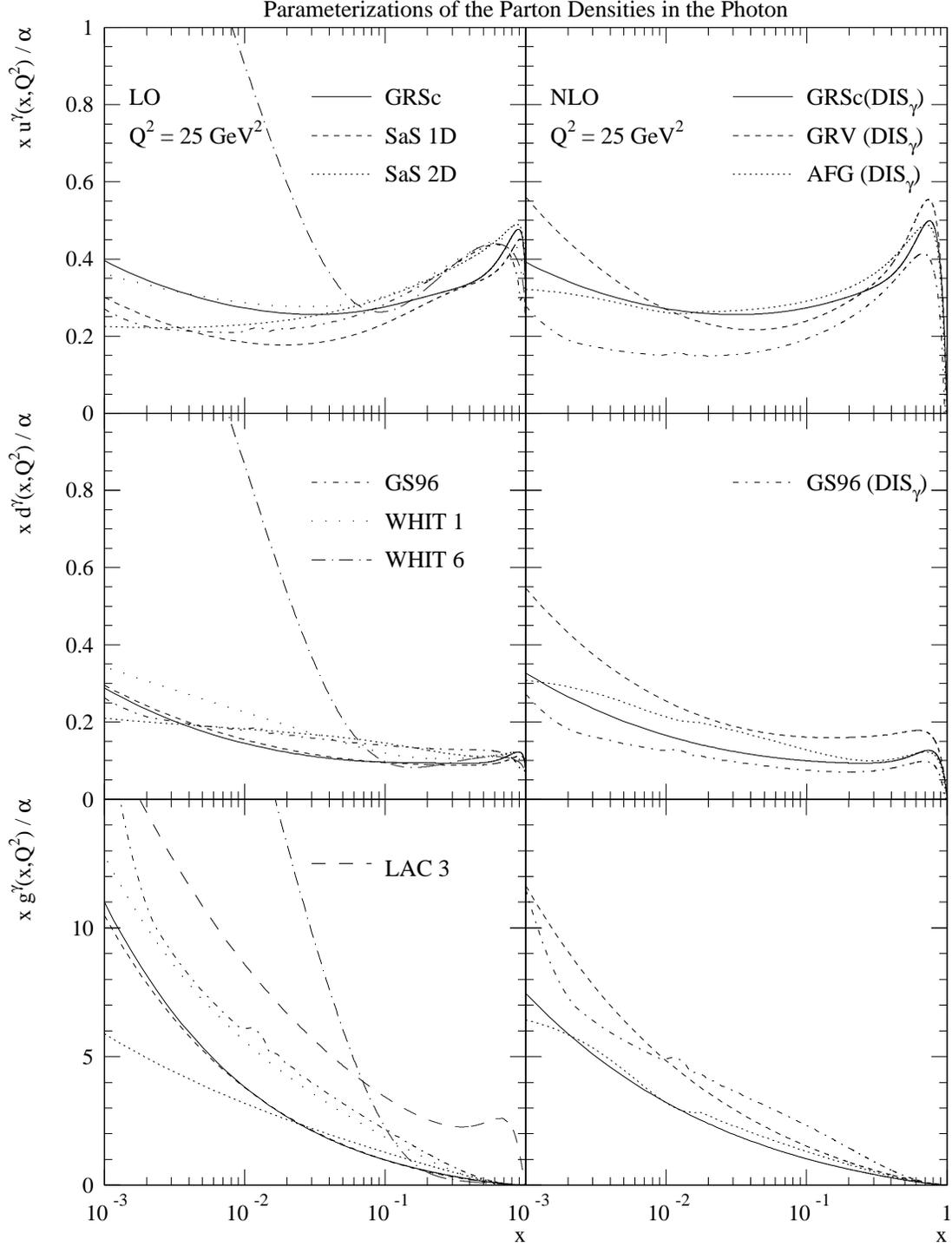,width=0.9\linewidth}
 \caption{\label{fig:02}
 Parameterizations of the up-quark (top), down-quark (center), and gluon
 (bottom) densities in the photon in LO (left) and NLO (right) at
 $Q^2=25$ GeV$^2$. The NLO AFG and GS96 parameterizations have been
 transformed from the $\ms$ to the $\dis$ scheme.}
 \end{center}
\end{figure}

Different LO and NLO parameterizations of the parton densities in the photon
are compared in Fig.\ \ref{fig:02} at a value of $Q^2=25$ GeV$^2$ relevant
for many hard photoproduction processes.
%
%
In LO (left) the ${\cal O}
(\alpha,\alpha_s)$ terms in Eq.\ (\ref{eq:f2gamma}) do not contribute and
$F_2^\gamma(x,Q^2)$ is
directly related to the quark densities. Thus the experimental constraints on
$F_2^\gamma(x,Q^2)$ lead to fairly good agreement of the different quark
density parameterizations, particularly at large $x$ $(x>0.1)$. Below this
value,
the WHIT~6 and also the similar WHIT~2-5 quark densities show a steep rise,
which is caused by a similar rise of the gluon density. While this behavior is
consistent with the large-$x$ data used in the fits, it is strongly disfavored
by recent small-$x$ data from LEP (Nisius, 2000). The parameterizations with a
high starting scale (WHIT, SaS~2, and GS96) have larger quark densities
at large $x$ than those with a low starting scale (GRSc and SaS~1). The sets
SaS~1M and SaS~2M, defined by absorbing $C_\gamma$ into pointlike initial
quark distributions even in LO, turn out similar to the sets SaS~1D and SaS~2D
after subtracting the $C_\gamma$-term.
If the vector-meson states are added in a coherent fashion (GRSc and SaS), the
$d$-quark input density in the photon is suppressed by a factor of four with
respect to the $u$-quark density. In this case the low-$x$ behaviors
differ, and the momentum sum rule Eq.\ (\ref{eq:sumrule2}) can be
fulfilled. With an incoherent superposition, the $u$-quark and $d$-quark
densities are identical at low $x$ (GS96 and WHIT),
and the momentum sum rule is
easily violated. The gluon density does not
enter directly in Eq.\ (\ref{eq:f2gamma}) in LO. It enters only through a
rather weak coupling to the singlet in the evolution equations and is
consequently badly constrained. Only a very steep rise at low $x$ (LAC~1-2,
WHIT~2-6) and a very hard gluon (LAC~3) can be ruled out by $F_2^\gamma$ data
from LEP and jet or particle production data from HERA. On the other hand, the
HERA data indicate that the SaS~1D and the almost identical GRSc gluon
densities may be too low (Nisius, 2000).

The NLO parton densities are shown on the right hand side of Fig.\
\ref{fig:02} in the $\dis$ scheme. The AFG and GS96 quark densities have been
properly transformed, while the GRSc and GRV densities are given directly in
the $\dis$ scheme. The quark densities do therefore not exhibit the $\ms$
singularity at $x=1$, but have shapes similar to their LO counterparts.
Furthermore, the GRSc and AFG densities are very similar, since both groups
use a coherent VMD ansatz at a low starting scale and relate the $\rho$-meson
input to recent determinations of the pion densities. An incoherent
superposition and older pion parameterizations have been used in the GRV
distributions, resulting in a steeper rise for the identical $u$-quark and
$d$-quark sea distributions at small $x$.
The gluon distribution enters directly in $F_2^\gamma$ in NLO (see Eq.\
(\ref{eq:f2gamma})) and can therefore be better constrained.
Again the GRV distribution exhibits a steeper shape than GRSc and AFG, and the
GS96 result is substantially harder. By definition the NLO GS96 set should,
like the older NLO GS set, lead to the same $F_2^\gamma(x,Q_0^2)$ as in LO.
However,
the NLO quark parameterizations turn out too small over the full $x$-range.
The LO $F_2^\gamma(x,Q_0^2)$ results and the figures in the original papers
(Gordon and Storrow, 1992a, 1997) cannot be reproduced with the available
NLO parameterization, which is therefore not usable in its present form.

\begin{figure}
 \begin{center}
  \epsfig{file=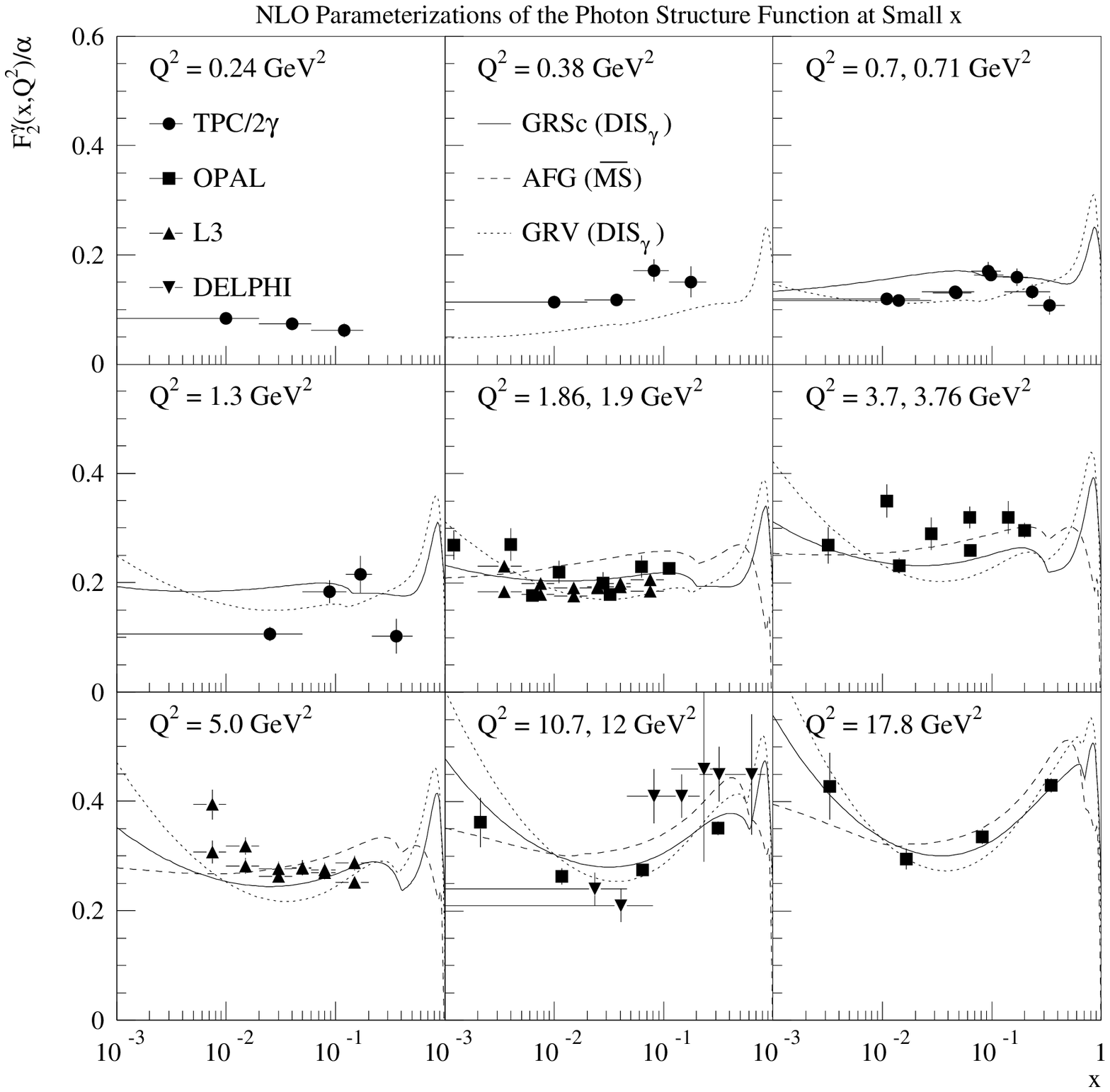,width=\linewidth}
 \caption{\label{fig:03}
 NLO parameterizations of the photon structure function at small $x$
 compared to the world data on $F_2^\gamma$ (Nisius, 2000).
 The GRSc and GRV parameterizations are given in the $\dis$ scheme,
 while the AFG parameterization is in the $\ms$ scheme.}
 \end{center}
\end{figure}

\begin{figure}
 \begin{center}
  \epsfig{file=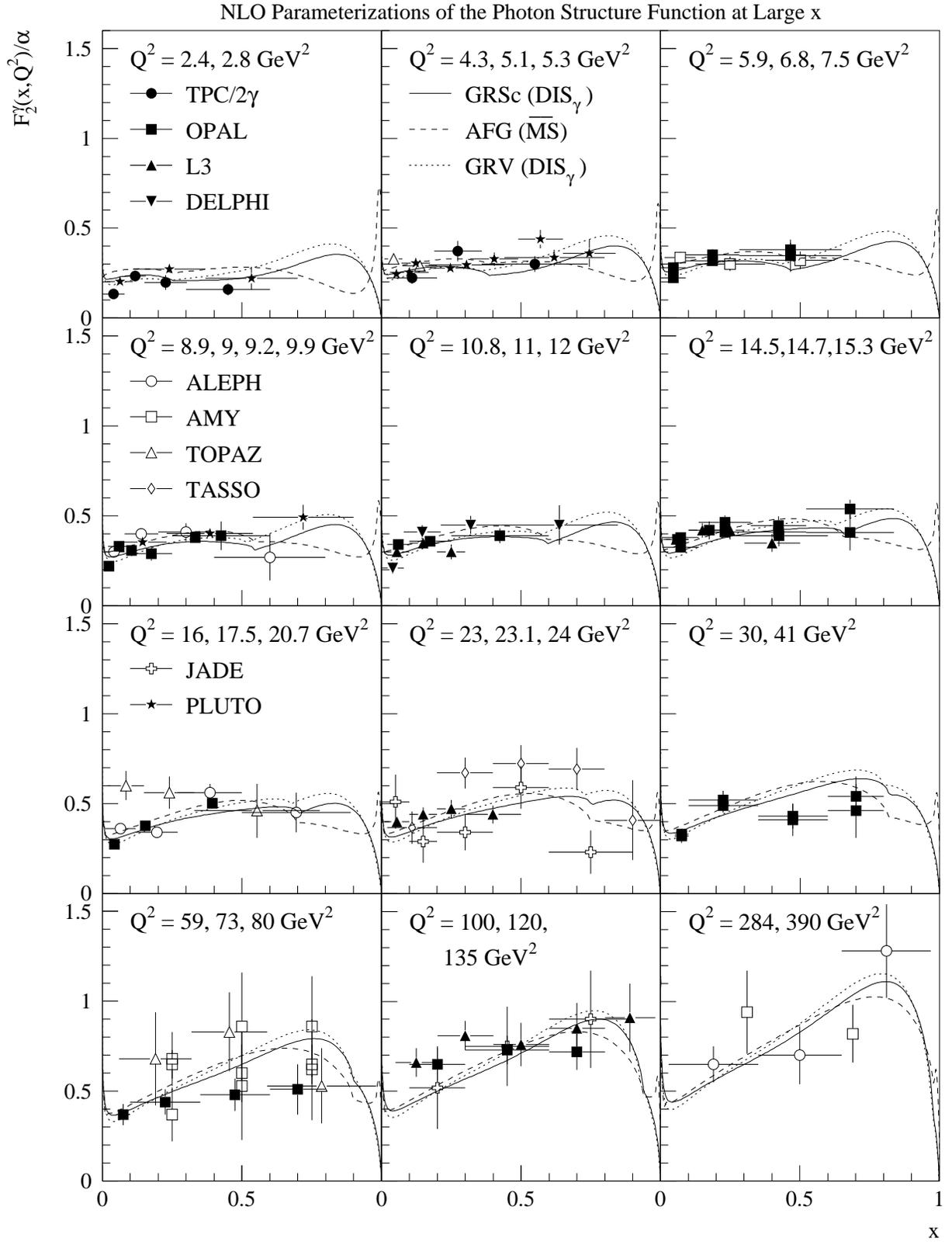,width=\linewidth}
 \caption{\label{fig:04}
 Same as Fig.\ \ref{fig:03}, but now for large $x$.}
 \end{center}
\end{figure}

The three usable NLO parameterizations of the photon structure function
(GRSc, AFG, and GRV) are confronted with the world data on $F_2^\gamma(x,Q^2)$
in Figs.\ \ref{fig:03} (small $x$) and \ref{fig:04} (large $x$).
None of
the parameterizations dare to describe the TPC/2$\gamma$ data at $Q^2=0.24$
GeV$^2$, which may be too close to $\Lambda^2\simeq 0.04-0.16$ GeV$^2$ to
allow for a perturbative treatment (see Fig.\ \ref{fig:03}).
%
%
The first NLO parameterization (GRV) starts at $Q_0^2=0.30$ GeV$^2$. While it
is still lower than the TPC/2$\gamma$ data at $Q^2=0.38$ GeV$^2$, it fits the
TPC/2$\gamma$ data rather well already at $Q^2=0.7$ and 1.3 GeV$^2$.
The low-$x$ and low-$Q^2$ data from OPAL and L3 are slightly underestimated.
The GRSc parameterization is evolved from $Q_0^2=0.40$ GeV$^2$. It describes
the LEP data at small $x$ best, due to the most recent determination of the
pion structure. AFG starts at $Q_0^2=0.50$ GeV$^2$, but this parameterization
should only be compared to data above $Q^2\simeq 2$ GeV$^2$. The theoretical
assumptions are very similar to GRSc, and so are the results for $F_2^\gamma$,
AFG being generally slightly larger.
At larger $x$ and $Q^2$ (Fig.\ \ref{fig:04}), the hadronic contribution
decreases, while the pointlike contribution increases.
%
%
Consequently the
different parameterizations converge, and they all describe the data very well.
The factorization scheme starts
to play a role at very large $x$, and thus the AFG $\ms$ prediction
differs strongly from the GRSc and GRV $\dis$ predictions.
The GRSc charm contribution is provided only by the direct and resolved
Bethe-Heitler terms in Eqs.\ (\ref{eq:f2bh}) and (\ref{eq:f2bhres}) for
$Q^2(1-x)/x > 4 m_c^2$, while GRV and AFG have included the charm quark also
in the massless evolution Eqs.\ (\ref{eq:evol_eq}). 
GRSc and AFG choose $m_c\simeq 1.4$ GeV, while GRV take $m_c=1.5$ GeV, so that 
the charm contribution sets on at slightly different values of $x$.

Early hopes to extract the strong coupling constant from the photon structure
function (Witten, 1977) were dashed shortly after by complications arising in
higher orders (Bardeen and Buras, 1979; Antoniadis and Grunberg, 1983; Rossi,
1983; Gl\"uck and Reya, 1983; Field, Kapusta, and Poggioli, 1987).
Since then it has been widely believed that the sensitivity of the photon
structure function to the strong coupling is low. Most of the parameterizations
use an approximate NLO solution of the renormalization group equation
\beq
 \alpha_s(Q^2) = \frac{4\pi}{\beta_0\ln(Q^2/\Lambda^2)}-\frac{4\pi\beta_1}
 {\beta_0^3}\frac{\ln\ln(Q^2/\Lambda^2)}{[\ln(Q^2/\Lambda^2)]^2}
\eeq
with the one- and two-loop $\beta$-functions $\beta_0=11-2N_f/3$ and
$\beta_1=102-38N_f/3$ and the fundamental QCD scale parameter
$\Lambda^{N_f=4}_{\ms}=200$ MeV. GS use the same $\Lambda$-value for all $N_f$,
while GRSc take $\alpha_s(M_Z)=0.114$ and solve the renormalization group
equation numerically. However, a new analysis shows that the now final PETRA,
TRISTAN, and LEP data lead to a competitive determination of $\alpha_s$ from a
single-parameter pointlike fit at large $x$ and $Q^2$ with
$\alpha_s(m_Z)=0.1183\pm0.0050$(exp.)$^{+0.0029}_{-0.0028}$(theor.) and from
a five-parameter full (pointlike and hadronic) fit at all $x$ and $Q^2$ with
$\alpha_s(m_Z)=0.1198\pm0.0028$(exp.)$^{+0.0034}_{-0.0046}$(theor.)
(Albino, Klasen, and S\"oldner-Rembold, 2002).

The hadronic structure of the photon is currently much less well known than
that of the proton, and current parameterizations rest largely on similar
theoretical assumptions like Vector Meson Dominance. To improve the situation,
the light quark and gluon densities have to be disentangled from various
observables in jet, hadron, and prompt photon production (see the following
Sections), and the heavy quark contributions have to be consistently resummed
according to variable flavor number schemes.


\pagebreak
\section{Jet production}
\label{sec:jetprod}
\setcounter{equation}{0}

Due to the confinement properties of QCD, only hadrons, and not partons, are
observable as asymptotic states. The hadronization process happens at low
scales, where the coupling $\alpha_s$ is large, and is therefore not calculable
in perturbation theory. It can, however, be modeled by combining several
partons or hadrons moving in the same direction into jets. At sufficiently
large transverse energies, the production cross section can then be calculated
perturbatively and jets can be related to partons.

Jets were first observed in event deformations at the $e^+e^-$ colliders SPEAR
(Hanson {\it et al.}, 1975) and DORIS (Berger {\it et al.}, 1978), where they
helped to identify the quark spin. At PETRA, they could then be seen even by
the naked eye. They established the existence of the gluon (Bartel {\it et
al.}, 1980; Brandelik {\it et al.}, 1980; Berger {\it et al.}, 1980; Behrend
{\it et al.}, 1982) and were also observed in photon-photon collisions
(Bartel {\it et al.}, 1981; Brandelik {\it et al.}, 1981; Althoff {\it et al.},
1984; Berger {\it et al.}, 1984a, 1985, 1987; Behrend {\it et al.}, 1991).
Jet production in photon-hadron collisions was subsequently observed at
Fermilab (Adams {\it et al.}, 1994) and at HERA (Abt {\it et al.}, 1993;
Derrick {\it et al.}, 1995a). In the following, various jet definitions and
jet cross sections will be discussed.

\subsection{Cone and cluster algorithms}
\label{sec:jetalgorithms}

In Sec.\ \ref{sec:phostr} we showed, how photons can split
into massless quark-antiquark pairs and exhibit a collinear singularity.
Collinear divergencies arise quite generally in scattering processes,
whenever a massless parton splits into two. In addition, infrared
divergencies are generated by the emission of soft particles.
Within the context of QED, it could be proven that soft (Bloch and
Nordsieck, 1937) and collinear (Kinoshita, 1962) divergencies
cancel between real and virtual emission in total cross sections. Lee and
Nauenberg (1964) demonstrated, that the cancellation happens also for
differential cross sections, if they are summed over degenerate initial and
final states. In QED the collinear divergency is regulated by the electron
mass, but in QCD it occurs even for non-zero quark masses, since
one massless gluon can split into two. Nevertheless, infrared and
collinear singularities are expected to cancel for all leading-twist QCD
observables. A rigorous proof of the Kinoshita-Lee-Nauenberg (KLN)
theorem does not exist for QCD, but in practical applications
counter-examples have never been found.

Shortly after the first experimental observation of jets, Sterman and Weinberg
(1977) realized, that beyond LO there can be no one-to-one correspondence
between jets and partons, but that jet definitions must be infrared safe
and satisfy the KLN theorem order by order in perturbation theory. In practice,
jet definitions should be simple to implement in experimental analyses and
theoretical calculations and be insensitive to hadronization effects.
In the Sterman-Weinberg definition, a final state in $e^+e^-$ annihilation
is classified as two-jet like, if all but an energy
$\epsilon$ is contained in a pair of cones of half-angle $\delta$, while all
other events are classified as three-jet like. This cone definition is
infrared safe, but it is not well-suited for events with more than three
jets, and it separates the phase space in an inefficient way.
A better solution is provided by clustering algorithms such as the one used by
the JADE collaboration (Bethke {\it et al.}, 1988), where two particles $i$ and
$j$ are combined into a jet and their four-momenta $p_{i,j}$ are added, if
their invariant mass
\beq
 (p_i+p_j)^2 = 2E_iE_j(1-\cos\theta_{ij})< yS
\eeq
is smaller than some fixed fraction $y$ of the overall center-of-mass energy
$S$.
Theoretically the JADE algorithm is not well suited to resum logarithms of
$\O(\alpha_s^n\ln^{2n}y)$, which can become large for small values of $y$.
This is easier in the $k_T$ or Durham algorithm (Catani {\it et al.}, 1991),
where two particles are combined, if their energies $E_{i,j}$ and the angle
$\theta_{ij}$ between them satisfy the condition
\beq
 2 \min (E_i^2,E_j^2) (1-\cos\theta_{ij})< yS.
\eeq

In photon and hadron collisions, the partonic system is boosted along the
beam axis, and the remnants of the beams must be separated from the hard jets.
It is therefore convenient to define outgoing particles $i$ in the transverse
plane with four-momenta 
\bea
 p_i&=&(E_i,p_{x_i},p_{y_i},p_{z_i}) \\
    &=& (m_{T_i}\cosh y_i,p_{T_i} \cos\phi_i,p_{T_i}\sin\phi_i,
        m_{T_i}\sinh y_i), \nonumber
\eea
transverse masses $m_{T_i}=\sqrt{p_{T_i}^2+m_i^2}$, transverse momenta
$p_{T_i}$, azimuthal angles $\phi_i$, and rapidities
\beq
 y_i={1\over 2}\ln\lr{E_i+p_{z_i}\over E_i-p_{z_i}}\rr.
\eeq
Experimentally the jet energies $E_i$ and scattering angles $\theta_i$ are
measured, and the $p_{T_i}$ are replaced with $E_{T_i}=E_i\sin\theta_i$. The
rapidities $y_i$ are additive under boosts along the $z$-direction.
For massless particles they coincide with 
the pseudorapidities $\eta_i=-\ln[\tan(\theta_i/2)]$.
According to the standardization of the 1990 Snowmass meeting
(Huth {\it et al.}, 1990), particles $i$ are added to a jet cone
$J$, if they have a distance
\beq
 R_i = \sqrt{(\eta_i-\eta_J)^2+(\phi_i-\phi_J)^2} < R
\eeq
from the cone center. $R=0.7...1$ is the jet radius in the $\eta-\phi$
plane and
\bea
 E_{T_J} = \sum_{R_i< R} E_{T_i}&,&
 \eta_J  = \frac{1}{E_{T_J}}\sum_{R_i< R} E_{T_i}\eta_i,\nonumber\\
{\rm and~~} \phi_J & = &\frac{1}{E_{T_J}}\sum_{R_i< R} E_{T_i}\phi_i
 \label{eq:recomb}
\eea
define the jet transverse energy and axis. In NLO QCD two partons, which may
be separated by as much as $2R$, can be combined, if they have equal
transverse energies. In this case, one could count the individual jets in
addition to the combined jet, but this double-counting should be avoided
(Ellis, {Kunszt}, and Soper, 1989b). The broad combined jet is difficult to
find experimentally, since it does not have a seed in its center. This led
Ellis, Kunszt, and Soper (1992) to propose an additional parameter
$R_{\rm sep}$, which restricts the distance
\bea
 R_{ij} &=& \sqrt{(\eta_i-\eta_j)^2+(\phi_i-\phi_j)^2} \nonumber\\
 &<& \min\le\frac{E_{T_i}+E_{T_j}}{\max (E_{T_i},E_{T_j})} R,
 R_{\rm sep}\re
\eea
of two partons $i$ and $j$. However, the value of $R_{\rm sep} = R...2R$ must
be fitted to different experimental implementations of the Snowmass
algorithm (Butterworth {\it et al.}, 1996; Klasen and Kramer, 1997b).
Furthermore, studies of three jet hadroproduction (Giele and Kilgore, 1997)
and jet shapes (Seymour, 1998) revealed, that the Snowmass cone algorithm is
not infrared safe in higher orders, unless the centers of jet pairs
are also considered as seeds (Akers {\it et al.}, 1994).

The deficiencies of the cone algorithm are remedied in the longitudinally
invariant $k_T$ clustering algorithm (Catani, Dokshitzer, and Webber, 1992;
Catani {\it et al.}, 1993; Ellis and
Soper, 1993), where one uses only the combination criterion
$R_{ij} < R_{\rm sep}$ with the $E_T$-weighted
recombination scheme as in Eq.\ (\ref{eq:recomb}). The results obtained
with the cluster algorithm in hadron-hadron collisions become similar to those
obtained with the cone algorithm, if one chooses $R_{\rm sep}\simeq 1.35 R$
(Ellis and Soper, 1993). This was also found to be the optimal value of
$R_{\rm sep}$ for the cone algorithm in photoproduction (Butterworth {\it et
al.}, 1996; Klasen and Kramer, 1997b).

\subsection{Single jets}
\label{sec:singlejets}

\begin{figure}
 \begin{center}
  \epsfig{file=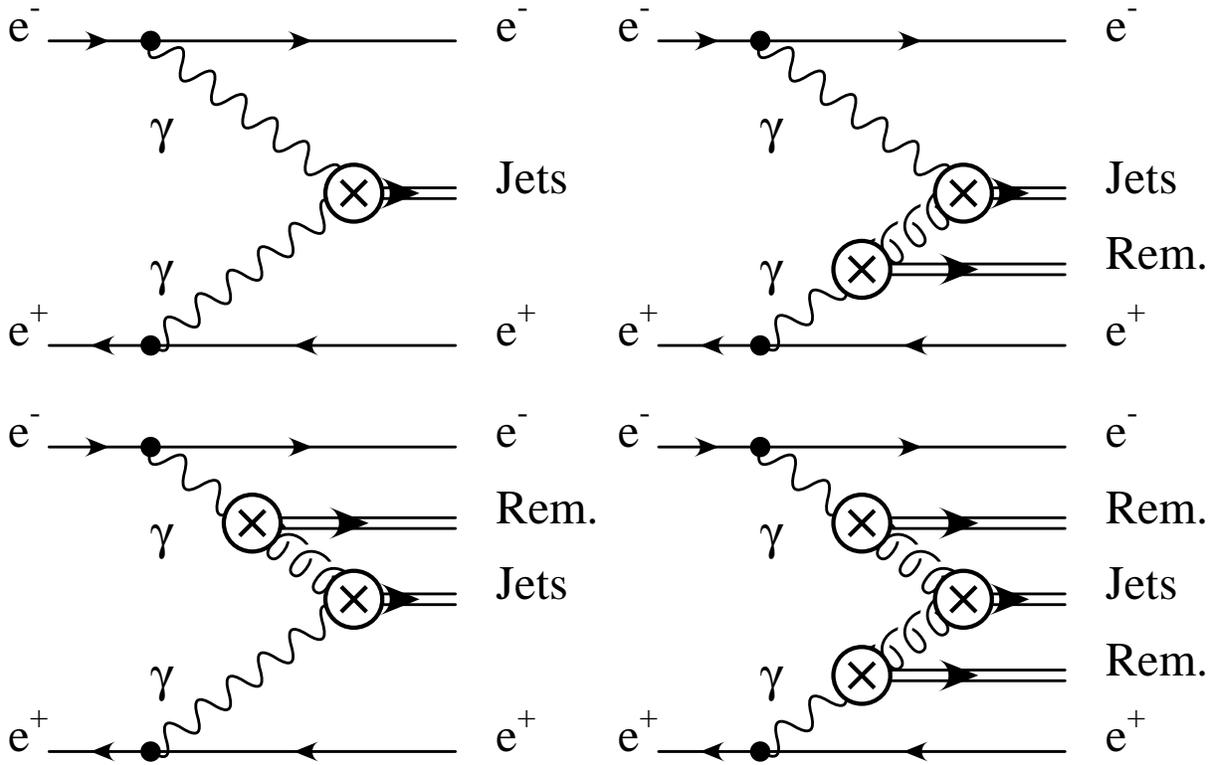,bbllx=60pt,bblly=290pt,bburx=280pt,bbury=430pt,%
          width=\linewidth}
 \caption{\label{fig:05}Factorization of photon-photon scattering into jets.}
 \end{center}
\end{figure}

\begin{figure}
 \begin{center}
  \epsfig{file=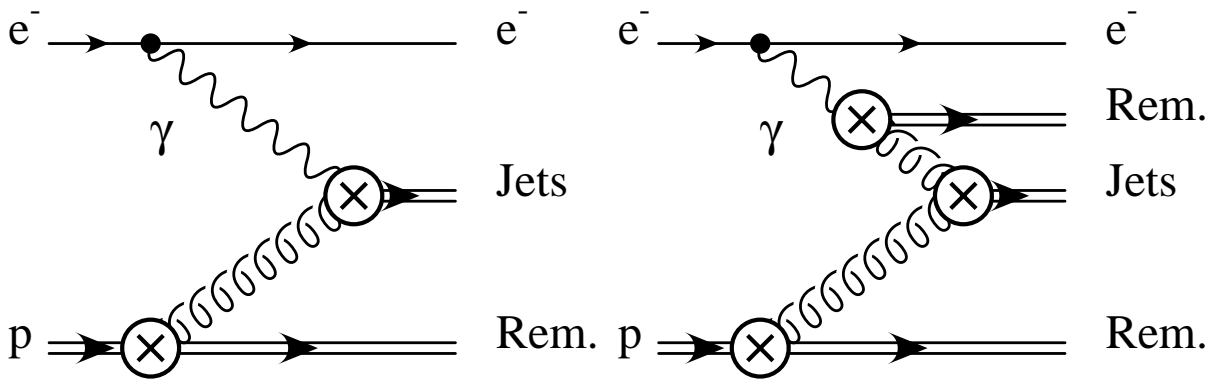,bbllx=60pt,bblly=360pt,bburx=280pt,bbury=430pt,%
          width=\linewidth}
 \caption{\label{fig:06}Factorization of photon-hadron scattering into jets.}
 \end{center}
\end{figure}

Since photons couple to charged quarks, they can participate directly in hard
scattering processes like jet production. On the other hand, they can also
exhibit a partonic structure (see Sec.\ \ref{sec:phostr}) and interact
indirectly through their constituents (quarks and gluons). In photon-photon
scattering, this leads to the direct,
single-resolved, and double-resolved processes shown in Fig.\ \ref{fig:05},
%
%
where the hard central jets are produced in association with zero, one, or two
forward-going photon remnant jets.
In photon-hadron scattering, the single- and double-resolved photon processes
described above correspond to the direct and resolved photon processes in
Fig.\ \ref{fig:06}, which
%
%
involve a proton remnant jet and, for resolved photoproduction, also a photon
remnant jet.

The LO partonic
cross section for the scattering of two massless partons $a$ and $b$
into two massless partons $1$ and $2$ is
\bea
 \frac{\d\sigma^B}{\d t}&=&\frac{1}{2s}
 \frac{1}{\Gamma(1-\varepsilon)}
  \left( \frac{4\pi s}{tu} \right) ^\varepsilon
  \frac{1}{8\pi s}
 \frac{g_{a,b}^2}{S_aS_bC_aC_b}
 |\M^B|^2
\label{eq:partonicxsec}
\eea
in $n=4-2\eps$ dimensions. $s = (p_a+p_b)^2,~t = (p_a-p_1)^2$, and
$u = (p_a-p_2)^2$ are the Lorentz-invariant Mandelstam variables,
$\Gamma(x)$ is the Euler $\Gamma$-function, and $g_{a,b}^2$ are the squared
coupling constants ($4\pi\alpha e_q^2$ for photons and $4\pi\alpha_s(\mu^2)$
for quarks and gluons). $S_{a,b}$ are the spin degrees
of freedom of the initial particles (2 for quarks and $n-2$ for photons and
gluons) and $C_{a,b}$ are the initial color degrees of freedom (1 for photons,
$N_C$ for quarks, and $N_C^2-1$ for gluons). While one can safely set $n=4$ or
$\eps=0$ in Eq.\ (\ref{eq:partonicxsec}) and in the squared matrix
element $|\M^B|^2$ in LO calculations (Brodsky {\it et al.}, 1978, 1979;
Owens, 1980; Fontannaz {\it et al.},
1980; Baer, Ohnemus, and Owens, 1989a),
the $\O(\eps)$ terms are needed in NLO in connection with
ultraviolet, infrared, and collinear divergencies (see below).
In jet cross sections, the
partonic masses $p_i^2=m_i^2\ll E_T^2$ can usually be neglected, and the
Mandelstam variables
\bea
  s =  4 x_a x_b E_A E_B&,&
  t = -2 x_a E_A E_T e^{-\eta},\nonumber\\
  {\rm and~~}  u &=& -2 x_b E_B E_T e^{\eta}
\eea
become simple functions of $E_T$, $\eta$, $E_{A,B}$, and $x_{a,b}$.

%
\begin{figure}
 \begin{center}
  \epsfig{file=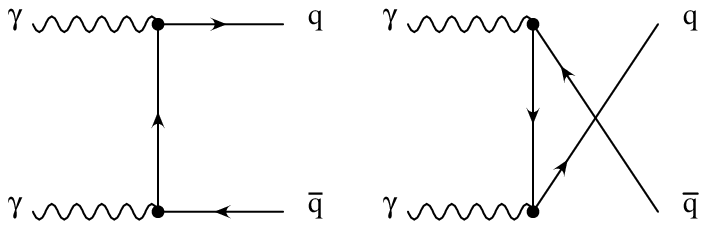,bbllx=60pt,bblly=360pt,bburx=266pt,bbury=431pt,%
          width=0.66\linewidth}
 \caption{\label{fig:07}Born diagrams for direct $\gamma\gamma$
          scattering.}
 \end{center}
\end{figure}
%
%
\begin{figure}
 \begin{center}
  \epsfig{file=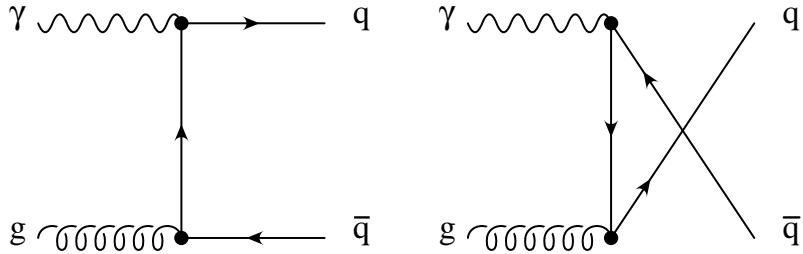,bbllx=60pt,bblly=360pt,bburx=266pt,bbury=431pt,%
          width=0.66\linewidth}
 \caption{\label{fig:08}Born diagrams for single-resolved $\gamma g$
           scattering.}
 \end{center}
\end{figure}
%
%
\begin{figure}
 \begin{center}
  \epsfig{file=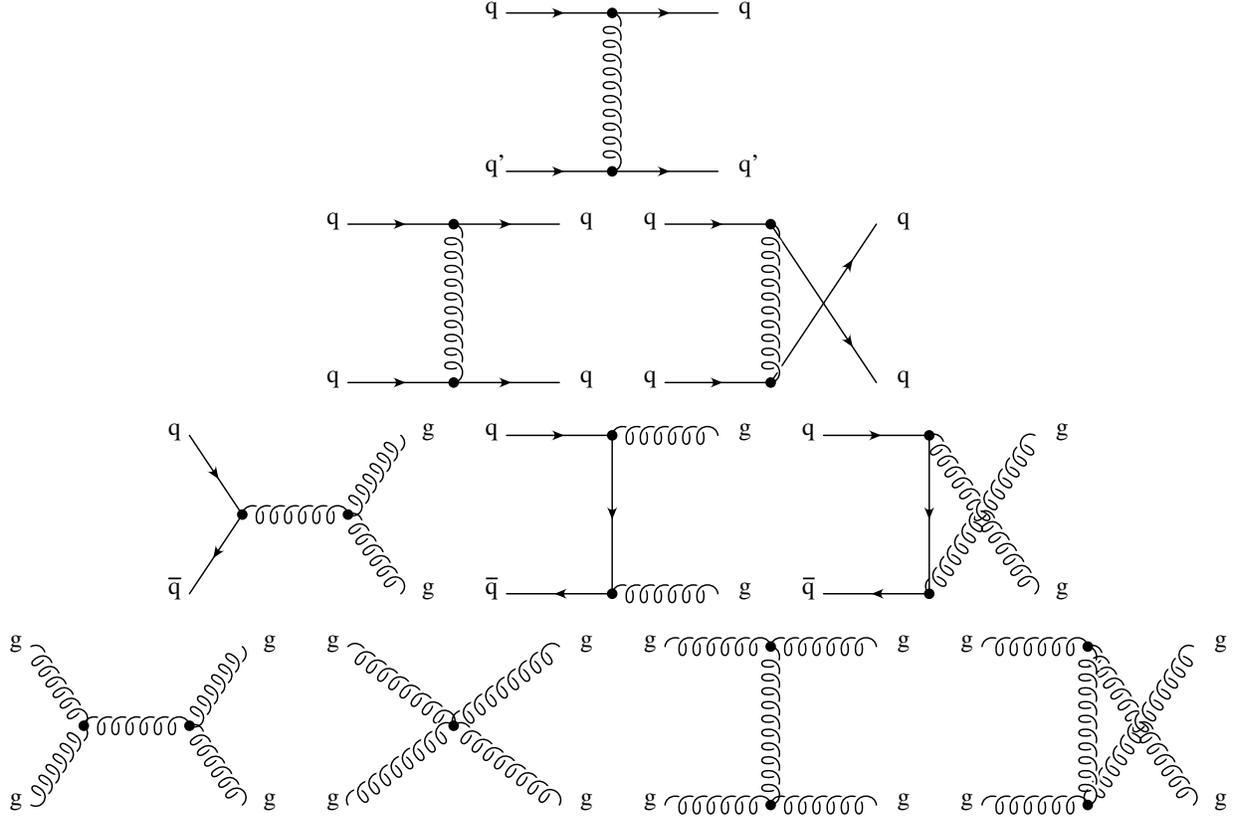,bbllx=60pt,bblly=146pt,bburx=483pt,bbury=431pt,%
          width=\linewidth}
 \caption{\label{fig:09}Born diagrams for double-resolved $qq',qq,q\bar{q}$,
          and $gg$ scattering.}
 \end{center}
\end{figure}
%
The massless Born diagrams for the direct process $\yytqb$, the single-resolved
resolved process $\ygtqb$, and the double-resolved processes $\qptqp,\,\qqtqq,
\,\qbtgg$, and $\ggtgg$ are shown in Figs.\ \ref{fig:07}, \ref{fig:08},
and \ref{fig:09}, and the corresponding squared matrix elements $|\M^B|^2$ are
summarized in Tab.\ \ref{tab:loxsec}. All other Born diagrams and squared
matrix elements can be obtained by crossing particle lines and Mandelstam
variables and including a factor of $(-1)$ for every crossed fermion line.
%
\begin{table}
\begin{center}
\begin{tabular}{|l|l|}
Process & LO matrix element squared $|\M^B|^2$ \\
\hline
$\yytqb$& $8N_C(1-\varepsilon)\le(1-\varepsilon) \lr\frac{u}{t}+
 \frac{t}{u}\rr-2\varepsilon\re$ \\
$\ygtqb$& $C_F|\M^B|^2_{\yytqb}(s,t,u)$ \\
$\qptqp$& $4N_CC_F\lr\frac{s^2+u^2}{t^2}-\varepsilon\rr$ \\
$\qqtqq$& $\le|\M^B|^2_{\qptqp}(s,t,u)+|\M^B|^2_{\qptqp}(s,u,t)\rp$ \\
        & $\lp-8C_F(1-\varepsilon)\lr\frac{s^2}{ut}+\varepsilon\rr\re/2!$ \\
$\qbtgg$& $\le 4C_F(1-\varepsilon)\lr\frac{2N_CC_F}{ut}-\frac{2N_C^2}
 {s^2}\rr(t^2+u^2-\varepsilon s^2)\re /2!$ \\
$\ggtgg$& $\le 32N_C^3C_F(1-\varepsilon)^2\lr 3-\frac{ut}{s^2}-
 \frac{us}{t^2}-\frac{st}{u^2}\rr\re /2!$ \\
\end{tabular}
\end{center}
\caption{\label{tab:loxsec}LO squared matrix elements $|\M^B|^2$ for massless
 $2\rightarrow 2$ parton processes involving two, one, and no photons
 in $n=4-2\eps$ dimensions.}
\end{table}
%

In NLO, the partonic cross section
\beq
 \frac{\d\sigma}{\d t}=\frac{\d\sigma^B}{\d t}
 +\frac{\d\sigma^V}{\d t}
 +\frac{\d\sigma^F}{\d t}
 +\frac{\d\sigma^I}{\d t}
 \label{eq:totalnloxsec}
\eeq
receives additional contributions from virtual loop corrections (V) and
real emission corrections in the final state (F) and initial state (I) with
\bea
 \frac{\d\sigma^{V,F,I}}{\d t}&=&\frac{1}{2s}
 \frac{1}{\Gamma(1-\varepsilon)}
  \left( \frac{4\pi s}{tu} \right) ^\varepsilon
  \frac{1}{8\pi s} \frac{g_{a,b}^2}{S_aS_bC_aC_b} \nonumber \\
 &\times&
 \frac{\alpha_s}{2\pi}
  \left( \frac{4\pi\mu^2}{Q^2} \right) ^\varepsilon
  \frac{\Gamma(1-\varepsilon)}{\Gamma(1-2\varepsilon)}
 |\M^{V,F,I}|^2.
 \label{eq:nloxsec}
\eea

\begin{figure}
 \begin{center}
  \epsfig{file=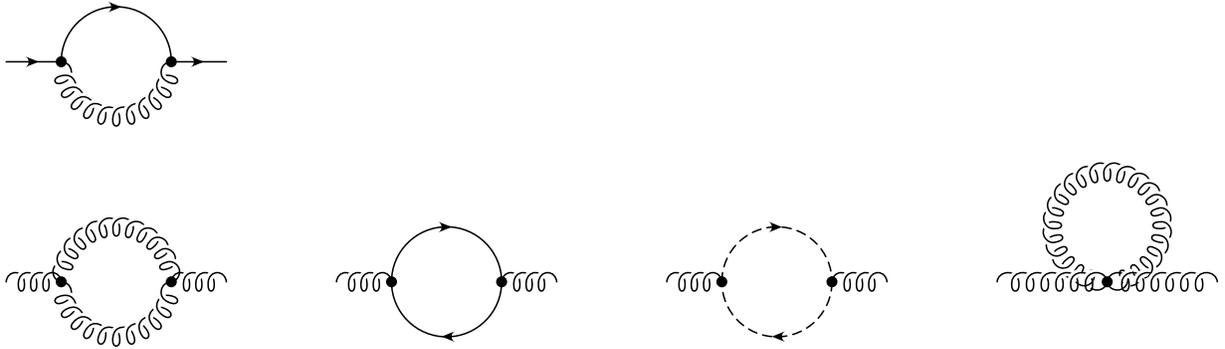,bbllx=66pt,bblly=290pt,bburx=472pt,bbury=418pt,%
          width=\linewidth,clip=}
 \caption{\label{fig:10}QCD quark (top) and gluon (bottom) self-energy
 diagrams. Faddeev-Popov ghosts are depicted as dashed lines.}
 \end{center}
\end{figure}

\begin{figure}
 \begin{center}
  \epsfig{file=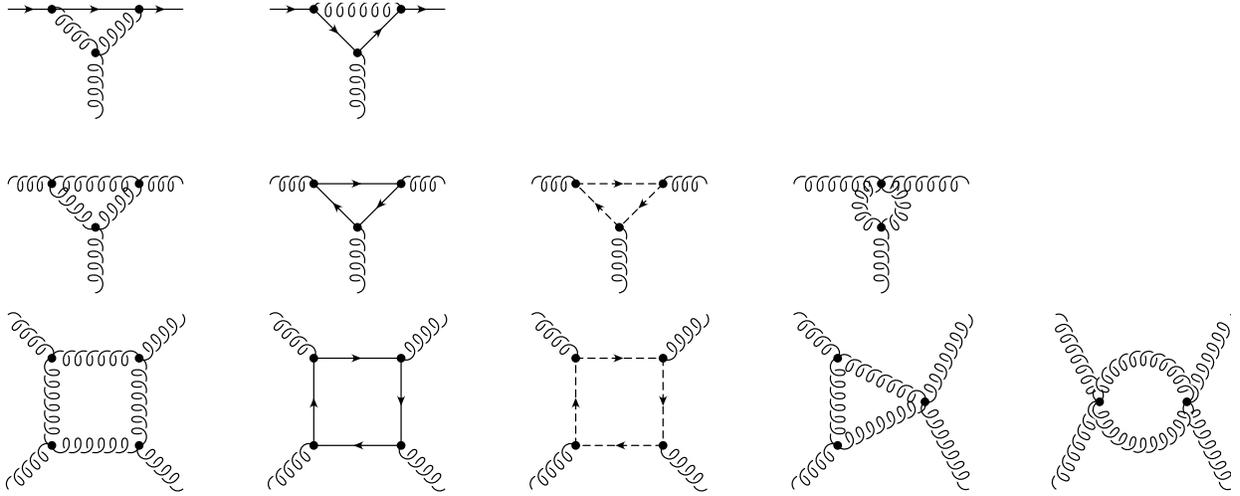,bbllx=66pt,bblly=204pt,bburx=578pt,bbury=418pt,%
          width=\linewidth,clip=}
 \caption{\label{fig:11}QCD corrections to the quark-gluon
 vertex (top), triple-gluon vertex (center), and four-gluon vertex (bottom).}
 \end{center}
\end{figure}

Virtual corrections arise in NLO by dressing all constituents of the LO
diagrams in Figs.\ \ref{fig:07}, \ref{fig:08}, and \ref{fig:09}
with one-loop particle exchanges. Fig.\ \ref{fig:10} shows the
%
%
self-energy diagrams for quarks (top) and gluons (bottom), which are
factorizable and thus independent of the
scattering process and have to be inserted in external and internal quark and
gluon lines. The ghost self-energy is obtained by replacing quark lines with
ghost lines in the upper diagram. While photon self-energies arise also at
one loop, they are of $\O(\alpha)$, not of $\O(\alpha_s)$, and thus not
included in NLO QCD calculations.
Vertex corrections are shown in Fig.\ \ref{fig:11} for the
%
%
quark-gluon (top), triple-gluon (center), and four-gluon vertices (bottom).
For the latter two, additional diagrams are generated by permutations of the
participating particles. The corrections to the ghost-gluon vertex are
of similar type as those for the quark-gluon vertex, and the photon-quark
vertex correction is obtained from the second diagram in the first row
of Fig.\ \ref{fig:11} by replacing the outgoing gluon with a photon.
Vertex corrections depend
on the four-momentum flow at the vertex and thus factorize only at the
amplitude, and not the squared amplitude, level.
A third type of loop diagrams is generated by box diagrams, which depend on
the types and momenta of all parton legs in a $2\to 2$ Born process and
factorize also only at the amplitude level. Typical
examples are the boxes generated by the four-gluon vertex correction in the
last row of Fig.\ \ref{fig:11}.
All loop diagrams with a closed fermion line receive a factor $(-1)$ and those
with identical particles (gluons) a statistical factor of 1/2. $|\M^V|^2$
in Eq.\ (\ref{eq:nloxsec}) denotes the interference terms of Born and
virtual one-loop matrix elements, which have been integrated over the loop
momentum $l$ with a measure $\d^n l/(2\pi)^n$. In this case only the real
parts of the loop integrals have to be kept. In loop diagrams with internal
fermion lines, the integration momentum $l$ appears not only in the propagator
denominator, but also in the tensor-valued numerator. In $2\to 2$ scattering
these integrals can involve between one and four vertices.
Tensor integrals can be reduced to scalar integrals by exploiting Lorentz
invariance and extending the well-known reduction procedure (Passarino and
Veltman, 1979) to $n$ dimensions. The
coefficients of the scalar integrals are then finite and have to be kept up to
$\O(\eps)$ or $\O(\eps^2)$. All divergences are contained in the scalar
integrals. They are of three
different types. Ultraviolet $1/\eps$ poles appear at the upper boundary
($\infty$) of the energy integration in the one- and two-vertex functions. They
are removed,
together with the universal finite terms $-\gamma_E+\ln(4\pi)$, by
renormalizing the bare strong coupling $\hat{g}_s$, wave functions, and masses
$\hat{m}_i$ (where applicable) in the $\ms$ scheme
(Bardeen {\it et al.}, 1978). Infrared
poles arise at the lower boundary (0) of the energy integration, when
massless particles are exchanged between real particles, and collinear poles
occur in the splitting of massless particles into two massless collinear
particles. Infrared and collinear poles and double poles appear in the two-,
three-, and four-vertex functions and in the derivatives of two-point
functions, which play a role in external self-energy corrections. These poles
are particularly cumbersome
in massless QCD calculations. In practice, the scalar integrals
can be calculated either with Feynman parameters ('t Hooft and Veltman, 1979)
or with Cutkosky cutting rules and dispersion relations ('t Hooft and Veltman,
1973). Analytical continuation of products of
logarithms can lead to numerically large terms of $\pi^2$.
The complete one-loop QCD corrections have been calculated in dimensional
regularization for massless parton scattering processes involving two
(Aurenche {\it et al.}, 1984a), one (Aurenche {\it et al.}, 1987), or no
photons (Ellis and Sexton, 1986).

\begin{figure}
 \begin{center}
  \epsfig{file=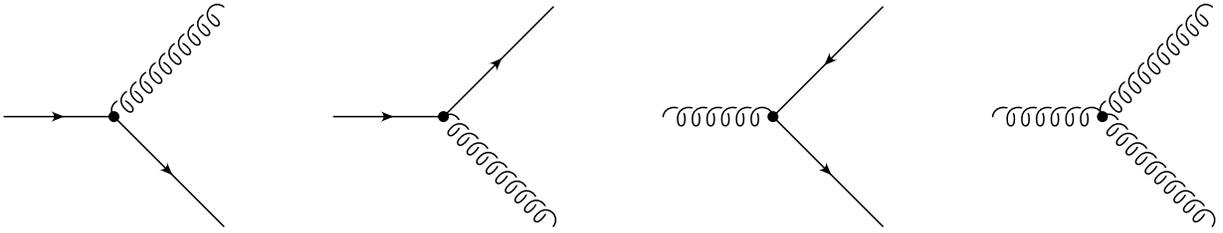,bbllx=66pt,bblly=274pt,bburx=473pt,bbury=367pt,%
          width=\linewidth,clip=}
 \caption{\label{fig:12}Real QCD corrections through $1\to 2$ parton
 splitting.}
 \end{center}
\end{figure}

Real corrections arise in NLO by the splitting of one particle into two as
shown in Fig.\ \ref{fig:12} for massless QCD partons.
%
%
These diagrams have to be attached at all possible places to the LO diagrams
in Figs.\ \ref{fig:07}, \ref{fig:08}, and \ref{fig:09} and
lead to $2\to 3$ processes with different kinematics.
The first three diagrams could also be applied to photons, but they would
constitute $\O(\alpha)$ corrections and should not be included at
$\O(\alpha_s)$. Only the $\gamma\to q\bar{q}$ diagram plays a special
role, since it provides a link at NLO between direct and resolved
photoproduction (see Sec.\ \ref{sec:phostr}).
Analytical expressions for the massless $2\to 3$ scattering matrix elements
have been computed in $n$ dimensions for photon-photon (Aurenche {\it et al.},
1984a), photon-parton (Aurenche {\it et al.}, 1987), and parton-parton (Ellis
and Sexton, 1986) scattering.

For single-particle production (see Secs.\ \ref{sec:hadprod} and
\ref{sec:promptyprod}), it is possible to
integrate the squared $2\to 3$ matrix elements $|\M^{F,I}|^2$ analytically
over the solid angle
\beq
 \int\d\Omega=\int_0^\pi\d\theta_1\sin^{1-2\eps}\theta_1\int_0^\pi
 \d\theta_2\sin^{-2\eps}\theta_2
 \label{eq:solidangle}
\eeq
of the two unresolved final (F) or initial (I) state particles in a
frame, where they have no net three-momentum. However, the integration over
the invariant mass $s_{ij}=(p_i+p_j)^2$ of the two unobserved partons with
four-momenta $p_i$ and $p_j$ has to be performed
numerically (Ellis {\it et al.}, 1980). These integrations are
implied in the cross section expression of
Eq.\ (\ref{eq:nloxsec}). The method described above can be applied to
direct (Aurenche {\it et al.}, 1994a), single- (Aurenche, Fontannaz, and
Guillet, 1994b) and double-resolved jet photoproduction (Aversa {\it et
al.}, 1989, 1990), if the phase space in Eq.\ (\ref{eq:solidangle})
is restricted to allow for an observed jet with a small cone of angular
size $\delta$ similar to the Sterman-Weinberg criterion.
For comparisons with experimental data, finite cone size
corrections have to be implemented by integrating the $2\to 3$
matrix elements numerically in the finite region between the small technical
cone of size $\delta$ and the larger experimental cone of size $R$
(Gordon and Storrow, 1992b; Greco and Vicini, 1994; Aversa {\it et al.}, 1991;
Kramer and Salesch, 1994a; Aurenche {\it et al.}, 1994a; B\"odeker, Kramer,
and Salesch, 1994; Aurenche, Fontannaz, and Guillet, 1994b; Klasen, Kramer,
and Salesch, 1995). The double-resolved contributions in these calculations
were all based on the same results by Aversa {\it et al.} (1989, 1990).
With the exception of the first two calculations, they were found to agree
numerically with later
calculations using different methods (Klasen, 1996b; Aurenche {\it et al.},
2000). Similar methods have been applied quite early in single-resolved
calculations (Baer, Ohnemus, and Owens, 1989b; Gordon and Storrow, 1992b;
B\"odeker, 1992a, 1992b, 1993) and will be described in detail in
Sec.\ \ref{sec:dijets}.

If the radiated gluons in Fig.\ \ref{fig:12} become soft and/or two of
the three particles become collinear, singularities appear
in the propagators of the intermediate particles and have to be extracted
analytically as $1/\eps$ or $1/\eps^2$ (double-)poles.
The interference of different amplitudes can furthermore
involve multiple singularities that have to be separated by partial
fractioning. According to the KLN theorem,
all soft and final state collinear singularities from the real corrections
have to cancel against corresponding virtual singularities in inclusive
jet cross sections. Only universal collinear singularities
\beq
 |\M^I|^2_{ab\rightarrow 123} = -\ede P_{c\leftarrow a}(x)
  |\M^B|^2_{cb\rightarrow 12}+\O(\eps^0)
\label{eq:photonsing}
\eeq
remain, which correspond to the parton splittings in Fig.\ \ref{fig:12}
attached to the initial state of the Born diagrams. They are proportional to
the Altarelli-Parisi splitting functions in Eq.\ (\ref{eq:altparsplit})
and are absorbed, together with scheme-dependent finite terms, into the
parton densities in the proton or in the photon (see Sec.\ \ref{sec:phostr}).
For the production of jets in direct, single-resolved, and double-resolved
photon-photon collisions,
the soft and collinear singular parts of the NLO squared matrix elements
$|\M^{V,F,I}|^2$ are collected in Tabs.\ \ref{tab:irdir}, \ref{tab:irsres},
and \ref{tab:irdres}. In all three cases the sums are finite, since the
factorizable initial state singularities have already been subtracted.
The arguments of the logarithms $l(x)=\ln|x/Q^2|$ have been normalized to the
arbitrary scale $Q^2$, which appears also in Eq.\ (\ref{eq:nloxsec}) and
drops out in the total result.

%
\begin{table}
\begin{center}
\begin{tabular}{|l|l|l|l|}
\hline
 Process  & Color  & Correc-    & Singular Parts of  \\
          & Factor & tion       & $|{\cal M}^{V,F,I}|^2/|{\cal M}^B|^2$\\
\hline
\hline
 $\yytqb$ & $C_F$        & Virtual  & $\le-\frac{2}{\eps^2}-\frac{1}
  {\eps}(3-2l(t))\re  $\\
          &              & Final   & $\le+\frac{2}{\eps^2}+\frac{1}
  {\eps}(3-2l(t))\re  $\\
\hline
\end{tabular}
\end{center}
\caption{\label{tab:irdir}
 Cancellation of soft and collinear singularities from virtual and final
 state NLO corrections for direct photon scattering.}
\end{table}
%
%
\begin{table}[htbp]
\begin{center}
\begin{tabular}{|l|l|l|l|}
\hline
 Process  & Color & Correc-    & Singular Parts of  \\
          & Factor& tion       & $|{\cal M}^{V,F,I}|^2/|{\cal M}^B|^2$\\
\hline
\hline
 $\yqtgq$ & $C_F$        & Virtual  & $\le-\frac{2}{\eps^2}-\frac{1}
  {\eps}(3-2l(t))\re  $\\
          &              & Final     & $\le+\frac{1}{\eps^2}+\frac{1}
  {2\eps}(3-2l(t))\re  $\\
          &              & Initial   & $\le+\frac{1}{\eps^2}+\frac{1}
  {2\eps}(3-2l(t))\re  $\\
\cline{2-4}
          & $N_C$        & Virtual   & $\le-\frac{1}{\eps^2}-\frac{1}
  {2\eps}\lr\frac{11}{3}-2l(s/t)-2l(u)\rr\re  $\\
          &              & Final     & $\le+\frac{1}{\eps^2}+\frac{1}
  {2\eps}\lr\frac{11}{3}-l(s/t)-l(u)\rr\re  $\\
          &              & Initial   & $\le     +\frac{1}
  {2\eps}\lr-l(s/t)-l(u)\rr\re  $\\
\cline{2-4}
          & $N_f$        & Virtual   & $+\frac{1}{3\eps} 
  $\\
          &              & Final     & $-\frac{1}{3\eps} 
  $\\
\hline
\end{tabular}
\end{center}
\caption{\label{tab:irsres}Cancellation of soft and collinear singularities
 from virtual, final state, and initial state NLO corrections for
 single-resolved photon scattering and different color factors.}
\end{table}
%
%
\begin{table}
\begin{center}
\begin{tabular}{|l|l|l|l|}
\hline
 Process  & Color & Correc- & Singular Parts of  \\
          & Factor& tion    & $|{\cal M}^{V,F,I}|^2/|{\cal M}^B|^2$ \\
\hline
\hline
 $\qptqp$ & $C_F$        & Virtual   & $\le-\frac{4}{\eps^2}-\frac{1}
  {\eps}(6+8l(s/u)-4l(t))\re $\\
          &              & Final     & $\le+\frac{2}{\eps^2}+\frac{1}
  {\eps}(3+4l(s/u)-2l(t))\re  $\\
          &              & Initial   & $\le+\frac{2}{\eps^2}+\frac{1}
  {\eps}(3+4l(s/u)-2l(t))\re  $\\
\cline{2-4}
          & $N_C$        & Virtual   & $\le+\frac{1}{\eps}(4l(s)-2l(u)
  -2l(t))\re  $\\
          &              & Final     & $\le-\frac{1}{\eps}(2l(s)-~l(u)
  -~l(t))\re  $\\
          &              & Initial   & $\le-\frac{1}{\eps}(2l(s)-~l(u)
  -~l(t))\re  $\\
\hline
 $\qqtqq$ & $C_F$        & Virtual   & $\le-\frac{4}{\eps^2}-\frac{1}
  {\eps}(6+4l(s/t)-4l(u))\re  $\\
          &              & Final     & $\le+\frac{2}{\eps^2}+\frac{1}
  {\eps}(3+2l(s/t)-2l(u))\re  $\\
          &              & Initial   & $\le+\frac{2}{\eps^2}+\frac{1}
  {\eps}(3+2l(s/t)-2l(u))\re  $\\
\cline{2-4}
          & $N_C$        & Virtual   & $\le+\frac{2}{\eps}(2l(s)-l(t)
  -l(u))\re  $\\
          &              & Final     & $\le-\frac{1}{\eps}(2l(s)-l(t)
  -l(u))\re  $\\
          &              & Initial   & $\le-\frac{1}{\eps}(2l(s)-l(t)
  -l(u))\re  $\\
\hline
 $\qbtgg$ & $C_F$        & Virtual   & $\le-\frac{2}{\eps^2}-\frac{3}
  {\eps}\re  $\\
          &              & Initial   & $\le+\frac{2}{\eps^2}+\frac{3}
  {\eps}\re  $\\
\cline{2-4}
          & $N_C$        & Virtual   & $\le-\frac{2}{\eps^2}-\frac{11}
  {3\eps}\re  $\\
          &              & Final     & $\le+\frac{2}{\eps^2}+\frac{11}
  {3\eps}\re  $\\
\cline{2-4}
          & $N_f$        & Virtual   & $+\frac{2}{3\eps} 
  $\\
          &              & Final     & $-\frac{1}{3\eps} 
  $\\
          &              & Initial   & $-\frac{1}{3\eps} 
  $\\
\cline{2-4}
          & $1$          & Virtual   & $+\frac{1}{\eps}\;l(s)\lr\lr 4N_C^3
  C_F\!+\!\frac{4C_F}{N_C}\rr\!\frac{t^2+u^2}{tu}\!\rp    
     \lp-16N_C^2C_F^2\frac{t^2+u^2}{s^2}\rr/|{\cal M}^B|^2
  $ \\
          &              & Final     & $-\frac{1}{2\eps}l(s)\lr\lr 4N_C^3
  C_F\!+\!\frac{4C_F}{N_C}\rr\!\frac{t^2+u^2}{tu}\!\rp    
     \lp-16N_C^2C_F^2\frac{t^2+u^2}{s^2}\rr/|{\cal M}^B|^2
  $ \\
          &              & Initial   & $-\frac{1}{2\eps}l(s)\lr\lr 4N_C^3
  C_F\!+\!\frac{4C_F}{N_C}\rr\!\frac{t^2+u^2}{tu}\!\rp    
     \lp-16N_C^2C_F^2\frac{t^2+u^2}{s^2}\rr/|{\cal M}^B|^2
  $ \\
\cline{2-4}
          & $8N_C^3C_F$  & Virtual   & $\frac{1}{\eps}\; l(t)\lr\utx
  \!-\!\frac{2u^2}{s^2}\rr/|{\cal M}^B|^2\!+\!t\!\leftrightarrow\! u $\\
          &              & Final     & $\frac{-1}{2\eps} l(t)\lr\utx
  \!-\!\frac{2u^2}{s^2}\rr/|{\cal M}^B|^2\!+\!t\!\leftrightarrow\! u $\\
          &              & Initial   & $\frac{-1}{2\eps} l(t)\lr\utx
  \!-\!\frac{2u^2}{s^2}\rr/|{\cal M}^B|^2\!+\!t\!\leftrightarrow\! u $\\
\cline{2-4}
          & $8N_CC_F$    & Virtual   & $-\frac{1}{\eps}\lr\utx\!+\!\tux\rr
  (l(t)\!+\!l(u))/|{\cal M}^B|^2 $\\
          &              & Final     & $+\frac{1}{2\eps}\lr\utx\!+\!\tux\rr
  (l(t)\!+\!l(u))/|{\cal M}^B|^2 $\\
          &              & Initial   & $+\frac{1}{2\eps}\lr\utx\!+\!\tux\rr
  (l(t)\!+\!l(u))/|{\cal M}^B|^2 $\\
\hline
 $\ggtgg$ & $N_C$        & Virtual   & $\le-\frac{4}{\eps^2}-\frac{22}
  {3\eps}\re  $\\
          &              & Final     & $\le+\frac{2}{\eps^2}+\frac{11}
  {3\eps}\re  $\\
          &              & Initial   & $\le+\frac{2}{\eps^2}+\frac{11}
  {3\eps}\re  $\\
\cline{2-4}
          & $N_f$        & Virtual   & $+\frac{4}{3\eps}
  $\\
          &              & Final     & $-\frac{2}{3\eps}
  $\\
          &              & Initial   & $-\frac{2}{3\eps}
  $\\
\cline{2-4}
          & $32N_C^4C_F$ & Virtual   & $+\frac{1}{\eps} l(s) \lr 3
  -2\tus+\frac{t^4+u^4}{t^2u^2}\rr\!\!+{\rm\! cyc.} $\\
          &              & Final     & $-\frac{1}{2\eps}\!\! l(s) \lr 3
  -2\tus+\frac{t^4+u^4}{t^2u^2}\rr\!\!+{\rm\! cyc.} $\\
          &              & Initial   & $-\frac{1}{2\eps}\!\! l(s) \lr 3
  -2\tus+\frac{t^4+u^4}{t^2u^2}\rr\!\!+{\rm\! cyc.} $\\
\hline
\end{tabular}
\end{center}
\caption{\label{tab:irdres}Cancellation of soft and collinear singularities
 from virtual, final state, and initial state NLO corrections for
 double-resolved photon scattering and different color factors.}
\end{table}
%

The photoproduction cross section for inclusive jets
\bea
  \frac{\mbox{d}^2\sigma}{\mbox{d}E_T\mbox{d}\eta}
  &=& \sum_{a,b} \int_{x_{a,\min}}^1 \mbox{d}x_a x_a 
  f_{a/A}(x_a,M_a^2) x_b f_{b/B}(x_b,M_b^2)\nonumber\\
  &\times&\frac{4E_AE_T}{2x_aE_A-E_Te^{\eta}}
  \frac{\mbox{d}\sigma}{\mbox{d}t}
\label{eq:1jet}
\eea
depends on the transverse energy $E_T$ and pseudorapidity $\eta$ of the
observed
jet. In Eq.\ (\ref{eq:1jet}), $\eta$ is defined in the laboratory frame with
particle $A$ travelling in the positive $z$-direction. It is related to the
pseudorapidity in the hadronic center-of-mass frame
$\eta^\ast=\eta+1/2\ln(E_B/E_A)$ by a simple boost.
From the final state variables $E_T$ and
$\eta$ and from the energies $E_A$ and $E_B$
of the initial leptons or hadrons, the four-momenta of the incoming partons
$p_{a,b}=E_{A,B}(x_{a,b},0,0,\pm x_{a,b})$ cannot be fully reconstructed.
Instead the partonic
cross section $\d\sigma/\d t$ is integrated over $x_a$ with
\bea
  x_{a,\min} &=& \frac{E_BE_Te^{\eta}}{2E_AE_B-E_AE_Te^{-\eta}},\nonumber\\
  x_b        &=& \frac{x_aE_AE_Te^{-\eta}}{2x_aE_AE_B-E_BE_Te^{\eta}},
\eea
and convolved with the parton densities $f_{a/A}(x_a,M_a^2)$
and $f_{b/B}(x_b,M_b^2)$ at the factorization scales $M_{a,b}$.
The density of partons $i$ in initial leptons $l$ is given by the convolution
\beq
 f_{i/l}(x,M^2) = [f_{i/\gamma}(M^2)\otimes
 f_{\gamma/l}](x)
\label{eq:partonflux}
\eeq
of the photon spectrum $f_{\gamma/l}(x)$ as described in Sec.\
\ref{sec:phospec} with the parton densities in the photon $f_{i/\gamma}(x,M^2)$
(see Sec.\ \ref{sec:phostr}), so that the two can also not be uniquely
disentangled.

\begin{figure}
 \begin{center}
  \epsfig{file=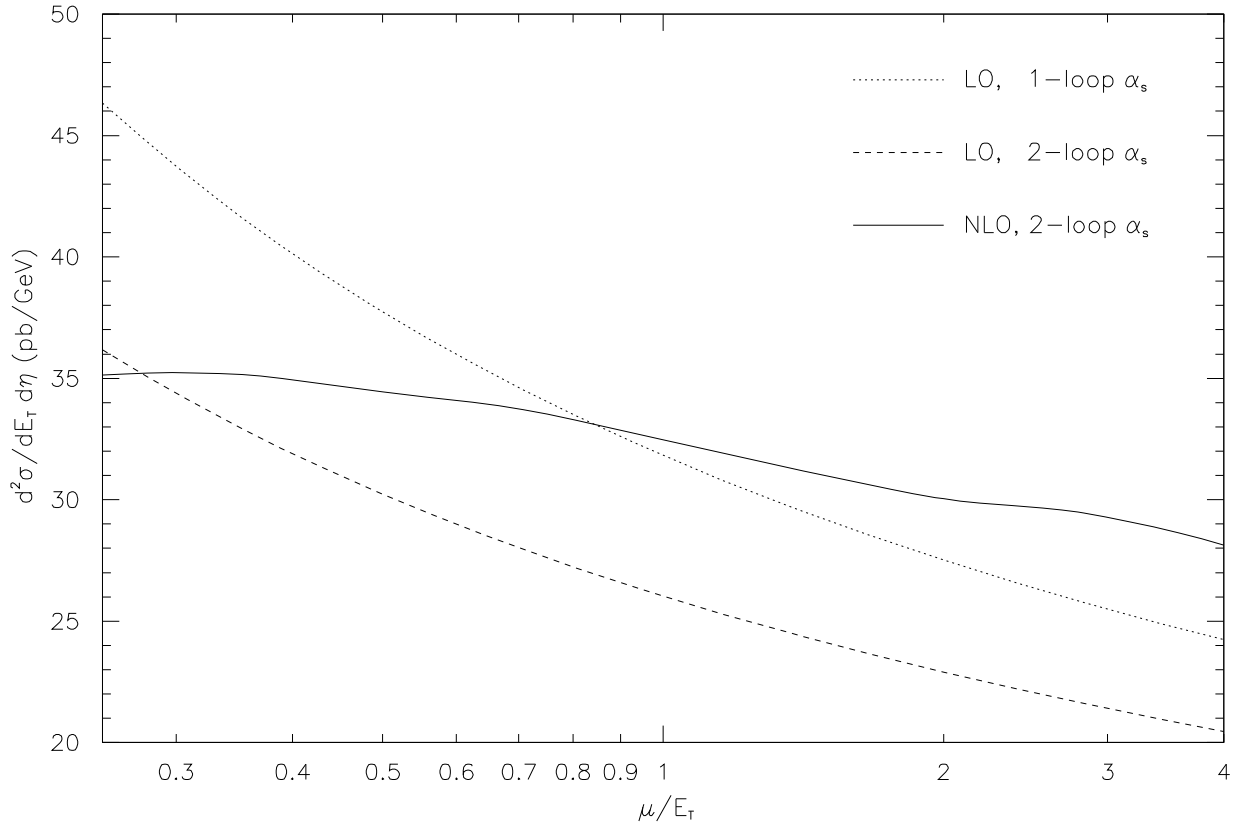,width=\linewidth}
 \caption{\label{fig:13}Renormalization scale dependence of the
 single-jet cross section d$^2\sigma/$d$E_T$d$\eta$ at HERA with $E_T=20$~GeV
 and $\eta=1$ in LO (dotted and dashed curves) and NLO (full curve).}
 \end{center}
\end{figure}

Due to the truncation of the perturbative series, fixed order predictions
depend on the scales, at which the strong coupling $\alpha_s(\mu)$ is
renormalized and at which the collinear initial state singularities are
factorized into the parton densities $f_{i/a}(x, M^2)$.
In LO, the logarithmic scale
dependence is generally quite strong.
It is reduced in NLO through the explicit logarithmic dependences in
the virtual and real initial state corrections
\bea
 |{\cal M}^V|^2_{ab\rightarrow 12\,} &=& \ln\!\lr{~\mu^2\,\over Q^2}\rr
 |{\cal M}^B|^2_{ab\rightarrow 12} \,\beta_0 \hspace*{9mm} \!\!+ ...\,, \\
 |\M^I|^2_{ab\rightarrow 123} &=&
 \ln\!\lr\frac{M^2}{Q^2}\rr
 |\M^B|^2_{cb\rightarrow 12} \,P_{c\leftarrow a}(x) \!\!+ ...\,.
\eea
For the renormalization scale $\mu$, this is demonstrated in Fig.\
\ref{fig:13}, where the total single-jet photoproduction cross section
Eq.\ (\ref{eq:1jet}) has been calculated using HERA kinematics ($E_p=820$ GeV
and $E_e=26.7$ GeV). Unless stated otherwise, NLO parton densities in the
$\ms$ scheme for the
%
%
proton (Lai {\it et al.}, 1995) and photon (Gl\"uck, Reya, and Vogt, 1992b)
have been used with $\Lambda_{\ms}^{N_f=4}=239$ MeV.
For a particular choice of $\mu$, the NLO corrections vanish
and the perturbative series converges apparently very fast (Grunberg, 1980).
If a one-loop $\alpha_s$ formula is used in the LO calculation, this happens
at $\mu\simeq M_p=M_\gamma=E_T$, while for two-loop $\alpha_s$ it happens
at $\mu\simeq E_T/4$. At this scale, the NLO curve exhibits also a local maximum and
has minimal sensitivity to scale variations (Stevenson, 1981a, 1981b).
However, the dependences on $M_p$ and $M_\gamma$ also have to be considered and
can spoil this simple picture. If all scales are varied simultaneously, the
Principles of Fastest Convergence and Minimal Sensitivity lead indeed to
unreasonably low scale choices (Klasen, 1996b). The situation becomes even more
complicated, if more than one scale is involved, as is the case in heavy quark
or DIS jet production.

\begin{figure}
 \begin{center}
  \epsfig{file=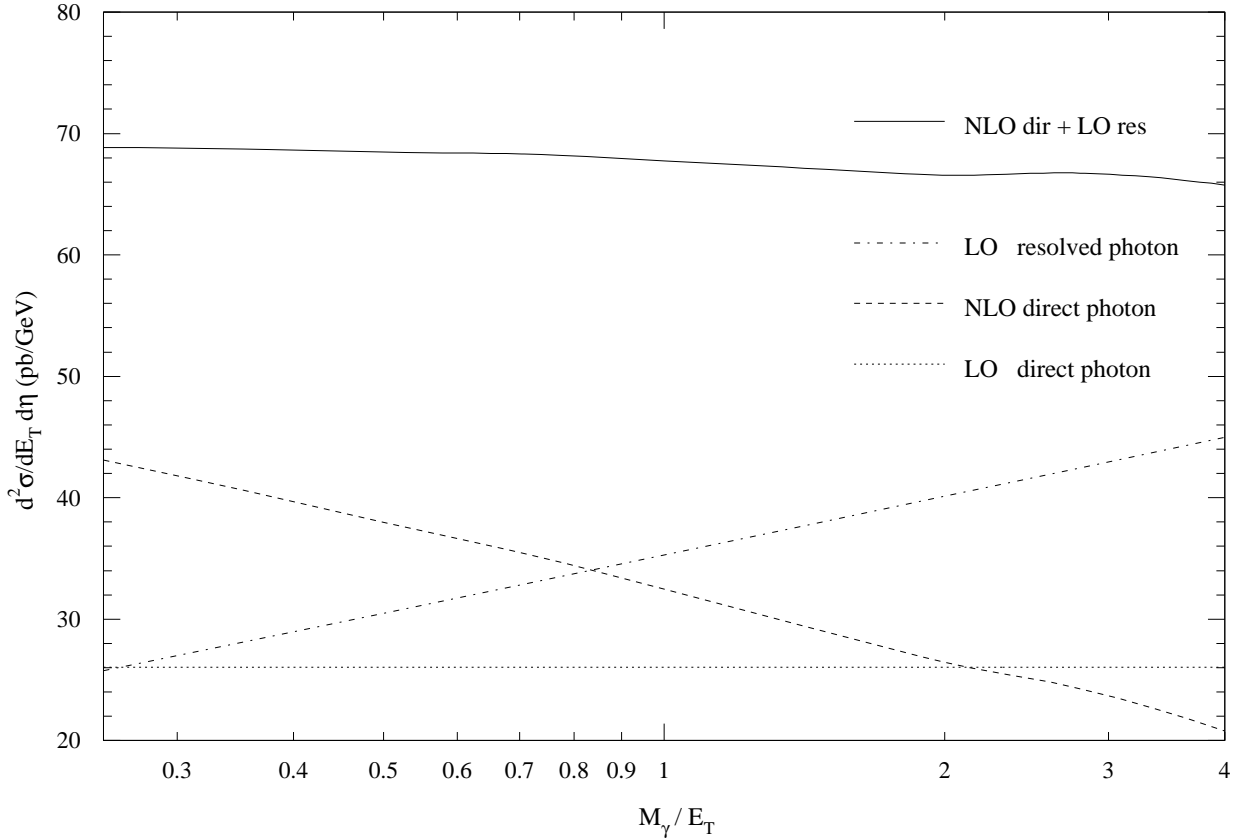,width=\linewidth}
 \caption{\label{fig:14}Photon factorization scale dependence of the
 single-jet cross section d$^2\sigma/$d$E_T$d$\eta$ at HERA with
 $E_T=20$~GeV and
 $\eta=1$. The LO direct curve (dotted) is independent of $M_\gamma$ as is
 the sum (full curve) of NLO direct (dashed) and LO resolved (dot-dashed)
 contributions.}
 \end{center}
\end{figure}

In photoproduction, the dependence on the photon factorization scale $M_\gamma$
is of particular interest, since
this scale defines the separation of direct and resolved processes.
While the direct LO cross section is manifestly independent of $M_\gamma$, the
resolved LO cross section depends logarithmically on $M_\gamma$ through the
LO parton densities in the photon. This dependence is canceled almost exactly
by the explicit logarithmic dependence of the NLO direct term, as can be
seen in Fig.\ \ref{fig:14} (B\"odeker, Kramer, and Salesch, 1994;
Klasen, 1996b; Klasen, Kleinwort, and Kramer, 1998).
%
%
However, the resolved cross section rises faster in NLO than in LO
(Klasen, 1996b). This reintroduces a (weaker) dependence in the total NLO
result, which would eventually be canceled by the direct NNLO contribution.

\begin{figure}
 \begin{center}
  \epsfig{file=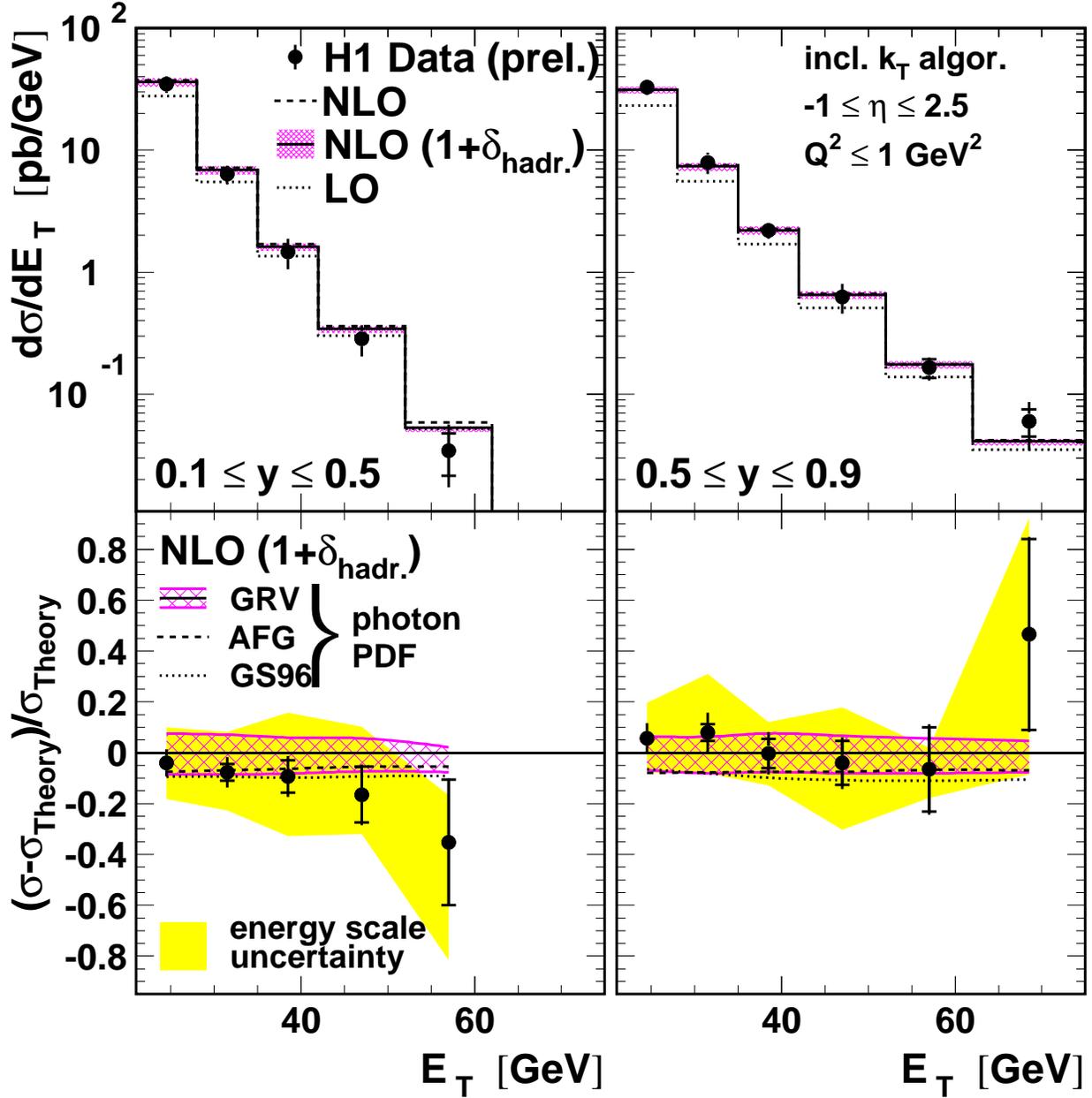,width=\linewidth}
 \caption{\label{fig:15}Transverse energy dependence of the total
 single-jet photoproduction cross section compared to preliminary H1 data
 (Adloff {\it et al.}, 2001a).}
 \end{center}
\end{figure}

\begin{figure}
 \begin{center}
  \epsfig{file=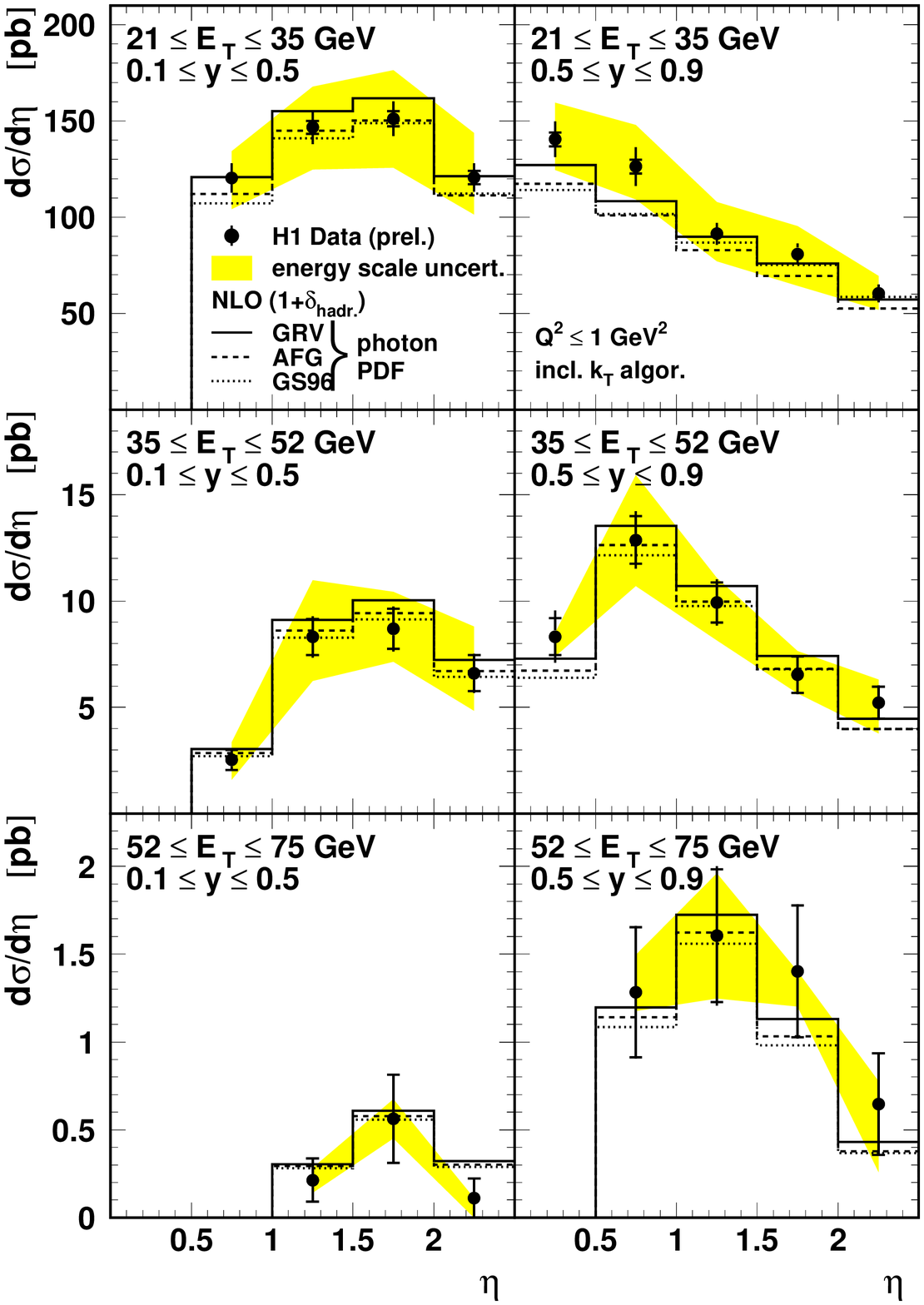,width=0.9\linewidth}
 \caption{\label{fig:16}Pseudorapidity dependence of the total
 single-jet
 photoproduction cross section compared to preliminary H1 data (Adloff
 {\it et al.}, 2001a).}
 \end{center}
\end{figure}

The HERA experiments H1 and ZEUS select photoproduction events by tagging and
anti-tagging the scattered electron, respectively, which limits the maximal
squared virtuality of the exchanged photon to $Q^2<0.01...1$ and $1...4$
GeV$^2$. The energy fraction $y$ transferred to the photon is determined from
the scattered electron energy $E_e'$ with $y=1-E_e'/E_e$ or with the
Jacquet-Blondel method
\beq
 y={1\over 2 E_e}\sum_i \lr E_i-p_{z_i} \rr
 \label{eq:jacblon}
\eeq
from the energies $E_i$ and longitudinal momenta $p_{z_i}$
of the final hadrons (Amaldi {\it et al.}, 1979). Single-jet production
has been analyzed with the cone algorithm and $R=1$ by H1 (Abt {\it et al.},
1993; Aid
{\it et al.}, 1996a) and ZEUS (Derrick {\it et al.}, 1995a; Breitweg {\it et
al.}, 1998a) and has been compared to various NLO calculations (Aurenche,
Fontannaz, Guillet, 1994b; Klasen, Kramer, and Salesch,
1995; Klasen, 1996a, 1996b, 1997a; Klasen, Kleinwort, and Kramer, 1998;
Harris and Owens, 1997, 1998). In these early comparisons at rather low
transverse energies of $\sim 10$~GeV, the theory overestimated the data in the
backward region, since no hadronization corrections had been applied, while
they underestimated the data in the forward region, triggering speculations
about additional soft interactions between the hard jets and the proton
remnant.
In more recent analyses (Breitweg {\it et al.}, 1998b; Adloff {\it et al.},
2001a), the invariant $k_T$ algorithm has been used. The measured H1 cross
section in Fig.\ \ref{fig:15} now extends out to $E_T<75$ GeV for large $y$.
%
%
It falls by three orders of magnitude and agrees
remarkably well with the NLO prediction (Frixione and Ridolfi, 1997). The
hadronization corrections $\delta_{\rm hadr.}$ are much smaller than the NLO
corrections, and the remaining theoretical uncertainty from varying the scale
$\mu=M_p=M_\gamma=E_T/2$ by a factor of two up and down is modest.
Different photon densities give slightly different normalizations, but the
experimental uncertainty is still too large to draw definite conclusions.
Single-jet pseudorapidity distributions are shown in Fig.\ \ref{fig:16}.
%
%
The maximum of the cross section is shifted in the electron direction (low
$\eta$) when increasing $y$ and lowering $E_T$. The H1 data are in general well
described by the NLO prediction, perhaps best with GRV photon densities, but
the experimental and theoretical (not shown) uncertainties are still quite
large. Direct photoproduction dominates at large $E_T$ and small
$\eta$, while resolved processes are more important at smaller $E_T$
and larger $\eta$.

\begin{figure}
 \begin{center}
  \epsfig{file=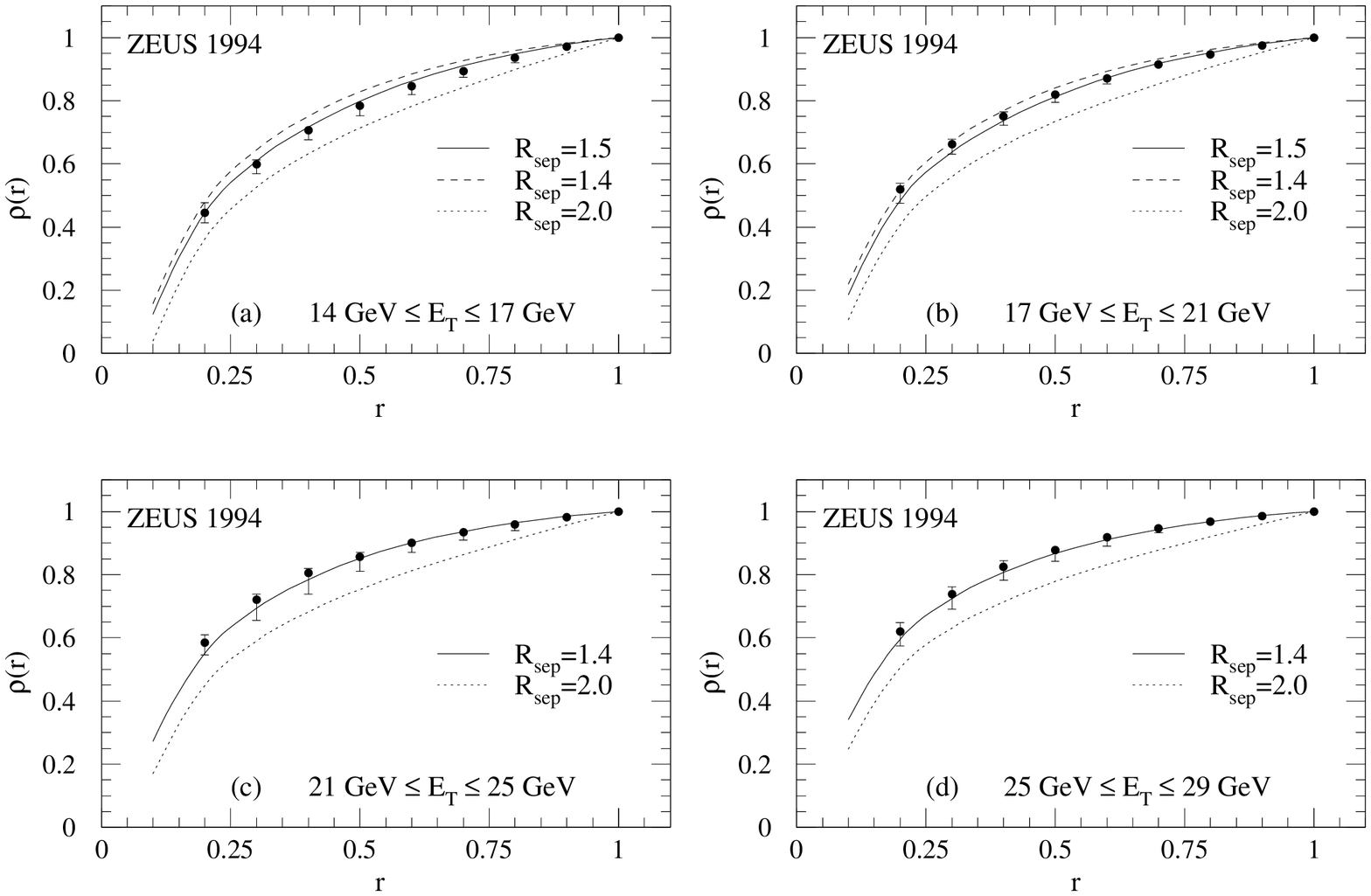,width=0.9\linewidth}
  \epsfig{file=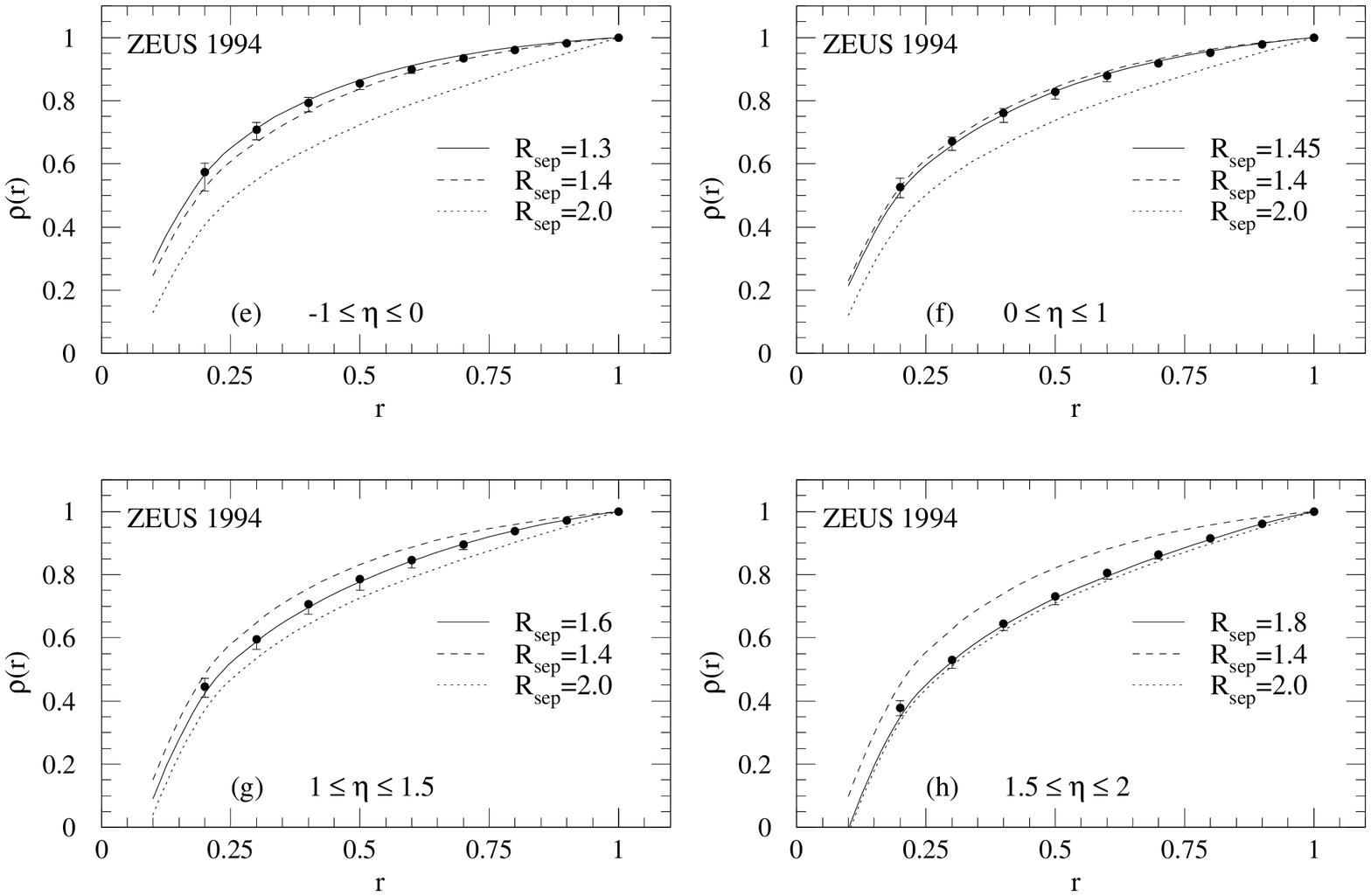,width=0.9\linewidth}
 \caption{\label{fig:17}Jet shapes $\rho(r)$ for
          single-jet photoproduction integrated over $-1<\eta<2$ and four
          different regions of $E_T$ (a)-(d) and integrated over $E_T>14$ GeV
          and four different regions of $\eta$ (e)-(h).
          1994 data from ZEUS (Breitweg {\it et al.}, 1998c)
          are compared to NLO results using the
          cone algorithm with $R=1$ and different values of
          $R_{\rm sep}$.}
 \end{center}
\end{figure}

A good measure for the width of a produced jet is the jet shape
\beq
 \rho(r\leq R) = 1-\frac{\int\d E_T\d\eta \,E_T\,\d^2\sigma^{\rm NLO}/(\d E_T\d\eta)}
                               {\int\d E_T\d\eta \,E_T\,\d^2\sigma^{\rm ~LO}/(\d E_T\d\eta)},
\eeq
which is the average fraction of a jet's transverse energy that lies inside
a concentric inner cone with radius $r\leq R$. It has been calculated in
resolved
(Kramer and Salesch, 1993, 1994b; Klasen and Kramer, 1997b; Klasen, 1997a) and
direct (Klasen and Kramer, 1997b; Klasen, 1997a) photoproduction by integrating
the $E_T$-weighted NLO $2\rightarrow 3$
cross section in the non-singular region between $r$ and $R$. In Fig.\
\ref{fig:17} jet shapes
for single-jet photoproduction are compared to ZEUS data
(Breitweg {\it et al.}, 1998c) for different
transverse energy and pseudorapidity intervals.
%
%
It is obvious that the data and the NLO predictions rise more steeply for large
values of $E_T$ and small values of $\eta$, where the jets become narrower.
Using LO Monte Carlo predictions, the ZEUS collaboration has recently been able
to associate thick and thin jets and with gluon and quark jets (Breitweg {\it
et al.}, 2000a).

\begin{figure}
 \begin{center}
  \epsfig{file=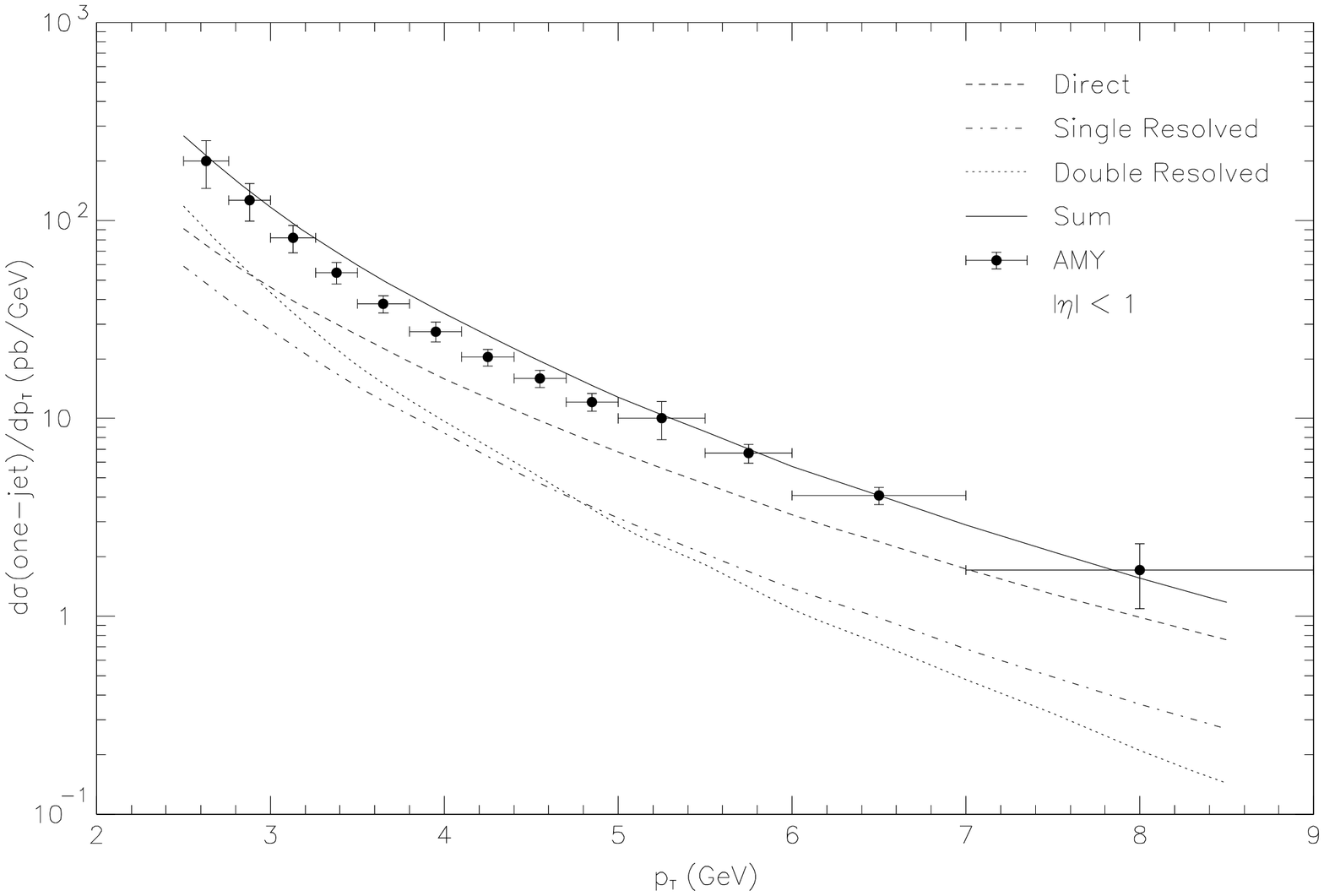,bbllx=105pt,bblly=85pt,bburx=518pt,bbury=716pt,%
          height=\linewidth,angle=270}
 \caption{\label{fig:18}Transverse momentum distribution of single jets
 produced in photon-photon collisions compared to AMY data (Kleinwort and
 Kramer, 1996a).}
 \end{center}
\end{figure}

Inclusive jet production in photon-photon scattering has been measured at the
$e^+e^-$ collider TRISTAN $(\sqrt{S}=58$ GeV) by the TOPAZ (Hayashii {\it et
al.}, 1993) and AMY collaborations (Kim {\it et al.}, 1994), where scattered
electrons were anti-tagged at small angles ($\theta_{\max}=$ 227 and 56 mrad).
OPAL (Ackerstaff {\it et al.}, 1997) performed a measurement at LEP1.5
$(\sqrt{S}=133$ GeV) with a maximal electron scattering angle of 33 mrad. All
experiments used the cone algorithm and $R=1$. As shown in Fig.\
\ref{fig:18}, the AMY data are well described by the sum of NLO direct,
%
%
single-, and double-resolved processes (Kleinwort and Kramer, 1996a) with
GRV photon densities. The same observation is made in NLO comparisons with the
TOPAZ (Aurenche {\it et al.}, 1994a; Kleinwort and Kramer, 1996a) and OPAL data
(Klasen, Kleinwort, and Kramer, 1998). At comparable values of $x_T=2E_T/
\sqrt{S}$, the jet width measured in photon-photon collisions is approximately
the same as in photon-hadron collisions (Adachi {\it et al.}, 1999).

\subsection{Dijets}
\label{sec:dijets}

NLO calculations for dijet photoproduction differ significantly from those for
single jets with respect to the real corrections. While the phase space of the
two unobserved partons could be integrated analytically for single jets (see
Eq.\ (\ref{eq:solidangle})), it is now restricted by the definition of the
second observed jet. The third unobserved parton momentum $p_3$ still has to
be integrated out.

In the phase-space slicing method, this is done analytically
in the soft and collinear regions, which can be defined by limiting the
invariant masses of two unresolved partons $s_{i3}=(p_i+p_3)^2<ys$ (Gutbrod,
Kramer, and Schierholz, 1984; Kramer and Lampe, 1989) or by applying separate
soft and collinear cut-offs on $E_3<\epsilon\sqrt{s}/2$ and $\theta_{i3}<
\delta$ (Fabricius {\it et al.}, 1981; Gutbrod, Kramer, and Schierholz, 1984),
on  $E_3<\delta_s\sqrt{s}/2$ and $s_{i3}<\delta_c s$ (Baer, Ohnemus, and Owens,
1989b; Harris and Owens, 2001), or on $E_{T_3}<p_{Tm}$ and $R_{i3}<R_c$
(Aurenche {\it et al.}, 2000). Outside the singular regions, the real
corrections are integrated numerically, so that the cross sections become
independent of the technical cut-offs and an experimental jet definition can be
implemented. While the $y$-cut method has been applied to direct (Kleinwort and
Kramer, 1996a, 1996b, 1997; Klasen, Kleinwort, and Kramer 1998),
single-resolved
(Klasen and Kramer, 1996a, 1996b, 1997a), and double-resolved (Klasen and
Kramer, 1997a) dijet photoproduction, the $\delta_{s,c}$ (Baer, Ohnemus, and
Owens, 1989b; Harris and Owens, 1997, 1998) and $p_{Tm},\,R_c$ methods
(Aurenche {\it et al.}, 2000) have only been applied to the single- and
double-resolved cases. All these results were found to be in good
agreement with each other (Harris, Klasen, and Vossebeld, 1999; Aurenche
{\it et al.} 2000).

In the subtraction method, $p_3$ is integrated
numerically also in the soft and collinear regions. The $2\to 3$ matrix
elements then have to be regulated by subtracting them in a finite phase-space
volume setting $p_3=0$ everywhere except in the singular propagator
denominators.
The subtracted terms are then integrated analytically in the same phase-space
volume and added to the virtual corrections to allow for an analytical
cancellation of the soft and collinear poles (Ellis, Ross, and Terrano, 1981).
Several solutions have been proposed to obtain numerical stability with this
method (Ellis, Kunszt, and Soper, 1989a, 1989b, 1990; Kunszt and Soper, 1992).
QCD
dipole factorization formulae allow for a straightforward computation of the
soft and collinear subtraction terms (Catani and Seymour, 1996, 1997). The
subtraction method has been applied to single-resolved (B\"odeker, 1992a,
1992b, 1993; Frixione and Ridolfi, 1997) and double-resolved (Frixione and
Ridolfi, 1997) dijet photoproduction. 

%
\begin{figure}
 \begin{center}
  \epsfig{file=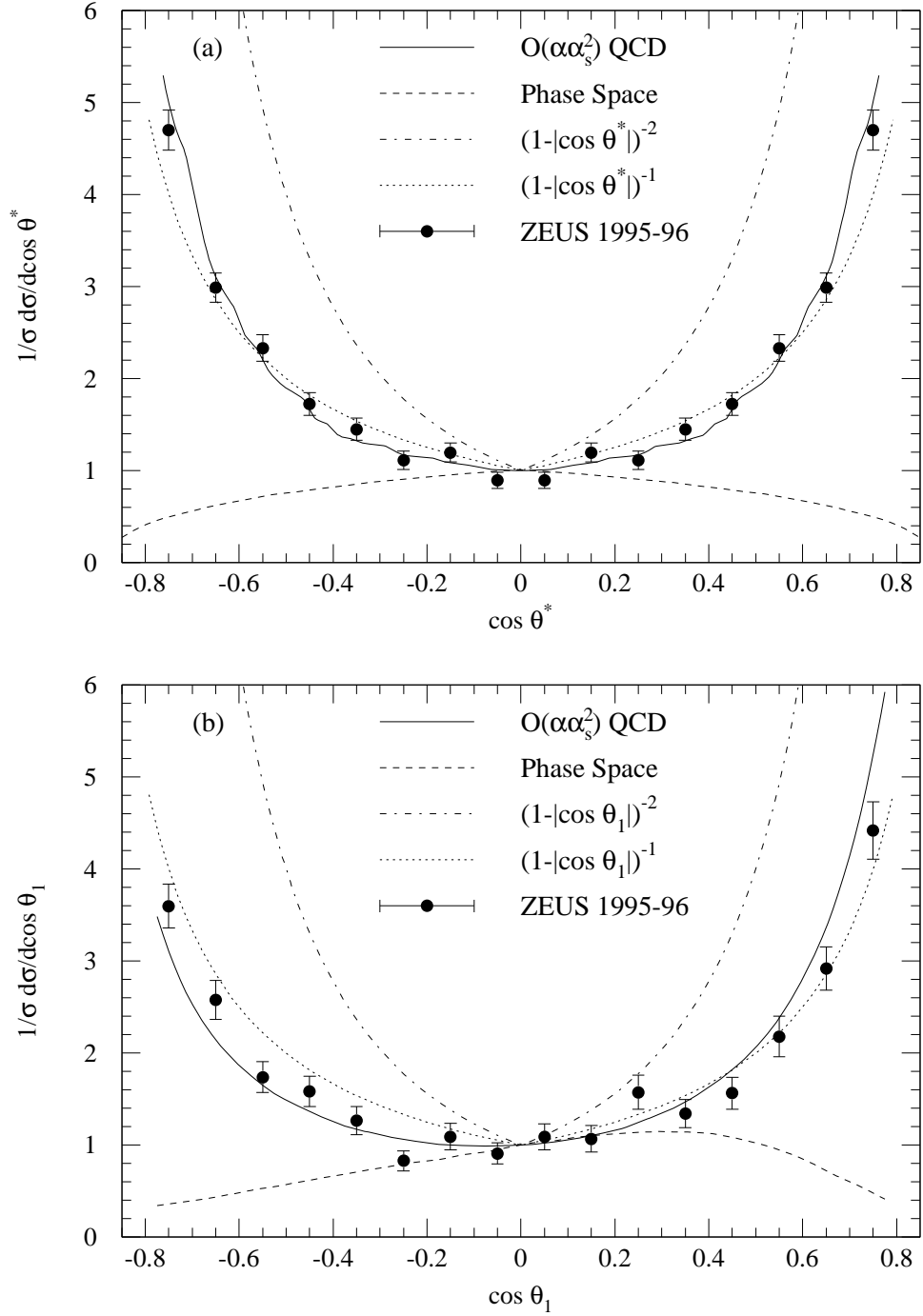,width=0.8\linewidth}
 \caption{\label{fig:19}
 Dependences of the (a) dijet and (b) three-jet cross sections on the fastest
 jet scattering angle, normalized at $\cos\theta=0$. The ZEUS dijet (Derrick
 {\it et al.}, 1997a) and three-jet (Breitweg {\it et al.}, 1998d) data are
 well described by the QCD predictions, but not by the pure
 phase-space distributions, and they favor the single $1/t$
 pole for
 massless fermion exchange over the double $1/t^2$ pole for
 massless boson exchange.}
 \end{center}
\end{figure}
%
In addition to the transverse energy $E_T$ and pseudorapidity $\eta_1$ of the
first jet, the inclusive dijet cross section
\beq
  \frac{\mbox{d}^3\sigma}{\mbox{d}E_T^2\mbox{d}\eta_1\mbox{d}\eta_2}
  = \sum_{a,b} x_a f_{a/A}(x_a,M_a^2) x_b f_{b/B}(x_b,M_b^2)
  \frac{\mbox{d}\sigma}{\mbox{d}t}
\label{eq:2jet}
\eeq
depends on the pseudorapidity of the second jet $\eta_2$. In LO only two jets
with equal transverse energies can be produced, and the observed momentum
fractions of the partons in the initial electrons or hadrons
\beq
 x_{a,b}^{\rm obs} = \sum_{i=1}^{2} E_{T_i}e^{\pm\eta_i} / (2E_{A,B})
\eeq
equal the true momentum fractions $x_{a,b}$. If the energy transfer
$y=E_\gamma/E_e$ is known (see Eq.\ (\ref{eq:jacblon})), momentum
fractions for the partons in photons $x_\gamma^{\rm obs}=x_{a,b}^{ \rm obs}/y$ 
can be deduced. In NLO, where a third jet can be present, the observed
momentum fractions are defined by the sums over the two jets with highest
$E_T$, and they match the true momentum fractions only approximately.
Furthermore, the transverse energies of the two hardest jets need no longer be
equal to each other. Even worse, for equal $E_T$ cuts and maximal azimuthal
distance $\Delta\phi=\phi_1-\phi_2=\pi$ the NLO prediction
becomes sensitive to the method chosen for the integration of soft and
collinear singularities, strongly scale dependent, and thus unreliable
(Klasen and Kramer, 1996a; Harris and Owens, 1997; Frixione and Ridolfi, 1997).
This sensitivity also propagates into the region of large observed momentum
fractions (Aurenche {\it et al.}, 2000). Two proposed solutions are to allow
for a small difference in the theoretical $E_T$ cuts or for a small technical
cut-off dependence, but it is preferable to cut instead on the average
$\bar{E}_T=(E_{T_1}+E_{T_2})/2$. Similarly one can define
$\Delta E_T=E_{T_1}-E_{T_2},\,\bar{\eta}=(\eta_1+\eta_2)/2,$ and
$\Delta\eta=\eta_1-\eta_2$, which is related
to the cosine of the center-of-mass scattering angle $\cos\theta^\ast =
\tanh(\Delta\eta/2)$. The dijet cross section then takes the form
\beq
 {\d^3\sigma\over \d x_a\d x_b\d\cos\theta^\ast} = {2\over s}
 \frac{\mbox{d}^3\sigma}{\mbox{d}E_T^2\mbox{d}\eta_1\mbox{d}\eta_2},
\eeq
which is particularly useful to determine the parton densities and scattering
processes.

%
\begin{figure}
 \begin{center}
  \epsfig{file=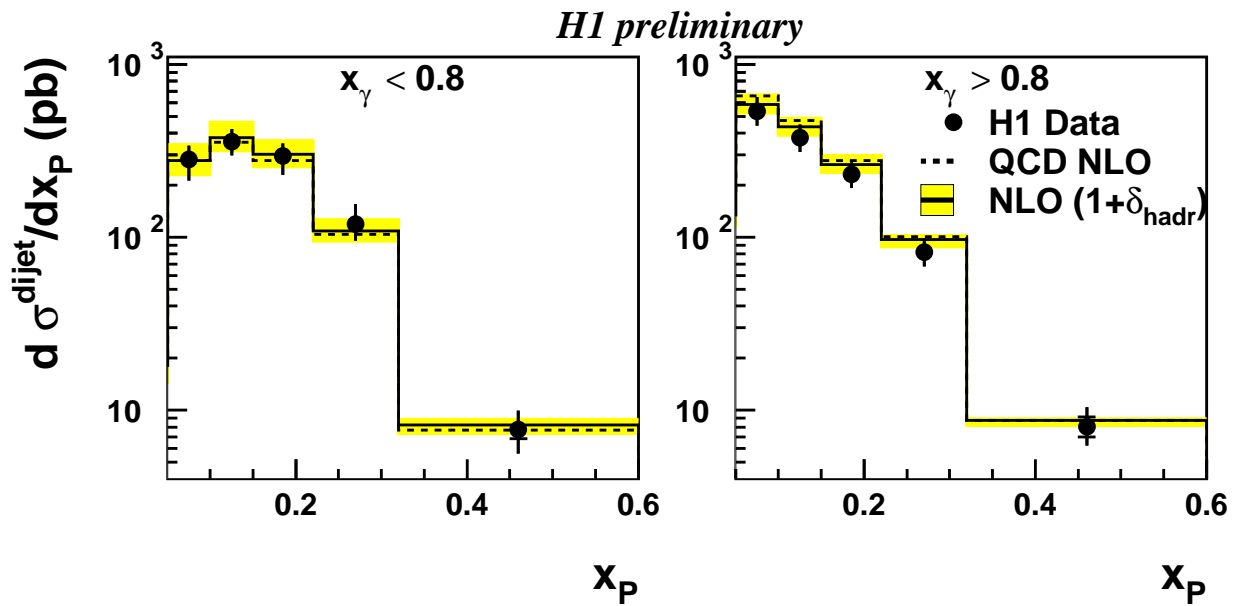,width=\linewidth,clip=}
 \caption{\label{fig:20}Dependence of the dijet photoproduction cross
 section on the observed parton momentum fraction in the proton compared to
 preliminary H1 data (Adloff {\it et al.}, 2001b). For ``resolved'' photons
 (left), the distribution is quark-like, while for ``direct'' photons
 (right) it is gluon-like, as is to be expected from the contributing
 parton scattering diagrams.}
 \end{center}
\end{figure}
%
The $\cos\theta^\ast$-dependence at HERA is shown in Fig.\ \ref{fig:19} (a)
for dijet masses $M_{12}=\sqrt{(p_1+p_2)^2}>47$ GeV and the $k_T$ algorithm.
The ZEUS data (Derrick {\it et al.}, 1997a) show good agreement with the QCD
prediction (full curve), but clearly disagree with the pure phase-space
distribution (dashed curve) and the Rutherford scattering form at small angle
$(1-|\cos\theta^{\ast}|)^{-2}$, which is characteristic for massless vector
boson exchange in the resolved processes. On the other hand, the data agree
very well with the less singular form $(1-|\cos\theta^{\ast}|)^{-1}$, which is
typical for massless fermion exchange in the direct processes,
indicating that the direct processes dominate over resolved processes in this
kinematic region. A similar behavior is observed in three-jet cross sections
with $M_{123}=\sqrt{(p_1+p_2+p_3)^2}>50$ GeV
for the scattering angle of the leading jet (see Fig.\ \ref{fig:19} (b))
(Breitweg {\it et al.}, 1998d; Klasen, 1999a).
NLO QCD also describes earlier ZEUS data using the cone algorithm at lower
values of $M_{12}$ $(M_{12}>23$ GeV) (Derrick {\it et al.}, 1996), which have been divided into a
flatter ``direct'' ($x_{\gamma}^{\rm obs} >0.75$) and a steeper ``resolved''
($x_{\gamma}^{\rm obs}<0.75$) sample (Harris and Owens, 1997).

Similar analyses have recently been carried out with the $k_T$ algorithm
by H1 (Adloff {\it et al.}, 2001b) and ZEUS (Breitweg {\it et al.}, 2000b), who
also measured distributions of the observed parton momentum fractions in the
proton (only H1) and photon with asymmetric cuts on $E_{T_1}>25$ GeV and
$E_{T_2}>15$ GeV (14 and 11 GeV in the ZEUS case).
The H1 $x_p^{\rm obs}$ distribution (Fig.\ \ref{fig:20}) is well
described by the CTEQ5M proton densities (Lai {\it et al.}, 2000), but it
is not sensitive to the poorly constrained gluon density at very small or
large $x_p$.
%
\begin{figure}
 \begin{center}
  \epsfig{file=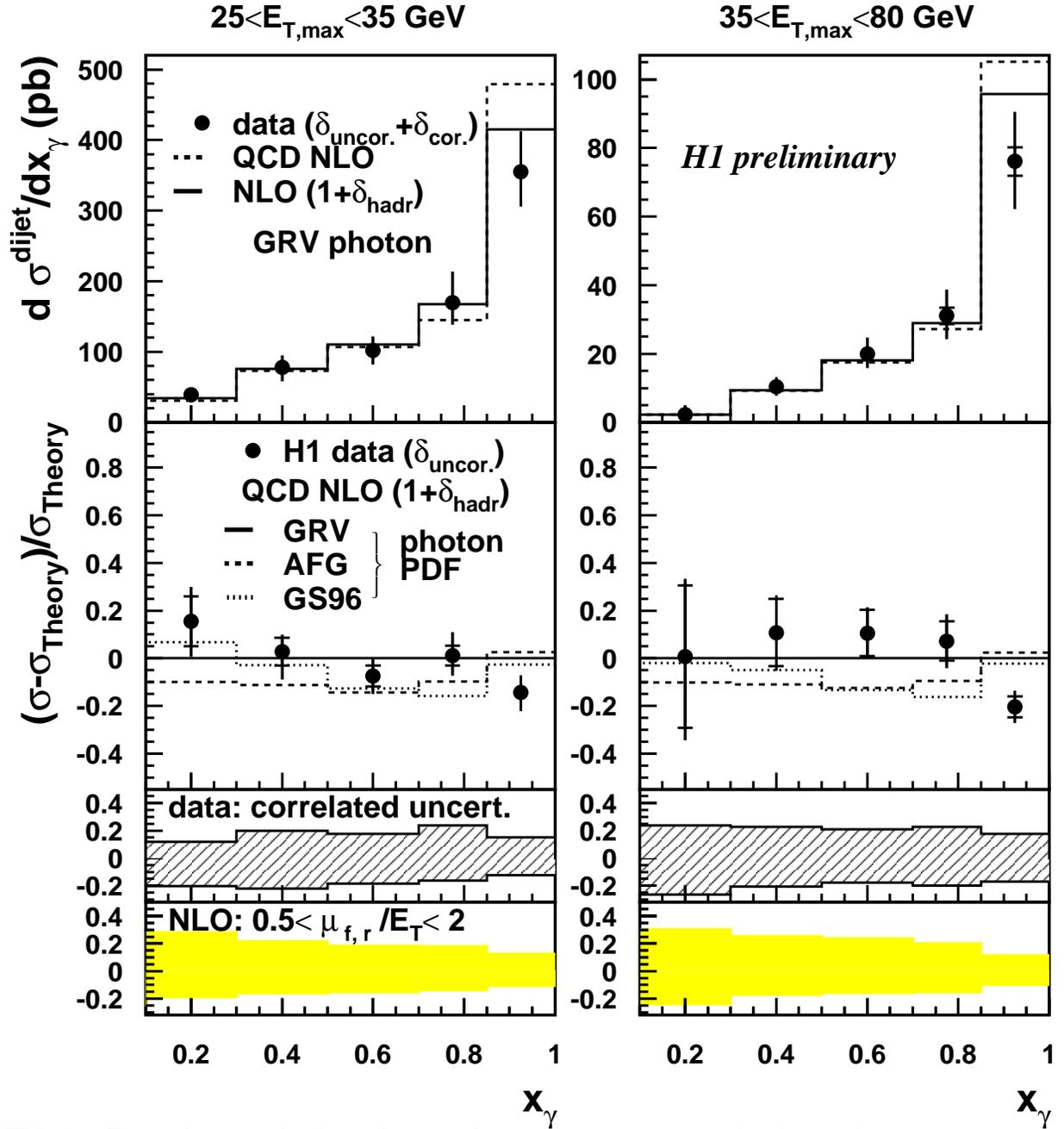,width=\linewidth,clip=}
 \caption{\label{fig:21}Dependence of the dijet photoproduction cross
 section on the observed parton momentum fraction in the photon compared to
 preliminary H1 data (Adloff {\it et al.}, 2001b).}
 \end{center}
\end{figure}
%
The H1 $x_\gamma^{\rm obs}$ distribution (Fig.\ \ref{fig:21}) shows a
slight preference for GS96 at lower $E_T$ and GRV at larger $E_T$ (see also
Adloff {\it et al.}, 1998), but the experimental and
theoretical uncertainties are still larger than the photon density variations.
H1 (Ahmed {\it et al.}, 1995; Adloff {\it et al.}, 2000) and ZEUS (Derrick
{\it et al.}, 1995b;
Breitweg {\it et al.}, 1998e) have earlier been able to rule out the LO LAC1 and
LAC3 photon densities in $x_\gamma^{\rm obs}$ and $\bar{\eta}$ distributions,
which are both sensitive to the photon densities (Forshaw and Roberts, 1993).
The ZEUS distributions in $\eta_2$ (Breitweg {\it et al.}, 1999a) do not allow
any firm conclusions, since sizeable hadronization corrections have not
been included in the NLO predictions (Harris, Klasen, and Vossebeld, 1999).
Dijet mass distributions have been measured up to 140 (Breitweg {\it et al.},
1999a) and 180 GeV (Adloff {\it et al.}, 2001b), respectively. They show no
deviations from NLO QCD.

\begin{figure}
 \begin{center}
  \epsfig{file=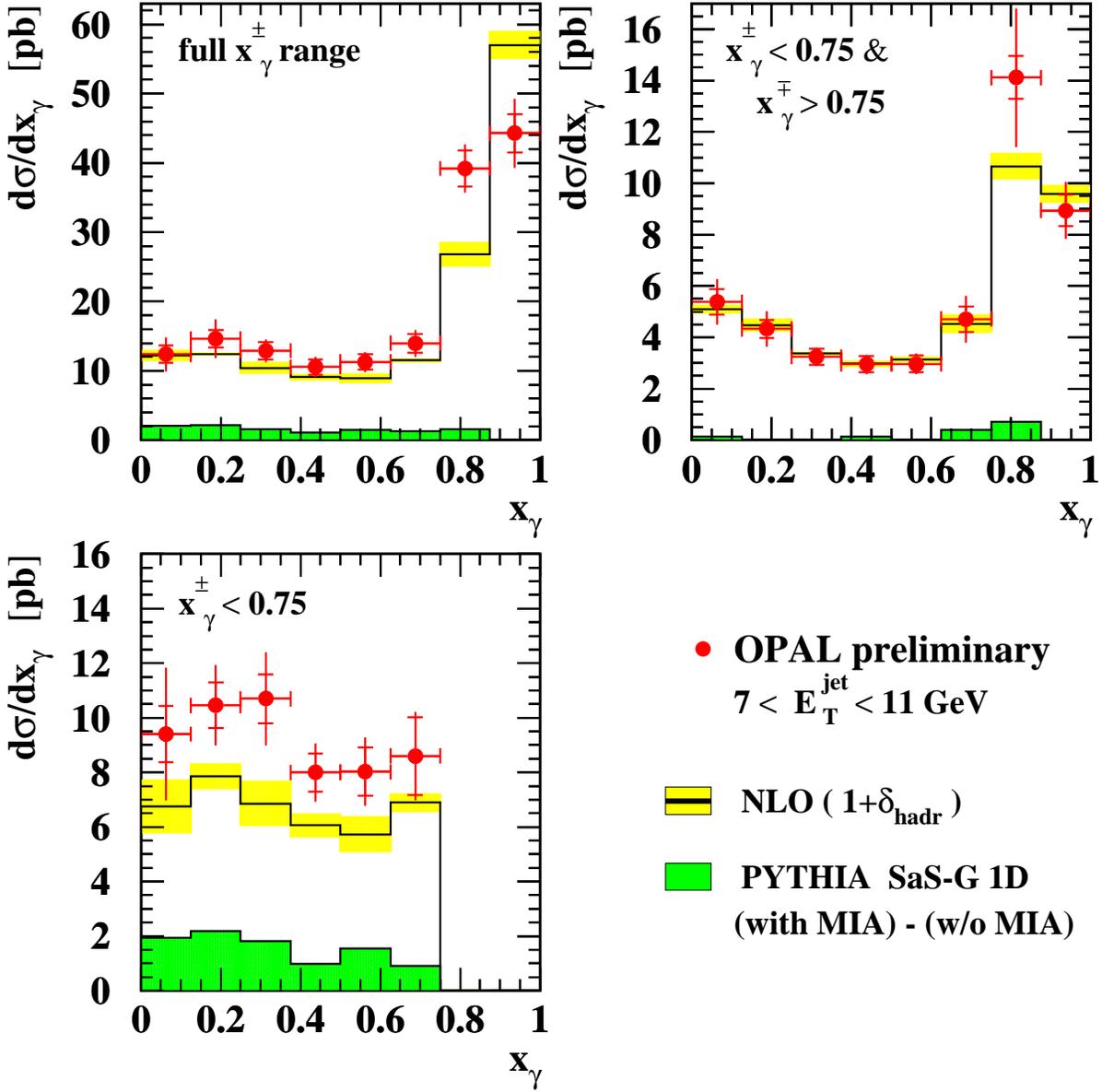,width=\linewidth}
 \caption{\label{fig:22}Dependence of the dijet cross section in
 photon-photon collisions on the observed parton momentum fraction in the
 photon compared to preliminary OPAL data (Wengler, 2002).}
 \end{center}
\end{figure}

In photon-photon collisions, dijet transverse energy distributions have been
measured with the cone algorithm by AMY (Kim {\it et al.}, 1994), TOPAZ
(Hayashii {\it et al.}, 1993), and OPAL (Ackerstaff {\it et al.}, 1997;
Abbiendi {\it et al.}, 1999). They are well described by NLO QCD (Kleinwort
and Kramer, 1996a, 1997; Klasen, Kleinwort, and Kramer, 1998), as are the OPAL
rapidity distributions (Abbiendi {\it et al.}, 1999). The $k_T$ algorithm has
been used by ALEPH (Barate {\it et al.}, 2000) and OPAL (Abbiendi {\it et
al.}, 2001c; Wengler, 2002)
to measure the transverse energy and $x_\gamma^{\rm obs}$ (only
OPAL) distributions. The former is well described by NLO QCD. However, the
infrared sensitivity of $x_\gamma^{\rm obs}$ is reflected in Fig.\
\ref{fig:22} in the large fluctuation of the NLO prediction in the
two highest bins.
%
%
When both photons interact with low $x_\gamma^{\rm obs}$, multiple parton
interactions (MIA) can lead to a larger observed cross section than predicted
by NLO QCD with GRV photon densities.

Higher center-of-mass energies may be reached at linear $e^+e^-$ colliders
like TESLA, ILC ($\sqrt{S}=500...1000$ GeV), or CLIC ($\sqrt{S}=1...3$ TeV) or
at a future $ep$ collider like THERA (TESLA$\times$HERA, $\sqrt{S}\sim 1$
TeV). At linear colliders, the photoproduction cross section
is enhanced due to beamstrahlung (see Sec.\ \ref{sec:beamstrahlung}) or laser
backscattering (see Sec.\ \ref{sec:laserback}). Fig.\ \ref{fig:23} demonstrates
that tests of QCD and determinations of the photon structure could be largely
extended:
%
\begin{figure}
 \begin{center}
  \epsfig{file=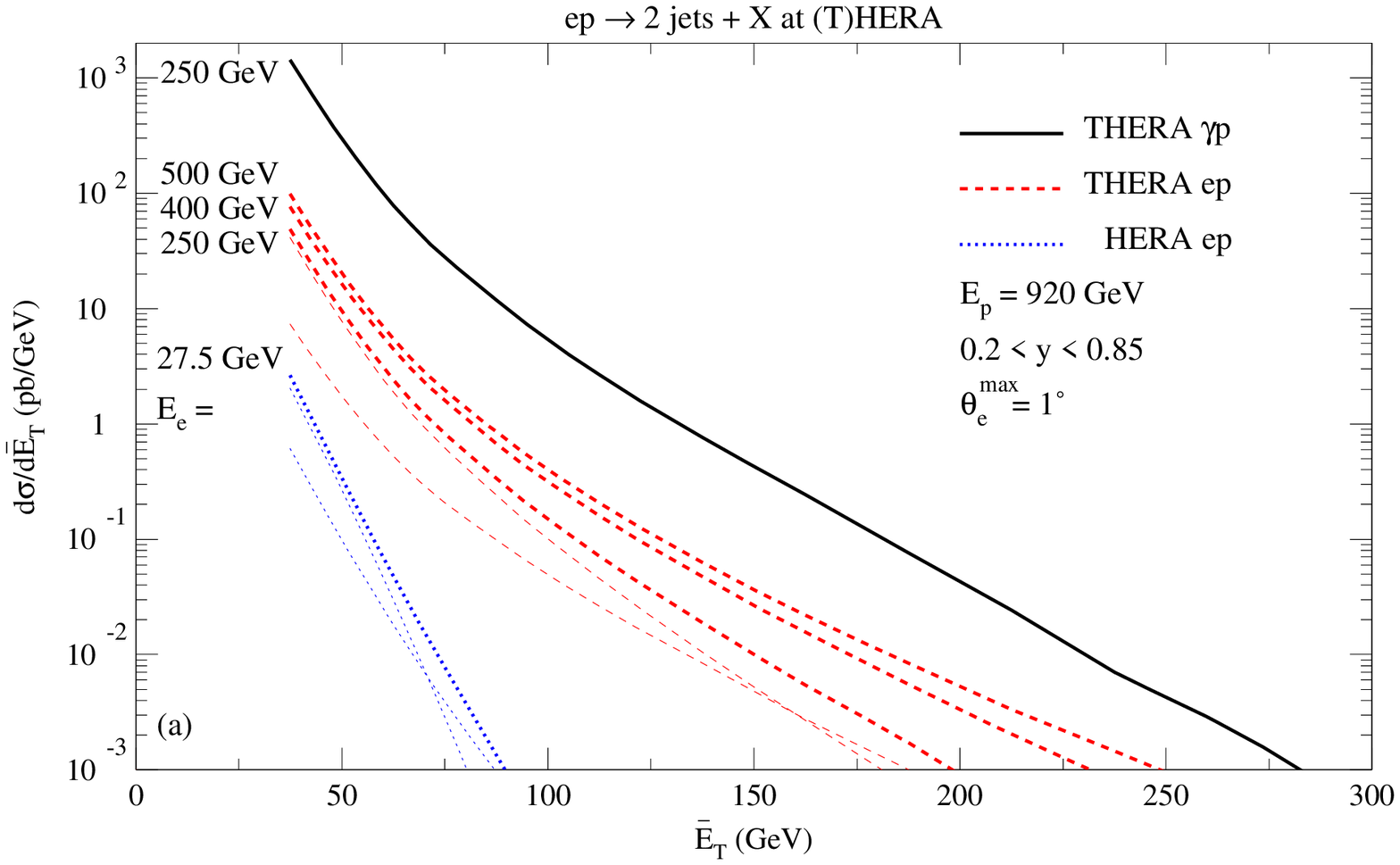,width=0.9\linewidth}
  \epsfig{file=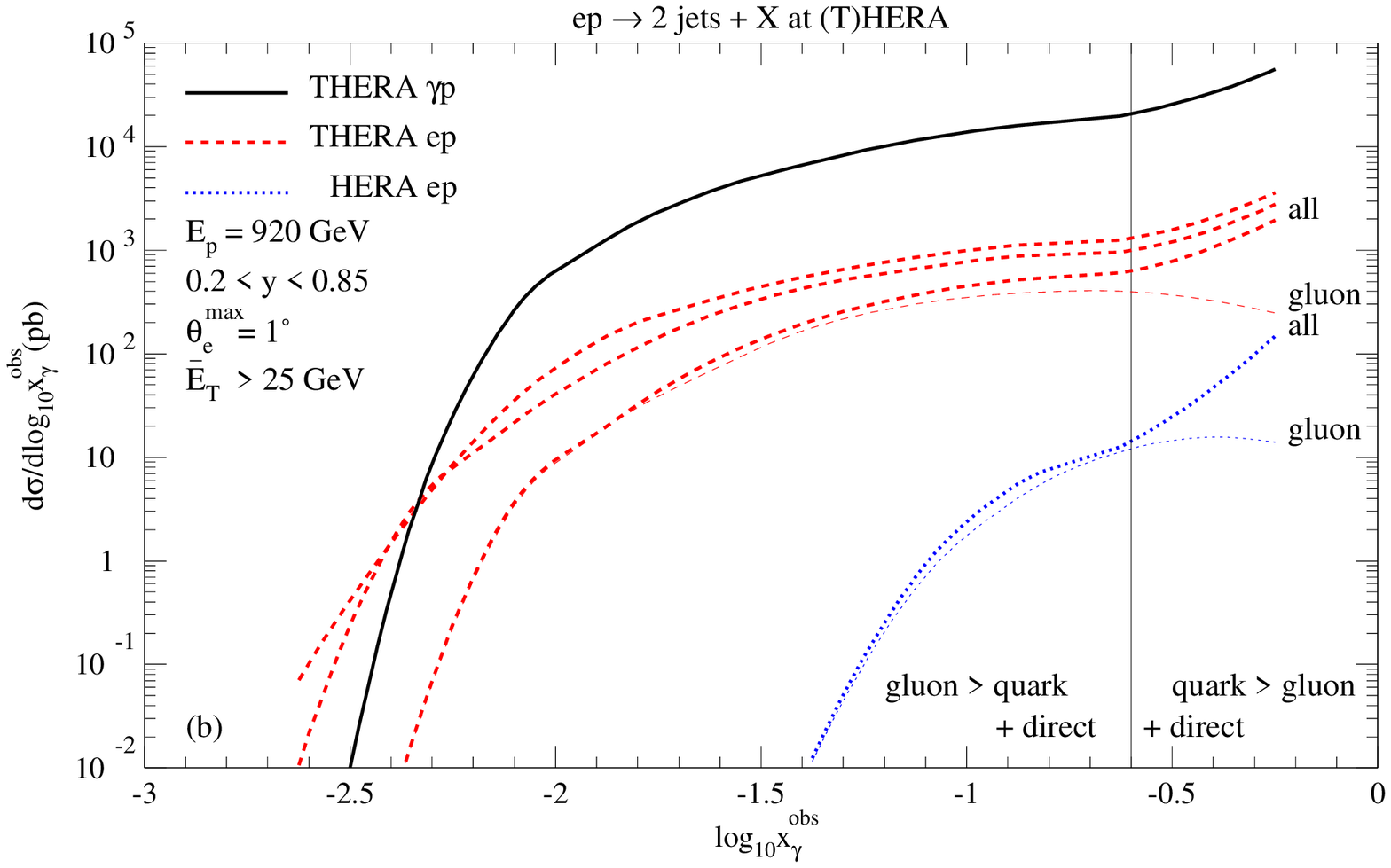,width=0.9\linewidth}
 \caption{\label{fig:23}
  Dijet photoproduction at a future THERA collider as a function of
  the average transverse energy
  of the two jets (a) and of the observed parton momentum fraction in the
  photon (b). The thin lines in (a) show the separate contributions from the
  resolved (direct) processes for HERA and THERA with $E_e=$250 GeV, which
  dominate at small (large) $\bar{E}_T$.}
 \end{center}
\end{figure}
%
Depending on the TESLA electron beam energy (250-500 GeV) and the collider
mode ($ep$ or $\gamma p$), the THERA range in the average transverse energy
of the two jets would be increased by a factor of 2-3 and the reach in
$x_\gamma^{\rm obs}$ by at least one order of magnitude. The reach in
$x_p^{\rm obs}$ is extended by about the same amount. It would thus become
possible to check the determinations of the gluon density in the proton
obtained in deep-inelastic scattering experiments and to measure
the gluon density in the photon down to low values of $x$ (Klasen, 2001c;
Wing, 2001). Similar studies have been performed for high-energy muon-proton
collisions, although
here bremsstrahlung is reduced and laser backscattering seems impossible
(Klasen, 1997b).

\subsection{Three jets}

For $N$ massless jets, one can choose $3N-4$ parameters that should span the
multijet parameter space. They should also facilitate a simple interpretation
within QCD and allow for a comparison of the $N-1$-jet to the $N$-jet cross
section. In the
case of $N=3$, the conventional choices are the three-jet mass $M_{123}
=\sqrt{(p_1+p_2+p_3)^2}$
and four dimensionless parameters. The Dalitz energy fractions
\begin{equation}
 x_i = \frac{2E_i}{M_{123}}
\end{equation}
specify, how the available energy is shared between the three jets. They are
ordered such that $x_1 > x_2 > x_3$. Since $x_1+x_2+x_3 = 2$, only $x_1$ and
$x_2$ are linearly independent. The third and fourth parameters are the
cosine of the scattering angle between the leading jet and the average beam
direction $\vec{p}_{\rm AV} = \vec{p}_a-\vec{p}_b$, where the incoming
parton $a$ is the one with the highest energy in the laboratory frame,
\begin{equation}
 \cos\theta_1=\frac{\vec{p}_{\rm AV}\vec{p}_1}{|\vec{p}_{\rm AV}||\vec{p}_1|},
\end{equation}
and the angle between the three-jet plane and the plane containing the
leading jet and the beam direction,
\begin{equation}
 \cos\psi_1=\frac{(\vec{p}_1\times\vec{p}_{\rm AV})(\vec{p}_2\times\vec{p}_3)}
                 {|\vec{p}_1\times\vec{p}_{\rm AV}||\vec{p}_2\times\vec{p}_3|}.
\end{equation}
In the soft limit, where $E_3 \rightarrow 0$ and $x_{1,2} \rightarrow 1$,
$\cos\theta_1$ approaches the $2\rightarrow 2$ center-of-mass scattering angle
$\cos\theta^{\ast}$, thus relating three-jet to dijet cross sections. 
The similarity of the distributions in the dijet and three-jet scattering
angle distributions has already been discussed (see Fig.\ \ref{fig:19}).
Of course, the third jet must not be too soft (or the hard jets not too hard)
to avoid soft singularities that would have to be absorbed into the
next-to-leading order dijet cross section. This can be achieved by a cut on
$x_1$, {\it e.g.} $x_1 < 0.95$. However, since the energy of a jet is always
larger than its transverse
energy, a cut like $E_{T,3} > 5$ GeV already insures the absence of soft
singularities. The three jets also have to be well separated in phase space
from each other and from the incident beams to avoid initial and final state
collinear singularities. This is insured by a cut like $|\cos\theta_1| < 0.8$,
by cuts on the pseudorapidities, and by the $k_T$ jet algorithm.

%
\begin{figure}
 \begin{center}
  \epsfig{file=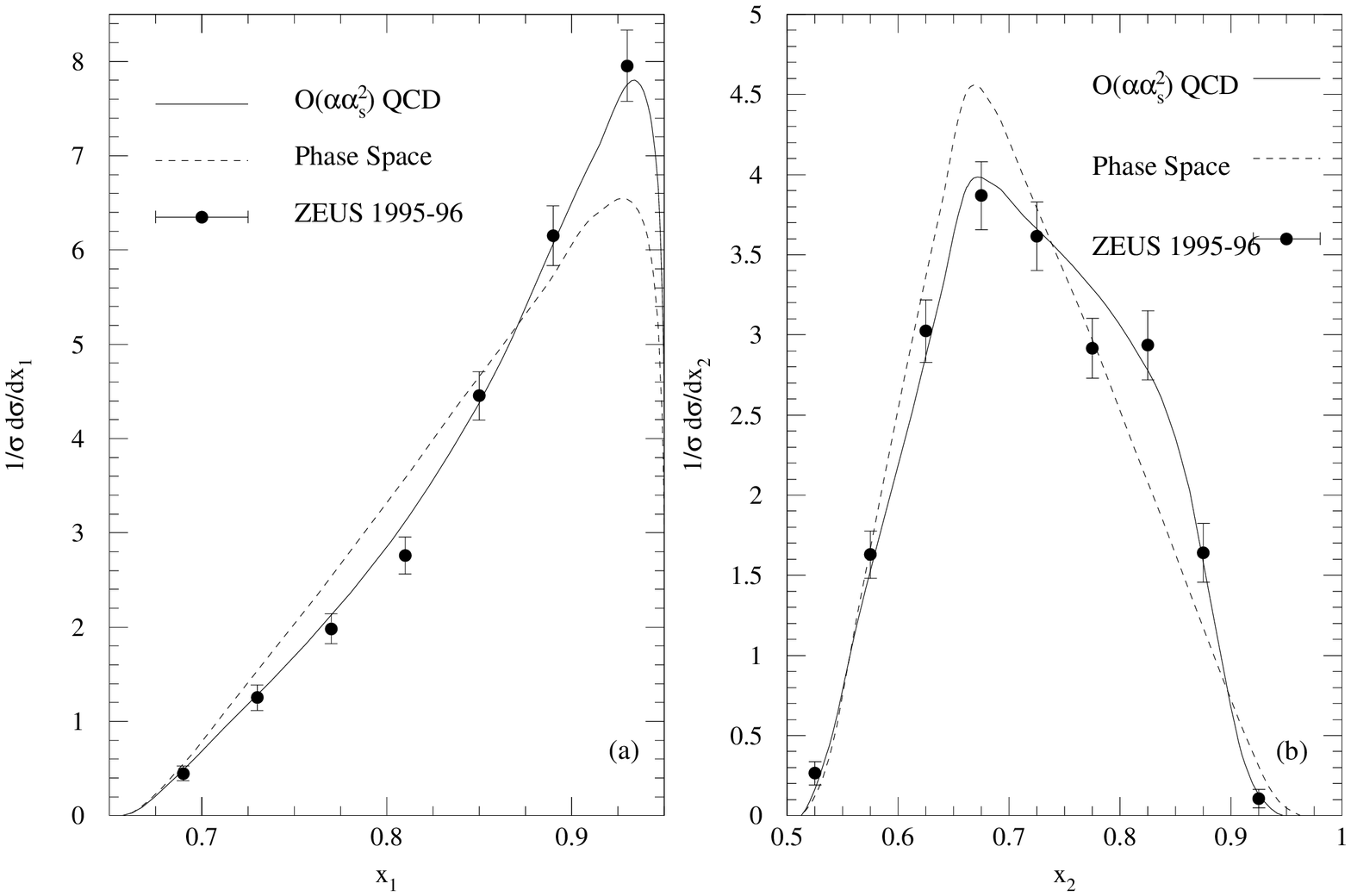,width=0.9\linewidth}
  \epsfig{file=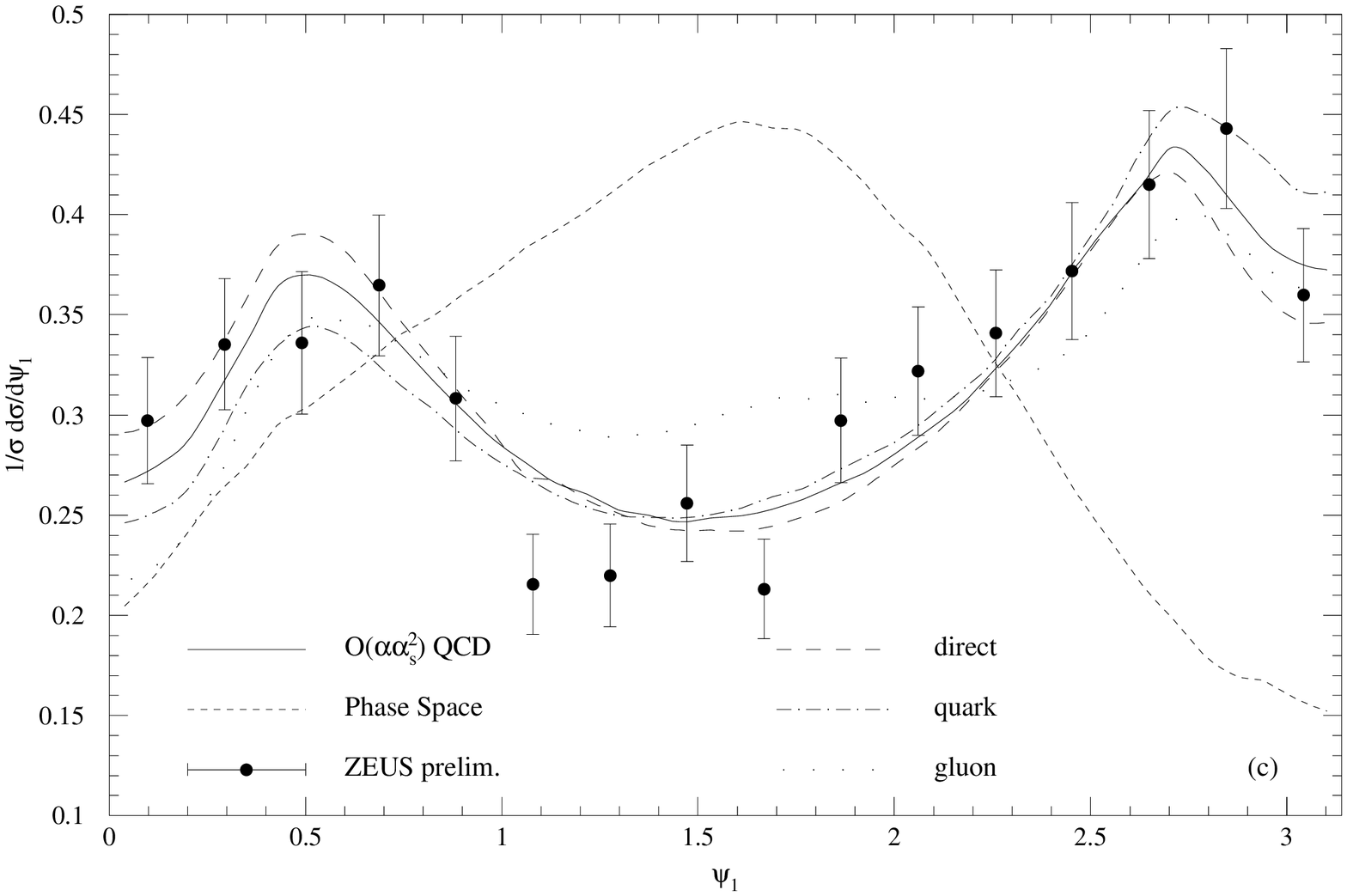,width=0.9\linewidth}
 \caption{\label{fig:24}
  Dependence of the three-jet cross section on the energy fractions
  $x_1$ (a) and $x_2$ (b) and on the angle between the three-jet plane and the
  plane containing the leading jet and the average beam direction (c).
  The ZEUS data (Breitweg {\it et al.}, 1998d) rule out the pure phase-space
 and pure gluon-initiated distributions.}
 \end{center}
\end{figure}
%
The three-jet photoproduction cross section
\bea
  \frac{\d^4\sigma}{\d x_1\d x_2 \d\cos\theta_1\d\psi_1}
  &=& \sum_{a,b} x_a f_{a/A}(x_a,M_a^2) x_b f_{b/B}(x_b,M_b^2)\nonumber\\
  &\times&\frac{\overline{|\M|}_{ab\rightarrow 123}^2}{1024\pi^4}
\label{eq:3jet}
\eea
has been calculated from the direct and resolved tree-level $2\to 3$ processes
by several groups with good mutual agreement (Baer,
Ohnemus, and Owens, 1989a; Klasen, 1999a, 1999b; Harris, Klasen, and
Vossebeld, 1999). Figure \ref{fig:24} shows the three-jet cross section at HERA
as a function of the energy fractions of the leading (a) and next-to-leading
(b) jet $x_1$ and $x_2$, normalized to the total cross section. The prediction
from the ${\cal O}(\alpha\alpha_s^2)$ QCD matrix elements are rather
similar to the pure phase-space distributions with constant matrix elements,
but the ZEUS data
(Breitweg {\it et al.}, 1998d) slightly prefer the QCD predictions.
The dependence on the angle between the three-jet plane and the plane
containing the leading jet and the average beam direction is shown in
Fig.~\ref{fig:24} (c).
The full QCD curve again agrees well with the data as do the contributions
from direct photons and quarks in the photon, whereas the pure phase space
has a completely different shape.

Unfortunately, a full NLO calculation for three-jet photoproduction has not yet
been performed. Such a calculation would allow, {\it e.g.}, for a determination
of $\alpha_s$ from the ratio of three-jet to dijet cross sections and for a
detailed study of the jet substructure. It could furthermore form the basis of
a NNLO dijet calculation with reduced renormalization and factorization scale
dependences. Higher order corrections can also be taken into
account approximately with resummation and parton shower techniques, leading
eventually to more precise Monte Carlo generators for use by the
experimental groups.

\subsection{Dijets with a leading neutron}

Recently H1 (Adloff {\it et al.}, 2001c) and ZEUS (Breitweg {\it et
al.}, 2000c) have presented data on dijet photoproduction with a leading
neutron, which is dominated by slightly off-shell pion exchange and can be
used to constrain the parton densities in the pion. These are not well
known, particularly at low $x_{\pi}$ and in the sea-quark and gluon
sectors. The pion structure carries important implications for the QCD
confinement mechanism and the realization of symmetries like isospin in nature.
It is also of practical importance for the hadronic input to the photon
structure at low scales (see Sec.\ \ref{sec:hadsol}).

Assuming factorization, the photoproduction cross section for two jets with a
leading neutron
\beq
   \frac{\d ^3\sigma}{\d E_T^2\d \eta_1\d \eta_2}
   = \sum_{a,b} x_a           f_{a/e}(x_a          ,M_\gamma^2) 
                  x_b           f_{b/p}(x_b          ,M_\pi^2   )
                  \frac{\d\sigma}{\d t}
\eeq
depends on the partonic cross section $\d\sigma/\d t$ (Eq.\
(\ref{eq:totalnloxsec})) and on the parton densities in the electron
$f_{a/e}$ (Eq.\ (\ref{eq:partonflux})) and proton
\beq
  f_{b/p}(x_b,M_\pi ^2  ) = \int_{x_b}^1\frac{\d (1\!-\!x_n)}
  {1\!-\!x_n}f_{b/\pi   }(x_\pi,M_\pi^2)f_{\pi/p}(1\!-\!x_n,t').
\eeq
The latter is a convolution of the parton densities in the pion $f_{b/\pi}$
with the pion flux in the $p \rightarrow n\pi$ transition
\bea
 f_{\pi/p}(1\!-\!x_n,t')&=& \frac{1}{4\pi}\frac{g^2}{4\pi}
  \frac{-t'}{(m_{\pi}^2-t')^2}(1-x_n)^{1-2\alpha_{\pi}'(t'-m_\pi^2)}\nonumber\\
 &\times& [F(x_n,t')]^2,
\eea
where $1/(m_{\pi}^2-t')^2$ is the squared pion propagator, $m_\pi$ is the
pion mass, and $(1-x_n)^{1-2\alpha_{\pi}'(t'-m_\pi^2)}$ accounts for the
virtuality and possible reggeization of the pion. In the ZEUS analysis, the
momentum fraction of the leading neutron $x_n$ and
the momentum transfer $t'$ are restricted to $x_n> 400$ GeV$/820$ GeV and
$t'=f(p_T)=-p_T^2/x_n-(1-x_n)(m_n^2-x_n m_p^2)/x_n$ with
$p_T<x_n\cdot 0.66$ GeV. The interaction term in the
pion-nucleon Lagrangian leads to the numerator $-t'$, and off-mass shell
effects of higher Fock states are modeled by a form factor
\beq
   F(x_n,t') = \left\{ \begin{array} {l} \exp [b(t'-m_{\pi}^2)],
  {\rm \hspace*{19.6mm} [Exponential]} \\ 
                                        \exp [R^2(t'-m_{\pi}^2)/(1-x_n)].
  {\rm \hspace*{4.5mm} [Light~Cone]}
                      \end{array} \right.
\eeq
Since the momentum transfer $t'$ is small, reggeization can be neglected
($\alpha'=0$), and a light cone form factor with $R=0.5$ GeV$^{-1}$ can be
chosen in good agreement with recent determinations
(Holtmann, Speth, and Szczurek, 1996; D'Alesio and Pirner, 2000).
The pion-nucleon coupling constant $g^2/(4\pi)=14.17$ can be taken from a
recent extraction from the Goldberger-Miyazawa-Oehme sum rule
(Ericson, Loiseau, and Thomas, 2000).

%
\begin{figure}
 \begin{center}
  \epsfig{file=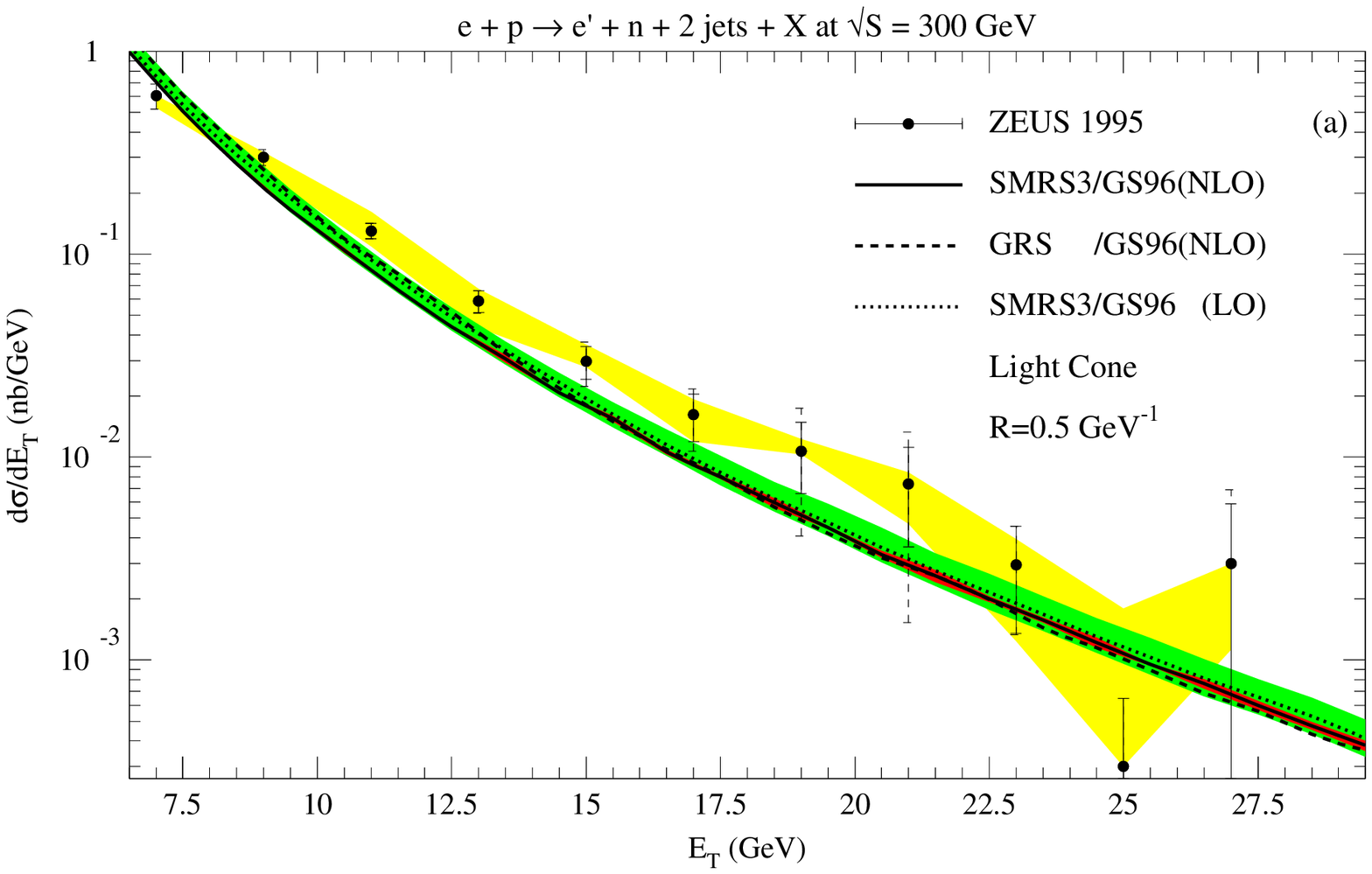,width=0.9\linewidth}
  \epsfig{file=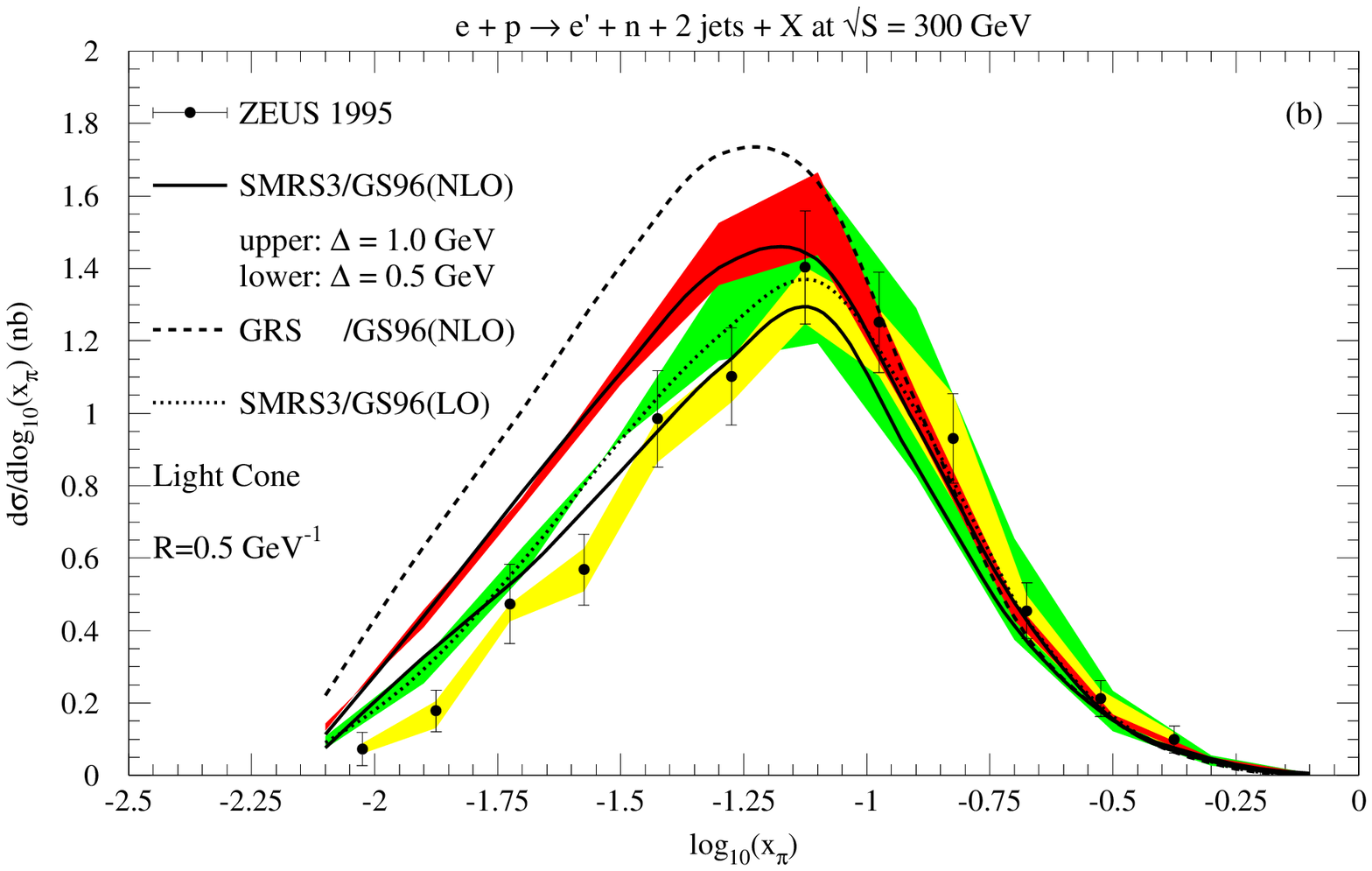,width=0.9\linewidth}
 \caption{\label{fig:25}
 Dependence of the dijet photoproduction cross section with a leading
 neutron on the transverse jet energy (a) and on the observed parton
 momentum fraction in the pion (b) compared to ZEUS data (Breitweg
 {\it et al.}, 2000c). The error bands show the theoretical scale uncertainty
 (medium: LO, dark: NLO) and the experimental energy scale uncertainty
 (light).}
 \end{center}
\end{figure}
%
In Fig.\ \ref{fig:25}, LO and NLO QCD predictions (Klasen and Kramer, 2001;
Klasen, 2001a, 2001b) with SMRS3 (Martin {\it et al.}, 1992) and GRS (Gl\"uck,
Reya, and Schienbein, 1999a) pion densities and GS96 photon densities are
compared to the ZEUS measurements (Breitweg {\it et al.}, 2000c).
In the transverse energy distribution (a), the scale dependence is reduced
considerably from LO to NLO. The
normalization is sensitive to the chosen pion flux factor. If one includes the
Regge trajectory ($\alpha'=1$ GeV$^{-2}$) and omits the form factor, the cross
section is reduced by $15\%$. The distribution in the observed momentum
fraction of the partons in the pion (b)
\beq
 x_{\pi}^{\rm obs} = \frac{E_{T_1}e^{ \eta_1}+E_{T_2}e^{ \eta_2}}
 {2 (1-x_n)E_p}
\eeq
suffers from the fact, that ZEUS have applied equal cuts on $E_{T_1},E_{T_2}>6$
GeV and $\eta_{1,2}\in[-2;2]$, which had to be relaxed to $E_{T,2}>6$ GeV$
-\Delta$
in the NLO calculation.
The best fit is obtained with GS96 photon densities and SMRS3 pion densities,
which have the lowest quark and gluon distributions, respectively, but
hadronization corrections may play an important role at small $x_{\pi}$.


\pagebreak
\section{Hadron production}
\label{sec:hadprod}
\setcounter{equation}{0}

In the last Section it was demonstrated, that jets provide an intuitive link
with the partonic scattering processes, yield large rates in experimental
measurements, and are well suited to determine the structure of the incoming
photons, protons, or pions. However, they do not provide detailed information
about the final state hadronization, which can be better studied in the
production of light and heavy hadrons and quark-antiquark bound states
(quarkonia).

Early measurements of inclusive light hadron production were performed in fixed
target collisions by the CERN NA14 experiment
(Auge {\it et al.}, 1986; Barate {\it et al.}, 1986) and 
in photon-photon collisions at DESY PETRA (Berger {\it et al.}, 1979; Brandelik
{\it et al.}, 1981). Heavy mesons were studied by the CERN
WA4 (Roudeau {\it et al.}, 1980) and  NA14/2
(Alvarez {\it et al.}, 1992, 1993), Fermilab E687 (Frabetti {\it et al.}, 1993,
1996; Moroni {\it et al.}, 1994) and E691 (Anjos {\it et al.}, 1989, 1990),
DESY JADE (Bartel {\it et
al.}, 1987) and KEK AMY (Aso {\it et al.}, 1995; Takashimizu {\it et al.},
1996), TOPAZ (Enomoto {\it et al.}, 1994a, 1994b), and VENUS experiments
(Uehara {\it et al.}, 1994, Ohyama {\it et al.}, 1997).

In this Section, the fragmentation of partons to light and heavy hadrons
and quarkonia and their production in photon-hadron and photon-photon
collisions will be discussed.

\subsection{Fragmentation}
\label{sec:hadfrag}

As mentioned in Sec.\ \ref{sec:jetalgorithms}, the KLN theorem guarantees,
that the soft and collinear singularities generated by real particle emission
beyond the LO cancel against those generated by virtual particle exchanges
after summation over degenerate final states. In semi-inclusive final states,
which contain identified particles, this cancellation is incomplete. Collinear
$1/\eps$-poles multiplying the time-like partonic splitting functions
$P_{j\leftarrow i}(x)$ in the transition functions
\bea
 \overline{\Gamma}_{j\leftarrow i}(x,M_f^2)&=&\delta_{ij}\,\delta(1-x)
 \!-\!\frac{1}
 {\eps}\frac{\alpha_s(\mu^2)}{2\pi}\frac{\Gamma(1-\eps)}{\Gamma(1-2\eps)}\!\lr
 \!\frac{4\pi\mu^2}{M_f^2}\!\rr^\eps\nonumber\\
 &\times& P_{j\leftarrow i}(x) 
 +{\cal O}(\eps,\alpha_s^2)\nonumber \\
 &=& \delta_{ij}\,\delta(1-x)
 \!-\!\le \frac{1}{\eps}\!-\!\gamma_E\!+\!\ln(4\pi)\!+\!\ln\frac{\mu^2}
 {M_f^2}\re\nonumber\\
 &\times&\frac{\alpha_s(\mu^2)}{2\pi}P_{j\leftarrow i}(x)
 +{\cal O}(\eps,\alpha_s^2)
\eea
remain. As in the case of parton distributions (see Sec.\ \ref{sec:phostr}),
the factorization theorem (Ellis {\it et al.}, 1979; Baier and Fey, 1979;
Altarelli {\it et al.}, 1979; Furmanski and Petronzio, 1980, 1982)
allows to absorb these singularities in the $\ms$-scheme
into renormalized fragmentation functions
\beq
 D_{H/i}(x,M_f^2)=\overline{D}_{H/i}(x)+\le
 \overline{\Gamma}_{j\leftarrow i}(M_f^2)\otimes\overline{D}_{H/j}\re (x),
 \label{eq:dqqren}
\eeq
where $x$ is the longitudinal momentum fraction of the hadron $H$ in the
parton $i$ and $\overline{D}_{H/i}(x),\,\overline{D}_{H/j}(x)$ are the bare
fragmentation functions. In this way one obtains important, yet incomplete,
perturbative information about the confinement of unobservable quarks and
gluons into observable hadrons. The fragmentation functions satisfy the
sum rules
\beq
 \sum_H\int_0^1\d xxD_{H/i}(x,M_f^2)=1,
 \label{eq:hadronsumrule}
\eeq
{\it i.e.} the momentum of the parton $i$ must be conserved after hadronization
into all available hadrons $H$.
Fragmentation functions constitute only the leading-twist contributions to
semi-inclusive hadron production in the operator product expansion.
Higher-twist, non-factorizable operators can also contribute to
the hadronization process. However, in the transverse/longitudinal and total
$e^+e^-$ cross sections they are suppressed by additional
factors of $1/Q$ and $1/Q^2$, respectively, and thus usually negligible
(Dasgupta and Webber, 1997).

The evolution of the fragmentation functions can be calculated by taking the
logarithmic derivative of Eq.\ (\ref{eq:dqqren}) with respect to the scale
$M_f^2\equiv Q^2$. This leads to the coupled homogeneous evolution
equations (Georgi and Politzer, 1978; Baier and Fey, 1979; Altarelli
{\it et al.}, 1979)
\bea
 \frac{\d D_{H/q}(Q^2)}{\d \ln Q^2} &=&
  \frac{\alpha_s(Q^2)}{2\pi}
  \left[ P_{q\leftarrow q} \otimes D_{H/q}(Q^2) \rp\nonumber\\
  &+&\lp      P_{g\leftarrow q} \otimes D_{H/g}(Q^2) \right] ,\nonumber\\
 \frac{\d D_{H/g}(Q^2)}{\d \ln Q^2} &=&
  \frac{\alpha_s(Q^2)}{2\pi}
  \left[ P_{q\leftarrow g} \otimes D_{H/q}(Q^2) \rp\nonumber\\
  &+&\lp      P_{g\leftarrow g} \otimes D_{H/g}(Q^2) \right],
 \label{eq:fragevoleq}
\eea
where the time-like splitting functions $P_{j\leftarrow i}$ are identical to
the space-like splitting functions in LO (Altarelli and Parisi, 1977), but
differ in NLO (Curci, Furmanski, and Petronzio, 1980; Furmanski and Petronzio,
1980; Floratos, Kounnas, and Lacaze, 1981).

The evolution equations can unfortunately not be solved analytically, and for
light hadrons an initial distribution similar to the
form $D_{H/i}(x,Q_0^2)=Nx^\alpha (1-x)^\beta$ has to be assumed at a low
starting scale $Q_0$. The free constants $N,\,\alpha$, and $\beta$ are then
fitted
to experimental data, usually from the process $e^+e^- \rightarrow \gamma^\ast
(q) \rightarrow H(p_H) X$, where the initial state and the total center-of-mass
energy
$Q^2=q^2$ are uniquely fixed. In the $\ms$ scheme,
the normalized NLO semi-inclusive hadron cross section is given by
(Baier and Fey, 1979; Altarelli {\it et al.}, 1979)
\bea
 \label{eq:inclhadxsec}
 \frac{1}{\sigma_{\rm tot}} \frac{\d\sigma(Q^2)}{\d x} &=& \sum_q 2
 \lg D_{H/q}(Q^2)+\sum_{i=T,L}
 \frac{\alpha_s(Q^2)}{2\pi} \rp \\
 &\times&\lp\le C_q^i \otimes D_{H/q}(Q^2)
 +   C_g^i \otimes D_{H/g}(Q^2)\re\rg, \nonumber
\eea
where
\beq
 \sigma_{\rm tot}=N_C\sum_qe_q^2\sigma_0\le1+\frac{\alpha_s(Q^2)}{\pi}\re
 \label{eq:tothadxsec}
\eeq
is the total hadronic cross section, $x=2(p_H\cdot q)/Q^2$ is the fraction
of the center-of-mass energy transferred to the observed hadron $H$, 
and $\sigma_0=4\pi\alpha^2/(3Q^2)$ is the total cross section for
$e^+e^-\rightarrow\mu^+\mu^-$. At larger $Q^2$, the couplings and propagators
in Eqs.\ (\ref{eq:inclhadxsec}) and (\ref{eq:tothadxsec}) are modified due to
additional $Z$-boson exchange.
The time-like transverse (T) and longitudinal (L) Wilson
coefficients are given by (Altarelli {\it et al.}, 1979)
\bea
 C_{q,T}^T &=& C_F\le\frac{3}{2}(1-x)-\frac{3}{2}\frac{1}{(1-x)}_+
  +2\frac{1+x^2}{1-x}\ln x\rp \nonumber \\
 &+&(1+x^2)\lr\frac{\ln(1-x)}{1-x}\rr_+
  +\lp\lr\frac{2\pi^2}{3~}-\frac{9}{2}\rr\delta(1-x)\re, \nonumber \\
 C_{q,L}^T &=& C_F, \nonumber \\
 C_{g,T}^T &=& C_F\le\frac{1+(1-x)^2}{x}\ln[x^2(1-x)]-\frac{2(1-x)}{x}\re ,
  \nonumber \\
 C_{g,L}^T &=& C_F\frac{2(1-x)}{x}.
 \label{eq:tlwilcoeff}
\eea
The transverse cross section starts at ${\cal O} (\alpha_s^0)$, while the
longitudinal cross section starts only at ${\cal O}(\alpha_s)$. Thus in NLO the
longitudinal ${\cal O}(\alpha_s^2)$ Wilson coefficients have to be included
for longitudinal cross sections, but not for total cross sections (Rijken and
van Neerven, 1996, 1997).

%
\begin{table}
\begin{center}
\begin{tabular}{|lll|cccl|}
 Group & Year & Hadron & $Q_0^2$ & Factor.\        & $N_f$ &
   $\Lambda_{\ms}^{N_f=4}$ \\
       &      &     & (GeV$^2$) &        Scheme & &  (MeV) \\
\hline
\hline
 BEP & 1979 &$\pi^{\pm,0},K^{\pm},p$ & 25.0 & LO  & 3             & 450/600  \\
 AKL & 1983 &$\pi^{\pm  },K^{\pm}  $ & 25.0 & LO  & 3             & 300/400  \\
 GR+ & 1993/5&$\pi^{\pm,0},\eta,K^{\pm,0}$& 900 & $\ms$ & 5  & 269/319 \\
 NW  & 1994  & $h^{\pm}$& $m_Z^2$ & $\ms$ & 5  &  344  \\
 BKK & 1995 &$\pi^{\pm  },K^{\pm}  $ &  2.00 & LO  & 5    & 190      \\
     &      &                        &    &$\ms$&         & 190      \\
     & 1995/8 &$\pi^{\pm  },K^{\pm,0},$&  2.00,  & LO & 5    & 146 (fit)\\
     &        &$D^{\ast\pm},B^{+,0 }$    &$4m_{c,b}^2$&$\ms$&           & 317 (fit)\\
 CGRT& 1997  & $D^{0},D^{\ast 0}$ & $m_c^2$ & $\ms$ & 5  &  151       \\
 KKP & 2000 &$\pi^{\pm},K^{\pm},p$   &  2.00& LO  & 5    & 121 (fit)\\
     &      &                        &   &$\ms$&            & 299 (fit)\\
Kretzer&2000&$\pi^{\pm},K^{\pm},h^{\pm}$   &  0.26& LO  & 5    & 175      \\
     &      &                              &  0.40&$\ms$&            & 246      \\
 BFGW& 2001 &$h^{\pm}$                     &  2.00   &$\ms$& 5    & 300      \\
\end{tabular}
\end{center}
\caption{\label{tab:fragfunc}Parameterizations of various hadron
 fragmentation functions.}
\end{table}
%
For a long time, only LO fragmentation functions for charged pions, kaons, and
protons were available, which had been fitted to $e^+e^-$ data from MARK I
(Baier, Engels, and Petersson [BEP], 1979) and had subsequently been updated
using deep-inelastic data from EMC (Anselmino, Kroll, and Leader [AKL], 1983).
Improved LO sets for pions, $\eta$'s, kaons, and $D$-mesons were obtained by
comparing HERWIG Monte Carlo predictions in the next-to-leading logarithmic
approximation with $e^+e^-$ data from TPC, PETRA, and later LEP
(Greco and Rolli, 1993, 1995;
Chiappetta {\it et al.}, 1994; Greco, Rolli, and Vicini [GR+], 1995; Nason and
Webber [NW], 1994; Cacciari {\it et al.} [CGRT], 1997).
Recently several independent sets have been obtained by a full NLO evolution of
a non-perturbative input at low starting scales and fits to precise
$e^+e^-$ data from MARK II, TPC, PETRA, AMY, SLD, and LEP
(Binnewies, Kniehl, and Kramer [BKK], 1995a, 1995b, 1996a, 1997,
1998a, 1998b; Kniehl, Kramer, and P\"otter [KKP], 2000a; Kretzer, 2000; Bourhis
{\it et al.} [BFGW], 2001). These data make it possible to
simultaneously determine $\alpha_s$ with a precision that is competitive with
the world average (Binnewies, Kniehl, and Kramer, 1995b; Kniehl, Kramer, and
P\"otter, 2000b). The extracted value of $\alpha_s$ could be in principle
correlated with
the size of power corrections (Nason and Webber, 1994), but the effect was
found to be negligible (Kniehl, Kramer, and P\"otter, 2000b).
According to the factorization theorem, fragmentation functions should be
universal, {\it i.e.} process independent. The validity of this assumption
has been verified by comparing inclusive hadron cross sections with $e^+e^-$
data not used for the fits at different scales $Q^2$ and with different hard
scattering processes like photon-photon, photon-hadron, and hadron-hadron
collisions (Binnewies, Kniehl, and Kramer, 1995b, 1996b, 1997, 1998a, 1998b;
Kniehl, Kramer, and P\"otter, 2001).
The parameterizations of the various fragmentation functions discussed
above are compared in Tab.\ \ref{tab:fragfunc}.
%
\begin{figure}
 \begin{center}
  \epsfig{file=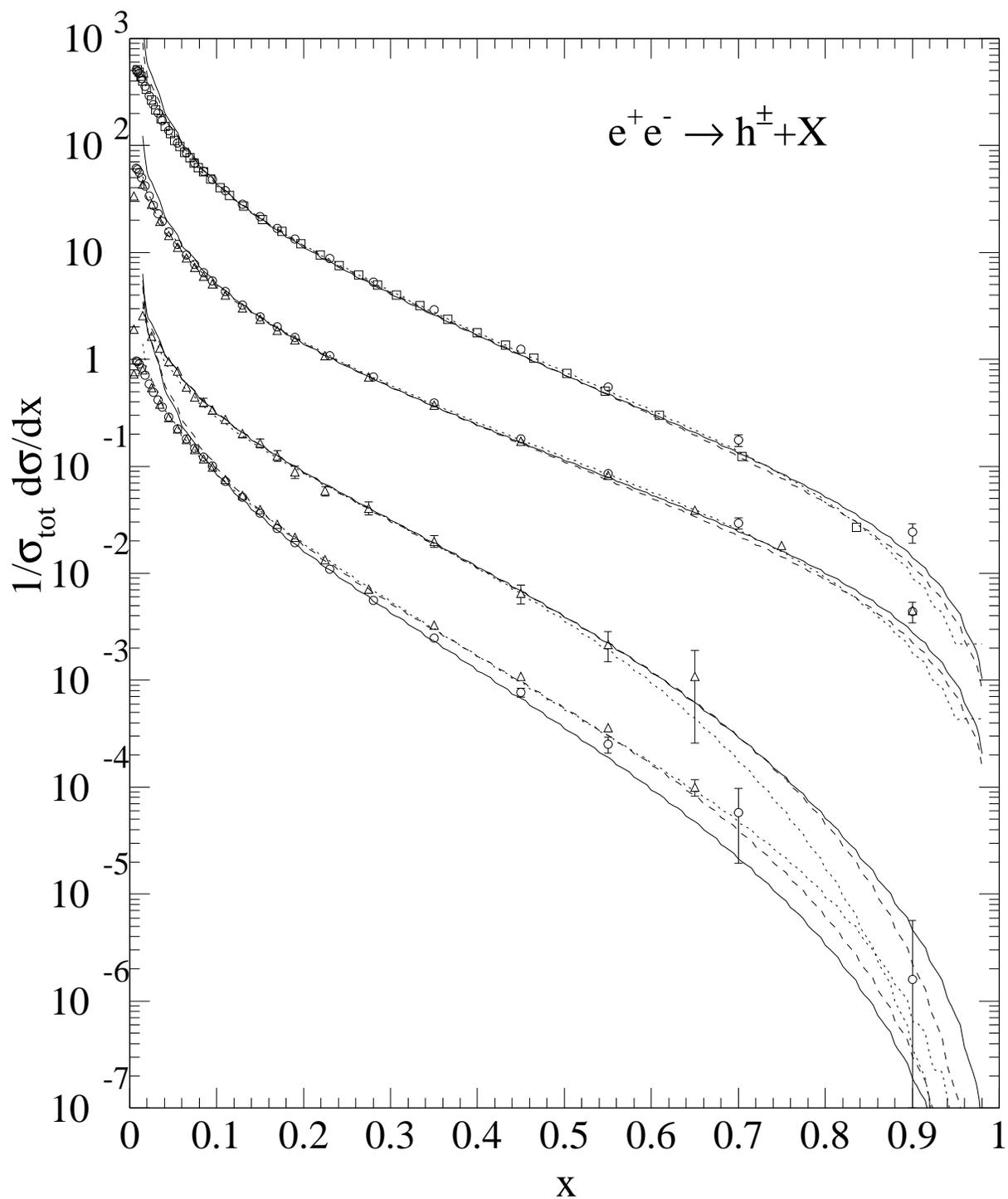,width=\linewidth}
 \caption{\label{fig:26}
 Dependence of the normalized inclusive hadron cross section at $\sqrt{S}=91.2$
 GeV on the scaled hadron momentum $x$. Data from DELPHI (triangles for 1991-93
 and circles for 1994 data) and SLD (squares) are compared with the NLO
 fragmentation functions of KKP (solid), Kretzer (dotted), and BFGW (dashed).
 Contributions from $b$, $c$, light,
 and all quarks (from bottom to top) are rescaled by factors of 1/5 relative
 to the nearest upper distribution (Kniehl, Kramer, and P\"otter, 2001).}
 \end{center}
\end{figure}
%
From Fig.\ \ref{fig:26} one observes that the three NLO fits
(KKP, Kretzer, and BFGW) are very similar for the fragmentation
of light quarks and all quarks, except for the theoretically unreliable
small-$x$ region ($x\leq 0.1$) and the experimentally badly constrained
large $x$-region ($x\geq 0.8$). For the fragmentation of $c$- and $b$-quarks
there are also differences at intermediate values of $x$. In the case of the
$b$-quarks the discrepancy is due to the fact that two incompatible data sets
from DELPHI (1991-93 and 1994 data) were used in the fits:
Table \ref{tab:fragfunccomp} shows that it is not possible to obtain good
values of $\chi^2$ per degree of freedom, $\chi^2_{\mathrm{DF}}$,
simultaneously for both data samples. Otherwise all three NLO
fits give satisfactory values of $\chi^2_{\mathrm{DF}}$.
%
\begin{table}
\begin{center}
\begin{tabular}{|r|c|c|r|r|r|r|}
Energy & Flavor & Experiment & \multicolumn{3}{c|}{FF Set} & No.\ of \\
\cline{4-6}
(GeV) & & & KKP  & Kretzer & BFGW & Points \\
\hline
29 & $uds$ & TPC  & $0.178^*$ & 0.159 & $0.167^*$ & 7 \\
   & $c$   &              & $0.876^*$ & 0.911 & $0.923^*$ & 7 \\
   & $b$   &              & $2.23^*$ & 1.21 & $1.14^*$ & 7 \\
\hline
91.2 & all   & DELPHI 94 & 1.28 & $1.51^*$ & 1.49 & 12 \\
     &       & SLD       & 1.32 & 0.370 & 0.421 & 21 \\
     & $uds$ & DELPHI 91-3  & $3.17^*$ & $0.990^*$ & 1.95 & 13 \\
     &       & DELPHI 94 & 0.201 & $0.588^*$ & $1.00^*$ & 12 \\
     & $c$   & DELPHI 91-3  & $0.473^*$ & $0.388^*$ & 0.401 & 11 \\
     & $b$   & DELPHI 91-3  & $28.9^*$ & $0.887^*$ & 1.03 & 12 \\
     &       & DELPHI 94 & 0.433 & $9.14^*$ & 8.74 & 12 \\
\hline
189  & all & OPAL  & $0.568^*$ & $0.250^*$ & $0.414^*$ & 11 \\
\end{tabular}
\end{center}
\caption{\label{tab:fragfunccomp}
$\chi^2_{\mathrm{DF}}$ values obtained in comparisons of the NLO fragmentation
functions by KKP, Kretzer, and BFGW to $e^+e^-$ data with $x>0.1$ at different
center-of-mass energies.
Data samples, which did not enter the respective fits, are marked by asterisks
(Kniehl, Kramer, and P\"otter, 2001).}
\end{table}
%

While the masses of light quarks and gluons can always safely be neglected
in the massless evolution approach, the masses $m_h$ of the heavier charm and
bottom quarks $h$ may become comparable to the physical scale $Q$.
The even heavier top-quark decays before it hadronizes. Bjorken (1978) argued,
that attaching a light antiquark $\bar{q}$ to a heavy quark $h$ should
decelerate the latter only slightly, so that it carries almost the same energy
as the hadron $H=(h\bar{q})$. As a consequence, a light quark fragmentation
function proposed by Field and Feynman (1978), which was peaked at rather low
$x$, was modified (Ali {\it et al.}, 1980a, 1980b). Later Peterson {\it et
al.} (1983) expanded the energies in the fragmentation amplitude
$1/(E_H-E_h+E_{\bar{q}})$ about the transverse particle masses with the result
\beq
 D_{H/h}(x)=\frac{N}{x[1-1/x-\eps_h/(1-x)]^2},
 \label{eq:peterson}
\eeq
where $\eps_h\propto m_q^2/m_h^2$ is a free parameter and $N$ is a
normalization factor constrained by the momentum sum rule Eq.\
(\ref{eq:hadronsumrule}). This fragmentation function peaks at
$x=1-\sqrt{\eps_h}$ and has a width $\sqrt{\eps_h}$ (Bodwin and Harris, 2001).
Alternative forms have been proposed by Kartvelishvili, Likhoded, and Petrov
(1978), Bowler (1981), Andersson {\it et al.} (1983), and Suzuki (1986).
Using Heavy Quark Effective Theory, Braaten, Cheung, and Yuan (1993a)
calculated the $\O(\alpha_s^2)$ fragmentation function
\bea
D_{B_c/b}(\!\!&x&\!\!,m_b+2m_c)=N\frac{rx(1-x)^2}{[1-(1-r)x]^6}
\left[6-18(1\!-\!2r)x\right.\nonumber\\
&+&(21\!-\!74r+68r^2)x^2-2(1\!-\!r)
(6\!-\!19r+18r^2)x^3\nonumber\\
&+&\left.3(1-r)^2(1-2r+2r^2)x^4\right]
\label{eq:bmesonfrag}
\eea
for $B_c$ mesons,
which depends on the mass ratio $r=m_c/(m_b+m_c)$. Like the non-perturbative
fragmentation functions discussed above, this result can also be used to
parameterize the input at the starting scale $Q_0$
for heavy-light mesons with the normalization $N$ and $r$ as free parameters
(Braaten {\it et al.}, 1995; Binnewies, Kniehl, and Kramer, 1998b). 
Eq.\ (\ref{eq:bmesonfrag}) peaks at large $x$ $(x\simeq 0.8)$ and vanishes
correctly like $(1-x)^2$ as $x\to 1$, whereas
a previous perturbative calculation by Collins and Spiller (1985) only vanishes
like $(1-x)$ and fails to describe recent ALEPH data (Heister {\it et al.},
2001). 

Matching a massless NLO calculation for $e^+e^-\to hX$ in the $\ms$ scheme to
a NLO calculation with $m_h$ as a regulator for the collinear singularity,
Mele and Nason (1991) calculated the functions
\bea
 D_{H/h}(x,Q^2)&=&\delta(1-x)\!+\frac{\alpha_s(Q^2)C_F}{2\pi}\le\frac{1+x^2}
 {1-x}\lr\ln\frac{Q^2}{m_h^2}\rp\rp\nonumber\\
 &-&\lp\lp 2\ln(1-x)-1\rr\re_+,\nonumber \\
 D_{H/g}(x,Q^2)&=&\frac{\alpha_s(Q^2)T_R}{2\pi}(x^2+(1-x)^2)\ln\frac{Q^2}{m_h^2}, \nonumber\\
 D_{H/q,\bar{q},\bar{h}}(x,Q^2)&=&{\cal O}(\alpha_s^2),
 \label{eq:melenason}
\eea
which reproduce the massive cross section when convolved with the massless hard
scattering cross section at the scale $Q$. For scales $Q\gg m_h$, however, the
logarithms in Eqs.\ (\ref{eq:melenason}) become large and have to be resummed.
In this case, these equations can be interpreted as boundary
conditions at a starting scale $Q_0$ and can be evolved to the scale $Q$.
Both approaches were found to give similar results in photoproduction
(Cacciari and Greco, 1996) and $e^+e^-$ annihilation (Nason and Oleari,
2000), but they yielded a satisfactory description of the data only when an
additional non-perturbative input $D_{H/i}(x,Q_0^2)=Nx^\alpha(1-x)^\beta$ 
(Cacciari {\it et al.}, 1997) or of the form in Eq.\ (\ref{eq:peterson})
(Cacciari and Greco, 1997; Nason and Oleari, 1999, 2000) was included.
An alternative solution consists in adding Eqs.\ (\ref{eq:melenason}) to
the massless cross section calculation with the interpretation of a change
of factorization scheme (Kniehl, Kramer, and Spira, 1997; Binnewies, Kniehl,
and Kramer, 1998a; Kramer, 1999). The fitted values of the non-perturbative
input depend, of course, on the treatment of the perturbative component.

The fragmentation of quarks and gluons into bound states of heavy quarks and
antiquarks $\Q=(h\overline{h})$ is based on the general factorization
analysis of Bodwin, Braaten, and Lepage (1995). It allows for the
fragmentation functions of quarkonia $\Q$ to be factorized into
calculable short-distance coefficients $d_i^n(x,Q^2)$, which describe the
production rate of a quark-antiquark pair with quantum numbers
$n=[\underline{a},^{2S+1}\!L_J]$ within a region of size $1/m_h$,
and long-distance operator matrix elements (OMEs) $\langle O^\Q(n) \rangle$,
which contain the non-perturbative dynamics responsible for the formation of
the bound state $\Q$ from the state $n$,
\beq
 D_{\Q/i}(x,Q^2)=\sum_n d_i^n(x,Q^2)\langle O^\Q(n) \rangle,\,\,\,
 i\in\{q,h,g\}.
\eeq
The size of the OMEs
may be estimated by how they scale with the relative velocity $v$
of the quarks inside the quarkonium state
(Lepage {\it et al.}, 1992). In the case of the 
charmonium state $J/\Psi$ the leading OME
\beq
 \OTSOO = \frac{9}{2\pi}|R_{J/\Psi}(0)|^2
\eeq
is related to the $J/\Psi$ radial wave function at the origin
$R_{J/\Psi}(0)$ (Braaten, Fleming, and Yuan, 1996). The subleading OMEs
$\OTSOE$, $\OOSZE$, and $\OTPJE$ follow at the relative order $\O (v^4)$.
Of particular importance is the transverse gluon coefficient
(Braaten and Yuan, 1994; Braaten and Lee, 2000; Ma, 1995)
\bea
 d_g^{[\underline{8},^3S_1]}(\!\!&x&\!\!,Q^2)=
 \frac{\pi\alpha_s(Q^2)}{24m_h^3}\lg \delta(1-x)
 +\frac{\alpha_s(Q^2)}
 {\pi} \rp\\
 &\times&\le\lr\beta_0\lr\ln\frac{Q}{2m_h}+\frac{13}{6}\rr
 +\frac{2}{3}-\frac{\pi^2}{2}+8\ln 2\rp\rp\nonumber\\
 &+&\lp 6\ln^22\rr\delta(1-x) 
 +\lr\ln\frac{Q}{2m_h}-\frac{1}{2}\rr P_{g\leftarrow g}(x)\nonumber\\
 &+&\lp\lp 6(2-x+x^2)\ln(1-x)-\frac{6}{x}\lr\frac{\ln(1-x)}{1-x}\rr_{\!+}\re\rg, 
 \nonumber
\eea
since it starts already at $\O(\alpha_s)$, while the gluon coefficient for the
leading OME $\OTSOO$ starts only at $\O(\alpha_s^3)$
(Braaten and Yuan, 1995). Since in the charmonium system
$v^2\simeq\alpha_s(m_c^2)\simeq 0.3$ (Quigg and Rosner, 1979), the
$\OTSOE$ contribution is
no longer suppressed with respect to the $\OTSOO$ contribution.
The perturbative coefficient for the fragmentation of heavy quarks into
the leading $[\underline{1},^3S_1]$ state is of $\O (\alpha_s^2)$
(Braaten, Cheung, and Yuan, 1993b; Ma, 1994).

For the $P$-wave $\chi_{c,J}$ states, the coefficients of the leading OMEs
$\langle O^{\chi_{c,J}}_1(^3P_J)\rangle$ with $J=0,\,1,\,2$
start at $\alpha_s^2$ for both gluons (Braaten and Yuan, 1994; Ma,
1995) and quarks (Ma, 1996a). The coefficients for the
subleading OMEs $\langle O^{\chi_{c,J}}_8(^3S_1)\rangle$
start at $\O (\alpha_s)$ for gluons (see above)
and $\O (\alpha_s^2)$ for quarks (Ma, 1996a), as do the quark (Braaten,
Cheung, and Yuan, 1993b; Ma, 1994) and gluon (Braaten and Yuan, 1993; Ma,
1994) coefficients for the leading OME in $\eta_c$ production
$\langle O^{\eta_c}_1(^1S_0)\rangle$.

\subsection{Light hadrons}
\label{sec:lighthad}

\begin{figure}
 \begin{center}
  \epsfig{file=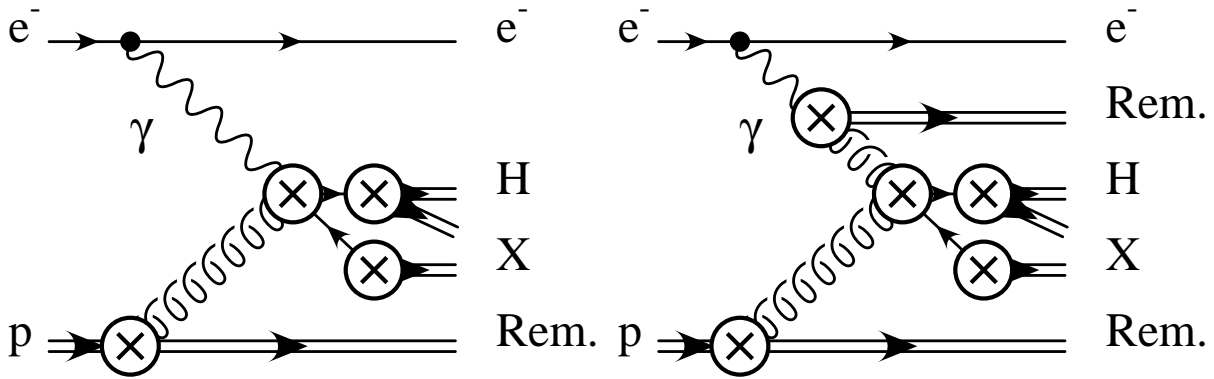,bbllx=60pt,bblly=360pt,bburx=280pt,bbury=430pt,%
          width=\linewidth}
 \caption{\label{fig:27}Factorization of photon-hadron scattering into
 hadrons.}
 \end{center}
\end{figure}

Like jets, hadrons can be produced in direct,
single-resolved, and double-resolved photon-photon scattering and in direct and
resolved photon-hadron scattering. As can be seen in Fig.\ \ref{fig:27} for
%
%
the photoproduction case, there is a new factorization in the final state
for the transition of partons $c$ into observed hadrons $H$, which is described
by fragmentation functions $D_{H/c}$. In the hadronic cross section
\bea
 \frac{\d^2\sigma}{\d p_T^2\d y}&=&
 \sum_{a,b,c}\int \d x_a \d x_b {\d z\over z^2}\,
 f_{a/A}(x_a,M_a^2)f_{b/B}(x_b,M_b^2)\nonumber\\
 &\times& D_{H/c}(z,M_c^2)
 \frac{\d\sigma}{\d t},
 \label{eq:hadronxsec}
\eea
the fragmentation function is convolved at a factorization scale $M_c$ with
the partonic cross section Eq.\ (\ref{eq:totalnloxsec}) and the initial state
parton densities $f_{a/A}$ and $f_{b/B}$, which were already encountered in
the jet production case.

%
\begin{figure}
 \begin{center}
  \epsfig{file=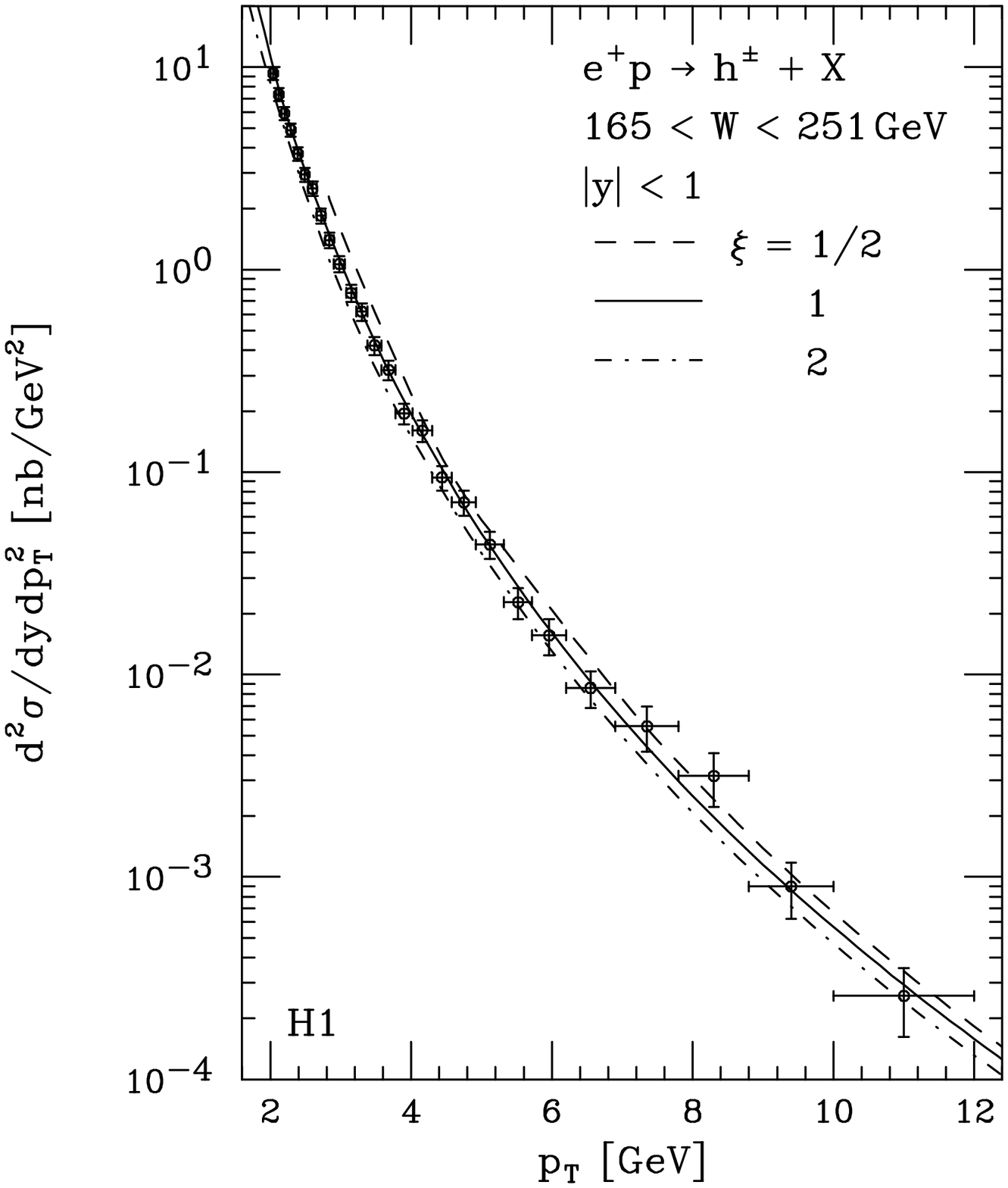,width=0.495\linewidth}
  \epsfig{file=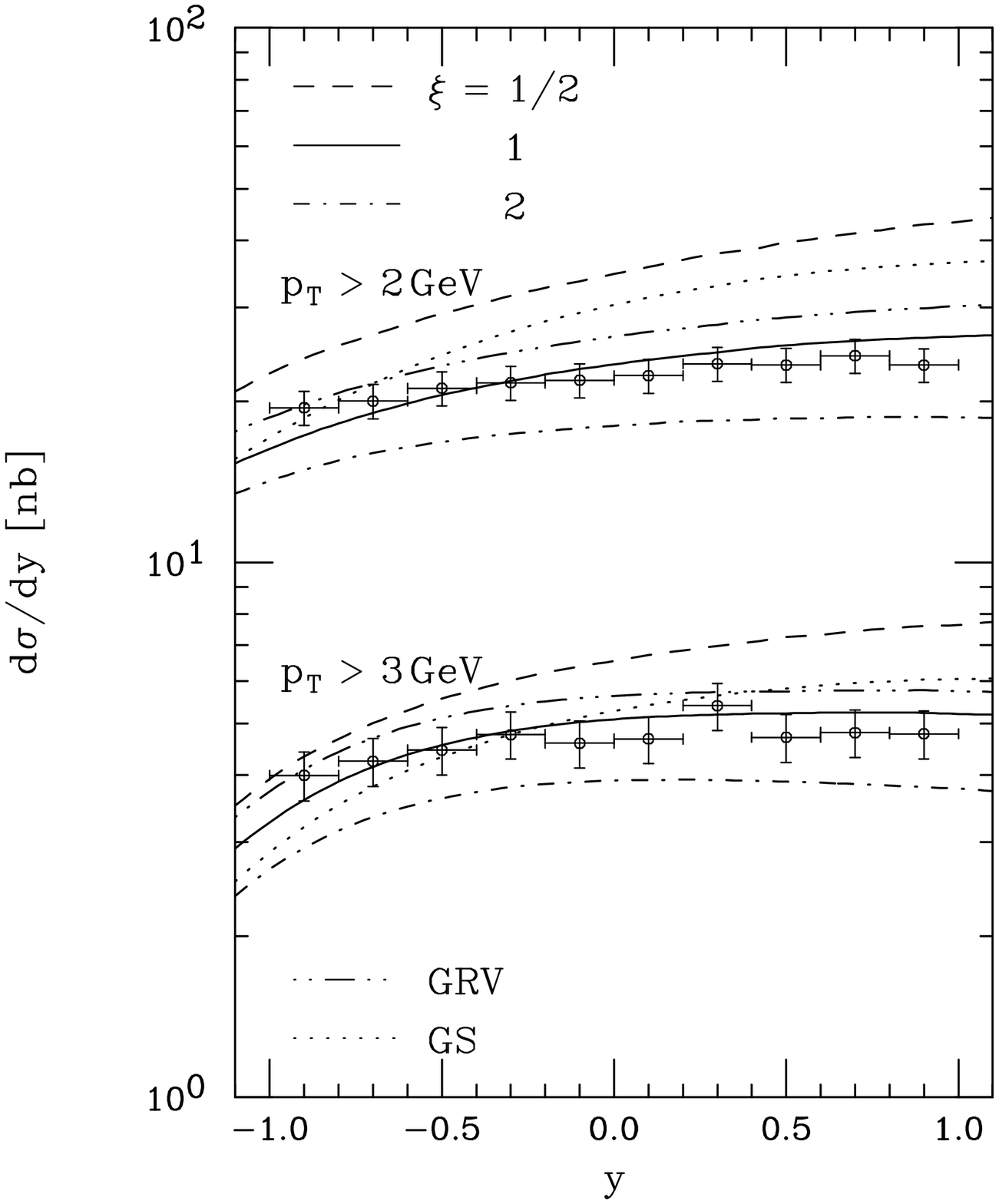,width=0.49\linewidth}
 \caption{\label{fig:28}
Transverse momentum (left) and rapidity (right) dependence of the NLO
inclusive charged hadron photoproduction cross section (Kniehl, Kramer, and
P\"otter, 2001) compared to H1 data (Adloff {\it et al.}, 1999a).}
 \end{center}
\end{figure}
%
The potential of inclusive hadron production for distinguishing the direct and
resolved contributions was stressed already in early LO calculations for
photon-photon (Brodsky {\it et al.}, 1979) and photon-hadron
scattering (Fontannaz {\it et al.}, 1980). For light hadrons, the calculation
of Born, virtual loop, and real emission processes proceeds as described in
Sec.\ \ref{sec:singlejets}. There is, however, an important difference. Since
the degeneracy of the final state is incomplete due to the observation of an
identified hadron, the collinear singularity associated with the real
emission of a massless parton $3$ from the observed final parton line
$2^\ast$
\beq
 |\M^F|^2_{ab\rightarrow 123} = -\ede P_{2\leftarrow 2^\ast}(x)
  |\M^B|^2_{ab\rightarrow 12^\ast}+\O(\eps^0)
\label{eq:hadronsing}
\eeq
remains uncancelled and has to be factorized into the fragmentation function
(see Sec.\ \ref{sec:hadfrag}). NLO calculations based on the formalism of
Ellis {\it et al.} (1980) have been performed for direct (Khalafi, Landshoff,
and Stirling, 1983; Aurenche {\it et al.}, 1985a; Gordon, 1994),
single-resolved (Aurenche {\it et al.}, 1984b, 1987; Gordon, 1994), and
double-resolved (Aversa {\it et al.}, 1989) light hadron production, and their
dependences on the renormalization and factorization scales were found to be
considerably reduced with respect to the LO calculations (Borzumati, Kniehl,
and Kramer, 1993; Kniehl and Kramer, 1994; Gordon, 1994;
Greco, Rolli, and Vicini, 1995). The NLO calculations
were then used to demonstrate the universality of NLO fragmentation functions
for charged hadrons, neutral pions, and neutral kaons (Binnewies, Kniehl, and
Kramer, 1995b, 1996a, 1996b; Binnewies, 1997; Kniehl, 1997; Kniehl, Kramer, and
P\"otter, 2001)  by comparing them with photoproduction data from H1 (Abt {\it
et al.}, 1994; Linsel, 1995; Adloff {\it et al.}, 1997, 1999a) and ZEUS
(Derrick {\it et al.}, 1995c; Breitweg {\it et al.}, 1998f) and
photon-photon data from
TASSO (Brandelik {\it et al.}, 1981), MARK II (Cords {\it et al.}, 1993), L3
(Achard {\it et al.}, 2001), and OPAL (Ackerstaff {\it et al.}, 1999).

In Fig.\ \ref{fig:28}, NLO calculations of the inclusive charged hadron
photoproduction cross section at HERA with AFG photon and CTEQ5M
(Lai {\it et al.}, 2000) proton densities and three different
choices of the scale factor $\xi=\mu/p_T=M_\gamma/p_T=M_p/p_T$ (Kniehl, Kramer,
and P\"otter, 2001) are compared to H1 data (Adloff {\it et al.}, 1999a). The
transverse momentum spectrum (left) agrees well with the data in shape and
normalization. The rapidity spectrum (right) is sensitive to variations of the
photon densities (AFG, GRV, or GS96), particularly for the lower $p_T$ cut,
but the scale uncertainty is still large.

\begin{figure}
 \begin{center}
  \epsfig{file=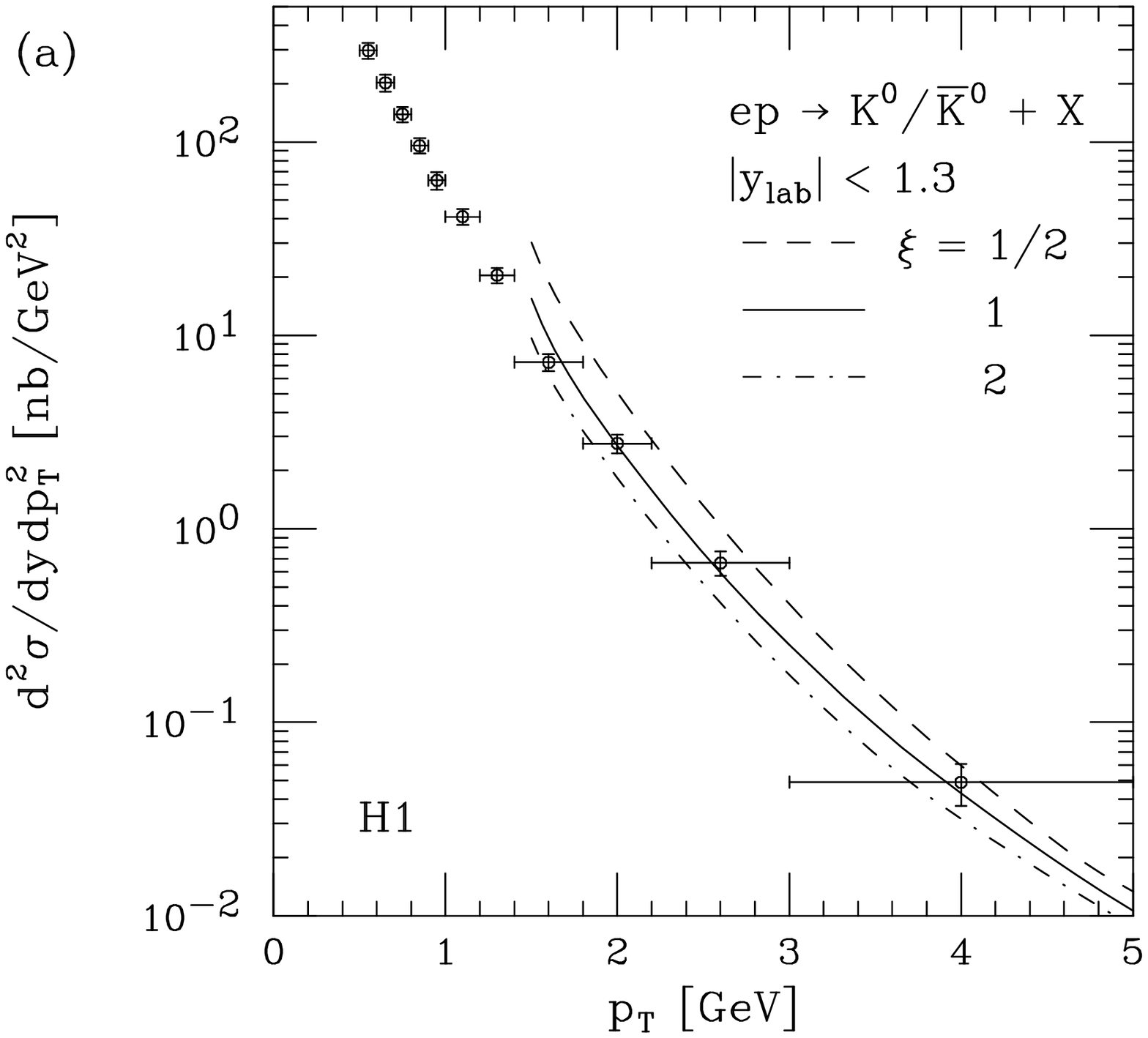,width=0.49\linewidth}
  \epsfig{file=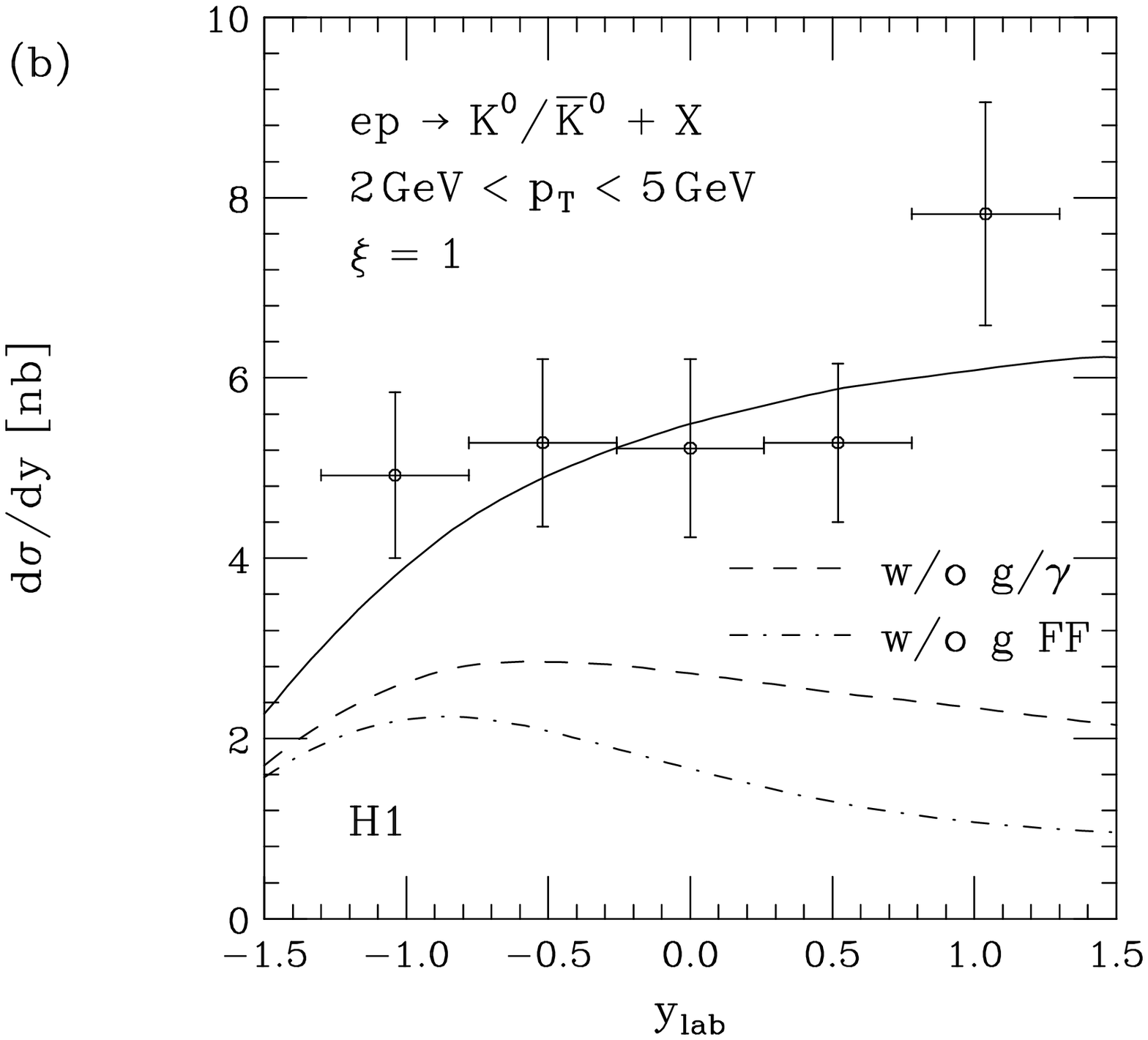,width=0.495\linewidth}
 \caption{\label{fig:29}
Transverse momentum (left) and rapidity (right) dependence of the NLO
inclusive neutral kaon photoproduction cross section (Kniehl, 1997)
compared to H1 data (Adloff {\it et al.}, 1997).}
 \end{center}
\end{figure}

A similar comparison is made in Fig.\ \ref{fig:29} for the photoproduction of
%
%
%
\begin{figure}
 \begin{center}
  \epsfig{file=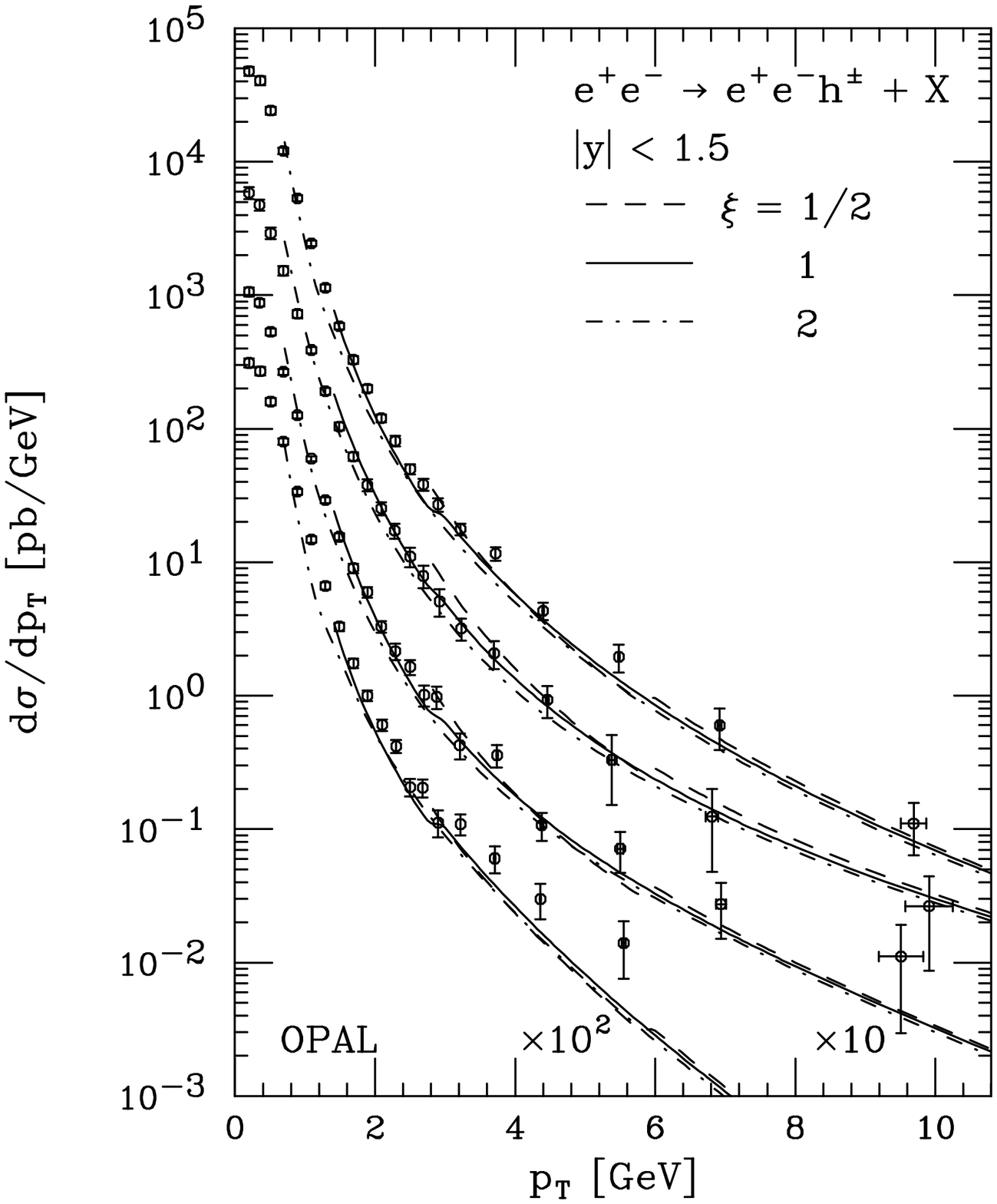,width=0.485\linewidth}
  \epsfig{file=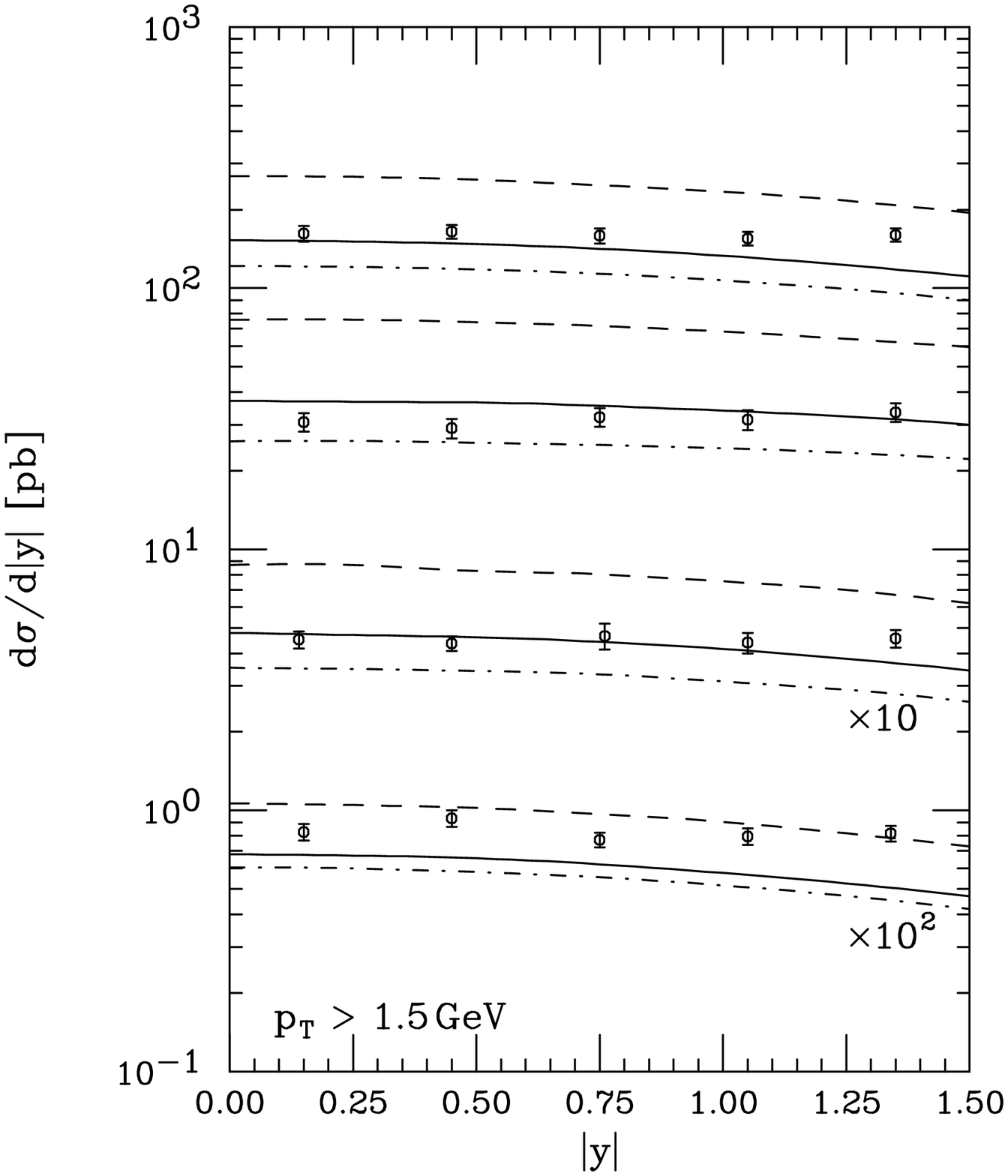,width=0.50\linewidth}
 \caption{\label{fig:30}
Transverse momentum (left) and rapidity (right) dependence of the NLO
inclusive charged hadron cross section in photon-photon collisions (Kniehl,
Kramer, and P\"otter, 2001), integrated over $\gamma\gamma$ invariant-mass
intervals $10<W<30$~GeV, $30<W<55$~GeV, $55<W<125$~GeV, and $10<W<125$~GeV
(from bottom to top in this order) and compared to OPAL data (Ackerstaff {\it
et al.}, 1999).}
 \end{center}
\end{figure}
%
neutral kaons. Due to the larger kaon mass, the massless NLO calculation
describes the measured $p_T$ spectrum (left) only at larger $p_T$
$(p_T\!>\!1.5$ GeV). The cut on $p_T$ for the rapidity spectrum (right,
$p_T\!>\!2$ GeV) may still be too low
for quantitative conclusions, but the potential to constrain $f_{g/\gamma}$
and $D_{K^0/g}$ is clearly visible.

NLO charged hadron cross sections in photon-photon collisions (Kniehl, Kramer,
and P\"otter, 2001) are shown in Fig.\ \ref{fig:30} and compared to
OPAL data (Ackerstaff {\it et al.}, 1999). Apart from the lowest $p_T$ points,
which are below the starting scale of the fragmentation functions, the
massless calculations describe the transverse momentum spectra (left) quite
well. The scale variations in the rapidity spectra (right) are again
significantly larger than the experimental errors and the sensitivity
to the photon structure (not shown).

\subsection{Heavy hadrons}

If the transverse momentum $p_T$ of a produced hadron is much larger than
its mass, the formalism described in the last Section can be applied
also to the production of heavy hadrons. The heavy quark is then considered to
be one of the active flavors in the evolution of parton densities (see Sec.\
\ref{sec:phostr}) and fragmentation functions (see Sec.\ \ref{sec:hadfrag}),
and the heavy quark mass $m_h$ is neglected in the hard scattering cross
section. It is only kept as the starting scale in the evolution
of the heavy quark parton
densities and fragmentation functions. NLO calculations in the massless
scheme have been performed for photoproduction (Cacciari and Greco, 1996,
1997; Kniehl {\it et al.}, 1995; Kniehl, Kramer, and Spira, 1997; Binnewies,
Kniehl, and Kramer, 1997, 1998; Kramer, 1999) and photon-photon collisions
(Cacciari {\it et al.}, 1996; Kramer and Spiesberger, 2001) with different
assumptions on the perturbative and non-perturbative components of the
fragmentation function (see Sec.\ \ref{sec:hadfrag}). They  have been
compared with $D^\ast$ photoproduction data from H1 (Aid {\it et al.}, 1996b)
and ZEUS (Breitweg {\it et al.}, 1997, 1999b; Derrick {\it et al.}, 1997b) and
photon-photon data from L3 (Acciarri {\it et al.},
1999a) and OPAL (Abbiendi {\it et al.}, 2000).

\begin{figure}
 \begin{center}
  \epsfig{file=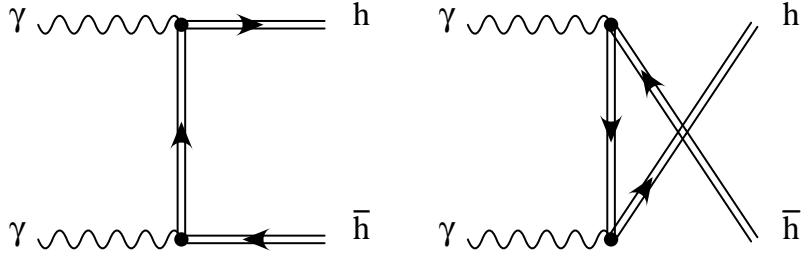,bbllx=60pt,bblly=360pt,bburx=266pt,bbury=431pt,%
          width=0.66\linewidth}
 \caption{\label{fig:31}Born diagrams for direct heavy-quark production.}
 \end{center}
\end{figure}

\begin{figure}
 \begin{center}
  \epsfig{file=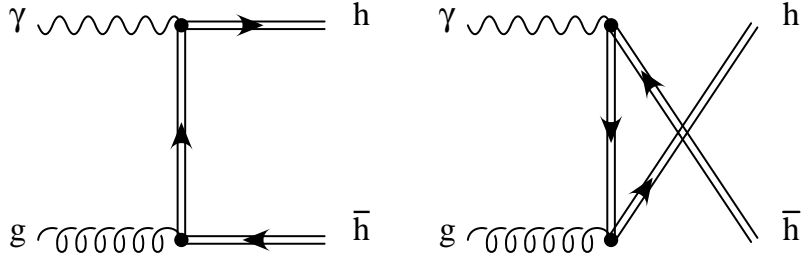,bbllx=60pt,bblly=360pt,bburx=266pt,bbury=431pt,%
          width=0.66\linewidth}
 \caption{\label{fig:32}Born diagrams for single-resolved heavy-quark
           production.}
 \end{center}
\end{figure}

\begin{figure}
 \begin{center}
  \epsfig{file=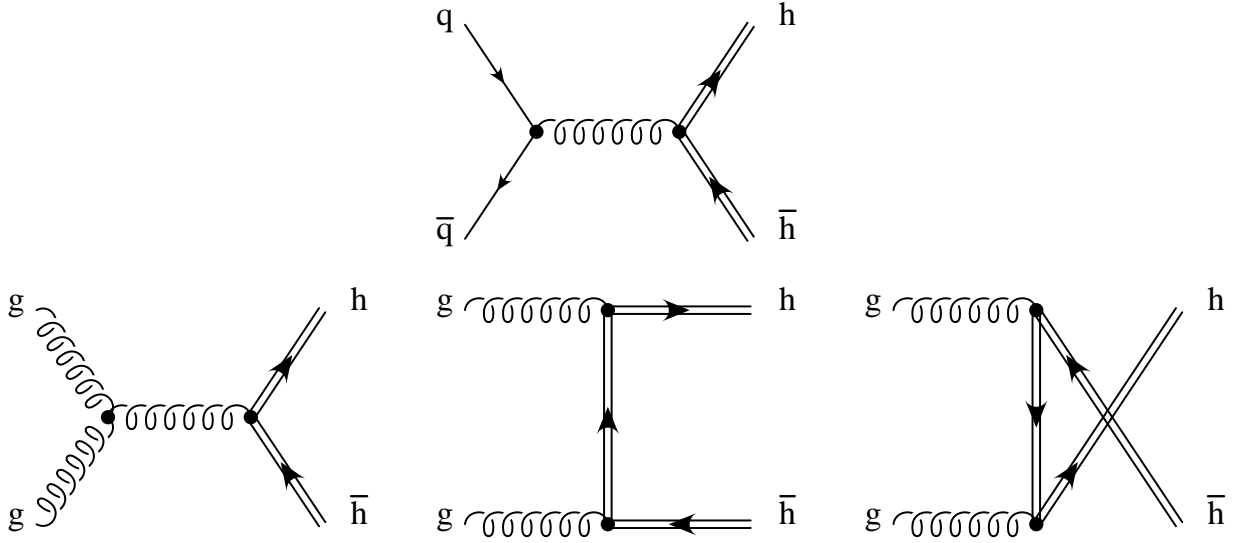,bbllx=60pt,bblly=287pt,bburx=373pt,bbury=431pt,%
          width=\linewidth}
 \caption{\label{fig:33}Born diagrams for double-resolved heavy-quark
           production.}
 \end{center}
\end{figure}

If, on the other hand, the transverse momentum $p_T$ is of the same order as
the mass of the produced hadron, the latter has to be kept in the partonic
cross section
\bea
 \frac{\d\sigma^B}{\d t}&=&\frac{1}{2s}
 \frac{1}{\Gamma(1-\varepsilon)}
  \left( \frac{4\pi s}{(t-p_2^2)(u-p_2^2)-p_2^2s} \right) ^\varepsilon
  \frac{1}{8\pi s} \nonumber \\
 &\times& \frac{g_{a,b}^2}{S_aS_bC_aC_b}
 |\M^B|^2.
\label{eq:massivepartonicxsec}
\eea
In this case, the heavy flavors are not active in the parton densities and
fragmentation functions, and only the Born diagrams in Figs.\
\ref{fig:31}, \ref{fig:32}, and Fig.\ \ref{fig:33} for direct,
single-, and double-resolved photon-photon scattering contribute.
%
%
%
%
%
%
The corresponding matrix elements are collected in Tab.\ \ref{tab:hqloxsec}.
%
\begin{table}
\begin{center}
\begin{tabular}{|l|l|}
Process & LO matrix element squared $|\M^B|^2$ \\
\hline
$\yythh$& $8N_C\le\frac{t_2}{u_2}+\frac{u_2}{t_2}+\frac{4p_2^2s}{t_2u_2}
 \lr 1-\frac{p_2^2s}{t_2u_2}\rr\rp $ \\
        & $+\eps\lr -1+\frac{s^2}{t_2u_2}\rr
           \lp +\eps^2\frac{s^2}{4t_2u_2}\re$ \\
$\ygthh$& $C_F|\M^B|^2_{\yythh}(s,t,u)$ \\
$\qbthh$& $4N_CC_F\le\frac{t_2^2+u_2^2}{s^2}+\frac{2p_2^2}{s}+\frac{\eps}{2}
\re $ \\
$\ggthh$& $\le\frac{N_CC_F}{2}\lr 1-2\frac{t_2u_2}{s^2}\rr-\frac{C_F}{2N_C}\re
 |\M^B|^2_{\yythh}(s,t,u)$ \\
\end{tabular}
\end{center}
\caption{\label{tab:hqloxsec}LO matrix elements squared $|\M^B|^2$ for
 $2\rightarrow 2$ parton processes involving two, one, and no photons
 in $n=4-2\eps$ dimensions. $t=t-p_2^2$ and $u=u-p_2^2$ are mass-subtracted
 Mandelstam variables.}
\end{table}
%
Figs.\ \ref{fig:32} and \ref{fig:33} and the respective matrix
elements apply also to direct and resolved photoproduction.
In LO calculations (Fritzsch and Streng, 1978) one can of course set $\eps=0$,
but not if NLO corrections
\bea
 \frac{\d\sigma^{V,F,I}}{\d t}&=&\frac{1}{2s}
 \frac{1}{\Gamma(1-\varepsilon)}
  \left( \frac{4\pi s}{(t-p_2^2)(u-p_2^2)-p_2^2s} \right) ^\varepsilon
  \frac{1}{8\pi s} \nonumber \\
 &&\hspace*{-13mm}\times \frac{g_{a,b}^2}{S_aS_bC_aC_b}
 \frac{\alpha_s}{2\pi}
  \left( \frac{4\pi\mu^2}{Q^2} \right) ^\varepsilon
  \frac{\Gamma(1-\varepsilon)}{\Gamma(1-2\varepsilon)}
 |\M^{V,F,I}|^2
 \label{eq:massivenloxsec}
\eea
are included. The heavy-flavor mass has to be kept in the denominators
of the virtual loop integrals and in the integrations over the solid angle
(Eq.\ (\ref{eq:solidangle})) and invariant mass
\beq
 \int_0^{s+t-p_1^2+{p_2^2s\over (t-p_2^2)}} \d s_{ij} {(s_{ij}-p_1^2)^{1-2\eps}\over s_{ij}^{1-\eps}}
\eeq
of the unresolved final state particles.
The presence of a mass usually complicates the integrals that have to be
computed, but the singularity structure is generally simpler than in
the massless case, since the collinear singularities are regulated by the
heavy-flavor
mass. The method of analytic integration was originally developed for
total and single-differential cross sections of heavy-flavor production in
hadron collisions (Dawson, Ellis, and Nason, 1988, 1989; Beenakker {\it et
al.}, 1989, 1991), but it was soon applied to photon-hadron (Ellis and Nason,
1989; Smith and van Neerven, 1992) and photon-photon collisions
(K\"uhn, Mirkes, and Steegborn, 1993; Drees {\it et al.}, 1993; Contogouris,
Kamal, and Merebashvili, 1995).
Later different methods based on numerical cancellations of divergencies
(Mangano, Nason, and Ridolfi, 1992) and phase-space slicing (Giele, Keller,
and Laenen, 1996) were designed for more differential cross sections in hadron
collisions. They were also subsequently extended to photon-hadron (Frixione
{\it et al.}, 1994a, 1994b, 1995; Frixione, Nason, and Ridolfi, 1995) and
photon-photon collisions (Kr\"amer and Laenen, 1996; Frixione, Kr\"amer, and
Laenen, 2000a, 2000b). The massless limits of the massive NLO results
have recently been compared with massless NLO calculations
(Cacciari, Frixione, and Nason, 2001; Kramer and Spiesberger, 2001).

After convolution with various non-perturbative
fragmentation functions, the massive NLO calculations were confronted with
measurements of total charm cross sections, $D$-meson $p_T$ and $y$
distributions, and $D\bar{D}$ azimuthal correlations by the Photon Emulsion
(Adamovich {\it et al.}, 1980, 1987), NA14/2 (Alvarez {\it et al.}, 1992,
1993),
E687 (Frabetti {\it et al.}, 1993, 1996; Moroni {\it et al.}, 1994), E691
(Anjos {\it et al.}, 1989, 1990), H1 (Aid {\it et al.}, 1996b), and ZEUS
(Breitweg {\it et al.}, 1997, 1999b, 2000e) photon-hadron experiments and by the
JADE (Bartel {\it et al.}, 1987), TASSO (Braunschweig {\it et al.}, 1990),
TPC/2$\gamma$ (Alston-Garnjost {\it et al.}, 1990),
AMY (Takashimizu {\it et al.}, 1996), TOPAZ (Enomoto {\it et al.}, 1994a,
1994b), ALEPH (Buskulic {\it et al.}, 1995), L3 (Acciarri {\it et al.}, 1999b,
2001a, 2001b), and OPAL (Abbiendi {\it et al.}, 2000, 2001a) photon-photon
experiments.

\begin{figure}
 \begin{center}
  \epsfig{file=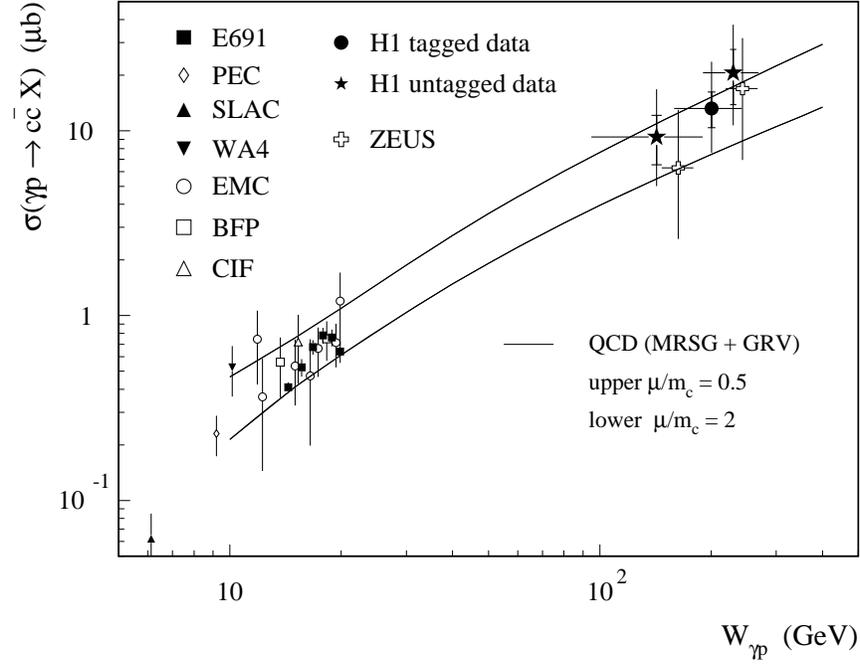,width=0.7\linewidth}
  \epsfig{file=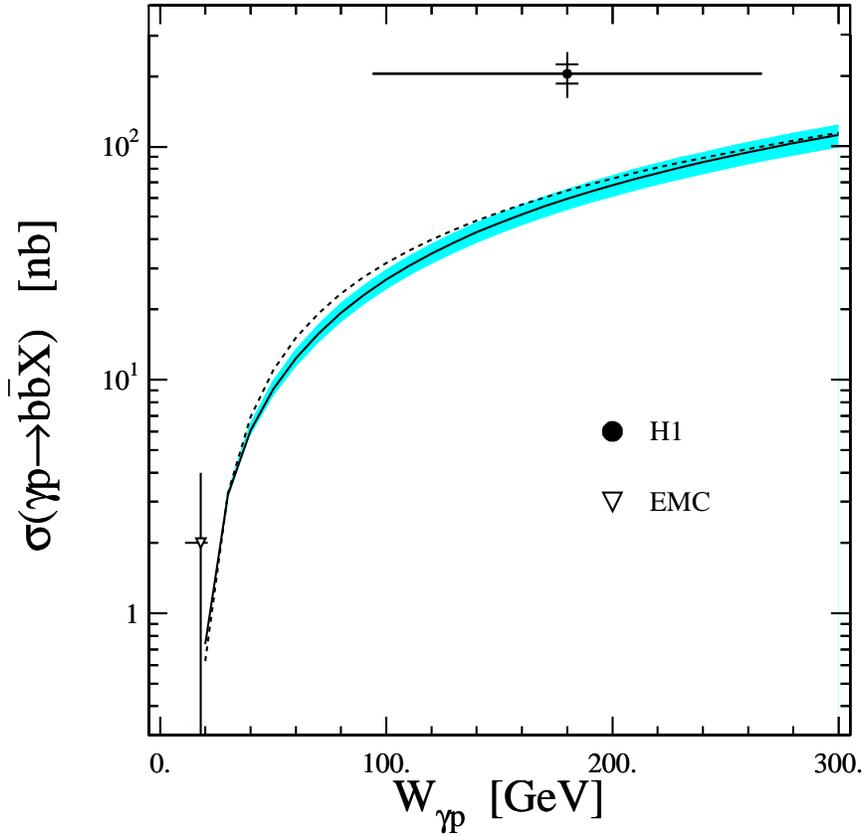,width=0.7\linewidth}
 \caption{\label{fig:34}Photon-proton center-of-mass energy
 dependence of the massive NLO total charm (top) and bottom (bottom) cross
 sections (Frixione {\it et al.}, 1994a) compared to fixed
 target and HERA data (Aid {\it et al.}, 1996b; Adloff {\it et al.}, 1999b).}
 \end{center}
\end{figure}

\begin{figure}
 \begin{center}
  \epsfig{file=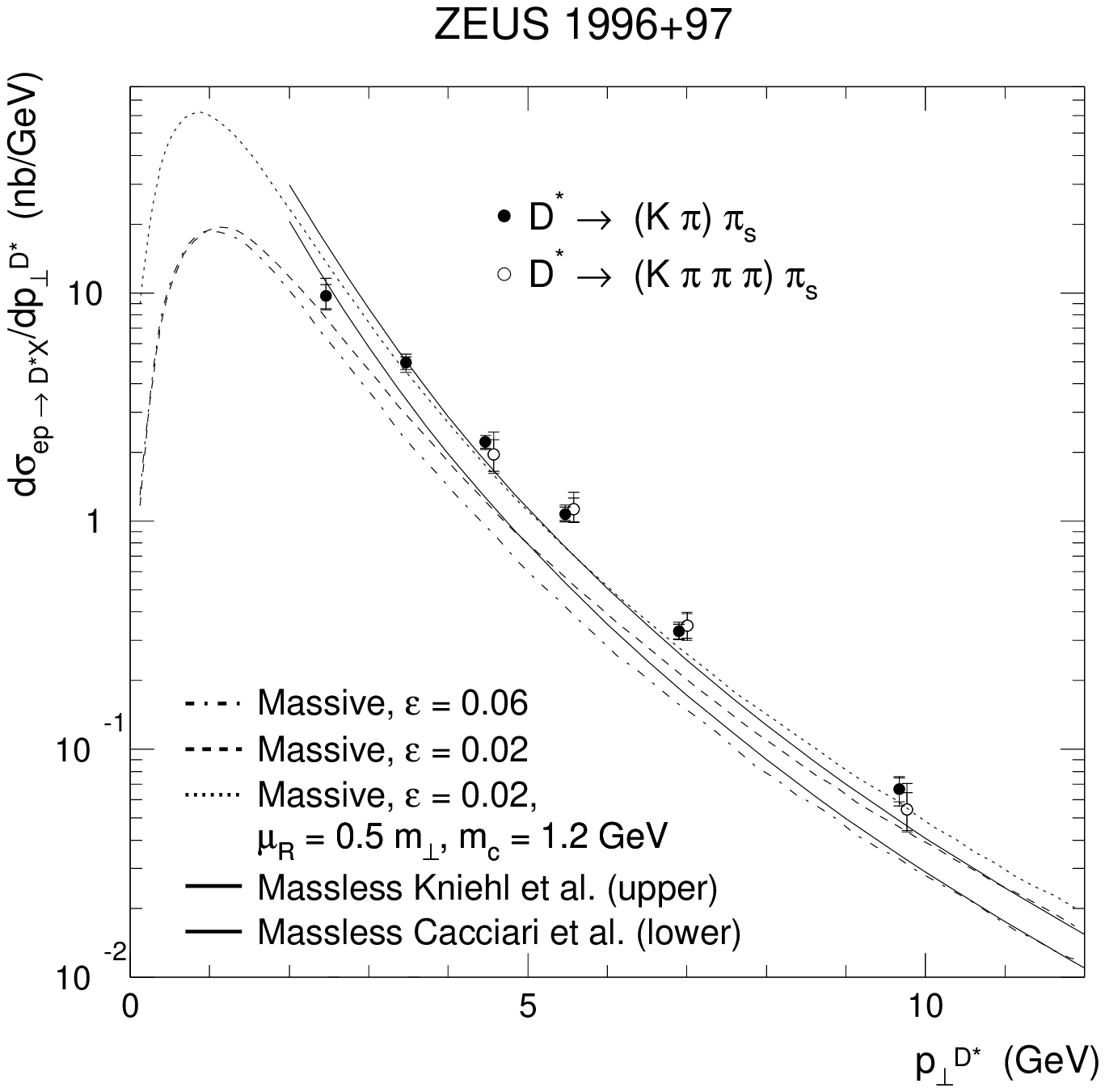,width=0.6\linewidth}
  \epsfig{file=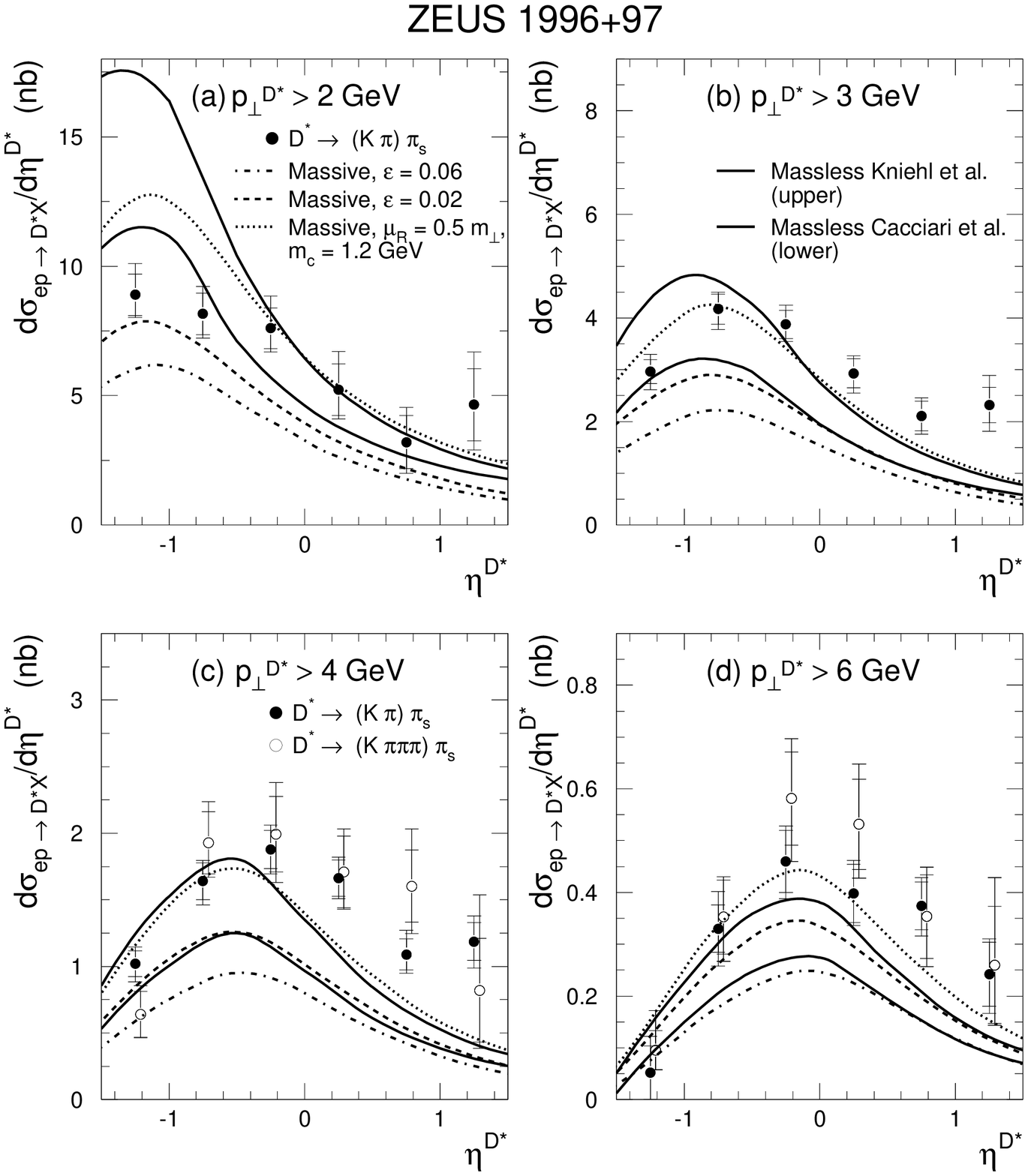,width=0.6\linewidth}
 \caption{\label{fig:35}Transverse momentum (top) and pseudorapidity
 (bottom) dependence of the massive (Frixione {\it et al.}, 1994a) and two
 different massless (Binnewies, Kniehl, and Kramer, 1997; Cacciari and Greco,
 1997) NLO $D^\ast$ photoproduction cross sections
 compared to ZEUS data (Breitweg {\it et al.}, 1999b).}
 \end{center}
\end{figure}

The massive NLO total charm cross section (Frixione {\it et al.}, 1994a)
shown in Fig.\ \ref{fig:34} (top)
%
%
describes the old fixed-target and the new HERA photoproduction data well
within the scale uncertainty (Aid {\it et al.}, 1996b). The H1 measurement
(Adloff {\it et al.}, 1999b) of the total bottom cross section in Fig.\
\ref{fig:34} (bottom) is, however, severely underestimated
by the massive NLO calculation, as are the corresponding ZEUS measurement
(Breitweg {\it et al.}, 2001) and the hadroproduction measurements from CDF
(Abe {\it et al.}, 1993, 1995) and D0 (Abbott {\it et al.}, 2000a, 2000b).
It should be noted that the NLO predictions are
rather sensitive to the photon (GRV) and proton (Martin, Roberts, and Stirling,
1993) parton densities, and to a lesser degree also to the fragmentation model
and the masses of the charm ($m_c=1.5$ GeV) and bottom quarks ($m_b=4.75$
GeV), but these variations cannot account for the discrepancy in the bottom
cross section (Blair {\it et al.}, 2000).
NLO massless calculations (Binnewies, Kniehl, and  Kramer, 1997; Cacciari and
Greco, 1997) cannot be applied to total cross sections due to the absence of
a hard scale. They can, however, be applied to $D^\ast$ transverse momentum
(top) and pseudorapidity (bottom) distributions as shown in Fig.\
\ref{fig:35}.
%
%
Massless distributions are usually larger than their massive counterparts, and
they describe the experimental data better (Breitweg {\it et al.}, 1999b),
particularly in the large $p_T$ region, where the massive calculation only fits
the data, if low values for $m_c=1.2$ GeV, the renormalization scale $\mu=0.5
\cdot m_T$, and the Peterson fragmentation function parameter $\eps=0.02$ are
chosen. The shapes of the measured pseudorapidity distributions are not well
described by either the massive or the massless calculations. This fact has
attracted considerable theoretical effort, from the consideration of ``drag
effects'' due to the beam remnants (Norrbin and Sj\"ostrand, 2000) to
$c\bar{c}$ vector meson components of the photon wave function (Berezhnoy
and Likhoded, 2001) and heavy-light quark recombination models inspired from
heavy quarkonia (Berezhnoy, Kiselev, and Likhoded, 2000; Braaten, Jia, and
Mehen, 2001), which may also have relevance for asymmetries in $B$-physics.
The cut on
$p_T>2$ GeV is probably too close to $m_c$ for the massless calculations to be
reliable. In the remaining $p_T$ regions, different charm densities in the
photon allow for a somewhat improved description of the data (Breitweg {\it et
al.}, 1999b).

%
\begin{figure}
 \begin{center}
  \epsfig{file=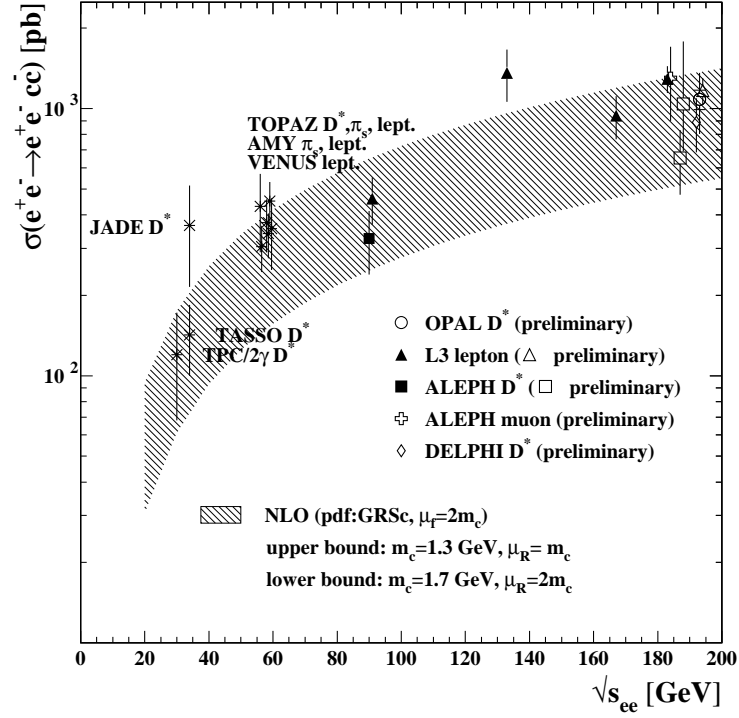,width=0.6\linewidth}
  \epsfig{file=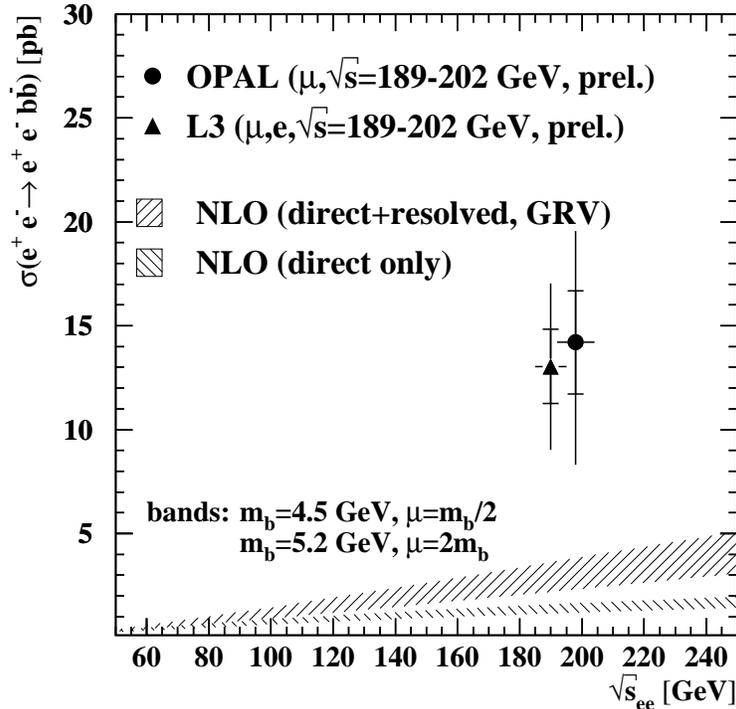,width=0.6\linewidth}
 \caption{\label{fig:36}$e^+e^-$ center-of-mass energy dependence
 of the massive NLO total charm (top) and bottom (bottom) cross sections in
 photon-photon collisions (Drees {\it et al.}, 1993) compared to
 TPC/2$\gamma$, PETRA, TRISTAN, and LEP data (Abbiendi {\it et al.}, 2001a,
 2001b).}
 \end{center}
\end{figure}
\begin{figure}
 \begin{center}
  \epsfig{file=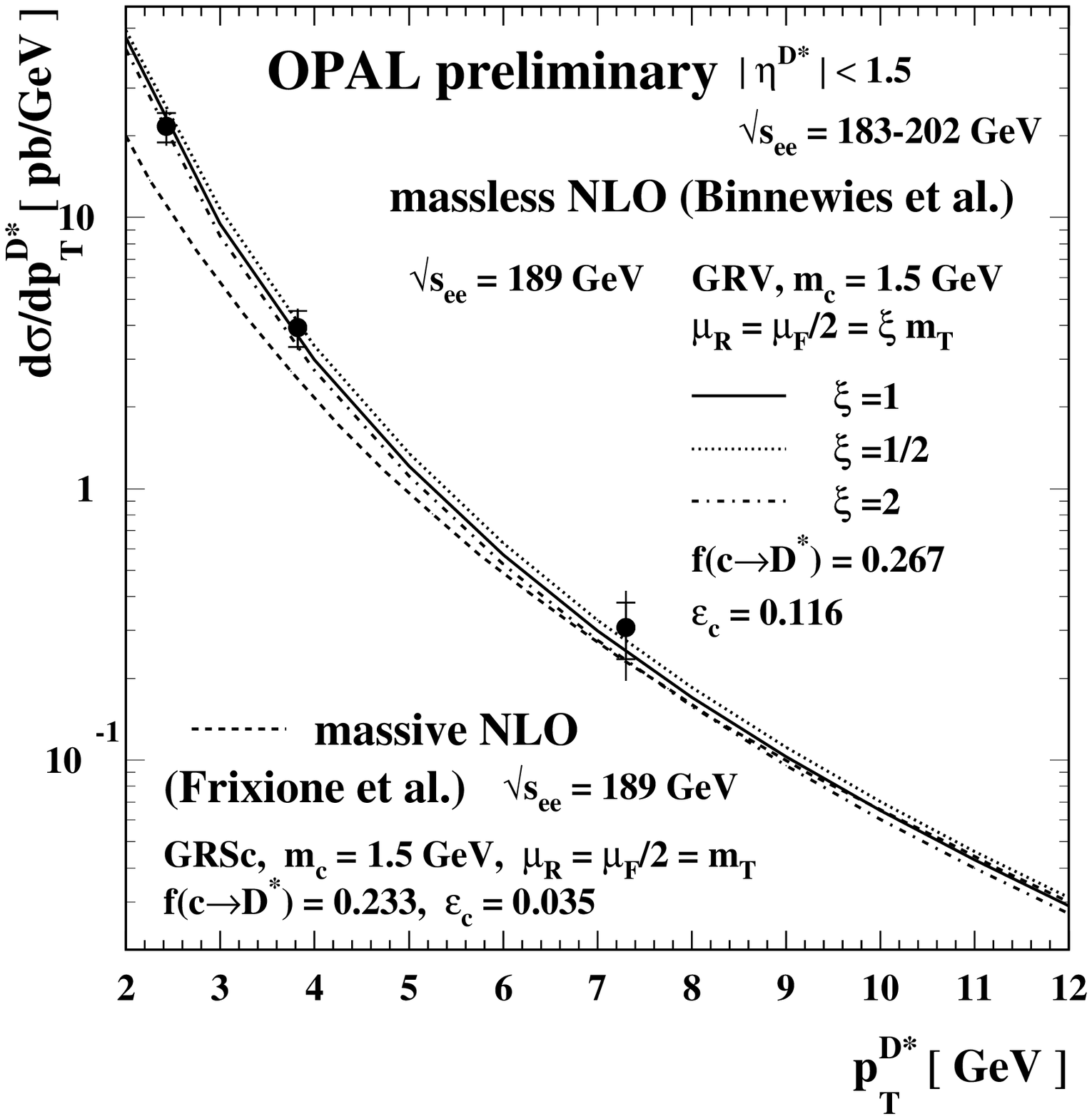,width=0.6\linewidth}
  \epsfig{file=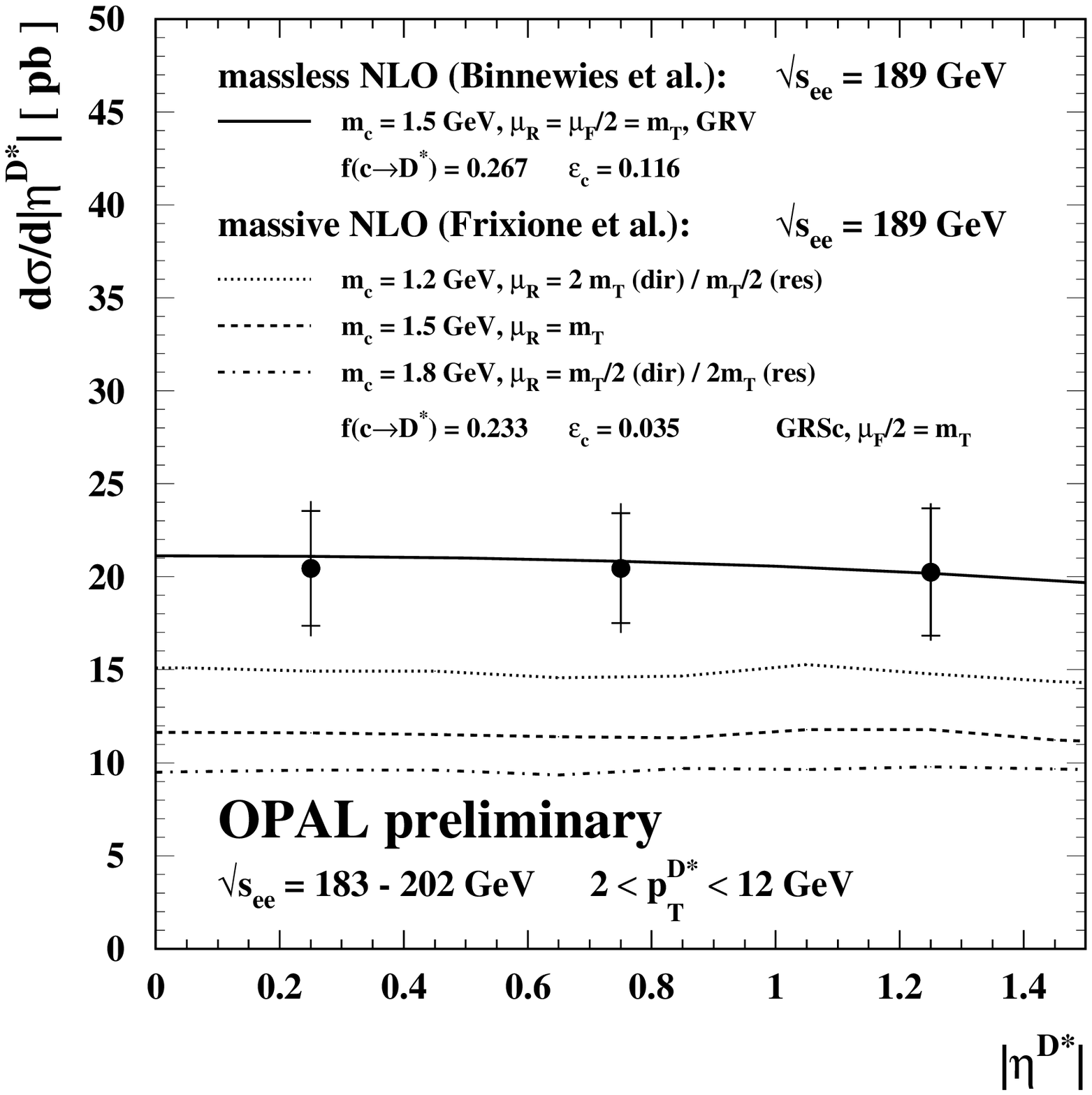,width=0.6\linewidth}
 \caption{\label{fig:37}Transverse momentum (top) and pseudorapidity
 (bottom) dependence of the massive (Frixione, Kr\"amer, and Laenen, 2000a)
 and a massless (Binnewies, Kniehl, and Kramer, 1996b, 1998a) NLO $D^\ast$
 photon-photon cross section compared to OPAL data (Abbiendi {\it et al.},
 2001a).}
 \end{center}
\end{figure}

%
The measured total charm cross sections in photon-photon collisions in Fig.\
\ref{fig:36} (top)
are again well described by massive NLO calculations (Drees {\it et al.},
1993), whereas the total bottom cross sections (bottom) are significantly
underestimated. Even though the experimental and theoretical uncertainties are
quite large, they cannot account for the discrepancy in the bottom case.
As for the $D^\ast$ transverse momentum (top) and pseudorapidity (bottom)
distributions shown in Fig.\ \ref{fig:37}
%
%
they are well described by the massless calculation, despite the low transverse
momenta studied ($p_T>2$ GeV). Surprisingly, the massive calculation agrees
with the data for larger $p_T$ ($p_T>3$ GeV), but not for small $p_T$, where it
should be more appropriate. Even low values of $m_c$ and $\mu$ cannot bring
the massive calculation into agreement with the measured pseudorapidity
distribution.

With future $e^+e^-$ and $ep$ colliders, it will not only be possible to
extend the kinematic ranges and production rates of charm and bottom quarks
(Kr\"amer and Laenen, 1996; Jankowski, Krawczyk, and Wing, 2001), but also 
to study the heavier top quark and its properties (K\"uhn, Mirkes, and
Steegborn, 1993; Bernreuther, Ma, and McKellar, 1995; Choi and Hagiwara, 1995).

\subsection{Quarkonia}

As already briefly discussed in Sec.\ \ref{sec:hadfrag}, the factorization
formalism of Non-Relativistic QCD (NRQCD) (Bodwin, Braaten, and Lepage, 1995)
provides a rigorous theoretical framework for the description of
heavy-quarkonium production and decay. This formalism implies a separation of
short-distance coefficients, which can be calculated perturbatively as
expansions in the strong-coupling constant $\alpha_s$, from long-distance
operator matrix elements (OMEs), which must be extracted from experiment.
The relative importance of the latter can be estimated by means of velocity
scaling rules (Lepage {\it et al.}, 1992), {\it i.e.} they are predicted to
scale with a definite power of
the heavy-quark velocity $v$ in the limit $v\ll1$. In this way, the
theoretical predictions are organized as double expansions in $\alpha_s$ and
$v$. A crucial feature of this formalism is that it takes into account the
complete structure of the $h\overline{h}$ Fock space, which is spanned by the
states $n=[\underline{a},^{2S+1}\!L_J]$
with definite spin $S$, orbital angular
momentum $L$, total angular momentum $J$, and color multiplicity
$\underline{a}= \underline{1}$, $\underline{8}$.
In particular, NRQCD predicts the existence of color-octet processes
in nature.
This means that $h\overline{h}$ pairs are produced at short distances in
color-octet states and subsequently evolve into physical (color-singlet)
quarkonia by the nonperturbative emission of soft gluons.
In the limit $v\to 0$, the traditional color-singlet model (CSM)
(Berger and Jones, 1981; Baier and R\"uckl, 1981) is recovered.
The greatest triumph of NRQCD was that it was able to correctly 
describe the cross section of inclusive charmonium hadroproduction
measured in $p\overline{p}$ collisions at the Fermilab Tevatron
(Braaten and Fleming, 1995; Cho and Leibovich,
1996a, 1996b), which had turned out to be more than one order
of magnitude in excess of the theoretical prediction based on the CSM
(Abe {\it et al.}, 1992, 1997).
An alternative model, which also predicts the existence
of color-octet processes,
is the duality or color evaporation model (CEM), where conservation of the
color quantum number is ignored in the hard interaction and assumed to be
recovered only through the exchange of {\em non-perturbative} soft gluons with
the underlying event (Fritzsch, 1977; Halzen, 1977; Gl\"uck, Owens, and Reya,
1978). The short-distance color-singlet and color-octet cross sections are
related to those of open heavy-quark production by statistical factors and
to those of the physically observed mesons by constant universal factors.
While this model can also account for the Tevatron data (Amundson {\it et al.},
1997), it lacks the status of NRQCD as an effective field theory.
Similar objections may be raised to a model, which allows to flip spin and
color of the heavy-quark pair through a {\em perturbative} gluon exchange with
an isotropic classical color field generated by bremsstrahlung of the initial
state partons (Hoyer and Peign\'{e}, 1999; Hoyer, Marchal, and Peign\'{e},
2000). This ``hard comover scattering'' model can explain not only the observed
ratio of $\chi_{c,1}$ and $\chi_{c,2}$ hadroproduction cross sections, but also
the observed non-polarization of hadroproduced $J/\psi$ and $\psi'$ mesons.
In both of these cases the CSM and NRQCD have some difficulty in explaining the
data. Quarkonium production in $p\overline{p}$ and $ep$ collisions has
recently been reviewed by Kr\"amer (2001). 

%
\begin{figure}
 \begin{center}
  \epsfig{file=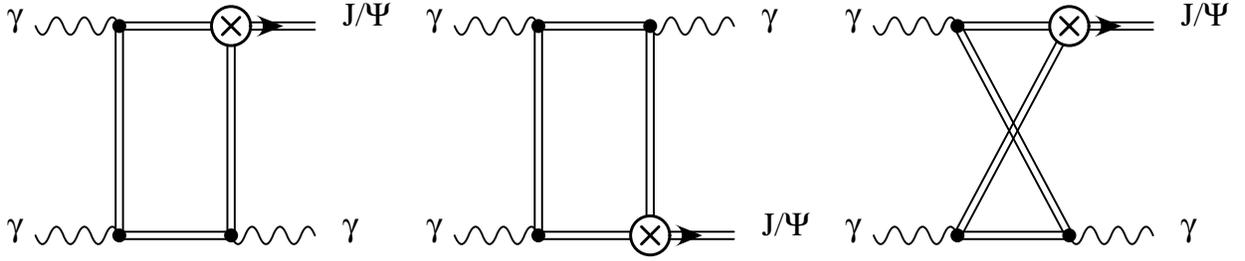,bbllx=60pt,bblly=360pt,bburx=380pt,bbury=431pt,%
          width=\linewidth}
 \caption{\label{fig:38}Born diagrams for direct color-singlet $J/\Psi
          +\gamma$ production. The charm quark lines in the loop run in both
          directions.}
 \end{center}
\end{figure}
%
%
\begin{figure}
 \begin{center}
  \epsfig{file=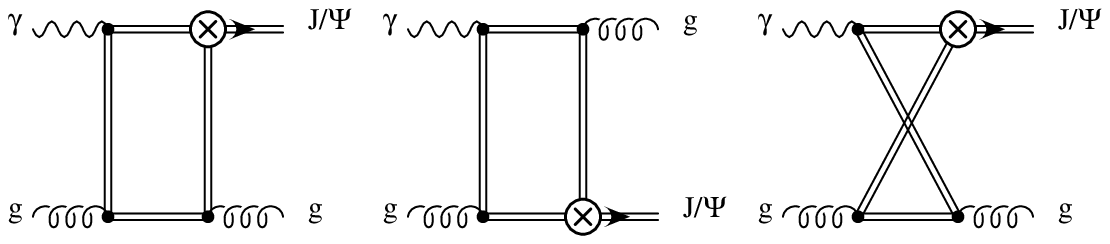,bbllx=60pt,bblly=360pt,bburx=380pt,bbury=431pt,%
          width=\linewidth}
 \caption{\label{fig:39}Born diagrams for single-resolved color-singlet
          $J/\Psi+$jet production.}
 \end{center}
\end{figure}
%
%
\begin{figure}
 \begin{center}
  \epsfig{file=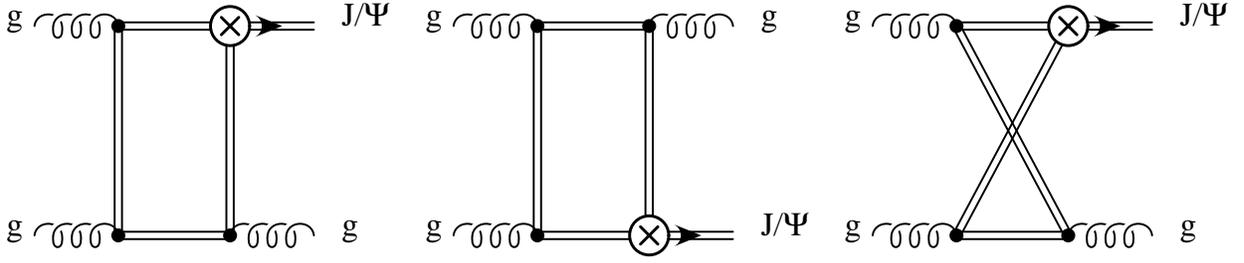,bbllx=60pt,bblly=360pt,bburx=380pt,bbury=431pt,%
          width=\linewidth}
 \caption{\label{fig:40}Born diagrams for double-resolved color-singlet
          $J/\Psi+$jet production.}
 \end{center}
\end{figure}
%
In order to convincingly establish the phenomenological significance of the
color-octet processes, it is indispensable to identify them in other kinds of
high-energy experiments as well.
In photon-photon and photon-hadron collisions, quarkonia can be
produced by direct, single-resolved, and double-resolved processes.
Which of the various photon, quark, and gluon initiated processes
contribute to the production of a given quarkonium state, depends on
the leading OMEs in the NRQCD velocity expansion
and on the additional particles in the final state. In inelastic
processes, the quarkonium state is produced with non-zero transverse
energy and a balancing jet or photon. In this case the inclusive
hadronic production cross section is given by Eq.\ (\ref{eq:hadronxsec})
with $D_{H/c}(z,M_c^2)=\delta(1-z)$.
$p_T$ and $y$ now relate to the transverse momentum and rapidity of
the observed quarkonium state, and the partonic cross section is given
by Eq.\ (\ref{eq:massivepartonicxsec}), where the mass $p_2^2$ is identified
with the quarkonium mass $M^2$ ($=4 m_h^2$ in the static limit).

Since its discovery in 1974, the $J/\Psi$ charmonium has remained in the
center of interest due to its relatively large production cross section
and easy experimental identification in the leptonic decay modes. The
diagrams contributing to the production of its leading color-singlet state
$[\underline{1},^3S_1]$ are shown in Figs.\ \ref{fig:38},
\ref{fig:39}, and \ref{fig:40}.
In photon-photon scattering, they relate to direct, single-resolved, and
double-resolved processes. In photoproduction, the last two apply to direct
and resolved processes, while in hadroproduction only the diagrams in Fig.\
\ref{fig:40} contribute. The corresponding hard scattering matrix
elements are related by simple transformations of couplings and color factors
and are listed in Tab.\ \ref{tab:onloxsec}.
%
\begin{table}
\begin{center}
\begin{tabular}{|l|l|}
Process & LO matrix element squared $|\M^B|^2$ \\
\hline
$\yytQy$& $\frac{3072 \OTSOO m_c}{3-2\eps}\cdot$ \\
        & $\frac{(2-5\eps)stu(s+t+u)+2(1-\eps)^2(s^2t^2+s^2u^2+t^2u^2)}
           {(s+t)^2(s+u)^2(t+u)^2}$ \\
$\ygtQg$& $\frac{C_F}{2N_C}|\M^B|^2_{\yytQy}$ \\
$\ggtQg$& $\frac{N_C^2-4}{4 N_C}|\M^B|^2_{\ygtQg}$ \\
\end{tabular}
\end{center}
\caption{\label{tab:onloxsec}LO matrix elements squared $|\M^B|^2$ for
 direct, single-resolved, and double-resolved color-singlet $J/\Psi$
 production.}
\end{table}
%
The diagrams in Figs.\ \ref{fig:38}, \ref{fig:39}, and \ref{fig:40}
contribute also to the production
of one of the leading color-octet states, $[\underline{8},^3S_1]$,
when the final photon in Fig.\ \ref{fig:38} is replaced with a gluon and
the color factors are changed appropriately. At the same order $v^4$
relative to $\OTSOO$, the OMEs $\OOSZE$ and $\OTPJE$ with
$J=0,\,1,\,2$ contribute in photon- and gluon-quark scattering
with final state quarks and in quark-antiquark annihilation with
final state gluons (Ma, McKellar, and Paranavitane, 1998; Japaridze
and Tkabladze, 1998; Godbole, Indumathi, and Kr\"amer, 2001;
Klasen, Kniehl, Mihaila, and Steinhauser, 2001a).

In the photon and gluon fusion processes described above, the two internal
quark propagators are off-shell by $p_T^2$, so that the differential cross
section falls like $1/p_T^8$. On the other hand, when $p_T\gg M$, the
quarkonium mass $M$ can be considered small and the cross section should scale
like any other single-particle inclusive cross section as $1/p_T^4$.
The dominant quarkonium production mechanism at sufficiently large $p_T$
must therefore be via fragmentation of a final state parton (see Sec.\
\ref{sec:hadfrag}) in the partonic processes shown in Figs.\ \ref{fig:07},
\ref{fig:08}, and \ref{fig:09} (Braaten and Yuan, 1993).

\begin{figure}
 \begin{center}
  \epsfig{file=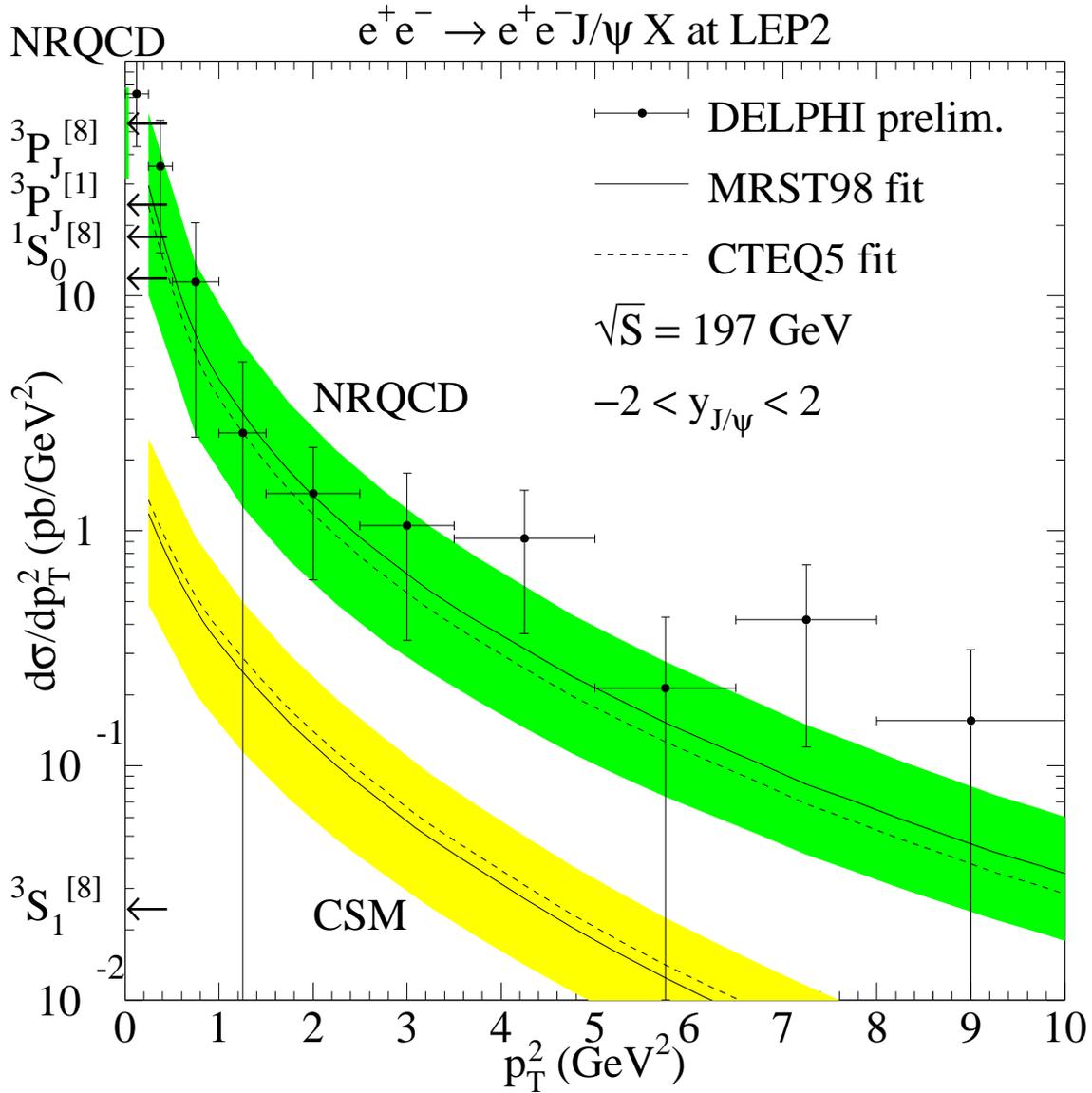,width=\linewidth}
 \caption{\label{fig:41}
 LO squared transverse momentum distribution of $J/\Psi$ mesons produced in
 photon-photon collisions at LEP2. The total NRQCD result describes the
 preliminary data from DELPHI (Todorova-Nova, 2001; Chapkin, 2002)
 well, while the CSM is an order of magnitude too low.}
 \end{center}
\end{figure}

In Fig.\ \ref{fig:41}, the LO 
color-singlet and the sum of color-singlet and color-octet contributions
%
%
are compared with preliminary data on $J/\Psi$ production in photon-photon
collisions from DELPHI (Todorova-Nova, 2001; Chapkin, 2002).
The agreement of the NRQCD prediction with the data is excellent, although the
statistical and theoretical uncertainties from the renormalization and
factorization scales, from $m_c=1.5$ GeV, and the numerical values of the
NRQCD operator matrix elements (Braaten, Kniehl, and Lee, 2000) are still
large (Klasen, Kniehl, Mihaila, and Steinhauser, 2001b).
Single-resolved processes are far more important than double-resolved and
direct processes (see also Godbole, Indumathi, and Kr\"amer, 2001).
The total color-singlet contribution is insufficient to describe the data, and
the size of the cross section is sensitive to the color-octet
operator values extracted at the Tevatron. The shapes of color-singlet and
color-octet contributions can, however, only be distinguished at larger
$p_T$.

\begin{figure}
 \begin{center}
  \epsfig{file=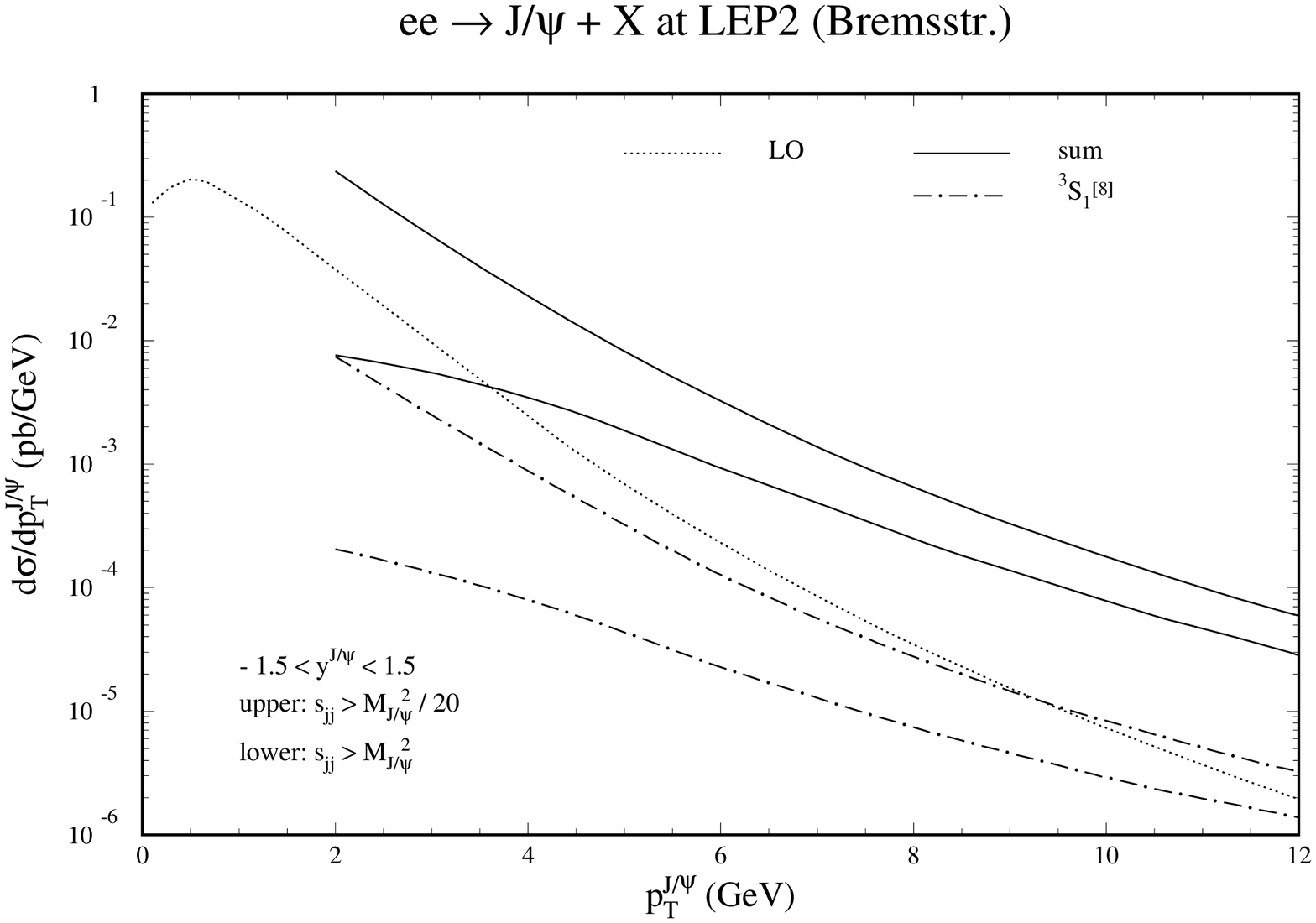,width=\linewidth}
 \caption{\label{fig:42}
 Transverse momentum spectrum for $J/\Psi$ production in direct photon-photon
 collisions at LEP2. At large $p_T$, the NLO real emission corrections
 enhance the LO cross section by an order of magnitude (Klasen, Kniehl,
 Mihaila, and Steinhauser, 2001a).}
 \end{center}
\end{figure}

At the scale of the charm mass, the strong coupling $\alpha_s$ is still
relatively large ($\simeq 0.3$) and NLO corrections can become
important. The real corrections to $J/\Psi$ production in direct
photon-photon collisions, which become dominant at large $p_T$, have recently
been calculated with an invariant mass cut-off for the associated dijet
system. Fig.\ \ref{fig:42}
%
%
shows that the LO cross section is enhanced by about an order of
magnitude at large $p_T$, where the missing virtual corrections are
no longer important and the uncertainty from the exact value of the
cut-off is reduced (Klasen, Kniehl, Mihaila, and Steinhauser, 2001a).
In a complete NLO calculation, there would of course be no dependence on
the cut-off.

\begin{figure}
 \begin{center}
  \epsfig{file=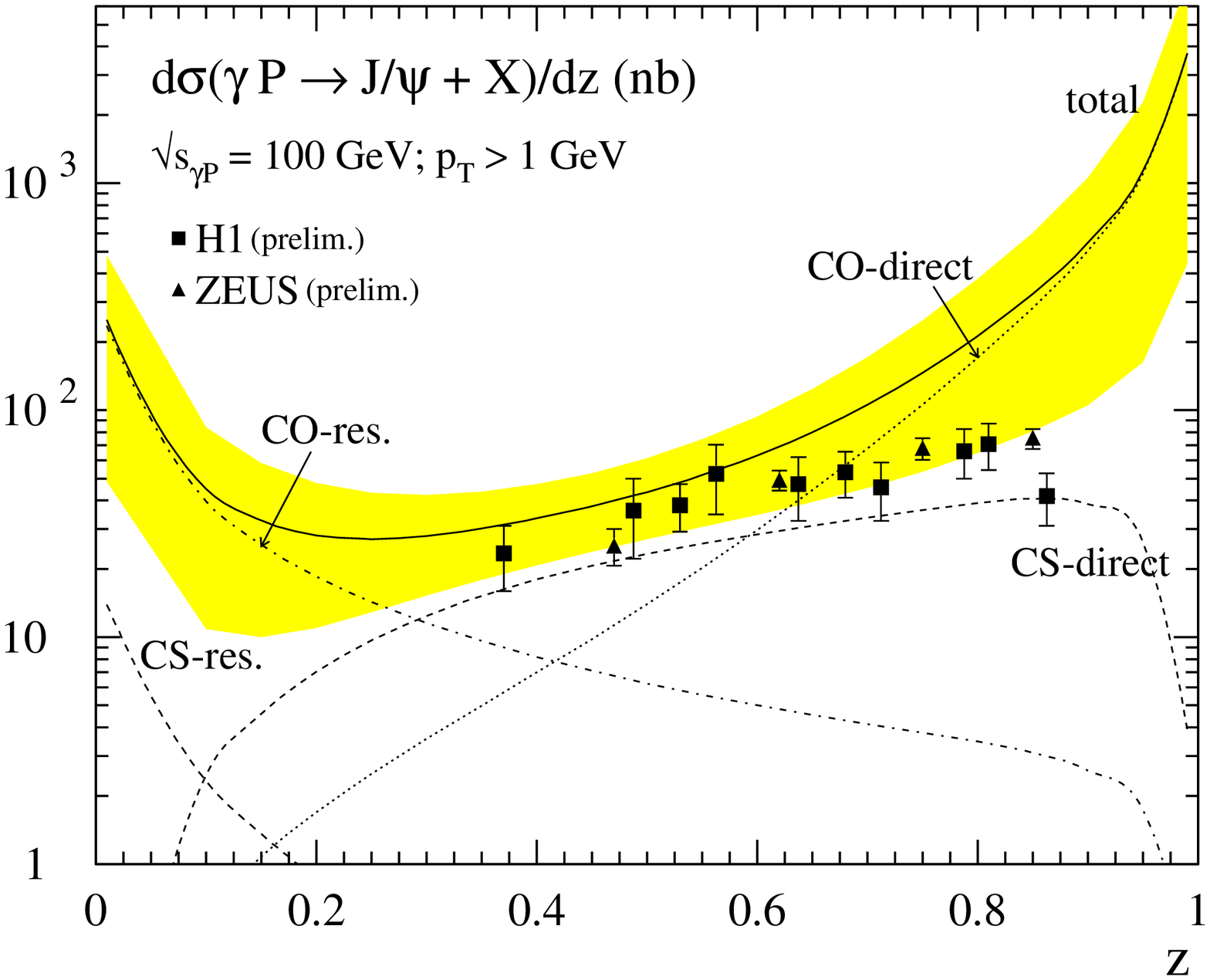,width=\linewidth}
 \caption{\label{fig:43}
 Direct and resolved contributions to the color-singlet and color-octet
 $J/\Psi$ energy distribution at HERA (Kr\"amer, 2001).}
 \end{center}
\end{figure}

A full NLO calculation is available for inelastic color-singlet $J/\Psi$
production in direct photon-hadron collisions (Kr\"amer {\it et al.},
1995; Kr\"amer, 1996). The direct process dominates
at the relatively large values of $z$ $(z=p_P\cdot p_{J/\Psi}/p_P\cdot
p_\gamma \simeq E_{J/\Psi}/E_\gamma\geq 0.3)$ currently observed at HERA.
Recent H1 data (Kr\"uger, 2000) show evidence also for resolved
photoproduction, which is closely connected to hadroproduction and
contributes mostly at low $z\leq 0.3$. This region is dominated
by color-octet processes in NRQCD (Beneke, Kr\"amer, and V\"anttinen, 1998;
Kniehl and Kramer, 1999), but also in the CEM (\'{E}boli, Gregores, and Halzen,
1999). As can be seen in Fig.\ \ref{fig:43}, the 
%
%
direct LO color-singlet prediction can almost account for the data. Within the
relatively large uncertainty of the theoretical parameters
(shaded band), the direct color-octet processes
improve the agreement at intermediate $z$, but show a stronger rise than the
data at large $z$
(Cacciari and Kr\"amer, 1996; Ko, Lee, and Song, 1996; Amundson, Fleming,
and Maksymyk, 1997). However, this region is close to the kinematic
limit and is influenced by  the non-perturbative soft transition of
the color-octet state into the physical $J/\Psi$ meson, so that LO
predictions are not reliable.
NLO corrections for color-octet photoproduction have unfortunately only been
calculated for the total cross section (Maltoni, Mangano, and Petrelli, 1998),
but higher-order terms have been resummed using non-relativistic shape
functions (Beneke, Rothstein, and Wise, 1997; Beneke, Schuler, and Wolf, 2000).
Relativistic corrections to the color-singlet state have been
evaluated, but there is unfortunately no agreement on their effects
in the literature (Greub et al., 1993; Khan and Hoodboy, 1996; Ma,
McKellar, and Paranavitane, 2000). Higher twist effects due to multiple
interactions with the proton or photon remnants should be suppressed as a
power of $\Lambda^2/(Q^2+p_T^2)$ with $Q=m_c,\,m_c v$, or $m_c v^2$
(Beneke, Kr\"amer, and V\"anttinen, 1998) and not as a power of $\Lambda^2/
(4 m_c^2)$ (Ma, 1997), since only the first result leads to the intuitively
expected $\Lambda^2/p_T^2$ behavior at large $p_T$. On the other hand,
the discrepancy between the NRQCD prediction and the HERA data can clearly
be reduced, if higher order effects of multi-gluon emission
are approximately taken into account in the extraction
of the operator matrix elements at the Tevatron (Kniehl and Kramer, 1999).

\begin{figure}
 \begin{center}
  \epsfig{file=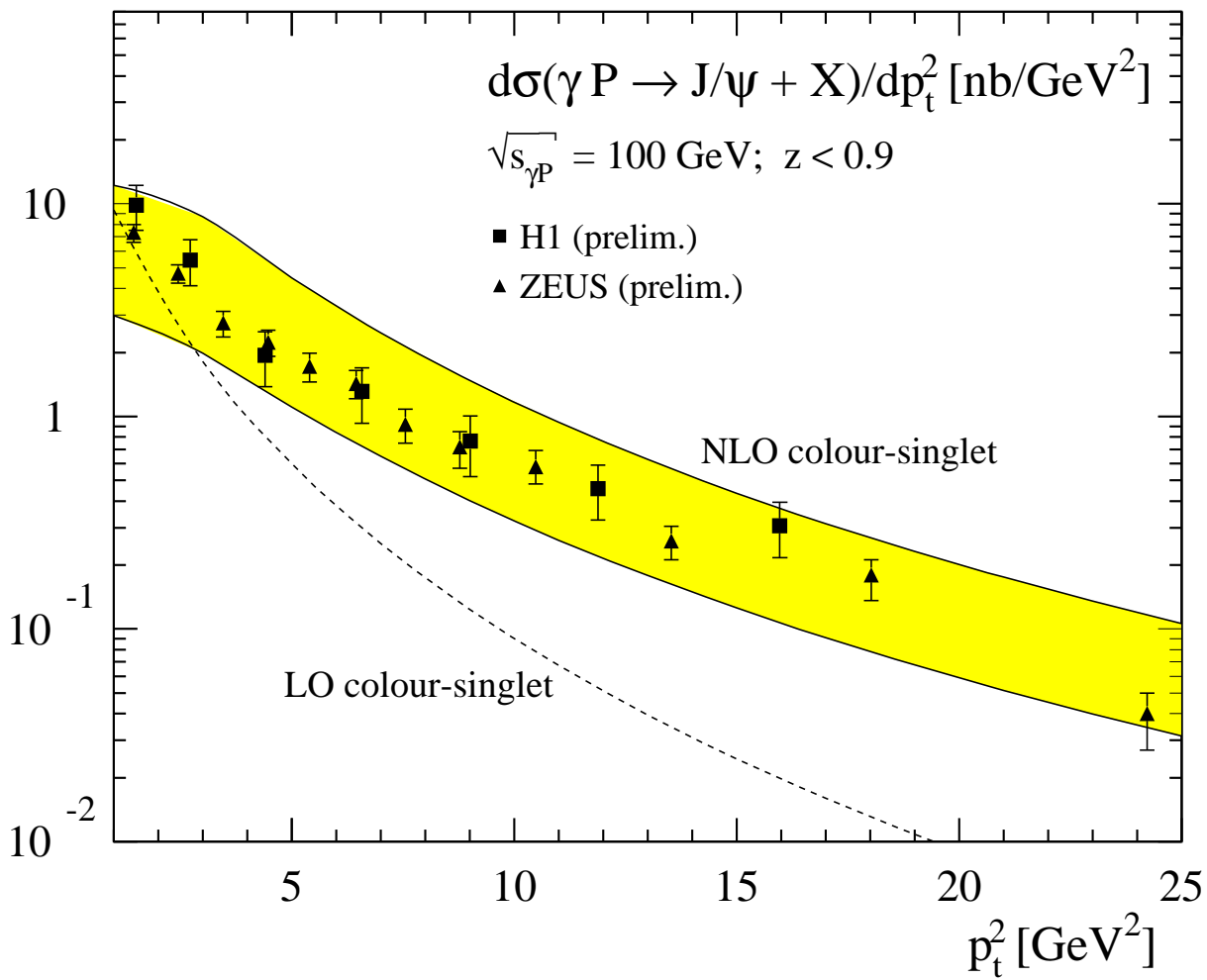,width=\linewidth}
 \caption{\label{fig:44}
 $J/\Psi$ transverse momentum distribution in LO and NLO direct color-singlet
 photoproduction (Kr\"amer, 2001).}
 \end{center}
\end{figure}

While the shape of the $J/\Psi$ energy distribution in Fig.\
\ref{fig:43} is already well described by the LO color-singlet
calculation, the shape of the LO transverse momentum distribution in Fig.\
\ref{fig:44}, which falls like $\alpha_s^2m_c^4/p_T^8$, differs markedly
%
%
from the HERA data. The data clearly require the NLO color-singlet
contribution, which is dominated by $t$-channel gluon exchange and falls only
like $\alpha_s^3m_c^2/p_T^6$ (Kr\"amer {\it et al.}, 1995; Kr\"amer, 1996).
A similar scaling is observed by contributions from charm quark
fragmentation in direct photoproduction, which exist at $\O(\alpha_s^3)$.
They are suppressed at low $p_T$, but exceed the color-singlet production via
photon-gluon fusion at $p_T\geq 10$ GeV according to LO (Godbole,
Roy, and Sridhar, 1996) and NLO calculations (Kniehl and Kramer, 1997a, 1997b).
Gluon fragmentation in resolved photoproduction exceeds the fusion process
already at $p_T\geq 5$ GeV due to a large enhancement by NLO corrections
(Kniehl and Kramer, 1997a, 1997b).

Further information about the underlying production mechanism (CSM, CEM, or
NRQCD) may be gained from the energy spectrum of $J/\Psi$'s produced in
polarized photon-hadron collisions (Chao {\it et al.}, 2000; Japaridze, Nowak,
and Tkabladze, 2000) and of $J/\Psi$'s produced in association with a photon,
where the color-octet contribution dominates already at intermediate $z$ (Kim
and Reya, 1993; Cacciari, Greco, and Kr\"amer, 1997; Mehen, 1997). Another
very sensitive observable is the energy spectrum of the polar angle in
polarized $J/\Psi$ decays
(Beneke, Kr\"amer, and V\"anttinen, 1998).
At the upgraded HERA collider one may furthermore hope
to observe and study the photoproduction of other quarkonia like the
$S$-wave states $\psi'$ (Morii, Roy, and Sudoh, 2000; Brugnera, 2001)
and $\eta_c$ (Chao, Hao, and Yuan, 1999) or the $P$-wave states
$\chi_c$ (Ma, 1996b) and $h_c$ (Fleming and Mehen, 1998).
Bottomonia have a smaller relative squared velocity $v^2\simeq 0.1$ than
charmonia with $v^2\simeq 0.3$ (Quigg and Rosner, 1979)
and should therefore be
better described by the NRQCD velocity expansion. Unfortunately, the
photoproduction cross section for $\Upsilon$ is more than two orders of
magnitude smaller than the one for $J/\Psi$ due to the smaller electric
charge and larger mass of the bottom quark.

The large conceptual and numerical uncertainties in quarkonium physics clearly
call for an increased theoretical effort, in particular for a full NLO
calculation of color-singlet production at hadron colliders and of color-octet
production at photon and hadron colliders. In open flavor production, the NLO
corrections should be calculated for the photoproduction of two light or heavy
hadrons and of hadrons in association with jets, which would help to
disentangle the flavor ({\it e.g.} charm) content of photons and protons and to
resolve the question, why massive calculations cannot describe the observed
bottom cross sections.


\pagebreak
\section{Prompt photon production}
\label{sec:promptyprod}
\setcounter{equation}{0}

The many facets, that photons exhibit in the initial state of hard scattering
processes, can be rediscovered in their production in the final state:
If produced inclusively, prompt photons are related to hadrons by
important fragmentation contributions; if produced in isolation, they exhibit
more perturbative aspects due to their pointlike coupling to quarks; if
produced in association with jets, they become powerful tools for constraining
the initial parton densities. All of these aspects will be discussed in this
Section.

\subsection{Fragmentation}
\label{sec:photfrag}

Since photons couple to charged quarks, they can be produced directly in
hard scattering interactions. If the photon is radiated from a massless
final-state quark, it exhibits a collinear $1/\eps$-singularity proportional
to the $P_{\gamma\leftarrow q}(x)$ splitting function and the underlying LO
matrix element. This $1/\eps$-pole has to be absorbed into a non-perturbative
fragmentation function, just like the collinear singularity of an
initial-state photon had to be absorbed into a non-perturbative parton
density. The photon fragmentation functions $D_{\gamma/i}$ are thus
intimately related to the parton densities in the photon $f_{i/\gamma}$.

The evolution equations of these fragmentation functions are of a
form similar to Eqs.\ (\ref{eq:evol_eq}) with the role of initial and final
particles reversed. They are given by (Koller, Walsh, and Zerwas, 1979)
\bea
 \label{eq:frag_evol}
 \frac{\d D_{\gamma/q}(Q^2)}{\d \ln Q^2} &=&
   \frac{\alpha       }{2\pi} P_{\gamma\leftarrow q}\otimes D_{\gamma/\gamma}(Q^2)
  +\frac{\alpha_s(Q^2)}{2\pi} \\
  &\times&\left[ P_{q\leftarrow q} \otimes D_{\gamma/q}(Q^2)
  +      P_{g\leftarrow q} \otimes D_{\gamma/g}(Q^2) \right] ,\nonumber\\
 \frac{\d D_{\gamma/g}(Q^2)}{\d \ln Q^2} &=&
   \frac{\alpha       }{2\pi} P_{\gamma\leftarrow g}\otimes D_{\gamma/\gamma}(Q^2)
  +\frac{\alpha_s(Q^2)}{2\pi}\nonumber\\
  &\times&\left[ P_{q\leftarrow g} \otimes D_{\gamma/q}(Q^2)
  +      P_{g\leftarrow g} \otimes D_{\gamma/g}(Q^2) \right] ,\nonumber\\
 \frac{\d D_{\gamma/\gamma}(Q^2)}{\d \ln Q^2} &=&
   \frac{\alpha       }{2\pi} P_{\gamma\leftarrow\gamma}\otimes D_{\gamma/\gamma}(Q^2)
  +\frac{\alpha}{2\pi}\nonumber\\
  &\times&\left[ P_{q\leftarrow\gamma} \otimes D_{\gamma/q}(Q^2)
  +      P_{g\leftarrow\gamma} \otimes D_{\gamma/g}(Q^2) \right] \nonumber
\eea
for a single quark flavor $q$. The time-like splitting functions
$P_{j\leftarrow i}$ are identical to the space-like splitting functions in LO
(Altarelli and Parisi, 1977), but differ in NLO (Curci, Furmanski, and
Petronzio, 1980; Furmanski and Petronzio, 1980; Floratos, Kounnas, and Lacaze,
1981). Like the evolution equations Eqs.\ (\ref{eq:evol_eq})
for photon densities, Eqs.\ (\ref{eq:frag_evol}) contain inhomogeneous terms:
The quark-photon splitting function can be obtained in LO from
$P_{g\leftarrow q}$ by the transformation $P_{\gamma\leftarrow q} =
e_q^2/C_F P_{g\leftarrow q}$. The gluon-photon splitting function enters only
in NLO and is given by (Aurenche {\it et al.}, 1993)
\bea
  P_{\gamma\leftarrow g} (\!\!&x&\!\!) =
 \frac{\alpha_s(Q^2)}{2\pi}\frac{e_q^2T_R}{2}
 \le -4+12x-\frac{164}{9}x^2+\frac{92}{9x}\rp\\
 &+&\lp\lr 10+14x+\frac{16}{3}x^2+\frac{16}
 {3x}\rr\ln x + 2(1+x)\ln^2x\re.\nonumber
\eea

Information about the photon fragmentation function can be obtained from
the process $e^+e^-\rightarrow \gamma X$. However, this process also has
a large contribution from the direct coupling of photons to quarks, which
dominates over the fragmentation contribution, particularly if the photon is
experimentally isolated (see Sec.\ \ref{sec:isolation}). The direct
contribution has been used extensively to extract the electroweak couplings of
up- and down-type quarks to the $Z$-boson by the ALEPH (Decamp {\it et al.},
1991; Buskulic {\it et al.}, 1993), DELPHI (Abreu {\it et al.}, 1995), L3
(Adriani {\it et al.}, 1992, 1993), and OPAL collaborations (Akrawy {\it et
al.}, 1990; Alexander {\it et al.}, 1991; Acton {\it et al.}, 1993).

In the $\ms$ scheme the NLO inclusive photon cross section is given by
(Altarelli {\it et al.}, 1979)
\bea
 \frac{1}{\sigma_0} \frac{\d\sigma(Q^2)}{\d x} &=& \sum_q 2 e_q^2 \lg
 D_{\gamma/q}(Q^2)
+\frac{\alpha}{2\pi} e_q^2 C_\gamma^T
 +\frac{\alpha_s(Q^2)}{2\pi}\rp\nonumber\\
 &\times&\lp \le C_q^T \otimes D_{\gamma/q}(Q^2)
 +   C_g^T \otimes D_{\gamma/g}(Q^2)\re
 \rg,
\eea
where $\sigma_0=4\pi\alpha^2N_C/(3Q^2)$ is $N_C$ times the pointlike cross
section for $e^+e^-\rightarrow\mu^+\mu^-$, the factor of two comes from
$D_{\gamma/q}(Q^2)=D_{\gamma/\overline{q}}(Q^2)$, and the time-like Wilson
coefficients for the transverse and longitudinal cross sections are
given by Eqs.\ (\ref{eq:tlwilcoeff}).
The photonic Wilson coefficients are $C_{\gamma,i}^T = C_{g,i}^T/C_F$.
$C_{\gamma,T}^T$ contains a
term $\propto\ln[x^2(1-x)]$ with a logarithmic singularity at $x=1$ and a
stronger
singularity at $x=0$. Gl\"uck, Reya, and Vogt (1993) defined a modified $\dis$
scheme for the photon fragmentation function by absorbing the transverse
photonic Wilson coefficient $C_{\gamma,T}^T$ into the quark fragmentation
function $D_{\gamma/q}^{\dis}(Q^2)=D_{\gamma/q}^{\ms}(Q^2)+\alpha/(2\pi)e_q^2
C_{\gamma,T}^T$. This affects again the NLO photon splitting functions
\begin{eqnarray}
P_{\gamma\leftarrow q}^{\dis} & = & P_{\gamma\leftarrow q}^{\ms} -
		e_q^2\, P_{q\leftarrow q} \otimes C_{\gamma,T}^T\nonumber,\\
P_{\gamma\leftarrow g}^{\dis} & = & P_{\gamma\leftarrow g}^{\ms} -
		2\, \sum_q  e_q^2\, P_{q\leftarrow g}\otimes C_{\gamma,T}^T\, .
\end{eqnarray}

In LO of $\alpha$, the third evolution equation in Eqs.\ (\ref{eq:frag_evol})
can be directly integrated with the result $D_{\gamma/\gamma}(x,Q^2)=\delta
(1-x)$. Furthermore in LO of $\alpha_s$ only the evolution equation of the
quark-photon fragmentation function
\beq
 \frac{\d D_{\gamma/q}(x,Q^2)}{\d\ln Q^2} =
 \frac{\alpha}{2\pi} P_{\gamma\leftarrow q}(x)
\eeq
survives, which can also be integrated with the result
(Gehrmann-de Ridder, Gehrmann, and Glover, 1997)
\beq
 D_{\gamma/q}(x,Q^2)=\frac{\alpha}{2\pi}P_{\gamma\leftarrow q}(x)\ln\frac{Q^2}
 {Q_0^2}+D_{\gamma/q}(x,Q_0^2).
 \label{eq:plsolfrag}
\eeq
The first term in Eq.\ (\ref{eq:plsolfrag}) is the perturbatively calculable
pointlike solution, while the second term is a hadronic boundary condition
which has to be fitted to experimental data, {\it e.g.} on $e^+e^-\rightarrow
\gamma+$jet. Following this approach the ALEPH collaboration obtained the LO
result (Buskulic {\it et al.}, 1996)
\beq
 D_{\gamma/q}(x,Q_0^2)=\frac{\alpha}{2\pi}\lr-P_{\gamma\leftarrow q}(x)
 \ln(1-x)^2-13.26\rr
\eeq
at the starting scale $Q_0=0.14$ GeV. For the fragmentation function of quarks
to photons with virtuality $P^2$, the purely perturbative LO result is (Qiu and
Zhang, 2001; Braaten and Lee, 2002)
\bea
 D_{\gamma/q}(x,Q^2,P^2) &=& e_q^2\frac{\alpha}{2\pi}\le\frac{1+(1-x)^2}{x}
 \ln\frac{xQ^2}{P^2}\rp\nonumber\\
 &-&\lp z\lr 1-\frac{P^2}{xQ^2}\rr\re.
\eea
The NLO evolution equation for the
quark-photon fragmentation function can also be integrated, if one neglects the
higher-order contributions from the gluon-photon fragmentation function, the
mixing between different quark flavors, and the running of $\alpha_s(Q^2)$. The
result
\bea
 D_{\gamma/q}(x,Q^2) &=& \frac{\alpha}{2\pi}
 P_{\gamma\leftarrow q}(x)\ln\left(\frac{Q^2} {Q_0^2}\right)
 + D_{\gamma/q}(x,Q_0^2) \nonumber \\
 &+& \frac{\alpha_s}{2\pi} P_{q\leftarrow q}(x)
 \ln\left(\frac{Q^2}{Q_0^2}\right) \\
 &\otimes& \left[\frac{\alpha}{2\pi}
 \frac{1}{2}P_{\gamma\leftarrow q}(x)
 \ln \left(\frac{Q^2}{Q_0^2}\right) +
 D_{\gamma/q}(x,Q_0^2) \right],\nonumber
\eea
is then exact at the fixed order ${\cal O}(\alpha\alpha_s)$, but does not take
into account the usual resummation of powers of $\ln(Q^2/Q_0^2)$. ALEPH have
fitted their $e^+e^-\to\gamma+$jet data in NLO, taking $\alpha_s=0.124$ in
order to reproduce the observed total rate of $e^+e^-\rightarrow$ hadrons,
and obtain (Buskulic {\it et al.}, 1996)
\bea
 D_{\gamma/q}(x,Q_0^2)&=&\frac{\alpha}{2\pi} \left(
 -P_{\gamma\leftarrow q}(x) \ln(1-x)^2 + 20.8(1-x) \rp\nonumber\\
 &-&\lp 11.07 \right)
\eea
at $Q_0 = 0.64$ GeV.

%
\begin{table}
\begin{center}
\begin{tabular}{|ccc|ccccc|}
 Group & Year&Set & $Q_0^2$         & Factor.\     & VMD     & $N_f$ &
     $\Lambda_{\ms}^{N_f=4}$      \\
       &     &    &        (GeV$^2$)&        Scheme &  Model              &        & (MeV)\\
\hline
\hline
 ACFGP & 1993  & NLO & 2.00          & $\ms$         &   coherent  & 4     &
   230 \\
 GRV & 1993  & LO  & 0.25            & LO            & incoherent  & 5     &
   200 \\
  &     & NLO & 0.30            & $\dis$        &                    &       &
   200 \\
 BFG & 1998 & NLO & 0.50            & $\ms$         &   coherent  & 4     &
   230 \\
\end{tabular}
\end{center}
\caption{\label{tab:photfrag}Parameterizations of the photon fragmentation
 functions.
 The coherence of the VMD model is determined by the coefficients $e_u$ and
 $e_d$ in Eq.\ (\ref{eq:vmd}).
 ACFGP and GRV relate the photon to the pion,
 while BFG fit $\rho$ production data from ALEPH and HRS.}
\end{table}
%
The fully resummed solution of the evolution equations consists of
pointlike and hadronic contributions $D_{\gamma/i}(Q^2)=D_{\gamma/i}^{\rm pl}
(Q^2)+D_{\gamma/i}^{\rm had}(Q^2)$. The full LO (Duke and Owens, 1982)
and NLO pointlike solutions (Aurenche {\it et al.}, 1993; Bourhis, Fontannaz
and Guillet, 1998)
\beq
 D^{\rm pl}_{\gamma/i}(Q^2)=\frac{\alpha}{2\pi}\le
 \frac{4\pi}{\alpha_s(Q^2)}a_i+b_i+{\cal O}(\alpha_s)\re.
 \label{eq:asymfrag}
\eeq
can only be calculated analytically in moment space.
The hadronic input can unfortunately not be determined from inclusive photon
production in $e^+e^-$ annihilation, since the experimental data (Buskulic
{\it et al.}, 1996; Ackerstaff {\it et al.}, 1998) are very limited and
furthermore dominated by the pointlike quark-photon fragmentation function.
Therefore all existing parameterizations (Aurenche {\it et al.} [ACFGP] 1993;
Gl\"uck, Reya, and Vogt [GRV], 1993; Bourhis, Fontannaz, and Guillet [BFG],
1998) use VMD to model the photon fragmentation at low scales. The different
input parameters are summarized in Tab.\ \ref{tab:photfrag}.
Heavy quarks
are included above their production thresholds with boundary conditions
$D_{\gamma/h}(x,m_h^2)=D_{\gamma/\bar{h}}(x,m_h^2)=0$.

\begin{figure}
 \begin{center}
  \epsfig{file=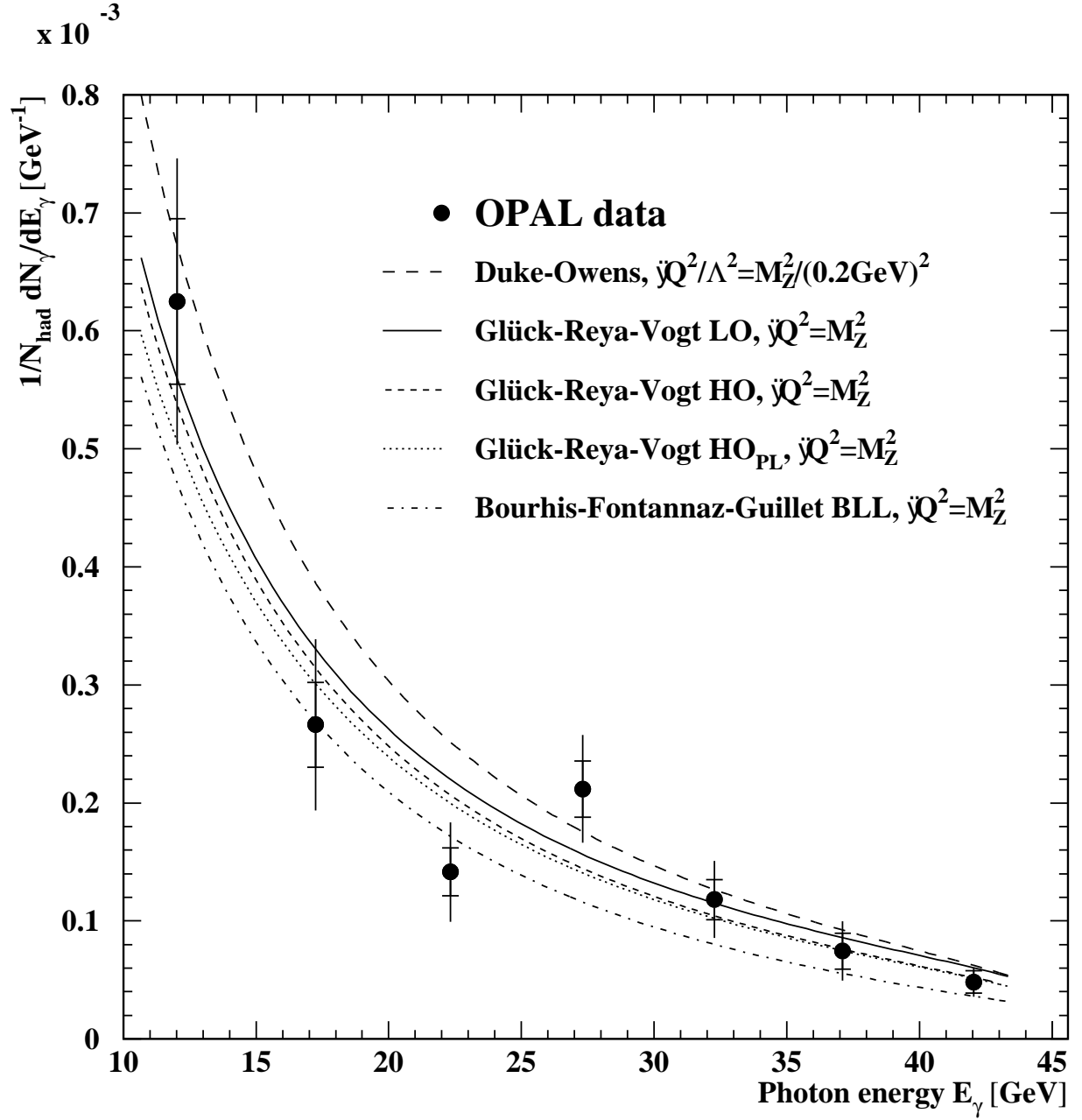,width=\linewidth}
 \caption{\label{fig:45}
 Parameterizations of the photon fragmentation function at $Q^2=M_Z^2$
 compared to OPAL data (Ackerstaff {\it et al.}, 1998).}
 \end{center}
\end{figure}

Fig.\ \ref{fig:45} demonstrates that the most recent data from OPAL
%
%
(Ackerstaff {\it et al.}, 1998) can already be described by the pointlike
fragmentation functions in LO (long-dashed curve) and NLO (dotted curve).
The data can thus not discriminate between the different LO (full curve)
and NLO (short-dashed and dot-dashed curves) assumptions for the
hadronic input.

\subsection{Isolation}
\label{sec:isolation}

\begin{figure}
 \begin{center}
  \epsfig{file=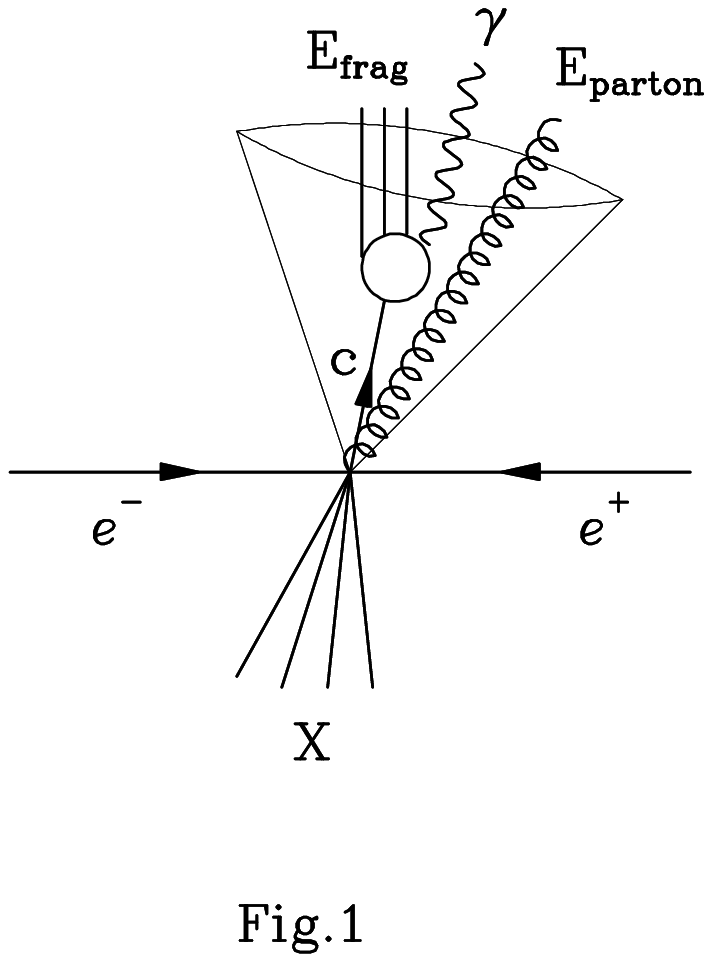,bbllx=222pt,bblly=300pt,bburx=432pt,bbury=531pt,%
          width=0.5\linewidth,clip=}
 \caption{\label{fig:46}
 Illustration of an isolation cone containing a parton $c$ that fragments
 into a photon $\gamma$ plus hadronic energy $E_{\rm frag}$. In addition,
 a gluon enters the cone and fragments giving hadronic energy
 $E_{\rm parton}$ (Berger, Guo, and Qiu, 1996).}
 \end{center}
\end{figure}

Photons produced via fragmentation usually lie inside hadronic jets, while
directly produced photons tend to be isolated from the final state hadrons.
The theoretical uncertainty coming from the non-perturbative fragmentation
function can therefore be reduced if the photon is isolated in phase space.
At the same time the experimental uncertainty coming from photonic decays of
$\pi^0$, $\eta$, and $\omega$ mesons is considerably reduced.
Photon isolation can be achieved by limiting the (transverse)
hadronic energy $E_{(T)}^{\rm had}$ inside a cone of size $R$ around the
photon to
\beq
 E_{(T)}^{\rm had}<\epsilon_{(T)} E_{(T),\gamma}.
\eeq
This is illustrated in Fig.\ \ref{fig:46}.
%
%
For small photon pseudorapidities $\eta$, the cone size $R=\delta/\cosh\eta$ is
approximately equal to the half angle of the cone $\delta$.
Since in NLO the direct and fragmentation processes are intimately linked
at the factorization scale $Q$, it is mandatory that the isolation
criterion does not interfere with the cancellation of soft and collinear
singularities. For certain differential cross sections like
d$\sigma/$d$x_\gamma$ in $e^+e^-$ annihilation, where $x_\gamma=2E_\gamma/Q^2$
is the fractional photon energy, spurious infrared singularities remain at
the point $x_\gamma=1/(1+\epsilon)$ (Berger, Guo, and Qiu, 1996). The cross
section
d$\sigma$/d$x_\gamma$ is therefore only defined as a distribution, which has to
be integrated with a test function over finite bin widths. So for
physical observables these integrable singularities disappear.
Nevertheless, the cross section contains logarithms of the type $\ln[1-
x_\gamma (1+\epsilon)]$, which can become large at $x_\gamma=1/(1+\epsilon)$
(Aurenche {\it et al.}, 1997; Catani, Fontannaz, and Pilon, 1998).
For physical observables it is therefore important to integrate over large
enough bins in $x_\gamma$. The situation is very similar to the infrared
sensitivity of the NLO dijet cross section at large values of the initial
state photon energy fraction $x_\gamma^{\rm obs}$. Recently an improved photon
isolation criterion
\beq
 \sum_i E^{\rm had}_{(T),i} \theta(\delta-R_{i}) < \epsilon_{(T)}
 E_{(T),\gamma} \lr{1-\cos\delta\over 1-\cos\delta_0}\rr,
\eeq
has been proposed, where $\delta\leq\delta_0$ and $\delta_0$ is now the
isolation cone (Frixione, 1998a). This procedure allows the fragmentation
contribution to vanish in an infrared safe way.

\subsection{Inclusive photons}

\begin{figure}
 \begin{center}
  \epsfig{file=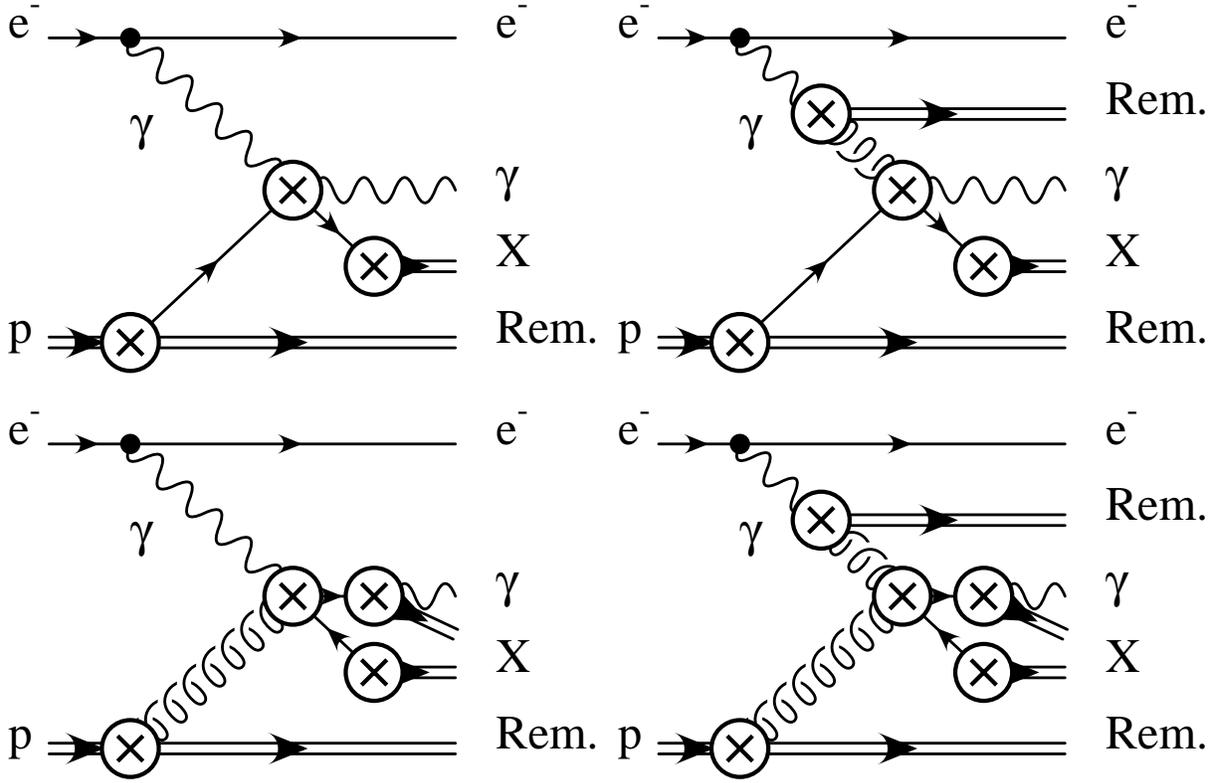,bbllx=60pt,bblly=290pt,bburx=280pt,bbury=430pt,%
          width=\linewidth}
 \caption{\label{fig:47}Factorization of lepton-hadron scattering into photons.}
 \end{center}
\end{figure}

The photoproduction of inclusive photons proceeds through direct and resolved
initial state photons and through direct and fragmentation production of
the final state photon. Thus four types of partonic subprocesses contribute
to $\gamma h\to\gamma X$, as shown in Fig.\ \ref{fig:47},
%
%
and eight types to $\gamma\gamma\to\gamma X$.
The hadronic cross section is given by Eq.\
(\ref{eq:hadronxsec}), where the hadron fragmentation function is replaced by
the photon fragmentation function $D_{\gamma/c}$. For direct photon production
the fragmentation function is, of course, simply $\delta(1-z)$, and the cross
section coincides with Eq.\ (\ref{eq:1jet}).

The partonic LO diagrams and cross sections can be obtained by crossing the
diagrams in Figs.\ \ref{fig:07}, \ref{fig:08}, and \ref{fig:09} and
the matrix elements in Tab.\ \ref{tab:loxsec}, and the NLO corrections are
calculated as outlined in Sec.\ \ref{sec:lighthad}. The only
additional subtlety occurs when a final state quark splits into a quark and a
photon. The corresponding collinear singularity
\bea
 \overline{\Gamma}_{\gamma\leftarrow q}(x,\mu^2)&=&\delta_{\gamma q}
 \,\delta(1-x)-\frac{1}
 {\eps}\frac{\alpha}{2\pi}\frac{\Gamma(1-\eps)}{\Gamma(1-2\eps)}\lr
 \frac{4\pi\mu^2}{Q^2}\rr^\eps \nonumber \\
 &\times& P_{\gamma\leftarrow q}(x)
 +{\cal O}(\eps,\alpha^2,\alpha\alpha_s)\nonumber \\
 &=& \delta_{\gamma q}\,\delta(1-x)
 \!-\!\le \frac{1}{\eps}\!-\!\gamma_E\!+\!\ln(4\pi)\!+\!\ln\frac{\mu^2}
 {Q^2}\re\nonumber \\
 &\times&\frac{\alpha}{2\pi}P_{\gamma\leftarrow q}(x)
 +{\cal O}(\eps,\alpha^2,\alpha\alpha_s)
 \label{eq:realphotfrag}
\eea
is absorbed into the photon fragmentation function
\beq
 D_{\gamma/q}(x,M_f^2)=\overline{D}_{\gamma/q}(x)+\le
 \overline{\Gamma}_{\gamma\leftarrow q}(M_f^2)\otimes\overline{D}_{\gamma/\gamma}\re (x)
 \label{eq:dqyren}
\eeq
at the factorization scale $M_f^2$.

NLO corrections to resolved-direct and resolved-fragmentation processes
have been calculated in the context of prompt photon production in hadronic
collisions by different groups (see Sec.\ \ref{sec:promptyhad}), while those
for the direct-fragmentation process have been calculated in the context of
inclusive hadron photoproduction (see Sec.\ \ref{sec:lighthad}). The NLO
corrections, which are specific only for the photoproduction of prompt photons,
{\it i.e.} the corrections to the direct-direct process, have been calculated
for inclusive photons by Duke and Owens (1982), Aurenche {\it et al.} (1984a),
Bawa, Krawczyk, and Stirling (1991), and Gordon and Storrow (1994).
Full NLO calculations for photon-hadron scattering have been performed by
Gordon and Vogelsang (1995), using a subtraction term in order to
account also for photon isolation, and by Fontannaz, Guillet, and
Heinrich (2001a) in a completely differential form.
Gordon and Storrow (1996) have furthermore calculated NLO corrections to
prompt photon production in photon-photon collisions.

\begin{figure}
 \begin{center}
  \epsfig{file=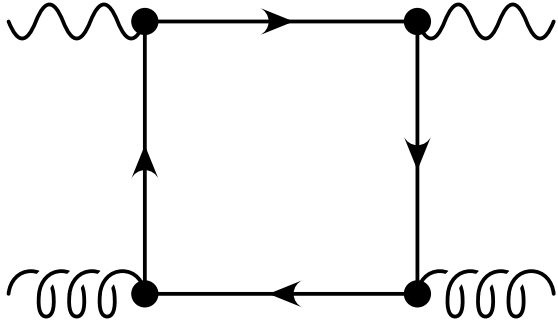,bbllx=66pt,bblly=299pt,bburx=148pt,bbury=351pt,%
          width=0.5\linewidth}
 \caption{\label{fig:48}
 The quark box contribution to direct photon photoproduction.}
 \end{center}
\end{figure}

The direct process $\gamma\gamma\to\gamma X$ can only proceed through a
quark box diagram with $X=\gamma$ and is of $\O(\alpha^4)$. The process
$\gamma g\to\gamma X$ with $X=g$ proceeds through the quark box 
shown in Fig.\ \ref{fig:48} and similar diagrams with reversed
and crossed fermion flow
%
%
and is of
$\O (\alpha^2\alpha_s^2)$. It thus represents a NNLO correction to the
tree-level subprocesses which contain at least one quark line and are of
$\O (\alpha^2)$. Since the box contribution is by itself gauge invariant and
can be numerically large, it can and has to
be included together with the NLO corrections of the tree-level processes
(Aurenche {\it et al.}, 1984a; Krawczyk and Zembrzuski, 2001a; Fontannaz,
Guillet, and Heinrich, 2001a).

%
\begin{figure}
 \begin{center}
  \epsfig{file=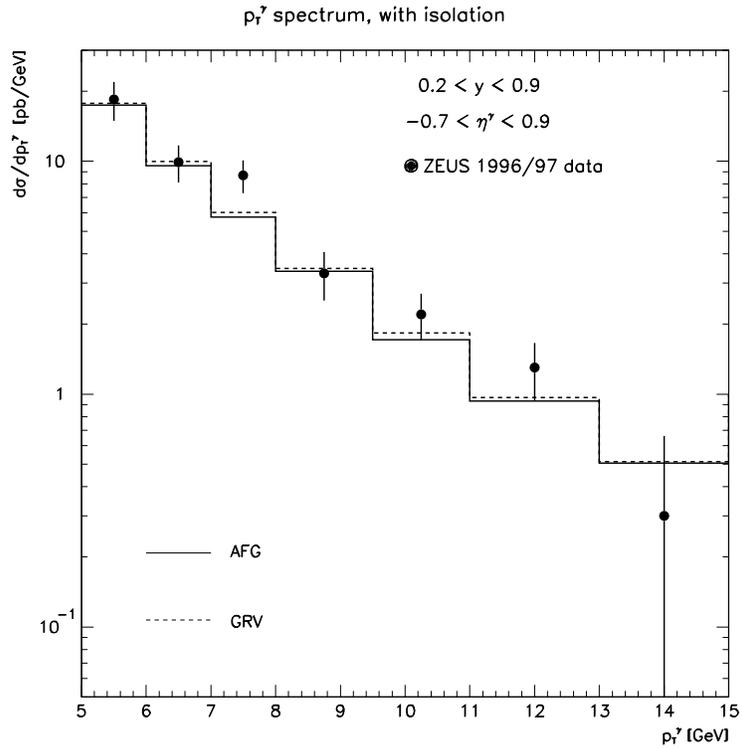,width=0.6\linewidth}
  \epsfig{file=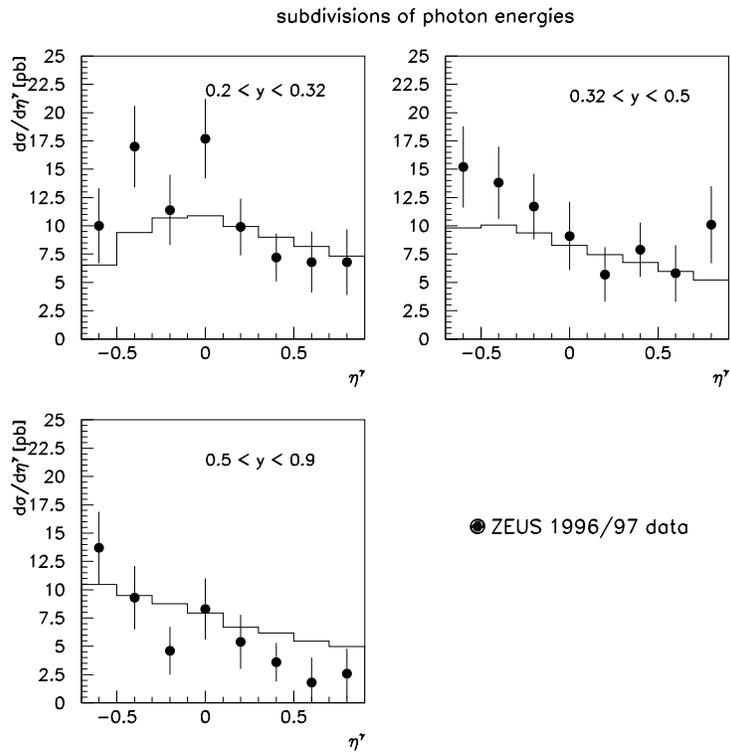,width=0.6\linewidth}
 \caption{\label{fig:49}
 Transverse momentum (top) and pseudorapidity (bottom) spectra of isolated
 photons in NLO compared to ZEUS data (Fontannaz, Guillet, and Heinrich,
 2001a).}
 \end{center}
\end{figure}
%
In Fig.\ \ref{fig:49} the only existing complete NLO calculation for
isolated photons, which also includes the NNLO box contribution (Fontannaz,
Guillet, and Heinrich, 2001a), is compared to recent data from ZEUS (Breitweg
{\it et al.}, 2000d). The transverse momentum ($p_T^\gamma$) distribution (top)
agrees quite well with the data, but the experimental errors are still too
large to distinguish between the AFG and GRV parton densities in the photon.
On the other hand, the NLO pseudorapidity ($\eta^\gamma$) distributions with
AFG photon densities (bottom), integrated over $p_T^\gamma\in[5;10]$ GeV,
show some discrepancies with the data:
The calculations underestimate the data for negative rapidities and small
fractional photon energies $y=E_\gamma/E_e$ and
overestimate them for positive rapidities and large photon energies. 
Similar results are obtained with the calculations of Gordon (1998) and
Krawczyk and Zembrzuski (2001a), which are partly based on LO cross sections.
The movement of the maximum towards the backward direction is caused by an
increase of the photon energy participating in the hard scattering process.
A possible explanation for the discrepancy in the forward direction might be
that there is more hadronic activity in this experimental region, which is
more strongly affected by the isolation cut than the theoretical parton level
simulation.
At a future THERA collider, where 250 GeV electrons would be collided with
920 GeV protons, the cross section in the ZEUS kinematic region would be
about three times larger (Krawczyk and Zembrzuski, 2001b).

\subsection{Photons and jets}

\begin{figure}
 \begin{center}
  \epsfig{file=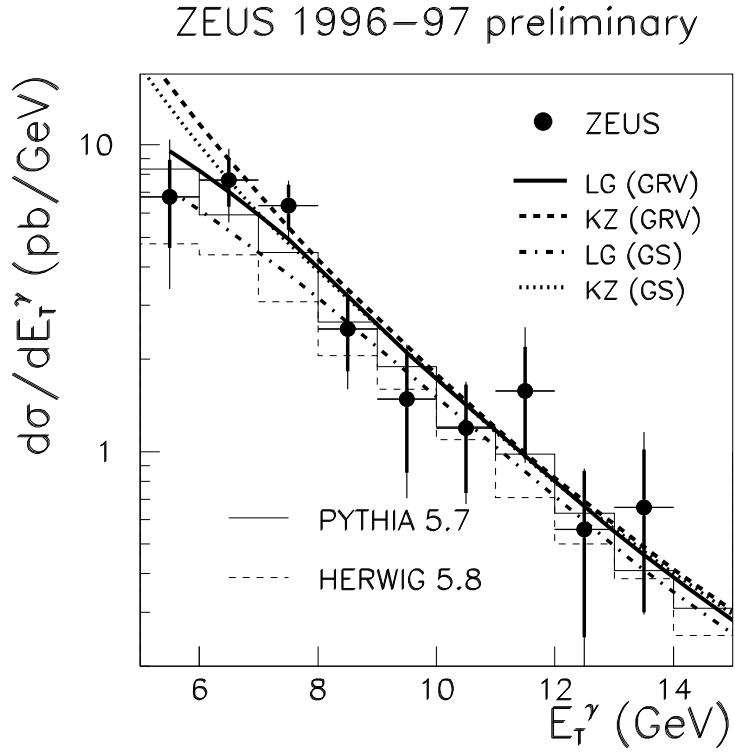,width=0.6\linewidth}
  \epsfig{file=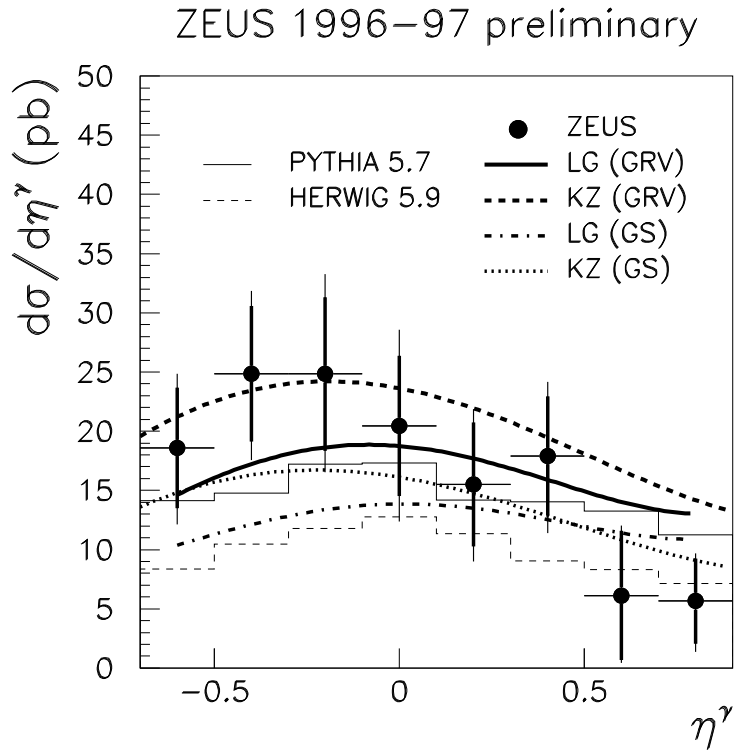,width=0.6\linewidth}
 \caption{\label{fig:50}
 Transverse energy (top) and pseudorapidity (bottom) spectra of isolated
 photons
 produced in association with a jet of $E_T>5$ GeV, $-1.5<\eta<1.8$, and cone
 size $R=1$ compared to ZEUS data (Lee, 2000).}
 \end{center}
\end{figure}

Prompt photon production in association with an
observed jet has been calculated by Gordon [LG] (1998) and by Krawczyk and
Zembrzuski [KZ] (2001a). However, both of these calculations use LO expressions
for the resolved-fragmentation contribution (the latter also for the
direct-fragmentation and resolved-direct contributions). 
Numerical results of these calculations, using GRV and GS parton densities in
the photon, are compared with ZEUS data (Lee {\it et al.}, 2000) and PYTHIA
(Sj\"ostrand,
1994) and HERWIG (Abbiendi {\it et al.}, 1992) Monte Carlo predictions in
Fig.\ \ref{fig:50}.
%
%
The shape of the transverse energy distribution (top),
integrated over $\eta^\gamma\in[-0.7;0.9]$, is described by all theoretical
predictions, but the magnitude is not quite reproduced by the Monte Carlo
predictions. The same observation can be made in the pseudorapidity
distribution
(bottom) integrated over $E_T^\gamma\in[5;10]$ GeV, but here the higher
order calculations with GS photon densities also fall below the data in the
backward region. The main difference between the two higher order calculations 
comes from the box contribution, which is included only in the calculation
of Krawczyk and Zembrzuski (2001a).

Fixed-target and hadron collider experiments indicate that prompt photon
production may be influenced by intrinsic transverse momenta of the scattering
initial
partons (see Sec.\ \ref{sec:promptyhad}). ZEUS have analyzed
isolated prompt photon photoproduction in association with a jet and observed
an excess over the LO QCD expectation in the distributions of perpendicular
momentum $p_\bot$, longitudinal momentum $p_\|$, and azimuthal angle
$\Delta\phi$ of the photon relative to the balancing
jet (Chekanov {\it et al.}, 2001).
They have then attributed these discrepancies to an effective $\langle k_T
\rangle$ of 1.69 $\pm 0.18^{+0.18}_{-0.20}$ GeV, which includes effects
coming from the initial-state parton showering as modeled within PYTHIA
(Sj\"ostrand, 1994).
This value of $\langle k_T \rangle$ seems to be consistent with determinations
in hadron collisions at different energies,
which are, however, obtained using a variety of methods. Within the LO Monte
Carlo comparison, the ZEUS data seem to support the trend that the effective
$\langle k_T \rangle$ in the proton rises with the available hadronic energy.
However,
the data have recently been confronted with a calculation for photon plus
jet photoproduction, that includes all NLO corrections and the NNLO box
diagram (Fontannaz, Guillet, and Heinrich, 2001b). The comparison is
complicated by the fact that equal cuts on the transverse energies of the
photon and recoiling jet have been used in the experiment ($E_T>5$ GeV),
which has to be relaxed in the
NLO calculation and induces some sensitivity on the size of the mismatch
$|E_{T,\min}^\gamma-E_{T,\min}^{\rm jet}|< 0.5$ GeV. The sensitivity within
this range is, however, smaller than the experimental uncertainty. Within
errors, the full
NLO calculation can describe the ZEUS data without any intrinsic $\langle k_T
\rangle$ effects.
The production of direct photons and jets in direct, single-, and
double-resolved photon-photon collisions has also recently been calculated in
NLO (Fontannaz, Guillet, and Heinrich, 2002), and good agreement with
preliminary OPAL data (Abbiendi {\it et al.}, 2001d) was found.
In the future, NLO calculations should be performed for the production of
two photons and photons in association with light and heavy hadrons in order
to obtain more information about the flavor content of photons and protons.


\pagebreak
\section{Related topics}
\label{sec:related}
\setcounter{equation}{0}

In this Section, the topics discussed so far will be extended in several
directions. First the photon virtuality, which was assumed to be negligible
so far, will be explicitly taken into account. Second, the initial photon
will be allowed to have a well-defined polarization, and third the production
of prompt photons in hadron collisions will be discussed, which is closely
related to photoproduction by crossing an initial state photon into the final
state.

\subsection{Virtual photons}

\begin{figure}
 \begin{center}
  \epsfig{file=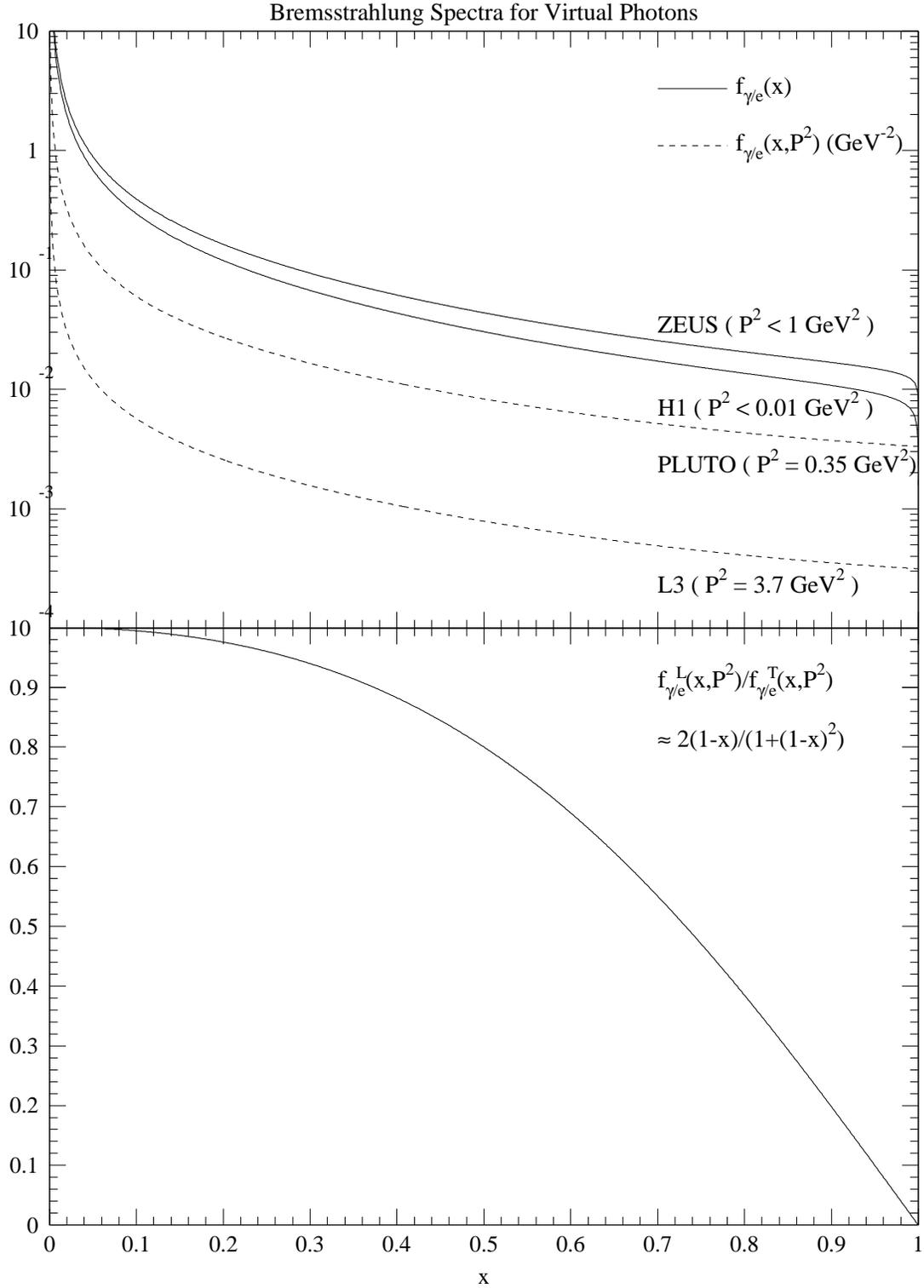,width=0.9\linewidth}
 \caption{\label{fig:51}
 Bremsstrahlung spectra for transverse photons with different virtualities
 (top) and the ratio of longitudinal and transverse photon fluxes (bottom).
 The latter is essentially independent of the photon virtuality.}
 \end{center}
\end{figure}

If the momentum transfer in a lepton scattering process $q=p_l-p_{l'}$ is
small, but not completely negligible, the corresponding cross section can be
calculated in the Equivalent Photon approximation using a differential form
of the transverse photon flux (Budnev {\it et al.}, 1974; Kessler, 1975)
\beq
 \frac{{\rm d}f_{\gamma/l}^{\rm brems}}{{\rm d}P^2}(x,P^2)=\frac{\alpha}
 {2\pi}\left[
 \frac{1+(1-x)^2}{x}\frac{1}{P^2}
 -\frac{2 m_l^2 x}{P^4}\right].
 \label{eq:trans_brems}
\eeq
The Weizs\"acker-Williams approximation (Eq.\ (\ref{eq:unpol_brems}))
is recovered by integrating Eq.\ (\ref{eq:trans_brems}) over the photon
virtuality $P^2=-q^2$. As $P^2$ becomes larger, the longitudinal photon flux
\beq
 \frac{{\rm d}f_{\gamma/l}^{\rm brems}}{{\rm d}P^2}(x,P^2)=\frac{\alpha}
 {2\pi}\left[ \frac{2(1-x)}{x}\frac{1}{P^2}\right]
 \label{eq:long_brems}
\eeq
also has to be taken into account. Bremsstrahlung spectra for transverse
photons with different virtualities are shown in Fig.\ \ref{fig:51} (top)
together with the ratio of longitudinal and transverse photon fluxes (bottom),
which is essentially independent of the photon virtuality.
%
%

Like real photons, photons with a non-zero virtuality $P^2$ can be probed for
their hadronic structure in deep-inelastic electron-photon scattering.
As $P^2$ approaches the virtuality $Q^2$ of the probing photon, the structure
of the virtual photon reduces to the perturbative splitting of the
photon into a collinear quark-antiquark pair with probability (Uematsu and
Walsh, 1981, 1982; Rossi 1984)
\bea
 f^{\rm box}_{q/\gamma}(x,Q^2,P^2)&=&3e_q^2\frac{\alpha}{2\pi}\left\{\left[
 x^2+(1-x)^2\right]\ln\frac{Q^2}{x^2P^2}\rp\nonumber\\
 &+&\lp 6x(1-x)-2\right\}.
 \label{eq:virtphotuwr}
\eea
This expression includes contributions from transverse and longitudinal
photons. It differs from the real photon case $f_{q/\gamma}(x,Q^2)=
e_q^2\alpha C_\gamma/(2\pi)$ with $C_\gamma$ given in Eq.\ (\ref{eq:cgamma}),
since the collinear singularity is now mass
regulated by the photon virtuality $P^2$. In the limit $P^2\ll Q^2$, the
structure of the virtual photon should, of course, coincide with that of the
real photon.

A simple phenomenological ansatz to suppress smoothly the quark and gluon
content in the virtual photon at high $P^2$ is (Drees and Godbole [DG], 1994)
\bea
 f_{q/\gamma}(x,Q^2,P^2)&=&f_{q/\gamma}(x,Q^2) \frac{\ln\frac{Q^2+P_c^2}
 {P^2+P_c^2}}{\ln\frac{Q^2+P_c^2}{P_c^2}}, \nonumber \\
 f_{g/\gamma}(x,Q^2,P^2)&=&f_{g/\gamma}(x,Q^2) \frac{\ln^2\frac{Q^2+P_c^2}
 {P^2+P_c^2}}{\ln^2\frac{Q^2+P_c^2}{P_c^2}},
 \label{eq:virtphotdg}
\eea
where the typical hadronic scale $P_c$ lies between the QCD scale $\Lambda$
and the proton mass. Since gluons are radiated from off-shell
quarks, they must be more suppressed than quarks (Borzumati and Schuler, 1993).
A similar ansatz has also been used by Aurenche {\it et al.} (1994b). It can
be applied to any parameterization for the parton densities
in the real photon, which are recovered in the limit $P^2\ll Q^2$. At large
$P^2$, the {\it evolved} parton densities are globally suppressed. This means,
however, that the perturbative result of Eq.\
(\ref{eq:virtphotuwr}) is not reproduced, since the logarithmic suppression
factors are independent of $x$.

A refined version of the above prescription has been formulated by
Schuler and Sj\"ostrand [SaS] (1995, 1996), who sum the real hadronic
contributions from the three dominant vector mesons $\rho,\omega,$ and $\phi$
after weighting them with factors $\eta(P^2)=(1+P^2/m_{\rho,\omega,\phi}^2)^
{-2}$ and evolving them from max($P^2,Q_0^2$). The pointlike contribution
is evolved with a continuous suppression factor $(1+P^2/k^2)^{-2}$, so that
it also starts effectively from max($P^2,Q_0^2$). In the
low- and high-$P^2$ limits, the real photon
densities and the logarithmically enhanced terms of the virtual pointlike box
are correctly reproduced, but not the $x$-dependent terms of the latter.

Gl\"uck, Reya, and Stratmann [GRSt] (1995) interpret Eq.\
(\ref{eq:virtphotuwr})
at $P^2=Q^2$ and $f_{g/\gamma}(x,P^2,P^2)=0$ as NLO pointlike boundary
conditions. In LO these conditions differ since here also $f_{q/\gamma}
(x,P^2,P^2)=0$. The pointlike boundary conditions are then added to the
hadronic VMD input for real photons $f^{\rm had}_{i/\gamma}(x,Q_0^2)$, adopted
from Gl\"uck, Reya, and Vogt [GRV] (1992b), with an interpolating factor
$\eta(P^2)=(1+P^2/m_\rho^2)^{-2}$, so that
\bea
 f_{i/\gamma}(\!\!&x&\!\!,\max(P^2,Q_0^2),P^2) = 
 \eta(P^2)f^{\rm had}_{i/\gamma}(x,\max(P^2,Q_0^2)) \nonumber \\
 &+&[1-\eta(P^2)] f^{\rm box}_{i/\gamma}(x,\max(P^2,Q_0^2),P^2).
 \label{eq:virtphotgrs}
\eea
The functional form of $\eta(P^2)$ can be derived from a dispersion integral in
the photon virtuality, which is dominated by the poles of the vector mesons,
particularly the $\rho$-meson, in the region of small $P^2$.
In the limit $P^2\ll Q^2$, the purely hadronic real photon input is recovered
in LO and NLO, as required by the $\dis$ scheme. As $P^2$ approaches $Q^2$,
the {\em unevolved} initial parton densities are globally suppressed and
the virtual pointlike result Eq.\ (\ref{eq:virtphotuwr}) is recovered.
This is, however, inconsistent with the original $\dis$ scheme, since the NLO
pointlike quark boundary condition differs now from zero, but is evolved
using the same {\em massless} $\dis$ splitting functions (Eqs.\
(\ref{eq:dissplit})) in the inhomogeneous evolution equations (Eqs.\
(\ref{eq:evol_eq})) from the new starting scale $\max(P^2,Q_0^2)$.
A parameterization of these virtual photon parton densities, which is
valid in the region $P^2<Q^2/5$, has been performed only in LO (Gl\"uck, Reya,
and Stratmann, 1996).

%
\begin{figure}
 \begin{center}
  \epsfig{file=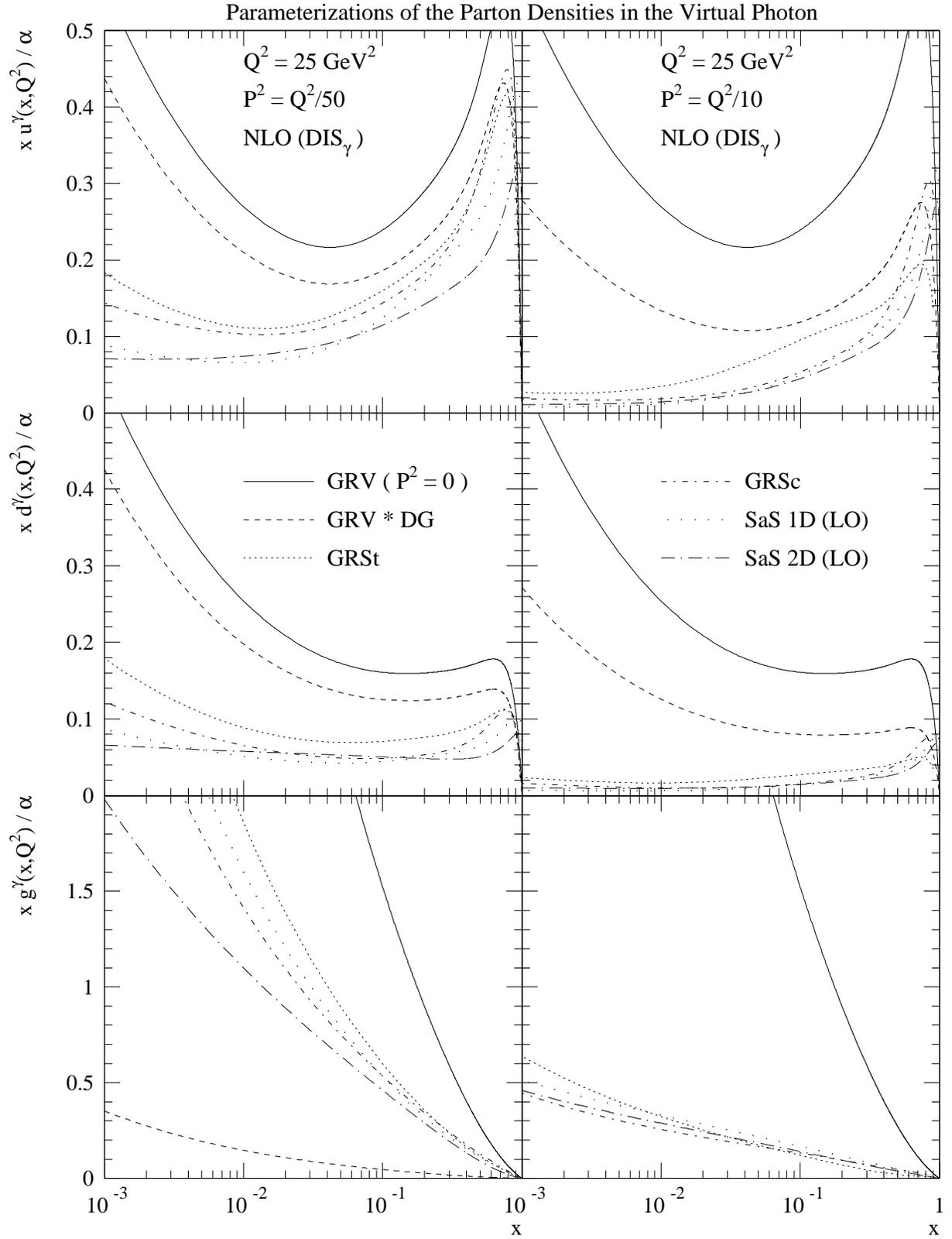,width=\linewidth}
 \caption{\label{fig:52}
 NLO parameterizations of the up-quark (top), down-quark (center), and gluon
 (bottom) densities in a photon with virtuality $P^2=0.5$ GeV$^2$ (left)
 and 2.5 GeV$^2$ (right) at $Q^2=25$ GeV$^2$.}
 \end{center}
\end{figure}
%
The inconsistencies with the $\dis$ scheme can be avoided if the pointlike term
in Eq.\ (\ref{eq:virtphotgrs}) is omitted (Gl\"uck, Reya, and Schienbein
[GRSc], 1999b,
2001). However, in this case the perturbative splitting Eq.\
(\ref{eq:virtphotuwr}) is not reproduced. Consequently, this virtual
parameterization is only applicable in the region of small $P^2$
$(P^2<Q^2/10)$,
where the virtuality of the pointlike term is of minor importance, and the
large-$P^2$ region should be calculated in fixed order. A parameterization
has also been given only in LO, although the NLO results are very similar.

The existing NLO parton densities in the transverse virtual photon
are compared in Fig.\ \ref{fig:52} at $Q^2=25$ GeV$^2$ and $P^2=0.5$ GeV$^2$
(left) and 2.5 GeV$^2$ (right) for the up-quark (top), down-quark (center),
and gluon (bottom) densities.
The suppression of the real photon GRV densities in the virtual photon 
is clearly visible. For the quarks (gluons) it is weakest (strongest) for the
global DG factors, for which $P_c^2=0.3$ GeV$^2$ has been used. At $P^2=Q^2/50$
the $x$-dependence is only weakly modified, while at $P^2=Q^2/10$ the
logarithmic singularity at $x=0$ in the perturbative box is significantly
weaker. The different parameterizations vary largely at small $P^2$,
particularly for the gluon, but much less at larger $P^2$ as expected.

Information about the parton densities in the virtual photon
can be gained from the total hadronic cross section in double-tagged $e^+e^-$
scattering in the Bjorken region $P^2\ll Q^2$. If the probing and target
photons are both soft, their transverse (T) and longitudinal (L) polarizations
contribute with equal weights to the effective virtual photon structure
function
\beq
 F_{\rm eff}^\gamma \simeq F_{\rm TT}^\gamma+F_{\rm TL}^\gamma
 +F_{\rm LT}^\gamma+F_{\rm LL}^\gamma
 \simeq F_2^{\gamma}+{3\over 2} F_L^{\gamma}.
\eeq
In the second equation, the relations
$F_{\rm TL}^\gamma\simeq F_{\rm LT}^\gamma\equiv F_L^\gamma$ and
$F_{\rm LL}^\gamma\simeq 0$ have been used, which apply strictly
speaking only to the virtual box contribution with
$F_L^\gamma = \sum_q 2x e_q^2 {\alpha\over 2\pi} e_q^2 N_C 4 x (1-x)$.
Like $F_2^{\gamma}$, $F_L^{\gamma}$ could be calculated by
inserting {\em longitudinal} virtual photon densities into Eq.\
(\ref{eq:f2gamma}) and employing the longitudinal Wilson coefficients
$C_{q,L}=8x/3$ and $C_{g,L}=2x(1-x)$. In practice, however,
only a pointlike parameterization for the longitudinal virtual photon is
currently available (Ch\'{y}la and Ta\v{s}evsk\'{y}, 2001), and it turns out
to be very close to the unevolved longitudinal box result.

\begin{figure}
 \begin{center}
  \epsfig{file=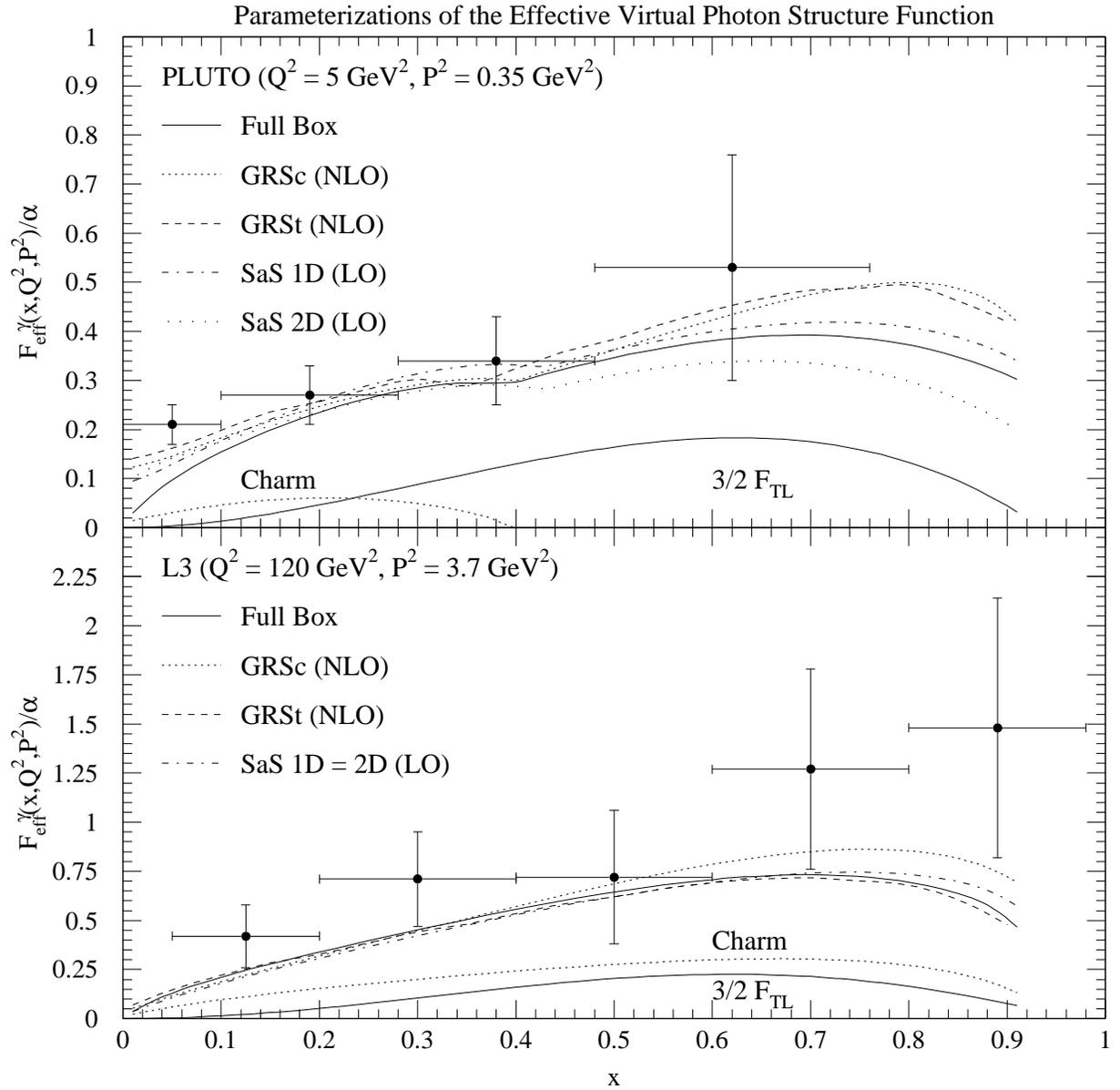,width=\linewidth}
 \caption{\label{fig:53}
 Parameterizations of the effective virtual photon structure function
 compared to data from PLUTO (top) (Berger {\it et al.}, 1984b) and L3 (bottom)
 (Acciarri {\it et al.}, 2000).}
 \end{center}
\end{figure}

In Fig.\ \ref{fig:53}, $F_{\rm eff}^\gamma$ measurements by the PLUTO
%
%
(top) (Berger {\it et al.}, 1984b) and L3 (bottom) (Acciarri {\it et al.},
2000) collaborations are compared with the virtual photon parameterizations
discussed above. In order to obtain $F_{\rm eff}^\gamma$ from $F_2^\gamma$, the
separately shown longitudinal virtual box contribution $3/2 F_{\rm TL}^\gamma$
has been added to all parameterizations. For the GRSc parameterization, which
is evolved with three massless flavors, the sum of direct and resolved heavy
charm quark contributions is also shown separately.
While the PLUTO and L3 data tend to be higher than all parameterizations and
the purely perturbative virtual box prediction, the theoretical curves all
describe the measurements within the errors, so that present data do not yet
convincingly establish the importance of the renormalization group improved
treatment
of parton densities in the virtual photon or the need for an intrinsic
hadronic component (Gl\"uck, Reya, and Schienbein, 2001).

The virtual photon fluxes and parton densities discussed above make it possible
to extend the calculations of real photoproduction processes to the region
$P^2>0$, as long
as $P^2$ is significantly smaller than the hard scattering scale $Q^2$
(typically the squared transverse energy $E_T^2$ of the observed jet, hadron,
or prompt photon). Therefore virtual photoproduction will also receive
contributions from direct and resolved photon processes. They can be
calculated by taking $P^2$ into account only in the photon flux and parton
densities, but neglecting it in the hard scattering. When $P^2$
approaches $Q^2$, however, the resolved
contribution vanishes logarithmically, the longitudinal photon cross section
has to be taken into account, and the DIS cross section has to be recovered.

As already discussed in Sec.\ \ref{sec:phostr}, direct and resolved real
photoproduction processes are related in NLO by the appearance of a collinear
$1/\eps$ singularity in the splitting of the photon into quarks with fractional
charge $e_q$ (Eq.\ (\ref{eq:realphotsplit})). This singularity is absorbed
into the
renormalized quark densities in the real photon (Eq.\ (\ref{eq:fqyren})) and
cancels its logarithmic factorization scale dependence.
A similar mechanism occurs for virtual photoproduction, where the singularity
is now mass regulated by $P^2$ and Eq.\ (\ref{eq:realphotsplit}) has to be
replaced with
\bea
 \overline{\Gamma}_{q\leftarrow\gamma^\ast}(x,P^2)&=&
 \delta_{q\gamma}\,\delta(1-x)
 - \frac{\alpha}{2\pi}\nonumber\\
 &\times&\le\ln\frac{P^2(1-x)}
 {Q^2}P_{q\leftarrow\gamma}(x)
 +N_Ce_q^2\re.
\eea
This defines an $\ms$ factorization scheme for virtual photons, which leaves
identical finite terms in the hard cross section as for real photons (Klasen,
Kramer, and P\"otter, 1998; P\"otter, 1999c). Since there is no $1/\eps$
singularity, the
subtraction is not mandatory to obtain finite cross sections, but only to
resum the logarithm which can become large for $P^2\ll Q^2$ and $x\to 1$.

%
\begin{figure}
 \begin{center}
  \epsfig{file=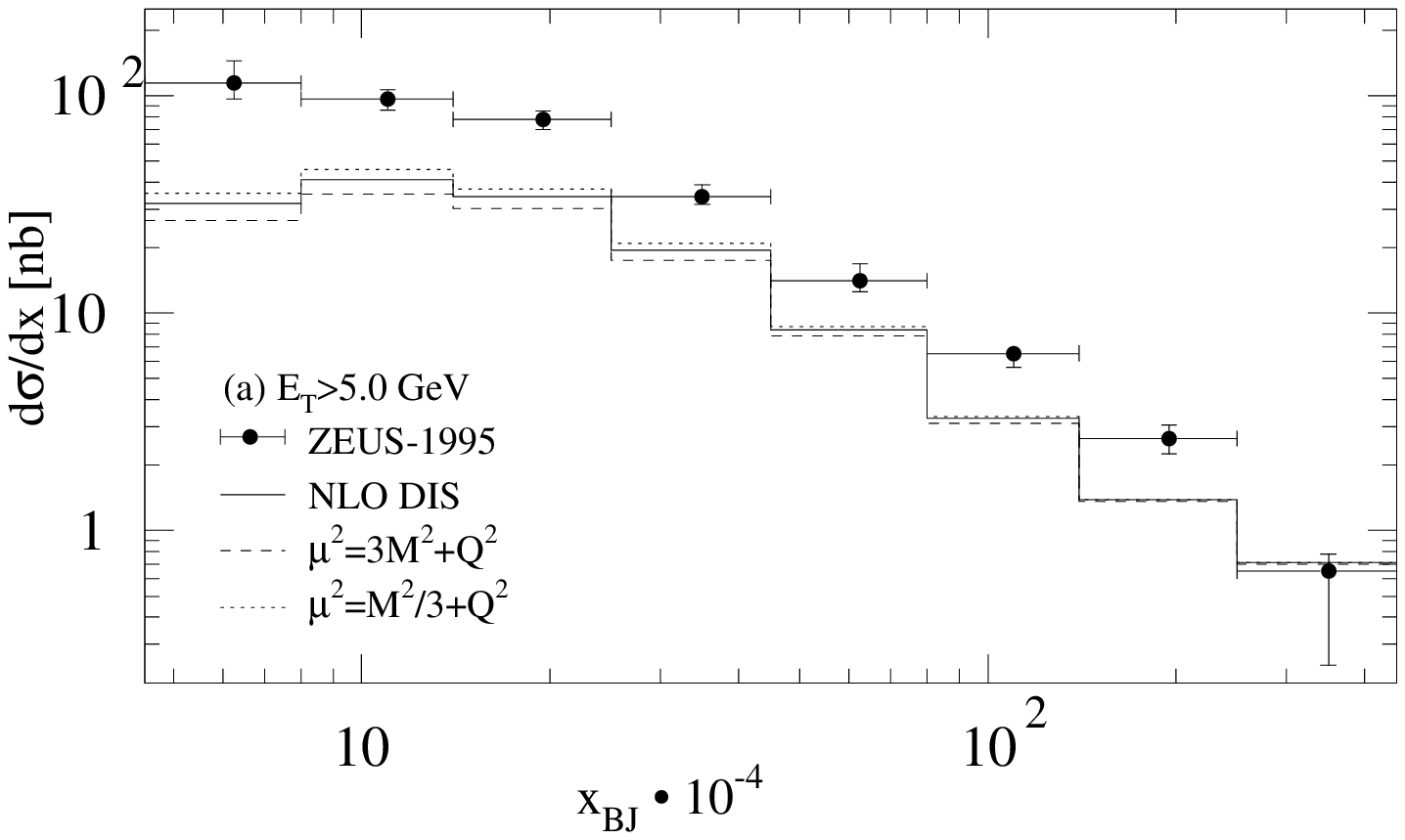,width=\linewidth}
  \epsfig{file=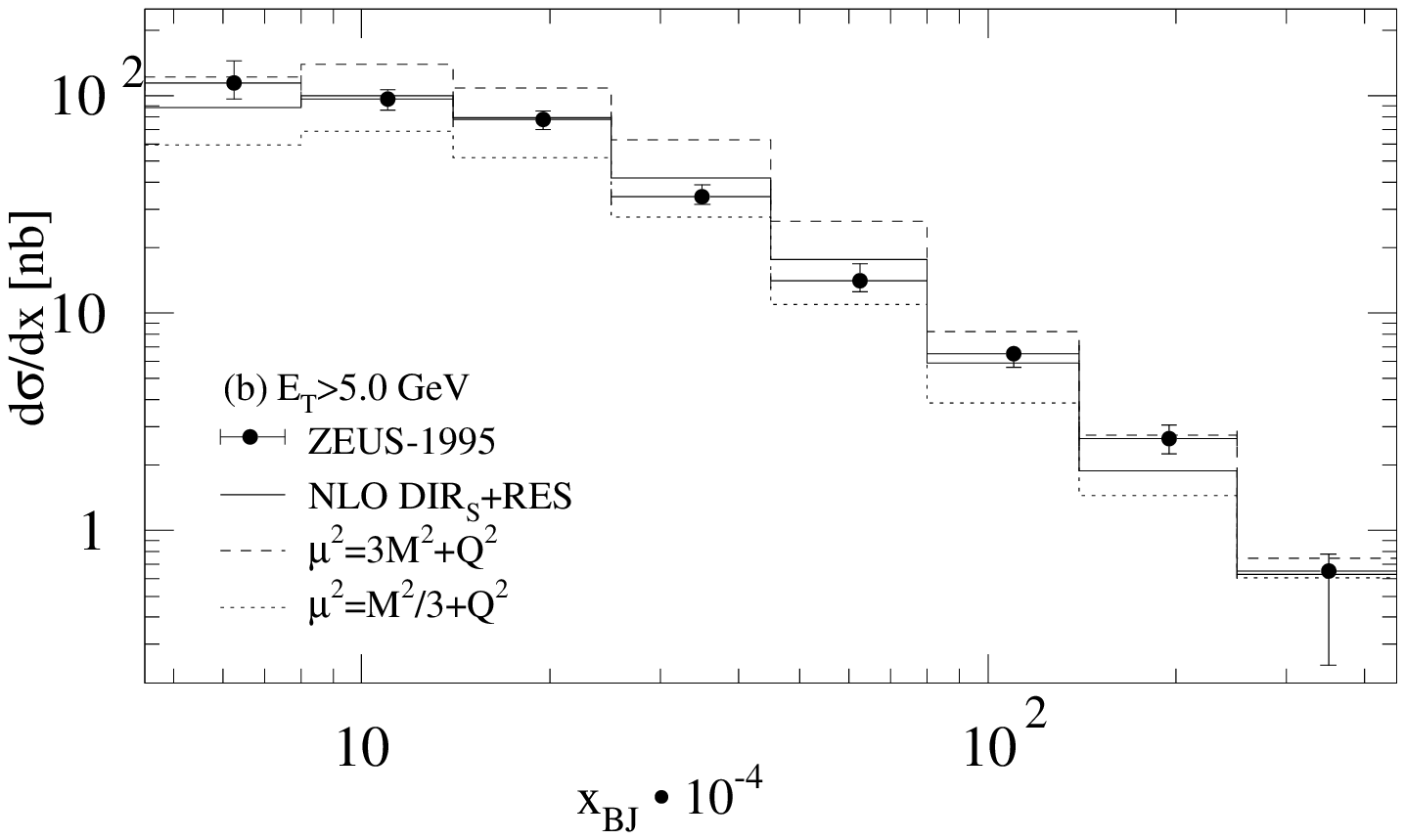,width=\linewidth}
 \caption{\label{fig:54}
 NLO predictions for forward dijet production with virtual photons compared to
 ZEUS 1995 data. In (a) only the DIS direct contribution is included, while
 in (b) direct and resolved virtual photon contributions have been added
 consistently (Kramer and P\"otter, 1999).}
 \end{center}
\end{figure}
%
In order to obtain more information about the structure of the virtual photon
it is useful to consider not only inclusive electron-photon scattering,
but also hard scattering processes like jet photoproduction. LO studies
indicate a significant contribution from resolved virtual photons at HERA,
if $P^2\ll E_T^2$ (Gl\"uck, Reya, and Stratmann, 1996; Gl\"uck, Reya, and
Schienbein, 2001). Virtual jet photoproduction has also been calculated in
NLO by adapting an existing NLO calculation for resolved real photons
(Klasen, Kleinwort, and Kramer, 1999) and using the $\ms$ subtraction for the
direct virtual photon singularity discussed above (Klasen, Kramer, and
P\"otter, 1998). It has then been extended to the DIS region of large $P^2$
and applied to jet production at HERA (Kramer and P\"otter, 1998, 1999) and
LEP (P\"otter, 1999a, 1999b).
The resolved contribution turns out to be particularly
important, if the jets are produced in the forward region.
The ZEUS (H1) collaborations have measured jets with $\eta > 2.6$
$(\eta\in[1.735;2.794])$ and $E_T>5$ GeV ($E_T>3.5$ GeV), respectively
(Breitweg {\it et al.}, 1999c; Adloff {\it et al.}, 1999c). In Fig.\
\ref{fig:54} the ZEUS data
are compared to NLO predictions. While the DIS direct calculation (a)
significantly underestimates the ZEUS data, they can be described if resolved
virtual photon contributions from the SaS 1D fit are added (b) (Kramer and
P\"otter, 1999).

In principle, the production of heavy quarks with mass $m_h$ is also sensitive
to the structure of the virtual photon. In this case resolved photon
contributions are applicable as long as $P^2 \ll m_h^2$. However a LO study
shows that they are numerically small at the relatively large values of $x$
currently probed at HERA (Gl\"uck, Reya, and Stratmann, 1996) and LEP energies
(Gl\"uck, Reya, and Schienbein, 2001).

Finally, a LO study for virtual photoproduction of prompt photons has shown
that the gluon content in the virtual photon may be tested if the photons are
produced in the forward direction at transverse energies $E_T\simeq 5$ GeV
(Krawczyk and Zembrzuski, 1998).

\subsection{Polarized photons}

\begin{figure}
 \begin{center}
  \epsfig{file=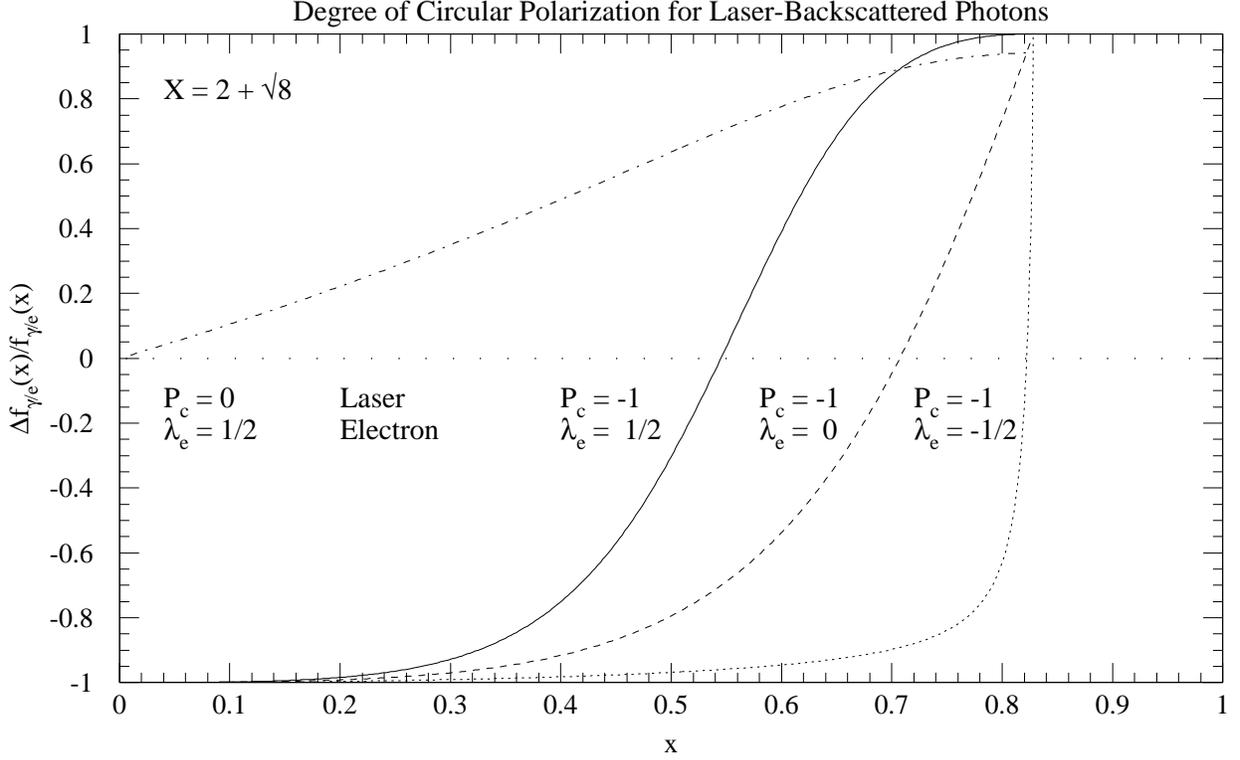,width=\linewidth}
 \caption{\label{fig:55}
 Degree of circular polarization for laser-back\-scattered photons. The
 dot-dashed curve applies also to brems- and beamstrahlung photons, but then
 it extends out to $x=1$ and $\Delta_{\gamma/e}f(x)/f_{\gamma/e}(x)=1$.}
 \end{center}
\end{figure}

Polarized photons can be produced by bremsstrahlung off lepton beams with
circular polarization $|\lambda_l|\leq 1/2$. The corresponding spectrum
$\Delta f^{\rm brems}_{\gamma/l}(x)=f^{\rm brems,+}_{\gamma/l}(x)-
f^{\rm brems,-}_{\gamma/l}(x)$ is given by (Philipsen, 1992;
De Florian and Frixione, 1999)
\beq
 \frac{{\rm d}\Delta f_{\gamma/l}^{\rm brems}}{{\rm d}P^2}(x,P^2)=\frac{\alpha}
 {2\pi}\left[
 \frac{1-(1-x)^2}{x}\frac{1}{P^2}
 -\frac{2 m_l^2 x^2}{P^4}\right]
\eeq
for transverse virtual photons and by 
\bea
 \Delta f_{\gamma/l}^{\rm brems}(x) &=& \frac{2\lambda_l\alpha}{2\pi}
 \left[\frac{1-(1-x)^2}{x} \ln \frac{Q^2_{\max} (1-x)}{m_l^2 x^2}\rp\nonumber\\
 &+&\lp 2 m_l^2 x^2\left(\frac{1}{Q^2_{\max}}-\frac{1-x}{m_l^2 x^2}\right)
 \right].
 \label{eq:pol_brems}
\eea
for almost real photons.
As in the unpolarized spectrum (Eq.\ (\ref{eq:unpol_brems})) a non-logarithmic
term is present, which is, however, not singular for $x\rightarrow 0$.
At $x\simeq 1$ the photons are completely polarized parallel to the incoming
lepton helicity, but at $x\simeq 0$, where most of the bremsstrahlung photons
are produced, they are completely unpolarized (see dot-dashed curve in Fig.\
\ref{fig:55}).
%
%
The lepton polarization is lost in the Lorentz transformation from the Breit
frame of the lepton-photon vertex to the center-of-mass frame of the
photon-target vertex.

The circularly polarized beamstrahlung spectrum has only been derived in an
integral form (Schroeder, 1990)
\bea
 \Delta f_{\gamma/e}^{\rm beam}(x) &=& \frac{5 \lambda_e}{2\sqrt{3}\Upsilon}
 \int_u^{\infty}{\rm d}v{\rm Ai}(v)\left[\left(\frac{2v}{u}-1\right)
 \rp\nonumber\\
 &\times&\lp\frac{1-(1-x)^2}{2(1-x)}+\frac{x^2}{2(1-x)}\right],
\eea
where the Airy-function Ai$(v)$ falls off exponentially at large $v$ and where
$u=[5x]^{2/3}/[4\sqrt{3}\Upsilon(1-x)]^{2/3}$. The polarization $\Delta
f_{\gamma/e}^{\rm beam}(x)/f_{\gamma/e}^{\rm beam}(x)$ depends only weakly on
the beamstrahlung parameter $\Upsilon$ and is very similar to the
bremsstrahlung polarization (Berge, Klasen, and Umeda, 2001). 

While the photon polarization at an $e^+e^-$ collider is thus rather limited, a
photon collider offers the additional possibility to control the helicity of
the laser photons $|P_c| \leq 1$.
The outgoing photons have a polarized spectrum (Ginzburg {\it et al.}, 1984)
\bea
 \Delta f_{\gamma/e}^{\rm laser}(x) &=& \frac{1}{N_c+2 \lambda_e P_c N_c'}
 \left\{ 2 \lambda_e \frac{x}{1-x}
 \left[ 1+(1-x) \rp\rp\nonumber\\
 &\times&\lp\left( 1-\frac{2 x}{(1-x) X}\right)^2\right] \\
 &+& \left. P_c \left( 1-\frac{2 x}{(1-x) X}\right) \left(1-x+\frac{1}{1-x}
 \right)
 \right\}. \nonumber
\eea
If only the electrons are polarized $(P_c=0)$, the $x$-dependence of the
polarization $\Delta f_{\gamma/e}^{\rm laser}(x)/f_{\gamma/e}^{\rm laser}(x)$
is similar to the brems- and beamstrahlung cases (dot-dashed curve in Fig.\
\ref{fig:55}). However if $P_c = \pm 1$, then the helicity of the
backscattered photons is opposite to that of the laser photons at $x =
x_{\max}$ (full, dashed, and dotted curves in Fig.\ \ref{fig:55}).
Therefore the choice $2 \lambda_e P_c = -1$
guarantees not only good monochromaticity (see Fig.\ \ref{fig:01}),
but also a high degree of
polarization of the produced photons. By switching the signs of $\lambda_e$
and $P_c$ simultaneously, one can switch the helicity of the outgoing photons
without spoiling the monochromaticity of the photon spectrum.

The polarized parton densities $\Delta f_{i/\gamma}(Q^2)$ of the photon obey
the same perturbative evolution Eqs.\ (\ref{eq:evol_eq}) as their unpolarized
counterparts, except that
polarized splitting functions $\Delta P_{j\leftarrow i}$ have to be used. These
have recently been calculated in NLO by Mertig and van Neerven (1996) and by
Vogelsang (1996a, 1996b). We review only the LO results (Altarelli and Parisi,
1977)
\bea
  \Delta P_{q\leftarrow q} (x) & = &
    C_F \left[ \frac{1+x^2}{(1-x)}_+ + \frac{3}{2} \delta (1-x)
    \right] +{\cal O}(\alpha_s) \nonumber\\
 &=& P_{q\leftarrow q} (x), \nonumber\\
  \Delta P_{g\leftarrow q} (x) & = &
    C_F \left[ \frac{1-(1-x)^2}{x} \right] +{\cal O}(\alpha_s), \nonumber\\
  \Delta P_{q\leftarrow g} (x) & = &
    T_R \left[ x^2-(1-x)^2 \right]+{\cal O}(\alpha_s), \nonumber \\
  \Delta P_{g\leftarrow g} (x) & = &
    N_C \left[ (1+x^4)\left(\frac{1}{x}+\frac{1}{(1-x)}_+\right)-
    \frac{(1-x)^3}{x} \re \nonumber \\
    &+& \left[ \frac{11}{6}N_C-\frac{1}{3}N_f\right] \delta (1-x)+{\cal O}
    (\alpha_s),
\eea
where the Dirac matrix $\gamma_5$ has been assumed to anticommute with the
Dirac matrices $\gamma_\mu$. Different results are obtained in other
$\gamma_5$-schemes as mandated in dimensionally regularized NLO calculations.
Some of them require finite renormalizations to arrive at the correct final
answer. The polarized photon-quark splitting function can be obtained in LO
by the
transformation $\Delta P_{q\leftarrow\gamma} = 2N_Ce_q^2 \Delta P_{q\leftarrow
g}$, and the polarized NLO photon-gluon splitting function is given by
(Stratmann and Vogelsang, 1996)
\bea
 \Delta P_{g\leftarrow\gamma} (x) &=& \frac{\alpha_s(Q^2)}{2\pi}
 N_CN_f\langle e^2\rangle C_F\le -2(1+x)\ln^2 x \rp\nonumber\\
 &+&\lp 2 (x-5)\ln x-10(1-x)\re.
\eea

As in the unpolarized case, the pointlike solution of the polarized evolution
equations dominates at large $x$ and $Q^2$.
It takes the same functional form as Eq.\
(\ref{eq:asymphot}), where $\Delta a_i$ and $\Delta b_i$ replace $a_i$ and
$b_i$ and have been explicitly calculated in moment space in LO (Irving and
Newland, 1980; Hassan and Pilling, 1981; Xu, 1984) and NLO (Stratmann and
Vogelsang, 1996). The position of the low-$x$ singularities in $\Delta a_i$ and
$\Delta b_i$ differs from the unpolarized case, since the polarized splitting
functions are now involved.

By combining the polarized parton distributions functions $\Delta
f_{q/\gamma}(Q^2)$ and $\Delta f_{g/\gamma}(Q^2)$ with the appropriate Wilson
coefficients (Ratcliffe, 1983; Bodwin and Qiu, 1990)
\begin{eqnarray}
\Delta C_q(\!\!&x&\!\!)  =  C_F\, \left[ (1+x^2)\lr\frac{\ln(1-x)}{1-x}\rr_+
 -\frac{3}{2}\frac{1}{(1-x)}_+\rp\nonumber\\
 &-&\lp\frac{1+x^2}{1-x}\ln x+2+x \right] 
 -C_F\left[\frac{9}{2}+\frac{\pi^2}{3}\right]\delta(1-x)
 ,\nonumber\\
 \Delta C_g(\!\!&x&\!\!)  =  T_R \left[ (2x\!-\!1)\lr\ln\frac{1\!-\!x}{x}\!-\!
 1\rr\!+2(1-x) \right]\!,
\end{eqnarray}
and (Stratmann and Vogelsang, 1996)
\bea
 \Delta C_{\gamma}(x)&=&
 2N_C\, \Delta C_g(x)=3\, \left[ (2x-1)\lr\ln\frac{1-x}{x}
 -1\rr\rp\nonumber\\
 &+&\lp 2(1-x) \right],
\label{eq:deltacgamma}
\eea
one obtains the NLO photon structure function in the $\ms$ scheme
\bea
 g_1^{\gamma}(&Q^2&)=\sum_q e_q^2\lg \Delta f_{q/\gamma}(Q^2)
 +\frac{\alpha}{2\pi}e_q^2\Delta C_{\gamma}+\frac{\alpha_s
 (Q^2)}{2\pi}\rp\nonumber\\
 &\times&\lp\le \Delta C_{q}\otimes \Delta f_{q/\gamma}(Q^2)+\Delta C_{g}
 \otimes \Delta f_{g/\gamma}(Q^2)\re\rg.
 \label{eq:g1gamma}
\eea
The $\ln(1-x)$-term in the Wilson coefficient of the polarized photon
causes similar stability problems in $g_1^\gamma$ as in $F_2^\gamma$ at large
$x$. Thus it should again be absorbed into the quark distributions, which
affects
also the polarized pointlike NLO splitting functions. The transformation is
given by Eqs.\ (\ref{eq:dissplit}) with all quantities replaced by their
polarized counterparts. This DIS$_{\Delta\gamma}$
factorization scheme leads then also to
perturbatively stable results and to purely hadronic boundary conditions
in LO and NLO.

\begin{figure}
 \begin{center}
  \epsfig{file=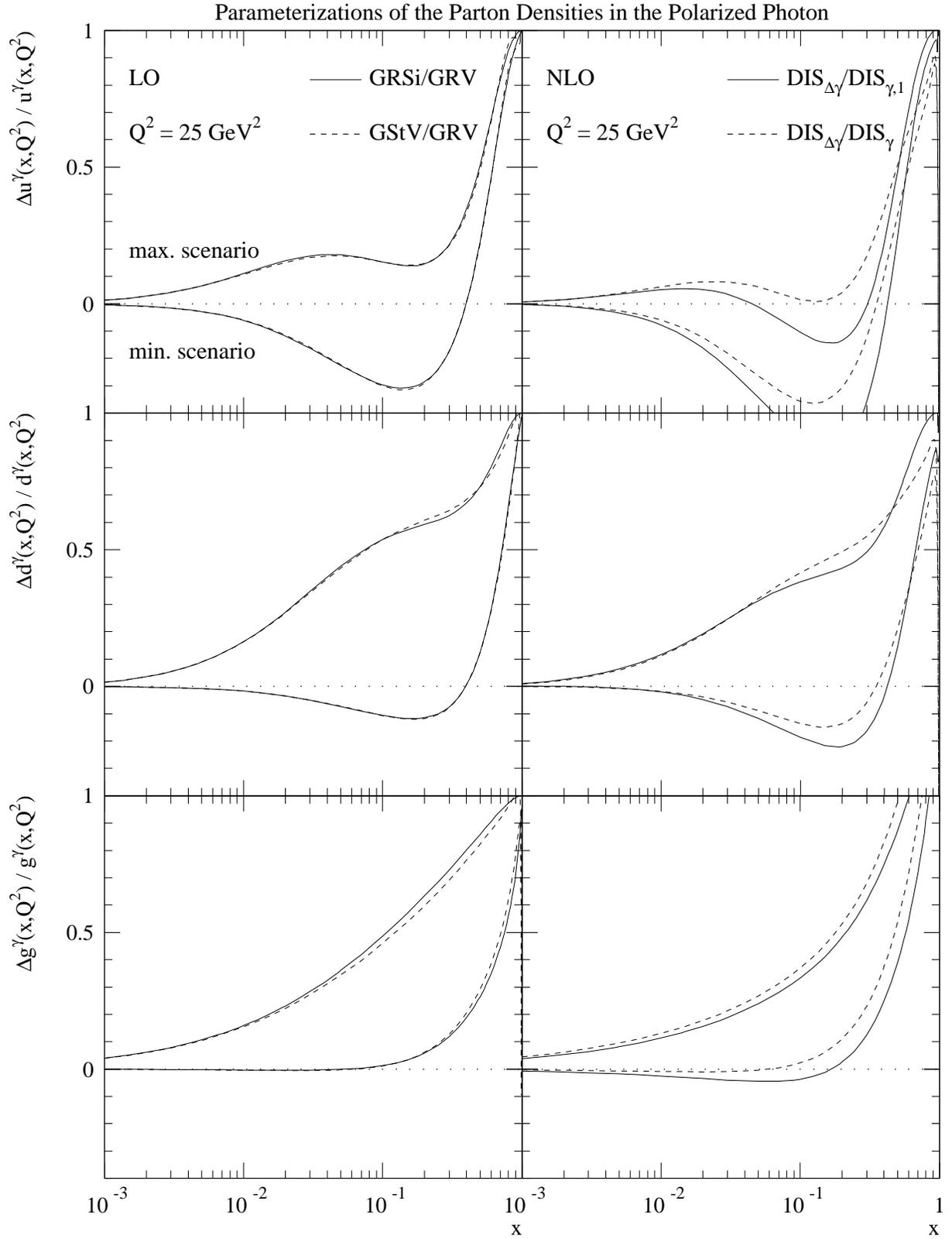,width=\linewidth}
 \caption{\label{fig:56}
 Parameterizations of the polarized up-quark (top), down-quark (center), and
 gluon (bottom) densities in the photon in LO (left) and NLO (right) at
 $Q^2=25$ GeV$^2$.}
 \end{center}
\end{figure}

For unpolarized photons, the hadronic input can be inferred from pionic
parton densities. These are, however, not known in the polarized case. Thus
Stratmann and Vogelsang (1996) were forced to make the two extreme assumptions
that either $\Delta f_{i/\gamma}^{\rm had}(Q_0^2)=f_{i/\gamma}^{\rm had}
(Q_0^2)$ or $\Delta f_{i/\gamma}^{\rm had}(Q_0^2)=0$, as had already been
suggested in the LO analysis of Gl\"uck, Stratmann, and Vogelsang [GStV]
(1994). Current conservation implies that the first moment of $g_1$
vanishes (Bass, 1992; Narison, Shore, and Veneziano, 1993), which can be
realized by demanding $\Delta f_{i/\gamma}^{{\rm had},~n=1}=0$. This is
trivial to fulfill in the second (minimal) scenario, but requires additional
assumptions on the low-$x$ behavior of the first (maximal) scenario.
In a previous LO analysis, Gl\"uck and Vogelsang (1992) considered these
two scenarios only for the gluon, while they fixed $\Delta f_{q/\gamma}^{\rm
had}(Q_0^2)=f_{q/\gamma}^{\rm had}(Q_0^2)$. 
The original $\dis$ scheme, which was constructed to avoid negative values of
$F_2^\gamma$ at large $x$, unfortunately causes a violation of the positivity
constraint $|g_1^\gamma|\leq F_1^\gamma=(F_2^\gamma-F_L^\gamma)/(2x)$. For
polarized photons it is thus preferable to absorb the photonic Wilson
coefficient of $F_1^\gamma$, $C_{1,\gamma}=C_\gamma-4N_Cx(1-x)$, into the
pointlike splitting functions Eqs.\ (\ref{eq:dissplit}) (Gl\"uck, Reya, and
Sieg [GRSi], 2001a). The maximal polarized boundary conditions in the
DIS$_{\Delta\gamma}$ scheme then have to be identified with the hadronic input
in the DIS$_{\gamma,1}$ scheme, and the minimal polarized boundary conditions
are given by $\Delta f_{i/\gamma}^{\rm had}(Q_0^2)e_q^2\alpha/(2\pi)
(C_{1,\gamma}-C_\gamma)$.
All the polarized photon densities discussed above are based on the
unpolarized GRV parameterization (Gl\"uck, Reya, and Vogt [GRV], 1992b), which
imposes an incoherent VMD input at a low starting scale and is evolved in
moment space with up to five massless flavors (see Sec.\ \ref{sec:hadsol}).
The different polarized parton densities in the photon are compared in
Fig.\ \ref{fig:56}.
%
%

Polarized photon densities can also be extended to non-zero photon
virtualities, either by multiplying the boundary conditions with a suppression
factor $\eta(P^2)=(1+P^2/m_\rho^2)^{-2}$ (Gl\"uck, Reya, and Sieg, 2001b) to
guarantee a smooth transition to real photons, or by imposing the polarized
virtual box result (Sasaki and Uematsu, 1999, 2000, 2001)
\beq
 \Delta f^{\rm box}_{q/\gamma}(x,Q^2,P^2)=3e_q^2{\alpha\over 2\pi}
 (2x-1)\lr\ln {Q^2\over x^2P^2} -2\rr
\eeq
at $Q^2=P^2$. As in the unpolarized case, the virtual box can be suppressed
at small $P^2$ by a factor $1-\eta(P^2)$.

\begin{figure}
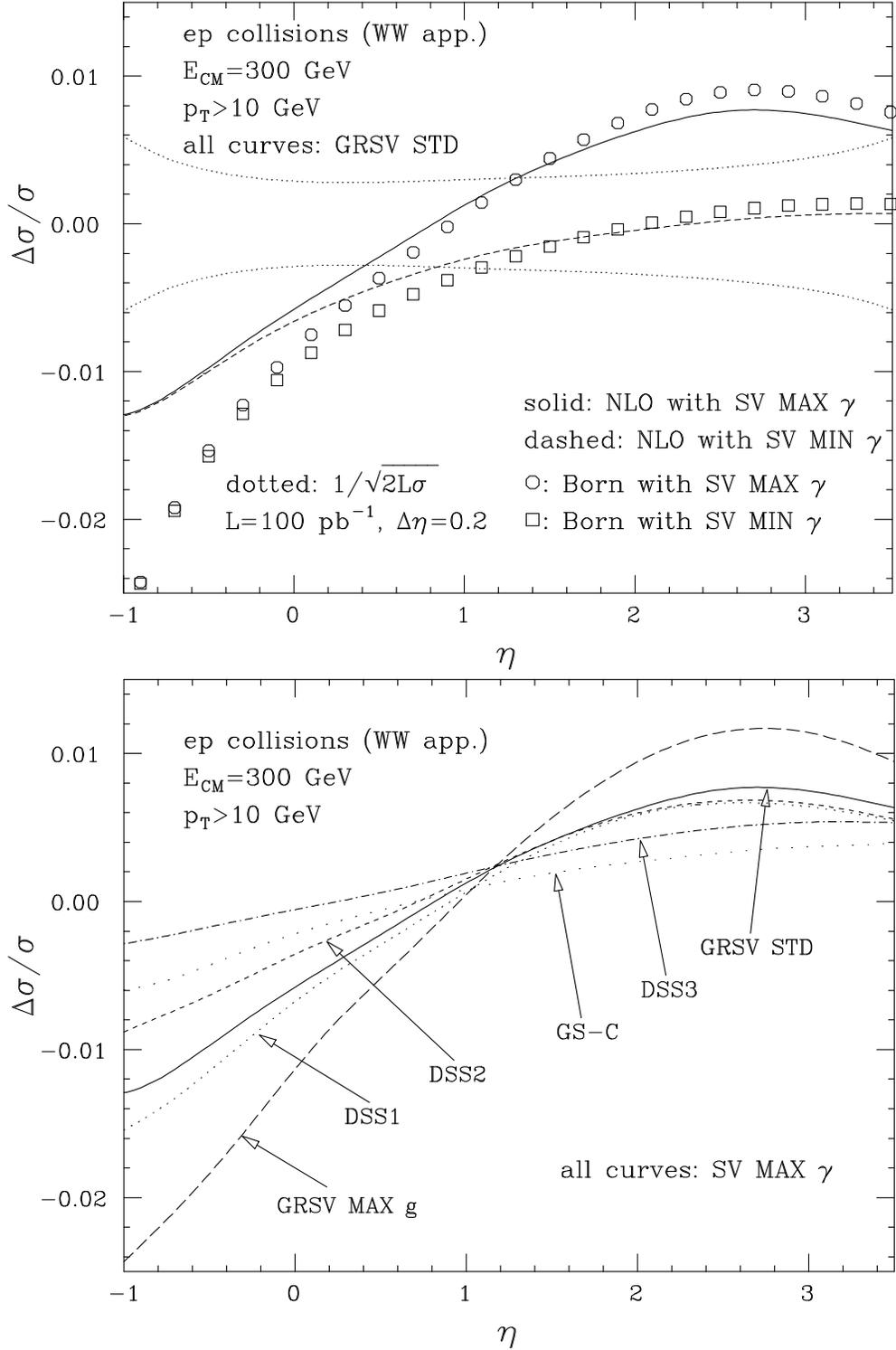

 \begin{center}
  \epsfig{file=fig_57a.eps,width=0.8\linewidth}
  \epsfig{file=fig_57b.eps,width=0.8\linewidth}
 \caption{\label{fig:57}
 Pseudorapidity dependence of the single-jet asymmetry in polarized
 photoproduction
 for different polarized photon (top) and proton (bottom) densities (De
 Florian and Frixione, 1999).}
 \end{center}
\end{figure}

The polarized cross sections for single-, two-, and three-jet photoproduction
can be obtained from Eqs.\ (\ref{eq:1jet}), (\ref{eq:2jet}), and
(\ref{eq:3jet}) by replacing the unpolarized photon spectra, parton densities,
and partonic cross sections with their polarized counterparts
$\Delta f_{\gamma/l}(x)$, $\Delta f_{i/\gamma}(x,Q^2)$, and
\beq
 \frac{\d\Delta\sigma^B}{\d t}=\frac{1}{2s}
  \frac{1}{8\pi s}\frac{2g_{a,b}^2}{S_aS_bC_aC_b}
 \lr|\M^B_{++}|^2-|\M^B_{+-}|^2\rr,
\eeq
where parity invariance guarantees that $|\M^B_{++}|^2=|\M^B_{--}|^2$
and $|\M^B_{+-}|^2=|\M^B_{-+}|^2$. Similarly the polarized cross section for
the photoproduction of light hadrons can be obtained from Eq.\
(\ref{eq:hadronxsec}). The massless Born diagrams are the same as those in
Figs.\ \ref{fig:07}, \ref{fig:08}, and \ref{fig:09}. The purely
partonic (double-resolved) diagrams were evaluated by Babcock, Monsay, and
Sivers (1979) using the projection operators
\bea
 u(p,h)\bar{u}(p,h)&=&{1\over 2}(1+h\gamma_5)p\!\!\!/\nonumber\\
 v(p,h)\bar{v}(p,h)&=&{1\over 2}(1-h\gamma_5)p\!\!\!/\nonumber\\
 \epsilon^\mu(p,h)\epsilon^{\ast\nu}(p,h)&=&{1\over 2}\lr -g^{\mu\nu}
 +{p^\mu n^\nu+p^\nu n^\mu\over p\cdot n}\rp\nonumber \\
 &&\lp -{ih\epsilon^{\mu\nu\alpha\beta}p^\alpha n^\beta\over p\cdot n}\rr
\eea
with an arbitrary four-vector $n$ ($n^2=0$)
for quarks and gluons with momentum $p$ and definite helicity $h$.
From their results, the direct and single-resolved matrix elements can easily
be obtained by appropriate changes of couplings and color factors. They can
then
be applied to jet production in a polarized collider mode of HERA (Stratmann
and Vogelsang, 1997) or to inclusive hadron photoproduction in fixed-target
collisions (Airapetian {\it et al.}, 2000; Contogouris, Grispos, and
Veropoulos, 2000). Polarized NLO calculations have been performed for direct
photoproduction of inclusive hadrons (De Florian and Vogelsang, 1998) and also
for resolved (De Florian {\it et al.}, 1999) and direct (De Florian and
Frixione, 1999) photoproduction of one or two jets. If large luminosities
could be accumulated in a polarized HERA or eRHIC collider, these processes
could be used to determine the polarized parton densities in the photon
and proton, in particular $\Delta f_{g/\gamma}$ and $\Delta f_{g/p}$. In this
way the information obtained from polarized structure functions in DIS would
be supplemented. Fig.\ \ref{fig:57} shows the pseudorapidity dependence
of the
%
%
single-jet asymmetry $\Delta\sigma/\sigma$ at HERA energies. The asymmetry is
clearly sensitive to the polarized photon (top) and proton (bottom) densities.
In addition, as a ratio of cross sections it is very stable with respect to
higher order corrections and variations of the factorization and
renormalization scales.

Polarized heavy quark photoproduction proceeds through the Born diagrams shown
in Figs.\ \ref{fig:31}, \ref{fig:32}, and \ref{fig:33}, which
applied already to the unpolarized production. Gl\"uck and Reya (1988) have
calculated the photon-gluon fusion squared matrix element, which is also
applicable to the photon-photon process after changing couplings and color
factors.
The resolved squared matrix elements have been evaluated by Contogouris,
Kamal, and Papadopoulos (1990) and by Karliner and Robinett (1994), but they do
not contribute significantly at the energies available at HERA and a possible
GSI collider with $E_e=5$ GeV and $E_p=50$ GeV (Stratmann and Vogelsang, 1997).
Due to this reduced sensitivity to the unknown polarized structure of the
photon, heavy quark photoproduction may be a useful tool to constrain the
polarized gluon density in the proton, despite the fact that fragmentation
effects have to be taken into account (Frixione and Ridolfi, 1996).
As expected the theoretical uncertainties are considerably reduced if NLO
corrections for the direct (Contogouris, Kamal, and Merebashvili, 1995; Jikia
and Tkabladze, 1996) and single-resolved (Bojak and Stratmann, 1998, 1999;
Contogouris, Grispos, and Merebashvili, 2000a, 2000b)
processes are included. For the double-resolved process the NLO corrections
are not yet available.

NLO corrections are also unavailable for polarized quarkonium production
in the color-singlet and in the color-octet channels. However LO studies
suggest that polarization provides an even more stringent test of NRQCD
factorization than unpolarized quarkonium production, since the color-octet
asymmetries are significantly different from the color-singlet asymmetries
and since the uncertainties from the quarkonium operator values cancel in the
asymmetries to a large extent. On the other hand, quarkonium production can
also be used to constrain the polarized gluon density in the proton, provided
that the color-octet operator values can be fixed in a different place
(Chao {\it et al.}, 2000; Japaridze, Nowak, and Tkabladze, 2000; Morii and
Sudoh, 2000).

\subsection{Prompt photons in hadron collisions}
\label{sec:promptyhad}

\begin{figure}
 \begin{center}
  \epsfig{file=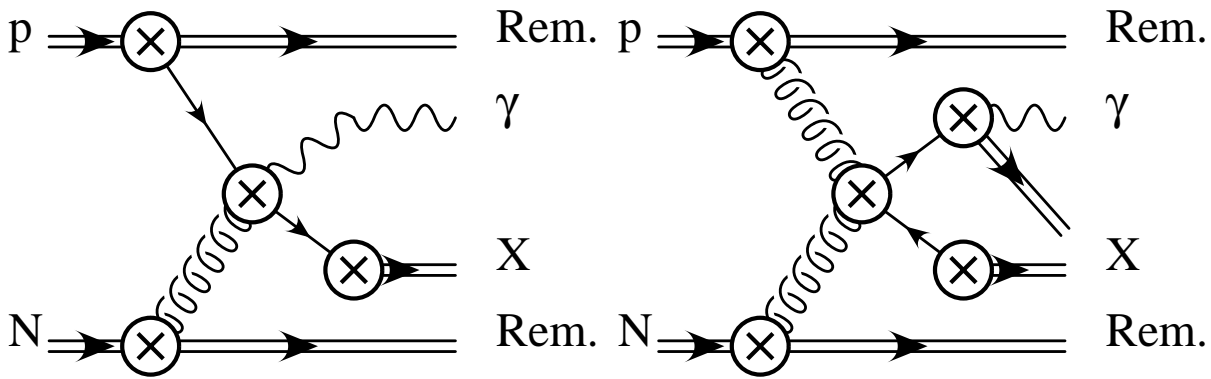,bbllx=60pt,bblly=360pt,bburx=280pt,bbury=430pt,%
          width=\linewidth}
 \caption{\label{fig:58}Factorization of hadron-hadron scattering into
 photons.}
 \end{center}
\end{figure}

The production of prompt photons in hadron collisions is intimately related
to photoproduction by crossing the initial state photon into the final
state. The contributing hadronic scattering processes are shown schematically
in Fig.\ \ref{fig:58}, where the photon is either produced directly in the
hard partonic subprocess (left) or via fragmentation of a final state parton
(right).
%
%
The inclusive cross section for the direct process Eq.\ (\ref{eq:1jet}) is a
convolution of the parton densities in the initial hadrons, which can be pions,
protons, antiprotons, or nuclei, with the partonic cross section.
The fragmentation cross section Eq.\ (\ref{eq:hadronxsec}) contains an
additional convolution with the photon fragmentation function $D_{\gamma/c}$.
The LO parton diagrams and matrix elements for
the fragmentation process are the same as
those in Fig.\ \ref{fig:09} and Tab.\ \ref{tab:loxsec}, while those for
the direct process are obtained by crossing the photon leg in Fig.\
\ref{fig:08} and Tab.\ \ref{tab:loxsec}. In these processes the photon
is balanced by an outgoing jet or hadron. Double prompt-photon production
can be calculated by taking into account the direct diagrams and matrix
elements in Fig.\ \ref{fig:07} and Tab.\ \ref{tab:loxsec} and a second
photon fragmentation function for the double-resolved processes discussed
above.

The calculation of the NLO corrections proceeds along the lines discussed in
Sec.\ \ref{sec:lighthad}. For the inclusive direct process, they have
been evaluated by Aurenche {\it et al.} (1984c, 1988, 1990) for unpolarized
and by Gordon and Vogelsang (1993, 1994a, 1994b) also for polarized prompt
%
\begin{figure}
 \begin{center}
  \epsfig{file=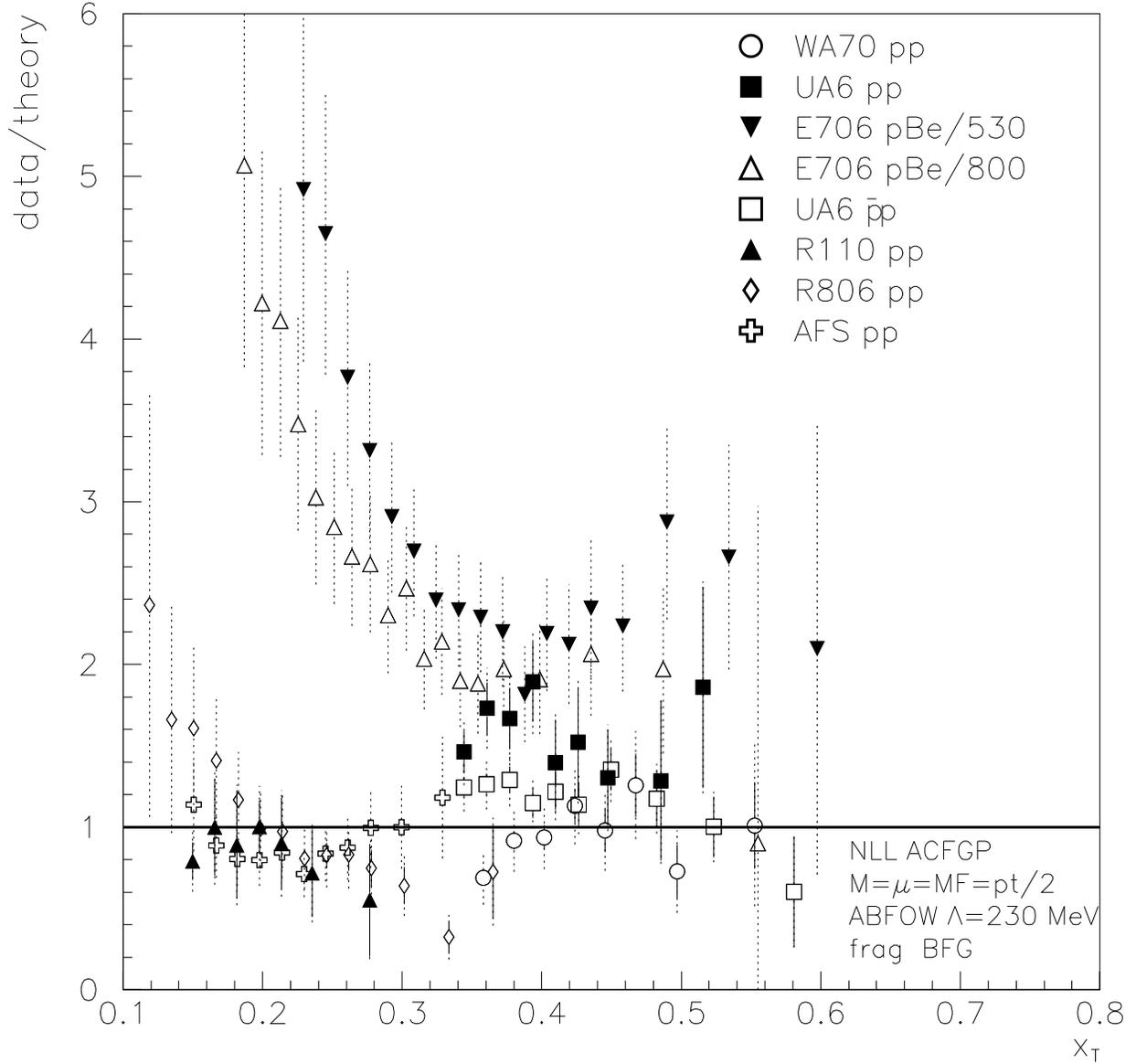,width=\linewidth}
 \caption{\label{fig:59}
 Prompt photon data from fixed-target and ISR experiments as a function of the
 scaled transverse momentum $x_T$ and normalized to NLO QCD predictions with
 ABFOW proton parton densities and BFG photon fragmentation functions
 (Aurenche {\it et al.}, 1999).}
 \end{center}
\end{figure}
%
photon production. The production of a direct photon in association with an
observed jet has been calculated in NLO by Baer, Ohnemus, and Owens (1990a,
1990b), by Gordon (1997a, 1997b), and by Frixione (1998b, 1999). The last two
calculations have again included beam polarization. Polarized and unpolarized
direct photon production with an additional final state charm quark has been
calculated in the massless scheme by Berger and Gordon (1996, 1998) and
by Bailey, Berger, and Gordon (1996). The direct NLO corrections to double
prompt-photon production have been evaluated by Aurenche {\it et al.}
(1985b), Bailey, Ohnemus, and Owens (1992), Corian\`{o} and Gordon (with
polarization) (1996a, 1996b), and Binoth {\it et al.} (2000, 2001).
NLO corrections to the fragmentation process have been evaluated by Aversa
{\it et al.} (1989) and were applied to prompt photon production by Aurenche
{\it et al.} (1993).
Fragmentation contributes typically less than 20 \% in fixed-target collisions,
but can become dominant at collider energies. An isolation cut can then help
to significantly reduce the theoretical and experimental uncertainties from the
fragmentation process (Berger and Qiu, 1990, 1991).

\begin{figure}
 \begin{center}
  \epsfig{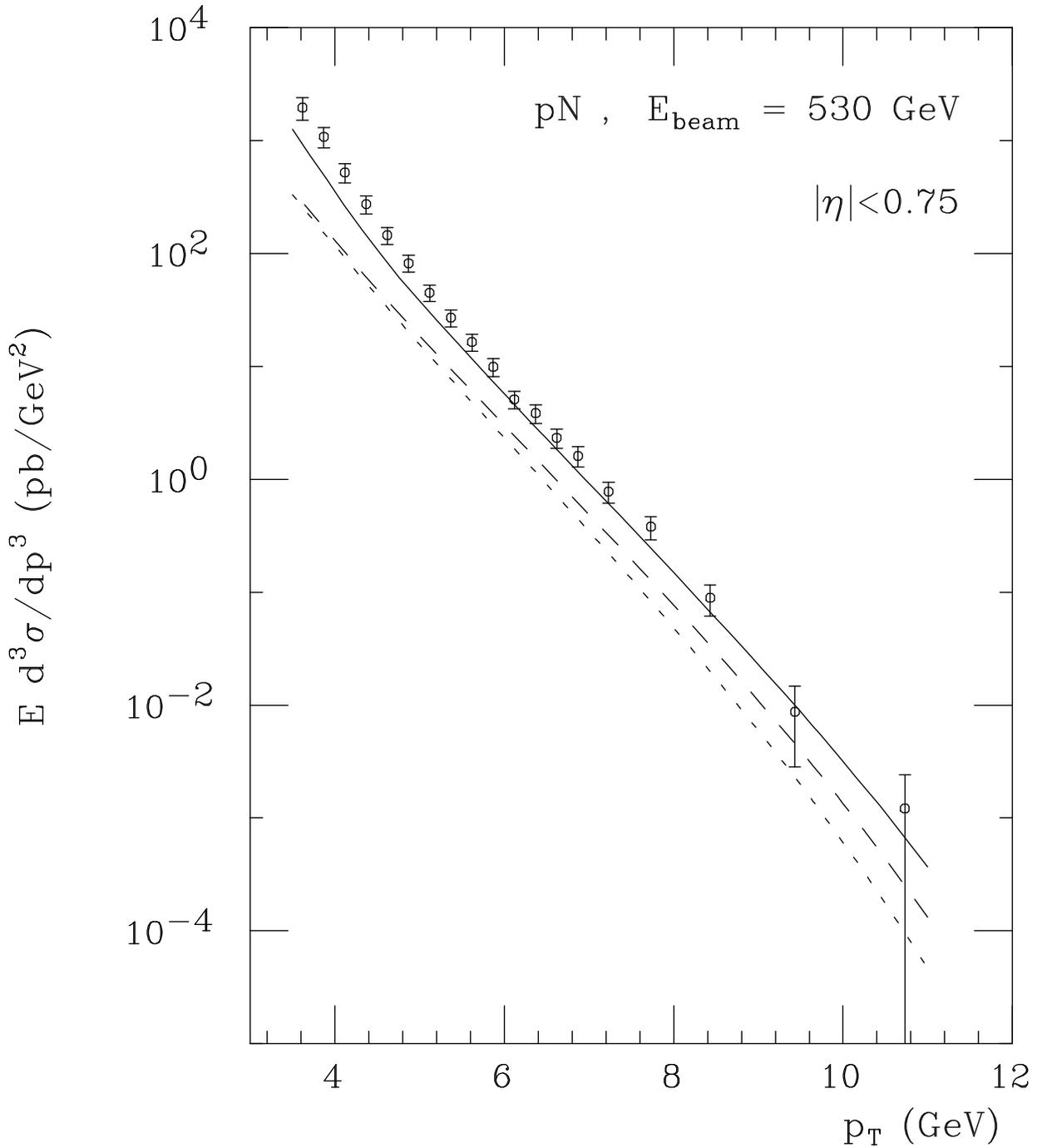}
 \caption{\label{fig:60}
 Transverse momentum distribution of prompt photons produced in $pN$
 collisions at $\sqrt{S}=31.5$ GeV. NLO (dotted), threshold resummed (dashed),
 and jointly resummed (full) results are compared to E706 data (Laenen,
 Sterman, and Vogelsang, 2000).}
 \end{center}
\end{figure}

At large values of transverse momentum $p_T$, the LO QCD Compton process
$qg\to q\gamma$ dominates over the competing $q\bar{q}\to g\gamma$ annihilation
process. This makes prompt photon production particularly sensitive to the
gluon density in the proton, which enters only in NLO in DIS. Prompt photon
production has indeed been used to constrain the unpolarized gluon density
in the proton at large $x$ (Vogelsang and Vogt, 1995; Martin {\it et al.},
2000) and could also be used to constrain the polarized gluon density in the
proton (Gordon and Vogelsang, 1996; Gordon, 1997c; Chang, Corian\`{o}, and
Gordon, 1998). Unfortunately, these determinations suffer from poorly known
fragmentation contributions, particularly of the gluon and at small $x$
(see Sec.\ \ref{sec:photfrag}), and potentially large logarithms in the
isolation criterion (see Sec.\ \ref{sec:isolation}). Even worse, the data from
fixed-target and ISR experiments cannot be described consistently by NLO
calculations (Aurenche {\it et al.}, 1999). As Fig.\ \ref{fig:59}
demonstrates, particularly the data from the Fermilab E706 experiment lie
a factor of two or more above NLO predictions (Apanasevich {\it et al.}, 1998).
They can only be described, if an intrinsic transverse momentum $\langle
k_T\rangle \simeq 1.3$ GeV of the incoming partons is taken into account.
This effect can either be generated by a simple Gaussian distribution or,
as shown in Fig. \ref{fig:60}, by considering multiple soft-gluon
%
%
radiation and simultaneously resumming the large logarithms at small values of
transverse momentum and at the partonic threshold $x_T=2 p_T/\sqrt{S}\simeq 1$
(Catani, Mangano, Nason, Oleari, and Vogelsang, 1999;
Laenen, Sterman, and Vogelsang, 2000). At collider energies the disagreement
with NLO QCD is less severe and exists only at low transverse momenta, but it
still leaves room for speculation on intrinsic $\langle k_T\rangle$ effects
(Abe {\it et al.}, 1994; Abbott {\it et al.}, 2000c). Due to the various
uncertainties discussed above, prompt photon data have currently been dropped
from global determinations of the gluon density at large $x$ (Martin {\it et
al.}, 2001; Lai {\it et al.}, 2000). Instead high-$E_T$ jet data are used,
which in turn suffer from a large factorization scheme dependence (Klasen and
Kramer, 1996c; Anandam and Soper, 2000).

%
\begin{figure}
 \begin{center}
  \epsfig{file=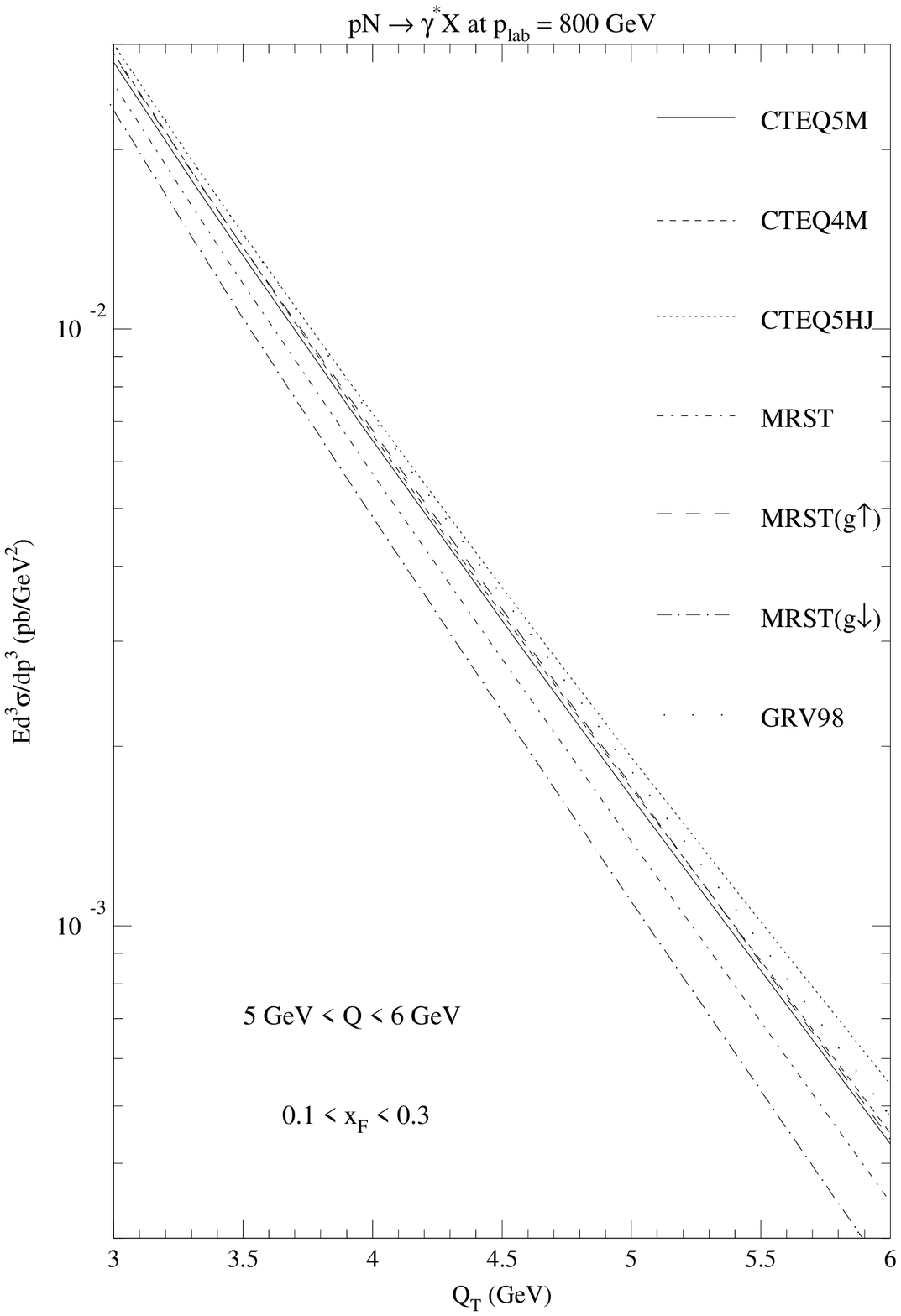,bbllx=60pt,bblly=100pt,bburx=495pt,bbury=725pt,%
          width=0.9\linewidth}
 \caption{\label{fig:61}
 Transverse momentum distribution of virtual photons produced in $pN$
 collisions at $\sqrt{S}=38.8$ GeV.}
 \end{center}
\end{figure}
%
Like real photon production, the production of photons with virtuality $Q$ is
also dominated by the QCD Compton process at large transverse momenta
$p_T>Q/2$ and thus sensitive to the proton's gluon distribution
(Berger, Gordon, and Klasen, 1998). However, the photon virtuality acts as a
mass regulator and eliminates the fragmentation contributions.
Isolation is also not necessary, since the virtual photon can be identified by
its muonic decay, and the presence of the scale $Q$ may reduce the importance
of large logarithms in $p_T$. In Fig.\ \ref{fig:61} we show that
virtual photon production in fixed-target collisions is indeed very sensitive
to the gluon density (Berger and Klasen, 2000). In a similar way the polarized
gluon distribution could be determined (Berger, Gordon, and Klasen, 2000a,
2000b).


\pagebreak
\section{Summary}
\label{sec:summary}
\setcounter{equation}{0}

From the fixed-target experiments in the 1980s to modern $e^+e^-$ and $ep$
colliders, hard photoproduction has been a fertile research ground
experimentally and theoretically. Precise measurements and calculations at
next-to-leading order of perturbative QCD have been performed for a large
variety of processes, and they have served a multitude of purposes. For one
thing, the factorization theorem for hard scattering processes has been
thoroughly tested. The photon energy spectra have been improved to meet the
precision mandated by current and future lepton colliders. The parton
densities of the photon, proton, and pion, and in particular the gluon
densities, are now better constrained, which reduces the theoretical
uncertainty at hadron and photon colliders. Ambiguities associated with jet
algorithms and kinematic regions of large hadronization corrections have been
identified and eliminated. Perturbative and non-perturbative aspects of parton
fragmentation into individual hadrons and photons and of quarkonium formation
have been disentangled, and the universality of fitted fragmentation functions
and quarkonium operator expectation values has been tested.

In spite of all these successes,
research into hard photoproduction is far from complete:
There is still large room for improvements, not only in even more precise
next-to-next-to-leading order calculations, but also in determinations of
the gluon density and spin structure of photons and protons, of the
transition from real to virtual photons, and of photon interactions with
new hypothetical particles as predicted {\it e.g.} by Supersymmetry.
Exciting possibilities may soon open up with a new linear $e^+e^-$
collider, which may even include a dedicated photon-photon experiment with
backscattered laser beams. Such an experiment would certainly initiate a whole
new era of hard photoproduction.


\section*{Acknowledgments}

I wish to thank I.\ Schienbein and C.\ Sieg for providing me with the virtual
and polarized parton distributions in the photon, B.A.\ Kniehl and G.\ Kramer
for carefully reading and commenting on the manuscript, and my family for their
strong support.
Financial support by the Deutsche Forschungsgemeinschaft through Grants No.\
KL~1266/1-1 and 1-2 and by the European Commission through Grant No.\
ERBFMRX-CT98-0194 is gratefully acknowledged.


\end{document}